\documentclass[a4paper,11pt]{article}
\usepackage{jinstpub} 
\usepackage{float}
\usepackage[section]{placeins}
\hypersetup{
 linkcolor=black
}
\usepackage{xspace}
\usepackage{multirow}
\usepackage{booktabs}
\usepackage{tabularray}
\usepackage[acronym,nonumberlist,shortcuts]{glossaries} 

\makeglossaries

\newacronym{AC}{AC}{Alternating Current}

\newacronym{BERT}{BERT}{Bertini Cascade Model} 

\newacronym{ADC}{ADC}{Analog-to-Digital Converter}

\newacronym{AliAOD}{AliAOD}{ALICE Analysis Object Data}

\newacronym{ALICE}{ALICE}{A Large Ion Collider Experiment}

\newacronym{ALPIDE}{ALPIDE}{ALICE PIxel DEtector}

\newacronym{AOC}{AOC}{Acive Optical Cable}

\newacronym{ASIC}{ASIC}{Application-Specific Integrated Circuit}

\newacronym{BC}{BC}{Bunch-Crossing}

\newacronym{CERN}{CERN}{Conseil Européen pour la Recherche Nucléaire}

\newacronym{CI}{CI}{Continuous Integration}

\newacronym{CMN}{CMN}{Common Mode Noise}

\newacronym{CMOS}{CMOS}{Complementary Metal Oxide Semiconductor}

\newacronym{CMS}{CMS}{Compact Muon Solenoid}

\newacronym{COTS}{COTS}{Commercial Off-The-Shelf}

\newacronym{CRU}{CRU}{Common Readout Unit}

\newacronym{CSP}{CSP}{Charge-Sensitive Preamplifiers}

\newacronym{CTP}{CTP}{Central Trigger Processor}

\newacronym{DAC}{DAC}{Digital-to-Analog Converter}

\newacronym{DAQ}{DAQ}{Data Acquisition}

\newacronym{DPG}{DPG}{Data Preparation Group}

\newacronym{DTP}{DTP}{Data Transfer Protocol}

\newacronym{EM}{EM}{Electromagnetic}

\newacronym{FCMD}{FCMD}{Fast Command}

\newacronym{FEC}{FEC}{Front-end-Card}

\newacronym{FF}{FF}{FireFly}

\newacronym{FIFO}{FIFO}{First In First Out}

\newacronym{FLP}{FLP}{First Level Processor}

\newacronym{FMC}{FMC}{FPGA Mezzanine Card}

\newacronym{FoCal}{FoCal}{Forward Calorimeter}

\newacronym{FoCal-E}{FoCal-E}{Electromagnetic Forward Calorimeter}

\newacronym{FoCal-H}{FoCal-H}{Hadronic Forward Calorimeter}

\newacronym{FR4}{FR4}{Flame Retardant}

\newacronym{FPC}{FPC}{Flexible Printed Circuit}

\newacronym{FTFP}{FTFP}{FriTioF parton model and Precompound model} 

\newacronym{FPGA}{FPGA}{Field-Programmable Gate Array}

\newacronym{FSM}{FSM}{Finite State Machine}

\newacronym{FWHM}{FWHM}{Full Width Half Maximum}

\newacronym{GBT}{GBT}{GigaBit Transceiver}

\newacronym{GBT-SCA}{GBT-SCA}{GBT Slow Control Adapter}

\newacronym{HB}{HB}{HeartBeat}

\newacronym{HBF}{HBF}{HeartBeam Frame}

\newacronym{HCAL}{HCAL}{Hadronic Calorimeter}

\newacronym{HG}{HG}{High Gain}

\newacronym{HGCal}{HGCal}{High Granularity Calorimeter}

\newacronym{HGCROC}{HGCROC}{High Granularity Circuit ReadOut Chip}

\newacronym{HGL}{HGL}{High Granularity Layer}

\newacronym{HIC}{HIC}{Hybrid Integrated Circuit}

\newacronym{HIJING}{HIJING}{Heavy Ion Jet Interaction Generator}

\newacronym{HLLHC}{HL-LHC}{High-Luminosity Large Hadron Collider}

\newacronym{HPK}{HPK}{Hamamatsu Photonics K.K.}

\newacronym{IB}{IB}{Inner Barrel}

\newacronym{I2C}{I2C}{Inter-Integrated Circuit}

\newacronym{IC}{IC}{Integrated Circuit}

\newacronym{IP}{IP}{Internet Protocol}

\newacronym{IR}{IR}{Infra Red}

\newacronym{ITS}{ITS}{Inner Tracking System}

\newacronym{langaus}{langaus}{A convoluted Landau and Gaussian distribution implemented in ROOT} 

\newacronym{LDO}{LDO}{Low-DropOut}  

\newacronym{LED}{LED}{Light Emitting Diode}

\newacronym{LG}{LG}{Low Gain}

\newacronym{LGL}{LGL}{Low Granularity Layer}

\newacronym{LHC}{LHC}{Large Hadron Collider}

\newacronym{LOI}{LOI}{Letter Of Intent}

\newacronym{LTU}{LTU}{Local Trigger Unit}

\newacronym{LVCMOS}{LVCMOS}{Low Voltage CMOS}

\newacronym{MAC}{MAC}{Medium Access Control}

\newacronym{MAPS}{MAPS}{Monolithic Active Pixel Sensor}

\newacronym{MC}{MC}{Monte Carlo}

\newacronym{MIP}{MIP}{Minimum Ionizing Particle}

\newacronym{MMCM}{MMCM}{Mixed-Mode Clock Manager}

\newacronym{MPV}{MPV}{Most Probable Value}

\newacronym{MSB}{MSB}{Most Significant Bit}

\newacronym{O2}{O2}{ALICE Online–Offline Computing System}

\newacronym{OB}{OB}{Outer Barrel}

\newacronym{OCDB}{OCDB}{Offline Conditions DataBase}

\newacronym{OMEGA}{OMEGA}{Organization for Micro-Electronics desiGn and Applications}

\newacronym{PCB}{PCB}{Printed Circuit Board}

\newacronym{pCT}{pCT}{proton Computed Tomography}

\newacronym{PDE}{PDE}{Photon Detection Efficiency}

\newacronym{PDF}{PDF}{Parton Distribution Functions}

\newacronym{PLL}{PLL}{Phase Locked Loop}

\newacronym{PS}{PS}{Proton Synchrotron}

\newacronym{QC}{QC}{Quality Control}

\newacronym{QGP}{QGP}{Quark Gluon Plasma}

\newacronym{QGSP}{QGSP}{Quark-Gluon String model and Precompound model} 

\newacronym{RU}{RU}{Readout Units}

\newacronym{RMS}{RMS}{Root Mean Square}

\newacronym{ROOT}{ROOT}{ROOT --- An Object-Oriented Data Analysis Framework}

\newacronym{SFP}{SFP}{Small Form-factor Pluggable}

\newacronym{SGMII}{SGMII}{Serial Gigabit Media-Independent Interface}

\newacronym{SiPad}{SiPad}{Silicon Pad}

\newacronym{SiPM}{SiPM}{Silicon Photomultiplier}

\newacronym{SNR}{SNR}{Signal-to-Noise Ratio}

\newacronym{SoC}{SoC}{System-on-Chip}

\newacronym{SPS}{SPS}{Super Proton Synchrotron}

\newacronym{SpTAB}{SpTAB}{Single point TAB}

\newacronym{SSTL}{SSTL}{Stub Series Terminated Logic}

\newacronym{TAB}{TAB}{Tape-Automated Bonding}

\newacronym{TB}{TB}{Trigger Board}

\newacronym{TC}{TC}{Transition Card}

\newacronym{TCP}{TCP}{Transmission Control Protocol}

\newacronym{TDC}{TDC}{Time-to-Digital Converter}

\newacronym{TDR}{TDR}{Technical Design Report}

\newacronym{TF}{TF}{Time Frame}

\newacronym{TIS}{TIS}{Trigger Input Switchboard}

\newacronym{ToA}{ToA}{Time-of-Arrival}

\newacronym{ToT}{ToT}{Time-over-Threshold}

\newacronym{TTC}{TTC}{Timing, Trigger and Control}

\newacronym{TTL}{TTL}{Transistor-Transistor Logic}

\newacronym{UDP}{UDP}{User Datagram Protocol}

\newacronym{ZIF}{ZIF}{Zero Insertion Force}

\newglossaryentry{5_pad_layer_gls}
{
    name=5-pad-layer board,
    description={5 silicon pad sensors wire-bonded to a PCB in a FoCal-E module. The size is roughly 450 mm x 80 x 5.5 mm}
}

\newglossaryentry{aggregator_gls}
{
    name=aggregator,
    description={Readout unit that can handle 18 5-pad-layer boards in one FoCal-E module}
}

\newglossaryentry{demonstrator_gls}
{
    name=demonstrator,
    description={A simple PCB serving as an intermediate connection between the 5-pad-layer boards in a FoCal-E module and the aggregator}
}

\newglossaryentry{HIC_gls}
{
    name=HIC,
    description={Hybrid Integrated Circuit. Sometimes referred to as a module. A flexible PCB that is wire-bonded to the ALPIDEs, mostly used in ITS2. Not the same as a stave}
}

\newglossaryentry{charge_injector_board_gls}
{
    name=charge injector board, 
    description={A PCB developed to test and characterize single-pad board before wire bonding the pad sensors} 
}

\newglossaryentry{LTU_gls}
{
    name=LTU,
    description={Local Trigger Unit. Hardware developed by the CTP group to distribute trigger and clock signal}
}

\newglossaryentry{module_gls}
{
    name=module, 
    description={A module in FoCal-E is the mechanical closure containing 18 pads and two pixel layers. The size is roughly ?}
}

\newglossaryentry{single_pad_board_gls}
{
    name=single-pad board, 
    description={One silicon pad sensor wire bonded to a PCB. Prototype for the development of the 5-pad-layer board. The size is roughly ?}
}

\newglossaryentry{stave_gls}
{
    name=stave,
    description={A stave in ITS2 combines several HICs, the cooling plate and the support structure. A HIC and a stave are not the same things}
}

\newglossaryentry{string_gls}
{
    name=string,
    description={A string is equivalent to the HIC in ITS2. However, the flexible PCB is made of aluminum and polyimide SpTABed to the chip cables of ALPIDEs}
}

\newglossaryentry{SUM_gls}
{
    name=SUM, 
    description={Difference between sums and SUM?}
}

\newglossaryentry{sums_gls}
{
    name=sums, 
    description={Difference between sums and SUM?} 
}

\newglossaryentry{tower_gls}
{
    name=tower, 
    description={The section in the z-direction of the FoCal-E covering the area of one pad sensor in all 20 layers. This configuration has been used for the first prototyping. Size is roughly 90 mm x 80 mm x 170 mm} 
}

\newglossaryentry{FTFP_old}
{
    name=FTFP, 
    description={Fritiof (FTF) parton model and Precompound (P) model} 
}

\newglossaryentry{BERT_old}
{
    name=BERT, 
    description={Bertini cascade model} 
}

\newglossaryentry{QGSP_old}
{
    name=QGSP, 
    description={Quark-gluon String (QGS) model and Precompound (P) model} 
}

\newglossaryentry{langaus fit}
{
    name=langaus fit, 
    description={A convoluted Landau and Gaussian fitting function implemented in ROOT} 
}

\newglossaryentry{langaus_old}
{
    name=langaus, 
    description={A convoluted Landau and Gaussian distribution implemented in ROOT} 
}

\newglossaryentry{IPBUS}
{
    name=IPBUS, 
    description={Control link by CERN} 
}

\newcommand{\com}[1]     {\relax}
\newcommand{\Fig}[1]     {Fig.~\ref{#1}}

\newcommand{\Figure}[1]  {Figure~\ref{#1}}
\newcommand{\Eq}[1]      {Eq.~\ref{#1}}

\newcommand{\Tab}[1]     {Tab.~\ref{#1}}

\newcommand{\Sec}[1]     {Sec.~\ref{#1}}

\newcommand{\App}[1]     {App.~\ref{#1}}
\newcommand{\red}[1]     {{\color{red}{#1}}}
\newcommand{\todo}[1]    {{\color{red}Todo: #1}}

\newcommand{\h}[1]       {\relax}
\def\geant               {\mbox{\textsc{Geant4}}\xspace} 

\newcommand{\pizero}       {\ensuremath{\pi^{0}}}

\newcommand{\kOhm}        {\ensuremath{\,\mathrm{k\Omega}}}
\newcommand{\pC}          {\ensuremath{\,\mathrm{pC}}}
\newcommand{\fC}          {\ensuremath{\,\mathrm{fC}}}
\newcommand{\pF}          {\ensuremath{\,\mathrm{pF}}}
\newcommand{\fF}          {\ensuremath{\,\mathrm{fF}}}
\newcommand{\nA}          {\ensuremath{\,\mathrm{nA}}}
\newcommand{\mV}          {\ensuremath{\,\mathrm{mV}}}
\newcommand{\GeV}         {\ensuremath{\,\mathrm{GeV}}}
\newcommand{\MeV}         {\ensuremath{\,\mathrm{MeV}}}
\newcommand{\ps}          {\ensuremath{\,\mathrm{ps}}}
\newcommand{\ns}          {\ensuremath{\,\mathrm{ns}}}
\newcommand{\us}          {\ensuremath{\,\mathrm{\mu s}}}

\newcommand{\um}          {\ensuremath{\,\mathrm{\mu m}}}
\newcommand{\mm}          {\ensuremath{\,\mathrm{mm}}}
\newcommand{\cm}          {\ensuremath{\,\mathrm{cm}}}
\newcommand{\MHz}         {\ensuremath{\,\mathrm{MHz}}}

\newcommand{\euler}      {e}
\newcommand{\dedx}       {d$E$/d$x$}

\newcommand{\Nhits}            {\ensuremath{N_{\text{hits}}}}
\newcommand{\Nhitsmin}         {\ensuremath{N_{\text{hits,min}}}}
\newcommand{\Nhitsfive}        {\ensuremath{N_{\text{hits},5}}}
\newcommand{\Nhitsten}         {\ensuremath{N_{\text{hits},10}}}
\newcommand{\woLseven}         {without-Layer-7}
\newcommand{\wLseven}          {with-Layer-7}

\def\dvers{v1.0}
\newif\ifcomment
\newif\ifextrafigs
\newif\ifdraft
\newif\ifgloss
\glosstrue
\ifdraft
\usepackage{lineno}
\linenumbers
\usepackage{fancyhdr}
\usepackage[us,12hr]{datetime}
\else
\extrafigsfalse
\fi
\title{Performance of the electromagnetic and hadronic prototype segments of the ALICE Forward Calorimeter}
\author[a]{M.~Aehle} 
\author[b]{J.~Alme}
\author[c]{C.~Arata} 
\author[d]{I.~Arsene} 
\author[e]{I.~Bearden}
\author[b]{T.~Bodova}
\author[f]{V.~Borshchov}
\author[c]{O.~Bourrion} 
\author[g]{M.~Bregant}
\author[h]{A.~van den Brink}
\author[i]{V.~Buchakchiev}
\author[e]{A.~Buhl}
\author[j,1]{T.~Chujo,\note{Corresponding author.}}
\author[e]{L.~Dufke}
\author[b]{V.~Eikeland}
\author[k]{M.~Fasel}
\author[a]{N.~Gauger} 
\author[l]{A.~Gautam}
\author[j]{A.~Ghimouz}
\author[m]{Y.~Goto}
\author[c]{R.~Guernane}
\author[n]{T.~Hachiya}
\author[o]{H.~Hassan}
\author[p]{L.~He}
\author[q]{H.~Helstrup}
\author[o]{L.~Huhta}
\author[r]{M.~Inaba}
\author[j]{T.~Inukai}
\author[l]{T.~Isidori}
\author[k,s]{F.~Jonas}
\author[j]{T.~Kawaguchi}
\author[t]{R.~Keidel} 
\author[m]{M.H.~Kim}
\author[i]{V.~Kozhuharov} 
\author[j]{T.~Kumaoka}
\author[a]{L.~Kusch} 
\author[k]{C.~Loizides}
\author[u]{Y.~Melikyan}
\author[j]{Y.~Miake}
\author[l]{N.~Minafra}
\author[b]{J.~Nystrand}
\author[j,k]{N.~Novitzky}
\author[b]{T.~{\O}kland}
\author[v]{K.~Oyama}
\author[j]{H.~Park}
\author[j]{J.~Park}
\author[e]{I.~Pascal}
\author[h]{T.~Peitzmann}
\author[f]{M.~Protsenko}
\author[o,u]{S.S.~R\"{a}s\"{a}nen}
\author[c]{F.~Rarbi} 
\author[b]{M.~Rauch}
\author[b]{A.~Rehman}
\author[b]{M.~Richter}
\author[b]{D.~R\"ohrich}
\author[d]{K.~R{\o}ed}
\author[k]{A.~Rusu}
\author[o]{H.~Rytk\"{o}nen}
\author[j]{S.~Sakai}
\author[j]{K.~Sato}
\author[t]{A.~Schilling} 
\author[m]{S.~Shimizu}
\author[n]{M.~Shimomura}
\author[i]{R.~Simeonov} 
\author[d]{E.~Solheim}
\author[w]{T.~Sugitate}
\author[x]{G.~Tambave}
\author[l]{D.~Tapia Takaki}
\author[c]{D.~Tourres} 
\author[f]{I.~Tymchuk}
\author[p]{J.~Yi}
\author[p]{Z.~Yin}
\author[b]{K.~Ullaland}
\author[b]{S.~Yang}
\author[j]{T.~Yokoo}
\author[p]{D.~Zhou}
\author[t]{S.~Zillien}

\affiliation[a]{University of Kaiserslautern-Landau (RPTU), Kaiserslautern, Germany}
\affiliation[b]{University of Bergen, Bergen, Norway}
\affiliation[c]{Universit\'e Grenobles Alpes, CNRS, Grenoble, France}
\affiliation[d]{University of Oslo, Oslo, Norway}
\affiliation[e]{University of Copenhagen, Copenhagen, Denmark} 
\affiliation[f]{Bogolyubov Inst.\ for Theoretical Physics of the Nat.\ Academy of Sciences of Ukraine (BITP), Kyiv, Ukraine}
\affiliation[g]{Universidade de São Paulo (USP), São Paulo, Brazil}
\affiliation[h]{Institute for Gravitational and Subatomic Physics (GRASP), Utrecht University, Netherlands}
\affiliation[i]{Faculty of Physics, University of Sofia, Sofia, Bulgaria}
\affiliation[j]{University of Tsukuba, Japan}
\affiliation[k]{ORNL, Oak Ridge, USA}
\affiliation[l]{The University of Kansas, Lawrence, USA}
\affiliation[m]{RIKEN, Japan} 
\affiliation[n]{Nara Women's University, Japan}
\affiliation[o]{University of Jyv\"askyl\"a, Jyv\"askyl\"a, Finland}
\affiliation[p]{Central China Normal University, Wuhan, China}
\affiliation[q]{Western Norway University of Applied Sciences, Bergen, Norway}
\affiliation[r]{Tsukuba University of Technology, Japan}
\affiliation[s]{University of Münster, Münster, Germany}
\affiliation[t]{Center for Technology and Transfer, (ZTT), University of Applied Sciences Worms, Worms, Germany}
\affiliation[u]{Helsinki Institute of Physics (HIP), Helsinki, Finland}
\affiliation[v]{Nagasaki Institute of Applied Science, Japan}
\affiliation[w]{Hiroshima University, Japan}
\affiliation[x]{Center for Medical and Radiation Physics, NISER, Jatni-752050, Odisha, India}

\emailAdd{tatsuya.chujo@cern.ch}

\abstract{
We present the performance of a full-length prototype of the \acs{ALICE} \acf{FoCal}. 
The detector is composed of a silicon-tungsten electromagnetic sampling calorimeter with longitudinal and transverse segmentation~(\acs{FoCal-E}) of about 20$X_0$ and a hadronic copper-scintillating-fiber calorimeter~(\acs{FoCal-H}) of about 5$\lambda_{\rm int}$.
The data were taken in various test beam campaigns between 2021 and 2023 at the \acs{CERN} \acs{PS} and \acs{SPS} beam lines with hadron beams up to energies of 350~GeV, and electron beams up to 300~GeV.
Regarding \acs{FoCal-E}, we report a comprehensive analysis of its response to minimum ionizing particles across all pad layers, employing various operational modes including different pre-amplifier and bias voltage settings. 
The longitudinal shower profile of electromagnetic showers is measured with a layer-wise segmentation of 1$X_0$. 
As a projection to the performance of the final detector in electromagnetic showers, we demonstrate linearity in the full energy range, and show that the energy resolution fulfills the requirements for the physics needs.
Additionally, the performance to separate two-showers events was studied by quantifying the transverse shower width.
Regarding \acs{FoCal-H}, we report a detailed analysis of the response to hadron beams between 60 and 350~GeV.
The results are compared to simulations obtained with a \geant model of the test beam setup, which in particular for \acs{FoCal-E} are in good agreement with the data.
The energy resolution of \acs{FoCal-E} was found to be lower than 3\% at energies larger than 100~GeV.
The response of \acs{FoCal-H} to hadron beams was found to be linear, albeit with a significant intercept that is about factor 2 larger than in simulations.
Its resolution, which is non-Gaussian and generally larger than in simulations, was quantified using the \acs{FWHM}, and decreases from about 16\% at 100~GeV to about 11\% at 350~GeV.
The discrepancy to simulations, which is particularly evident at low hadron energies, needs to be further investigated.

\ifdraft
\ifextrafigs
Extra info and figures --- for internal consumption --- are enabled
\fi
\vspace{0.5cm}
\color{red}DRAFT \dvers, \today\ at \currenttime\ UCT, INTERNAL ONLY\color{black}
\fi
}

\begin{document}
\maketitle
\flushbottom
\ifdraft
\pagestyle{fancyplain}
\fancyhead{}
\fancyhead[L,L]{\color{red}DRAFT \dvers}
\fancyhead[R,R]{\color{red}INTERNAL ONLY}
\fancyfoot[L,L]{\color{red}Compiled: \today}
\fancyfoot[R,R]{\color{red}at \currenttime\ UCT}
\fi
\section{Introduction}
\label{sec:intro}
A new instrument in the forward direction, the \ac{FoCal}~\cite{CERN-LHCC-2020-009}, was proposed for the \acs{ALICE}~\cite{ALICE:2008ngc} experiment at the \ac{LHC}.
Its main goal is to study the small-$x$ gluon structure of protons and nuclei by measuring direct photons, neutral hadrons, jets and their correlations at very forward rapidity~($3.4<\eta<5.5$ with coverage over the full azimuthal angle) in proton--proton and proton--lead collisions at \acs{LHC}, as well as J/$\psi$ production in ultra-peripheral heavy-ion collisions~\cite{ALICE:2023fov}. 

\ac{FoCal} consists of both electromagnetic and hadronic components: the \ac{FoCal-E} and the \ac{FoCal-H}, respectively. 
\ac{FoCal-E} has two independent subsystems embedded in longitudinally segmented modules, where 20~tungsten absorbers, each with about one radiation length~($1\,X_0$), are interleaved with active layers for shower particle measurements leading to a total of about $20\,X_0$.
Each module comprises 18 layers of Si pad detectors aimed at providing good energy resolution over a wide dynamic range for measuring the energy of \acs{EM} showers. 
Two layers of high granularity Si \ac{MAPS} are added to the stack at the 5th and 10th position to provide the required spatial resolution to identify photons from neutral-pion decays emitted with very small opening angle. 
The relative position of low and high granularity-layers were determined using simulations and result in good two-photon separation for shower energies from 0.1--1~TeV.
\ac{FoCal-H} consists of a modular transversely segmented spaghetti calorimeter based on commercially available capillary tubes, each containing a scintillating fiber. 
Groups of fibers are bundled together and read out by \acp{SiPM} making up a tower.
The performance of the proposed detector has recently been studied using \ac{MC} simulations~\cite{ALICE:2023rol}. 

To achieve the optimal detector design needed for the realization of the \ac{FoCal} physics program~\cite{ALICE:2023fov}, several performance studies of the detector prototypes have been carried out in the past decade.
In particular, for the \acs{EM} component, various prototype detector systems have been designed and tested. 
These included fully digital calorimeter designs using only pixel sensors~\cite{Nooren:2017jsz,Alme:2022hxo,Peitzmann:2022asy}, calorimeter designs using only 
pad-sensors~\cite{Muhuri:2014xva,Muhuri:2019hvg,Barthel:2023cut} and a combined pixel- and pad-sensor hybrid design~\cite{Awes:2019vfi}.
High-granularity calorimeter designs based on silicon sensors with analog readout have previously also been explored and tested by e.g.\ the CALICE collaboration~\cite{Poschl:2022lra}
and by the \acs{CMS} \acs{HGCal} collaboration~\cite{CMSHGCAL:2021nyx}.
In fact, the prototype as well as the final \ac{FoCal-E}, profits from dedicated readout electronics components developed in the context of the above projects. 
Fiber-scintillating calorimeters based on the capillary tubes have been designed and tested also by the IDEA collaboration~\cite{Giaz:2022puf}. 
The latter approach simultaneously uses scintillating and clear fibers to be able to measure both scintillation and Cherenkov light, with the goal to better constrain the electromagnetic component of hadronic showers. 
For \ac{FoCal-H}, the extra improvement in the energy resolution of hadronic showers that this may bring is not needed, so the present prototype uses only scintillating fibers.

In this publication, we focus on presenting scaled-down prototype systems closely representing individual \ac{FoCal} sub-detectors, which have been designed, constructed and tested since 2020 to benchmark the simulation studies~\cite{ALICE:2023rol}. 
Their functionality was first established in laboratory tests and their performance was then measured by collecting data at various particle beam lines. 
In particular, hadron and electron beams over a large range of energies provided by the \ac{PS} and \ac{SPS} at the East and North test beam areas at \acs{CERN} were used.
The tested \ac{FoCal} prototypes evolved with time closely connected with the development of the final detector design. 
Details for every prototype system are given in \Sec{sec:prototype}.
To model and understand the performance of the \acs{FoCal} prototypes, a dedicated \ac{MC} simulation chain was developed, where the realistic material budget of the prototype detectors was implemented.
Details are given in \Sec{sec:simulation}, and variations or additional information is provided when the results are discussed.
The evolution of the prototype systems through-out the test beam campaigns over the years are described in \Sec{sec:setup}, with an emphasis on the setup for November, 2022 and May, 2023, which provided the data for most of the presented results.
The results related to \ac{FoCal-E} are presented in \Sec{sec:focaleres} starting with a comprehensive analysis of the response to minimum ionizing particles across all pad layers, employing various operational modes including different pre-amplifier and bias voltage settings~(\Sec{subsec:MIP}). 
Details about the electron data and event selection of single electrons using the pixel layers are given in \Sec{subssec:electrondataset} and \ref{subsec:pixel-event-selection}, respectively.
The hit response of the pixel-layers to electrons and refinements to the standard transport model of the detector for the description of the resulting pixel hit distributions are discussed in \Sec{subsec:pixel-layer-hits}.
The pad channel calibration and electron-event selection for the pad-layer analysis are described in \Sec{subsec:padchannelcalib} and \ref{subsec:evtselpads}.
The measurement of the longitudinal shower profiles of electrons, mainly using the pad layers, but where possible including pixel-layer information, as well as comparison to simulations is presented in \Sec{subsec:shower_profiles}. 
The linearity of the response and the resolution\h{ of the 18-pad-layer stack, as well as a 17-pad-layer stack at higher energies} together with respective comparison to simulations is presented in \Sec{subsec:En_linearity}.
The performance of \ac{FoCal-H} obtained from hadron beams, where \ac{FoCal-E} was removed, is presented in \Sec{sec:focalhres}, starting with a brief description on the processing of the data, the determination of the pedestal, and the low-and-high gain matching provided by the readout in \Sec{subsec:pedestal_h}.
Linearity of \ac{FoCal-H} hadrons in data and simulations is discussed in \Sec{sec:HCal_calibration}, and used to calibrate data and simulations. 
The corresponding energy resolution in data and simulation, as well as the systematic uncertainties, are presented in \Sec{subsec:energy_res_hcal}.
A summary of the obtained results is given in \Sec{sec:summary}.
\App{appendix:pads_calibration} and \ref{appendix:hcal_calibration} give details on the calibration of the pad-layer readout and the inter-calibration of the \ac{SiPM} gains, respectively.
\section{Detector prototypes}
\label{sec:prototype}
In the following, detector prototypes developed since 2020 are described.
Even though the presented results focus on the later prototypes, we also mention earlier designs.
The \ac{FoCal-E} prototype is instrumented with low and high granularity Si sensors covering an area of 9x8x17cm, which corresponds to one tower out of the 5 towers foreseen in a module of the proposed \ac{FoCal-E} full setup~\cite{CERN-LHCC-2020-009}.
For the \ac{FoCal-H} a small proof-of-principle prototype and then a larger prototype with 9 modules were constructed covering about $10\times 10 \times 55\cm^3$ and $19.5\times19.5\times110\cm^3$, respectively.

\subsection{\acs{FoCal-E}: pad layers}
\label{subsubsec:Prototype_pads}

\begin{figure}[t!]
\begin{center}
\includegraphics[width=0.43\textwidth]{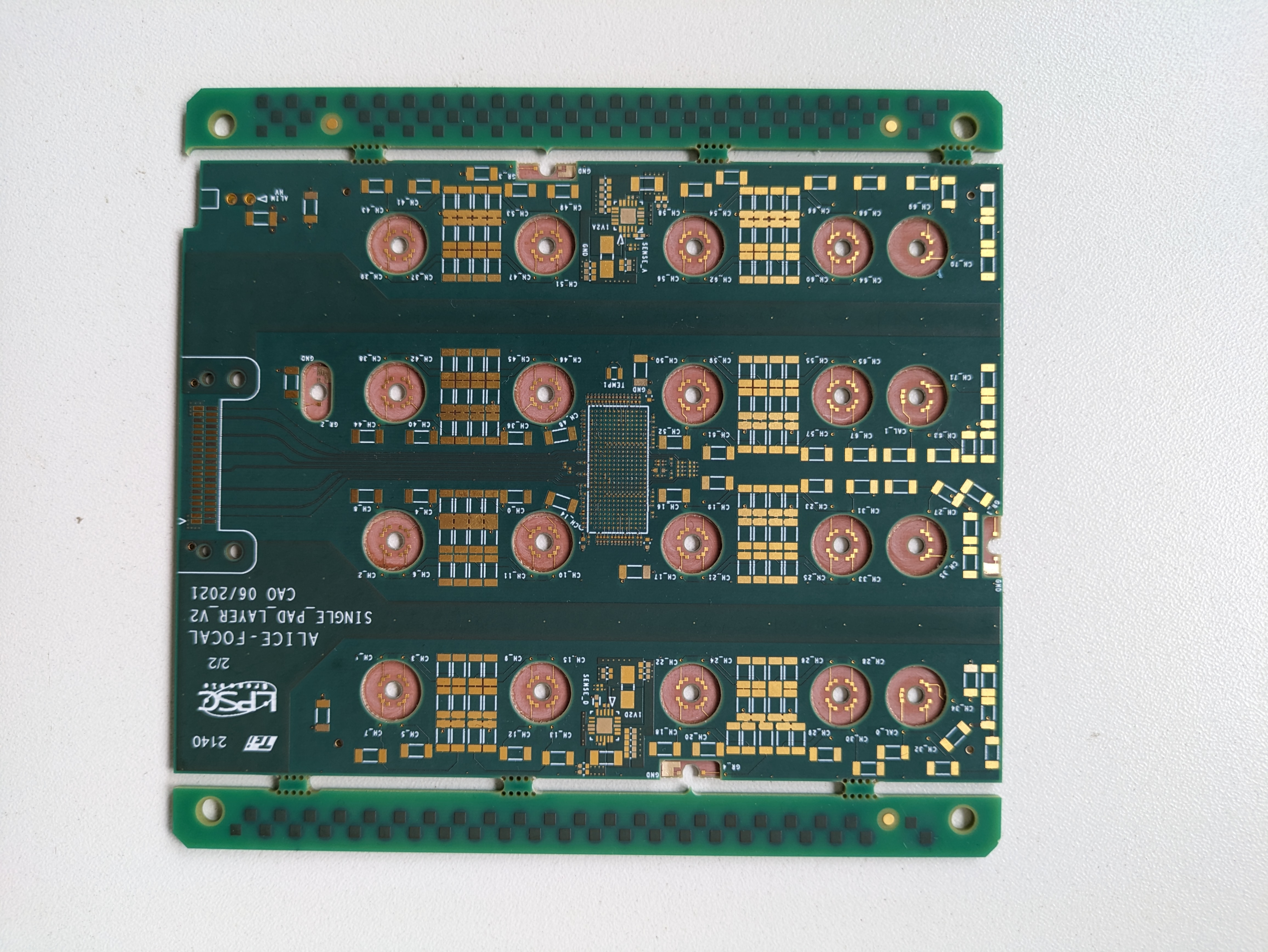}
\hspace{0.3cm}
\includegraphics[width=0.43\textwidth]{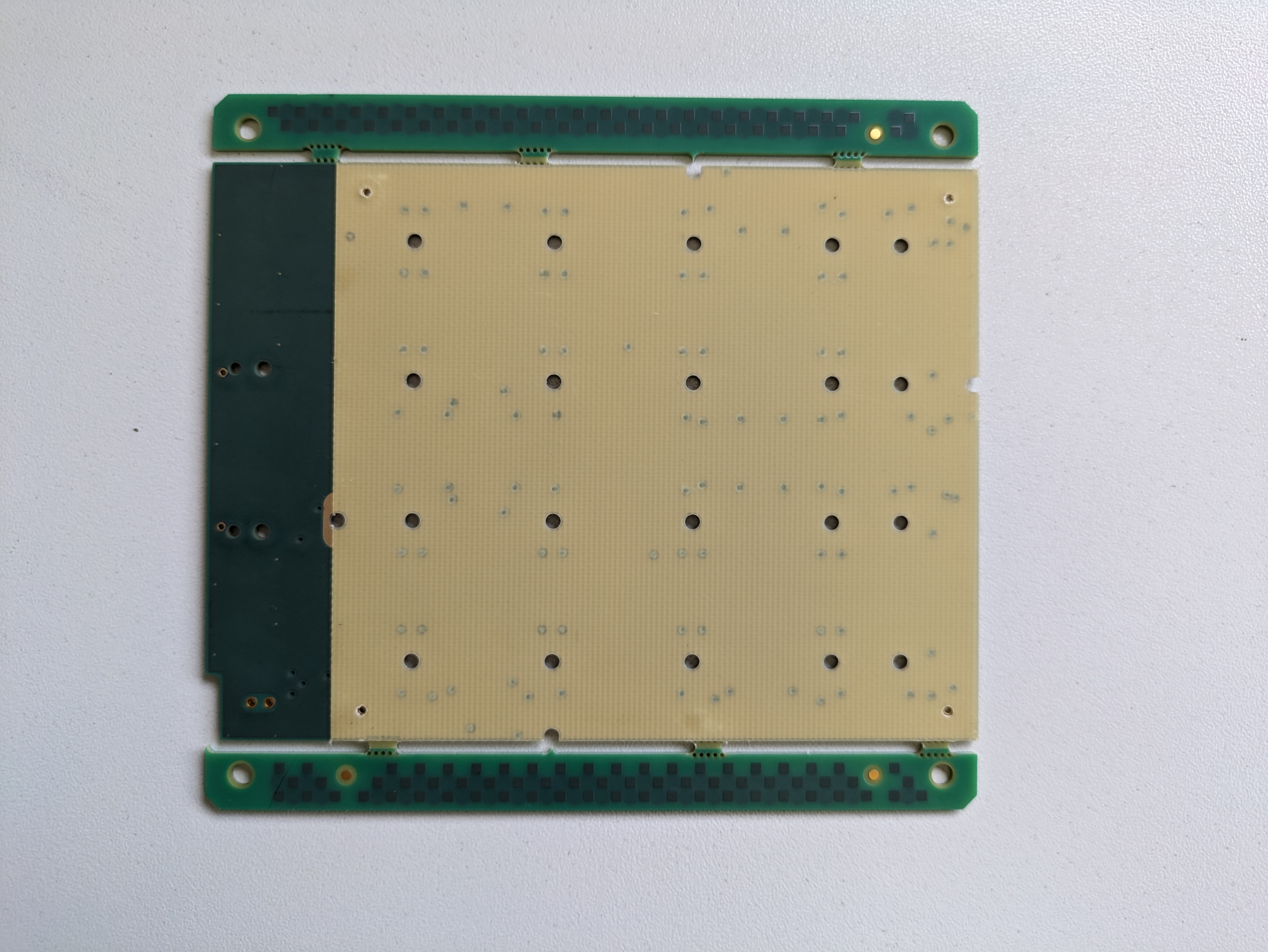}
\caption{Picture of the front, and back-plane of the single-pad board. 
The cavities are designed for wire bonding the individual silicon pads to the onboard electronics. 
At the center, some space is reserved for positioning the \ac{HGCROC} readout chip. 
The Si wafer is aligned and glued to the back-plane, and it receives the voltage for biasing the sensors through a connection routed by the board's edges. 
}
\label{fig:padpcbv2}
\end{center}
\end{figure}

The prototype uses custom-made boards for hosting the individual low-granularity layers~(\Fig{fig:padpcbv2}).
Each of the 9$\times$8\,cm$^{2}$ Si p-type sensors produced by \ac{HPK}  in the 6-inch wafer fabrication process, is segmented in a matrix of 72, $1\times1$\,cm$^{2}$, 320\,$\mu$m thick pads.
Two additional pads of smaller size are inserted in the array for calibration purposes. 
The front plane of the sensor is glued to the \ac{PCB} and wire bonded to the read-out electronics via through-holes carrying the lines to the back of the boards. 
Other cavities are used to connect the guard rings, high voltage, and ground connections. 
The depletion voltage is achieved by biasing the sensors through the gold-plated common cathode on the wafer back side. 
More details, as well as the systematic studies of the sensors' electrical properties, can be found in~\cite{Inaba:2020dng}. 

For the readout, each \ac{PCB} hosts the \ac{HGCROC} \ac{ASIC} originally developed by the \acs{OMEGA} group for the \acs{CMS} \ac{HGCal}~\cite{Thienpont:2020wau}. 
The energy deposit in each Si diode is converted into voltage pulses by the low-noise preamplification stage of the chip and later processed by a shape filter. 
The signal is then fed to the \ac{ADC} of the \ac{HGCROC} which provides a linear response up to 320\fC, equivalent to about 100\,\acp{MIP}.  
The signal is also discriminated and provided to two \acp{TDC} used to extend the dynamic range via measuring the \ac{ToT} of larger energy deposits. 
The \ac{ToT} is measured relatively to a precise \ac{ToA} information, granting a total dynamic range going from 0.2\fC\,to 10\pC.
These quantities are sampled every 25\,ns (40\,MHz) and stored in a 512 sample deep circular buffer which is streamed to the acquisition electronics via 1.28\,Gbit/s serial links when the \ac{HGCROC} receives a readout request through the \ac{FCMD} port. 
A buffer entry contains 72+2 measurements of \ac{ADC}, \ac{ToT}, and \ac{ToA}, the monitor of four common mode channels for coherent noise subtraction, and the digital sums of groups of nine pads~(eight sums for each sample) for trigger generation. 
The data streaming operation lasts a total of 1.075\,$\mu$s since the \acs{FCMD} signal is received by the chip. 
To retrieve the correct data from the buffer, the \ac{HGCROC} has to account for the delay between the readout command and the actual energy deposit in the detector. 
This can be set through the slow control as a global constant offset in the \ac{HGCROC} configuration that allows the probing of past recorded entries.  

The 18 pad layers sit on an interface card that connects the \ac{HGCROC}s to the front side of an aggregator board through differential high-speed serial links, and slow control lines shared between groups of five pads.
On the aggregator, a XILINX\textsuperscript{TM} XCKU035-2FFVA1156 \ac{FPGA} is responsible for controlling and monitoring the \ac{HGCROC} chips of the individual pads, as well as enabling bidirectional communication with the rest of the readout chain through \ac{SFP} modules hosting the optical and Ethernet transceivers. 
A detailed description of the \acs{PCB}, the pad-layer design and the aggregator readout card can be found in~\cite{Bourrion:2023cmf}.

\begin{figure}[t!]
\begin{center}
\includegraphics[width=1.0\textwidth]{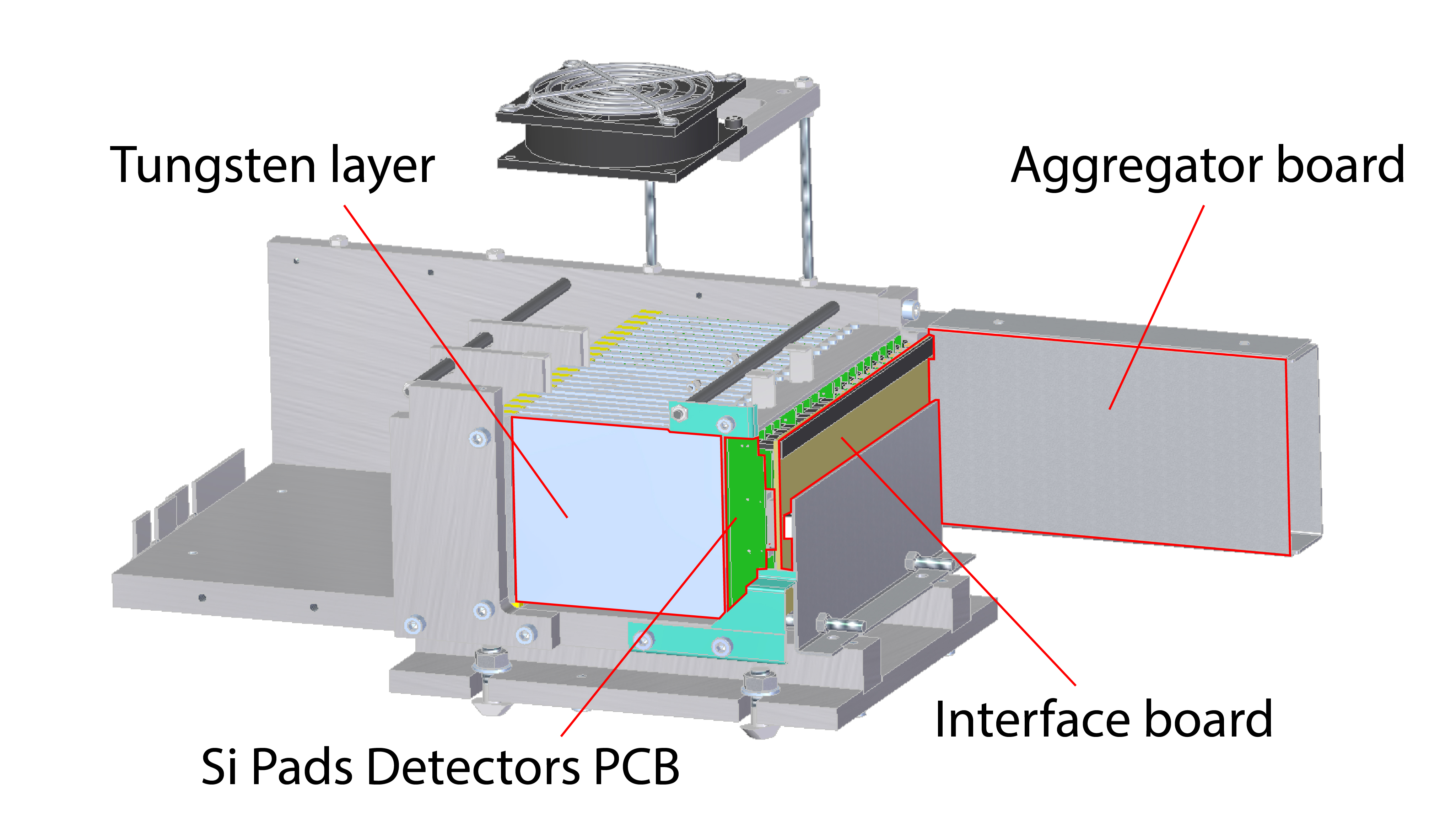}
\includegraphics[width=0.6\textwidth]{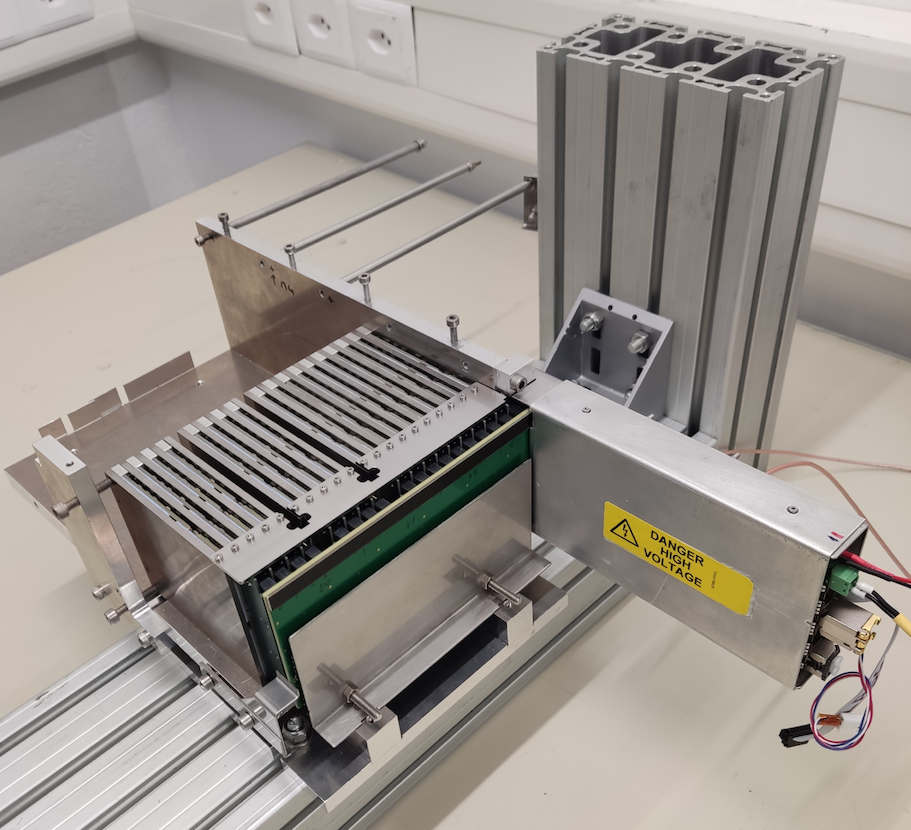}
\caption{Technical drawing~(top) and photo~(bottom) of the \ac{FoCal-E} pad prototype with the interface and aggregator cards on the right.
The slots for the pixel layers at position 5 and 10 in the stack are visible and not yet equipped.
}
\label{fig:focal_e_pads}
\end{center}
\end{figure}

The optical transceivers are allocated to interface the \ac{GBT} links to the \acs{ALICE} \ac{CRU}~\cite{Cachemiche:2016reb}. 
Here, the reception link is used to synchronize the detector with the \ac{CRU} \ac{DAQ} clock and deliver the \ac{CTP} trigger messages to the \ac{FPGA}. 
The \ac{PLL} locks to the recovered clock, and in turn provide the reference for the acquisition firmware. 
Packets of 80\,bits of data containing headers with unique event identifiers are sent to the \ac{CRU} at every clock cycle (40MHz) through the \ac{GBT} transmission link. 
When a physics trigger is received, the data and the trigger sum information of the 10 previous and 10 following bunch crossings from the \acp{HGCROC} are organized in payloads stored in \ac{FIFO} memory buffers containing fundamental information, such as the orbit and bunch crossing ID, the trigger type, and possible data drop flag. 
The output packets are built according to the \ac{O2} protocol~\cite{Buncic:2015ari}, and transmitted to the \ac{CRU}. 

An aluminum structure is designed to accommodate the 18 detector layers, as well as the interface and aggregator cards. The tungsten plates are glued to the back of the pad boards creating a rigid structure during the assembly. 
Empty slots in layer 5 and layer 10 were designed to host the high granularity detectors, described in \Sec{subsubsec:Prototype_pixels}.  
Two fans are used to mitigate the heat generated by the 20 detector layers and operate the detector at room temperature. This aims to replicate the final detector condition as it will be operated at a temperature of about $+20$\,$^{\circ}$C. 
The full stack is presented in \Fig{fig:focal_e_pads}. 

\subsection{\acs{FoCal-E}: pixel layers}
\label{subsubsec:Prototype_pixels}
The technology of the Si \ac{MAPS} used in the \ac{FoCal} high granularity layers is inherited from the \acs{ALICE} upgraded \ac{ITS}~\cite{Abelev:1625842}, which exploits their excellent spatial resolution and low material budget for vertex tracking. 
The \ac{ALPIDE} is \acs{CMOS} \ac{MAPS} produced with the TowerJazz 180\,nm imaging process~\cite{Senyukov:2013se}.    
The sensors have a pitch of about 29\,$\mu$m $\times$ 27\,$\mu$m, an epitaxial layer for charge collection of $25\um$, and a total thickness that ranges from 50\,$\mu$m to 100\,$\mu$m. 
In the \ac{ALPIDE} chips, the pixels are arranged in a matrix of 1024 $\times$ 512 for a total surface of about 30\,mm $\times$ 15\,mm. 

\begin{figure}[t!]
\begin{center}
\includegraphics[width=\textwidth]{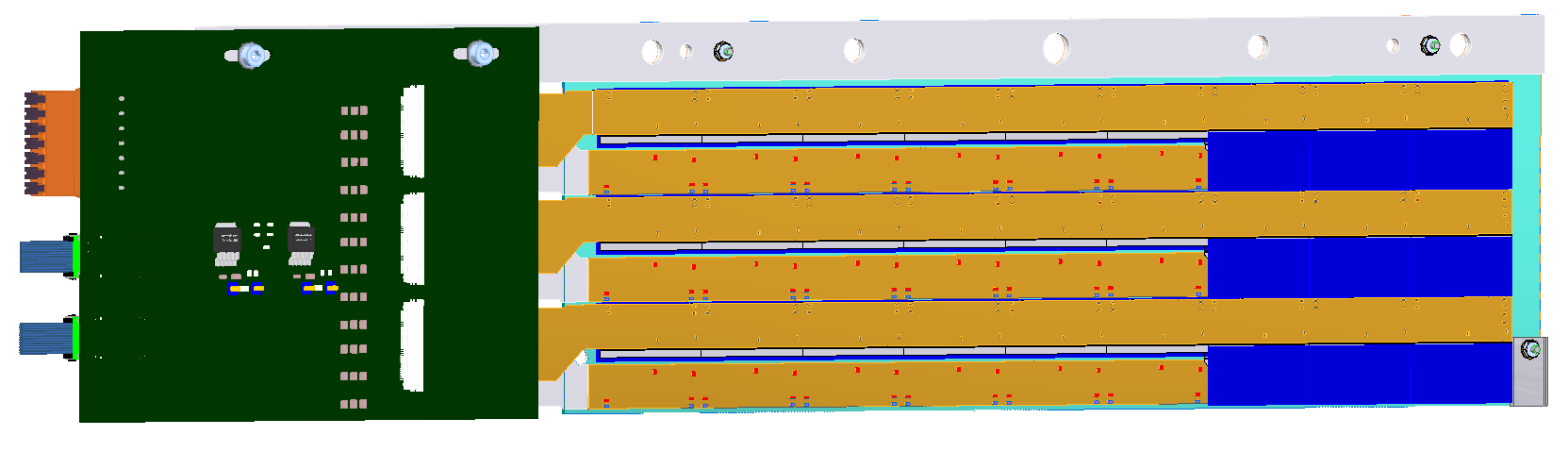}
\includegraphics[width=\textwidth]{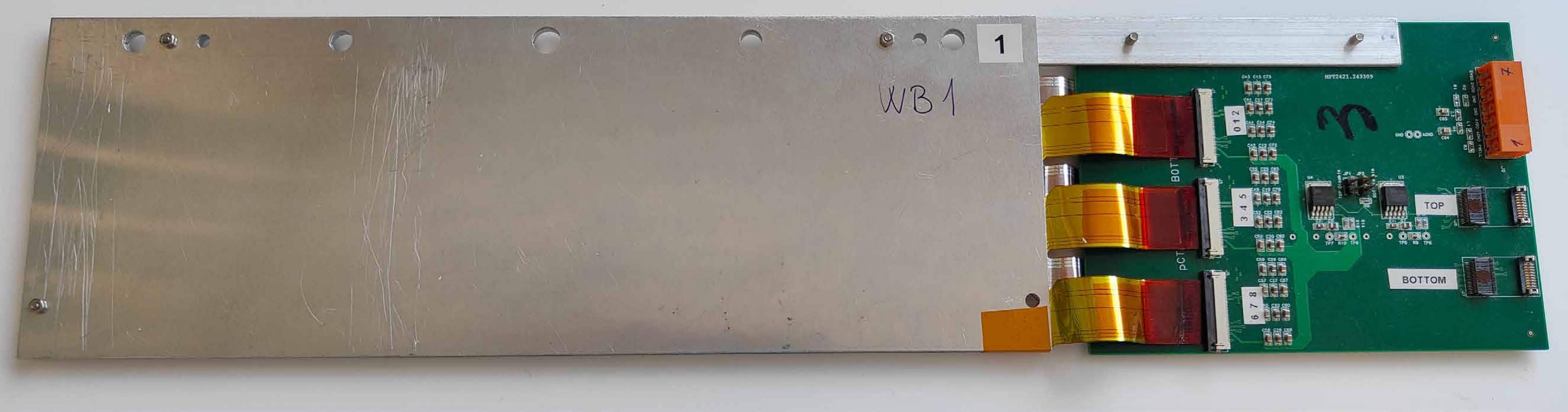}
\caption{Prototype of the \acs{pCT}-inspired \ac{FoCal-E} pixel layer; 3D drawing~(top panel) and photo of assembled layer~(bottom panel).}
\label{fig:pCT_cads}
\end{center}
\end{figure}

In the \ac{ITS}, the \ac{ALPIDE} sensors are hosted on polyimide \ac{FPC} referred to as \ac{HIC}. 
In each module, two \ac{HIC}~half-staves are placed in direct contact with high-thermal conductive carbon fiber sheets embedding polyimide cooling pipes, with rigid mechanical support provided by additional carbon fiber frames. 
There are two types of pixel sensors depending on their radial distance from the interaction point: \ac{IB} and \ac{OB}. 
The two differ in sensor thickness and data transfer rate capabilities provided by the embedded peripheral readout. 
A detailed description can be found in the \ac{ITS} \acrfull{TDR}~\cite{Abelev:1625842}. 

The \ac{FoCal} design concept uses both \ac{ALPIDE} readout modes, depending on the radial distance from the beam line, as the expected hit density grows rapidly as a function of increasing pseudorapidity. 
The central regions of the detector will require the use of \ac{IB}-mode chips that will benefit from the high-speed dedicated data link operating at a maximum of 1200~Mbps. 
The peripheral regions will be instrumented with \ac{OB}-mode chips, where the 400~Mbps data links are sufficient for the lower expected data rates.

As \ac{FoCal} encounters technical challenges different from the \ac{ITS}, the proposed mechanical and electrical structure of the pixel layers was entirely redesigned. 
\ac{FoCal} inherits its detector blueprint from the digital tracking calorimeter developed for the \ac{pCT}~\cite{pct_2020}, where the \acp{ALPIDE} are hosted on single and multilayered flexible microcables based on adhesiveless aluminium-polyimide foiled dielectrics. 
They are connected to the power, slow control, and data distribution lines using \ac{SpTAB} bonding technology.  

\begin{figure}[th!]
\begin{center}
\includegraphics[width=0.8\textwidth]{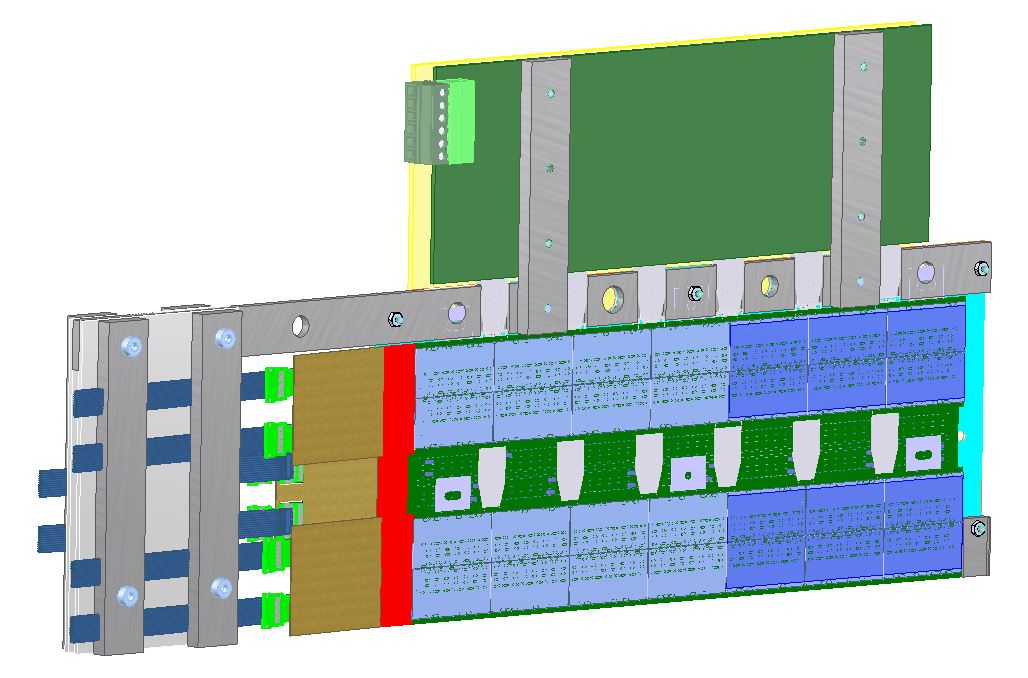}
\includegraphics[width=0.7\textwidth]{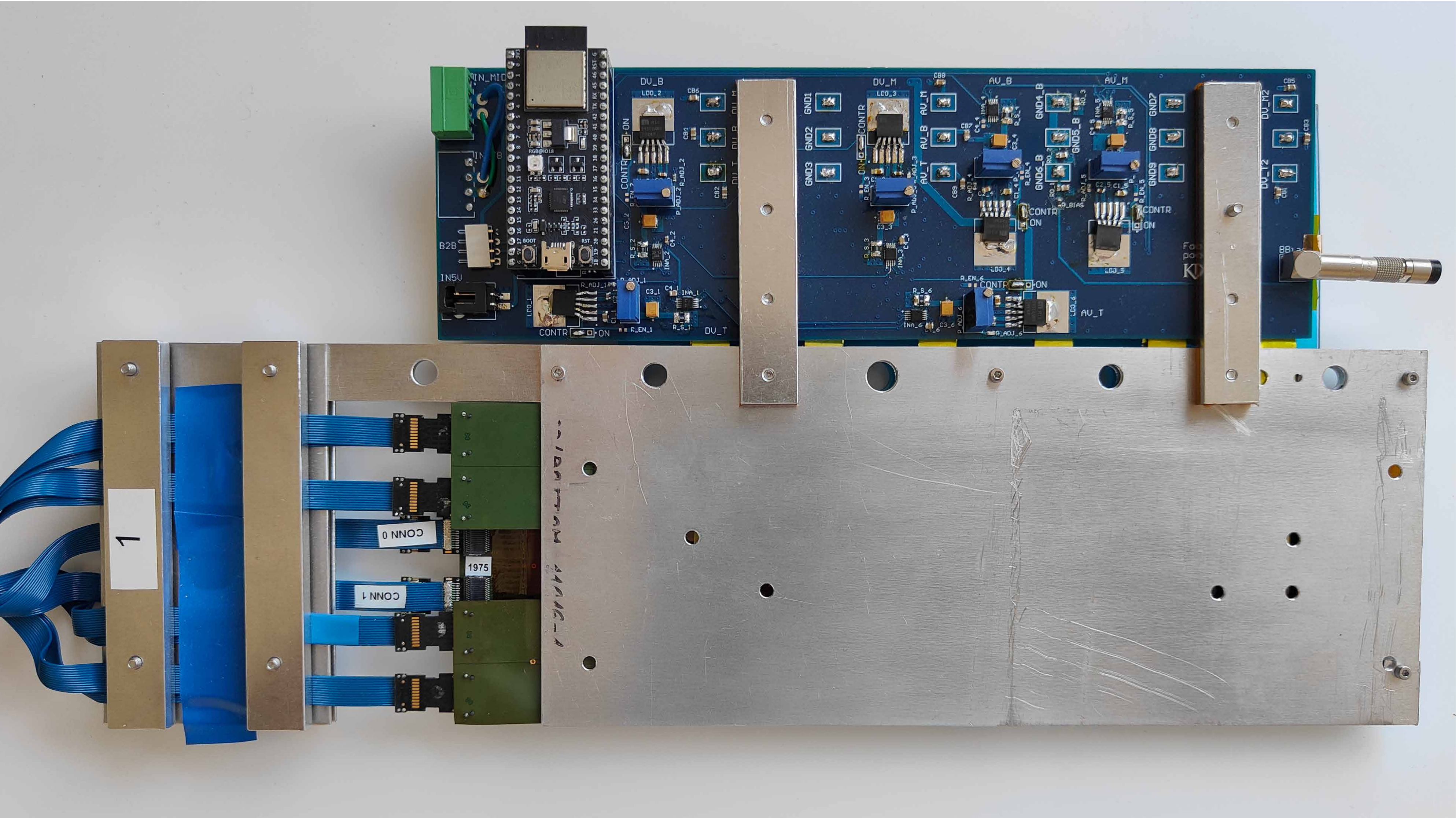}
\caption{Top: Realistic three-dimensional rendering of a single \ac{HIC} layer prototype. The flexible \acp{PCB} hosting the pixels are organized from top to bottom in strings composed by two \ac{ITS} half-layers. 
Each one is routed individually to a \acs{FF} connector. 
Bottom: Picture of one of the two \acs{FoCal-E} \ac{HIC}-based pixel prototypes. 
The three \acp{FPC} are connected to the \acs{FF} cables for the data, trigger, and slow control communication. The monitor and control of the power is enabled through the power board assembled in the aluminum enclosure. 
}
\label{fig:HIC_cads}
\end{center}
\end{figure}

Two pixel test prototypes were produced following these assembling techniques. 
Each layer is composed of two, complementary aluminum carriers hosting three adjacent strings of \ac{IB} \ac{ALPIDE} chips~(half-layers) glued to the supports. 
The carriers are then connected back-to-back to enclose and protect the sensors, providing a uniform active area that covers the same transverse acceptance of the pad detectors described in \Sec{subsubsec:Prototype_pads}. 
One of the assembled prototypes is shown in the bottom panel of \Fig{fig:pCT_cads}, while the top panel displays the 3D~drawing of the same pixel-layer prototype. 

The pixel strings terminate in \acrfull{ZIF} connectors that route the data, clock, and control lines to the off-detector electronics through an additional \ac{PCB}. 
These electronic boards, referred to as \ac{TC}, communicate with the \ac{ITS} \ac{RU}~\cite{Abelev:1625842} through a \ac{FF} differential copper cable assembly solution developed by SAMTEC.
The digital and analog power, regulated on the board via a \ac{LDO}, are also routed through the same connectors. 
The pixel prototypes are finally assembled inside the \ac{FoCal-E} stack, at nominal position~(5th and 10th layer), to provide the tracking information of the \acrfull{EM} showers. 

To verify the performance of \ac{OB} pixels at test beam lines, we developed two additional prototype layers based on \ac{ITS} \acp{HIC}. 
In this design, the aluminum carriers envelop three \ac{HIC}~half-layers~(a total of six strings) vertically ordered and labeled for simplicity: Top, Middle, and Bottom. 
In this version, the pixels closely resemble the \ac{ITS} stave structure, where the flexible \acp{PCB} are directly routed to the readout through a \ac{FF}. 
Gold-plated aluminum strips carry the digital and analog power to the pixels, through contacts directly bonded on the \ac{HIC} \acp{PCB}.
The \acp{HIC} are in turn connected to custom-made power boards controlled and monitored by ESP32 micro-controllers used to enable the power delivery and display the current absorption. 
\Figure{fig:HIC_cads} shows the 3D~drawing of an individual \ac{HIC} layer, as well as the corresponding assembled prototype layer.

\FloatBarrier

\subsection{\acs{FoCal-H} prototype modules}
\label{subsubsec:Prototype_Hcal}
We have built and tested two \ac{FoCal-H} prototypes each briefly described below.  
The main focus of the \ac{FoCal-H} prototypes was to test the conceptual design of the transversely segmented calorimeter based on commercially available copper capillary tubes, scintillating fibers, \acp{SiPM} and commercial readout electronics. 

\begin{figure}[ht!]
\centering
\includegraphics[width=0.48\textwidth]{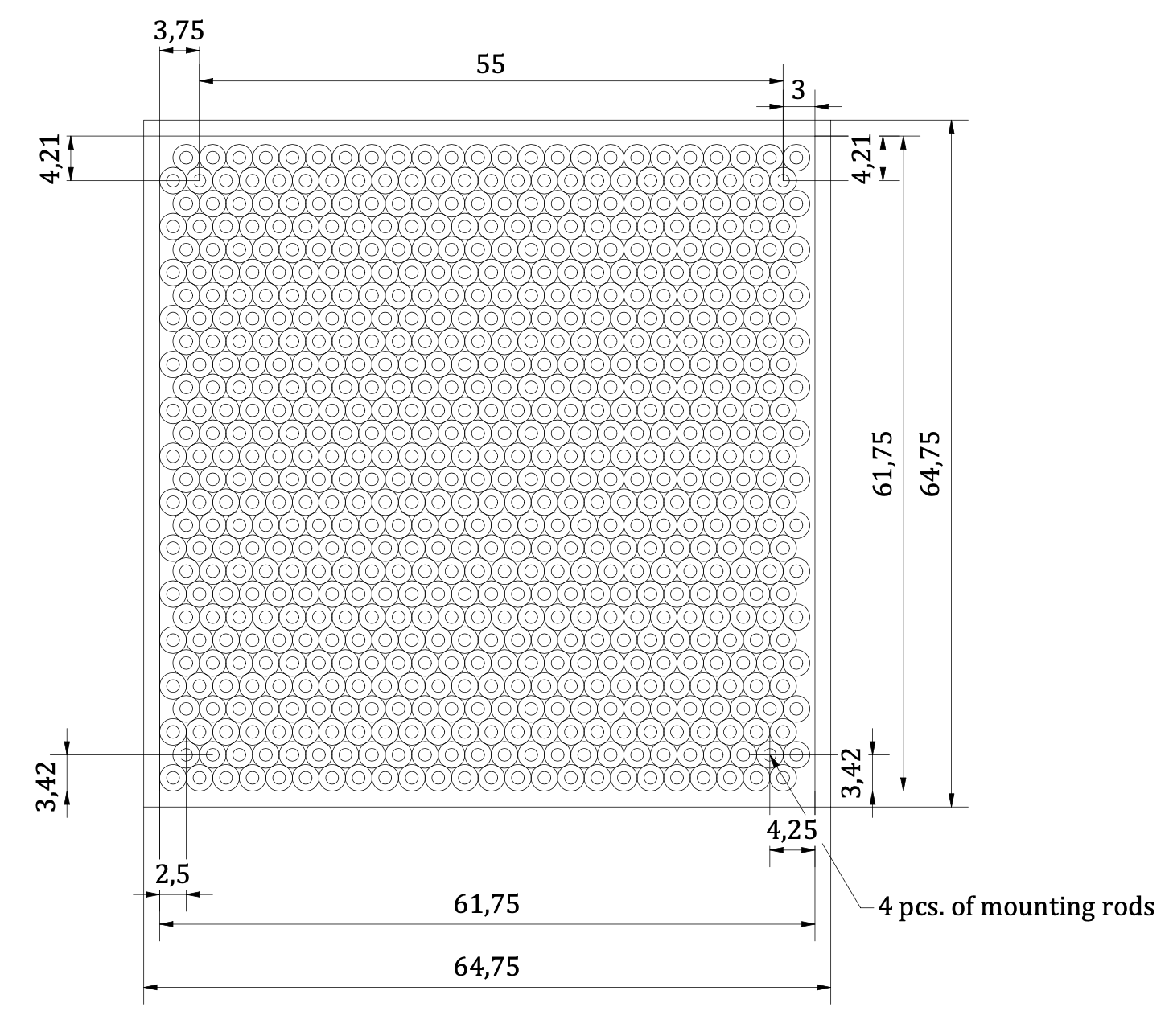}
\includegraphics[width=0.48\textwidth]{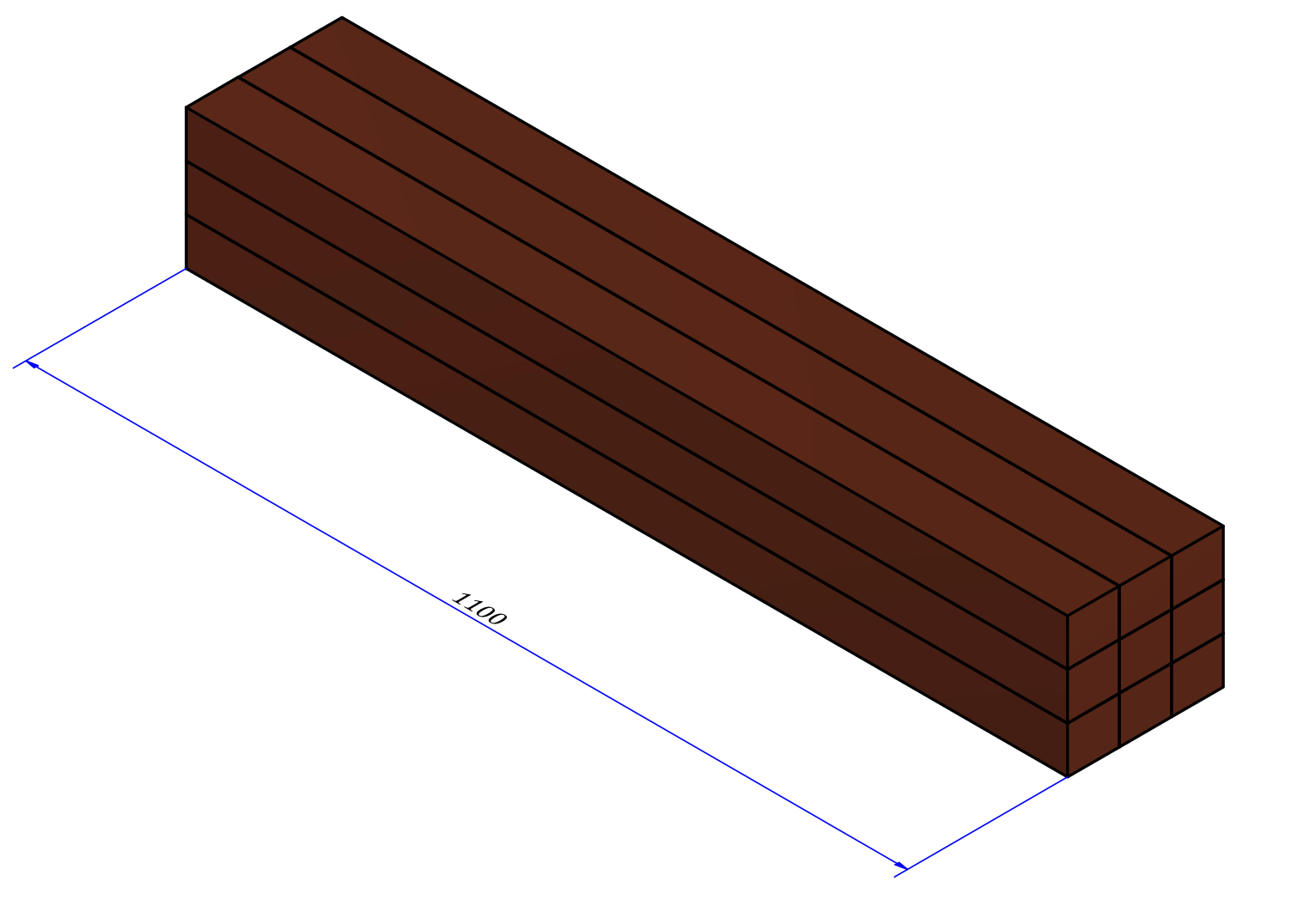}
\caption{\label{fig:P2_Design}
Left:~Schematic drawing of the design of the calorimeter modules for the FoCal-H second prototype. 
Right:~Placement of the 9 square calorimeter modules with respect to each other.
}
\end{figure}
\begin{figure}[ht!]
\includegraphics[width=1\textwidth]{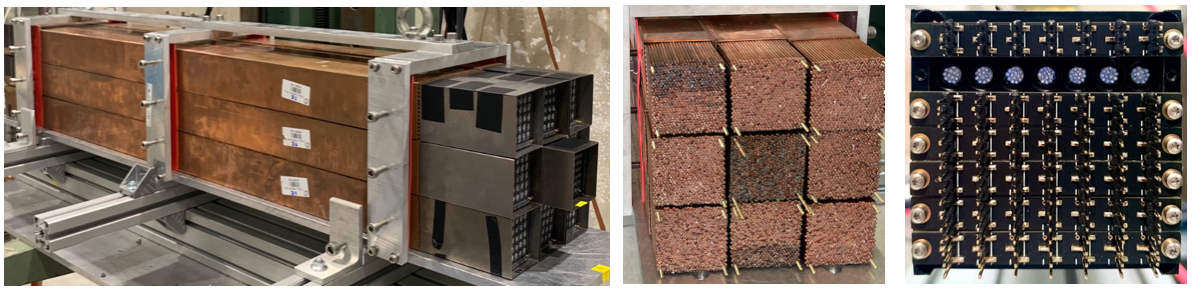}
\caption{\label{fig:P2_pic}
Left and middle: The assembled \ac{FoCal-H} prototype as a matrix of $3 \times 3$ modules.
Right: The \ac{SiPM} \ac{PCB} holders attached to the central module, which comprises $7\times7$ channels, with one \ac{PCB} removed to show the fiber bundles.
}
\end{figure}

The first prototype comprised 1440 copper capillary tubes of inner diameter $1.2$~mm and outer diameter $2.5$~mm. 
These were arranged in 40 rows of 36 capillary tubes, with every other row offset by the tube radius to allow the closest possible packing of the capillary tubes.    
The prototype had a size of 9.5x9.5x55~cm$^3$, i.e.\ a height and width of 9.5~cm and a length of 55~cm, half of the nominal length of \ac{FoCal-H}. 
The scintillating fibers, BCF10 from Saint-Gobain~\cite{Scint-fibers-datasheet}, used in this prototype had an outer diameter of 1~mm and were single cladded. 
Each of the capillary tubes contained a scintillating fiber, and thirty fibers were bundled together to be read out by a single \ac{SiPM}. 
This prototype used ON Semiconductor $6\times 6\,\mathrm{mm}^2$ MicroFC \acp{SiPM}.  
The reason for the large number of fibers per \ac{SiPM} was the difficulty in obtaining the \acp{SiPM} in 2020 and early 2021: only 48 were available.  
Two CAEN~A1702~\cite{CAEN-readout-datasheet} readout boards with each 32-channels were used to read the signal of the \acp{SiPM}. 
The prototype was tested on the H6 beam line at the \acs{CERN} \ac{SPS} with hadron beams from 20--80~GeV.  
Further details of this prototype and the results obtained in beam can be found in~\cite{Simeonov:2022yqy}.
The experimental energy distributions were in qualitative agreement with \ac{MC} simulations, demonstrating the viability of the capillary tube based design.

The second prototype was constructed from 9 modules of size $6.5 \times 6.5 \times 110$~cm$^3$ arranged in a stack of 3 by 3, as shown in \Fig{fig:P2_Design}.
Each module has 668 copper-tubes with a scintillating fiber inside, with a brass rod in each corner of a module to hold the collector plate for the fibers and \ac{SiPM} carrier boards. 
The scintillating fibers in the central calorimeter module were gathered in bundles of $\sim14$ fibers read out by 49 \acp{SiPM}, whereas the fibers in the outer calorimeter modules were gathered in bundles of $\sim27$ fibers read out by 25 \acp{SiPM}.
The distances between the \acp{SiPM} are reduced significantly compared to the first prototype with a separation of 2.5~mm between the edges of the \acp{SiPM} used for the outer modules, and 0.5~mm between the \ac{SiPM} used for the central module. 
The second prototype, hence, consisted of 249 optically isolated towers, each of which was instrumented with a $6\times 6\,\mathrm{mm}^2$ Hamamatsu S13360-6025PE \ac{SiPM}.
We used BCF12~\cite{Scint-fibers-datasheet} $1\,\mathrm{mm}$ diameter single cladded scintillating fiber. 
The performance of this scintillator is similar to BCF10 while the price per meter was slightly lower. 
\Figure{fig:P2_pic} shows the assembled prototype calorimeter.

\begin{figure}[thb!]
\includegraphics[width=0.33\textwidth]{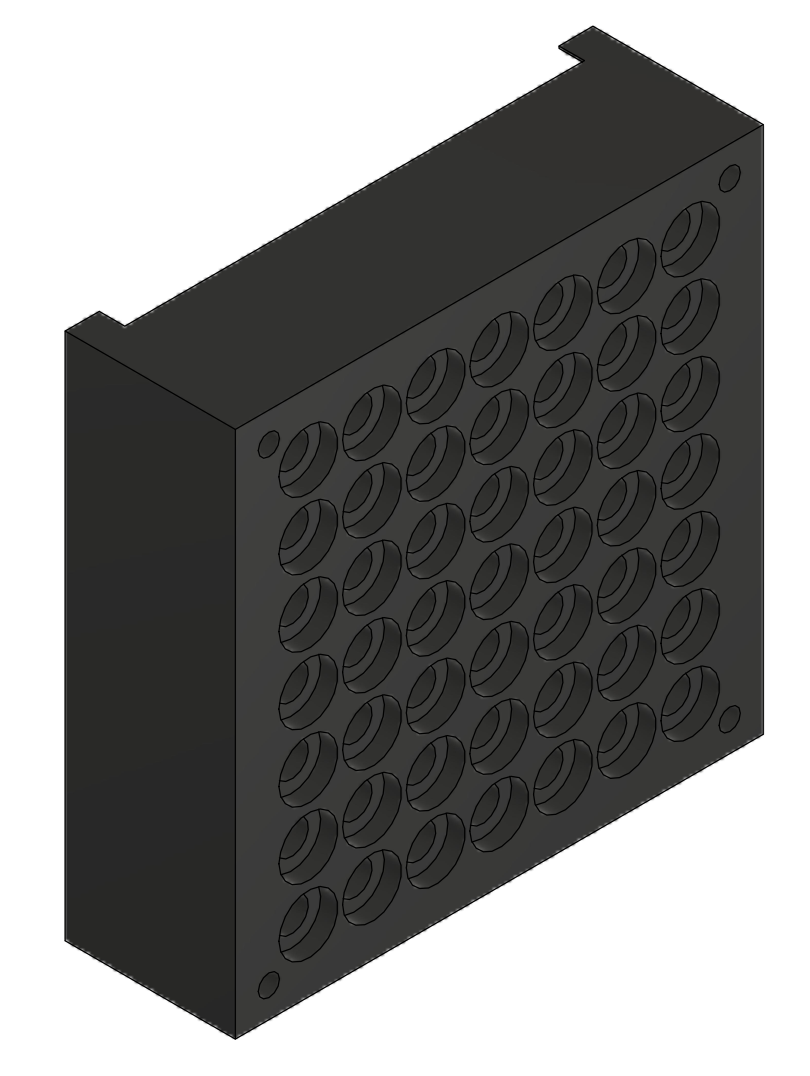}
\hspace{0.75cm}
\includegraphics[width=0.36\textwidth]{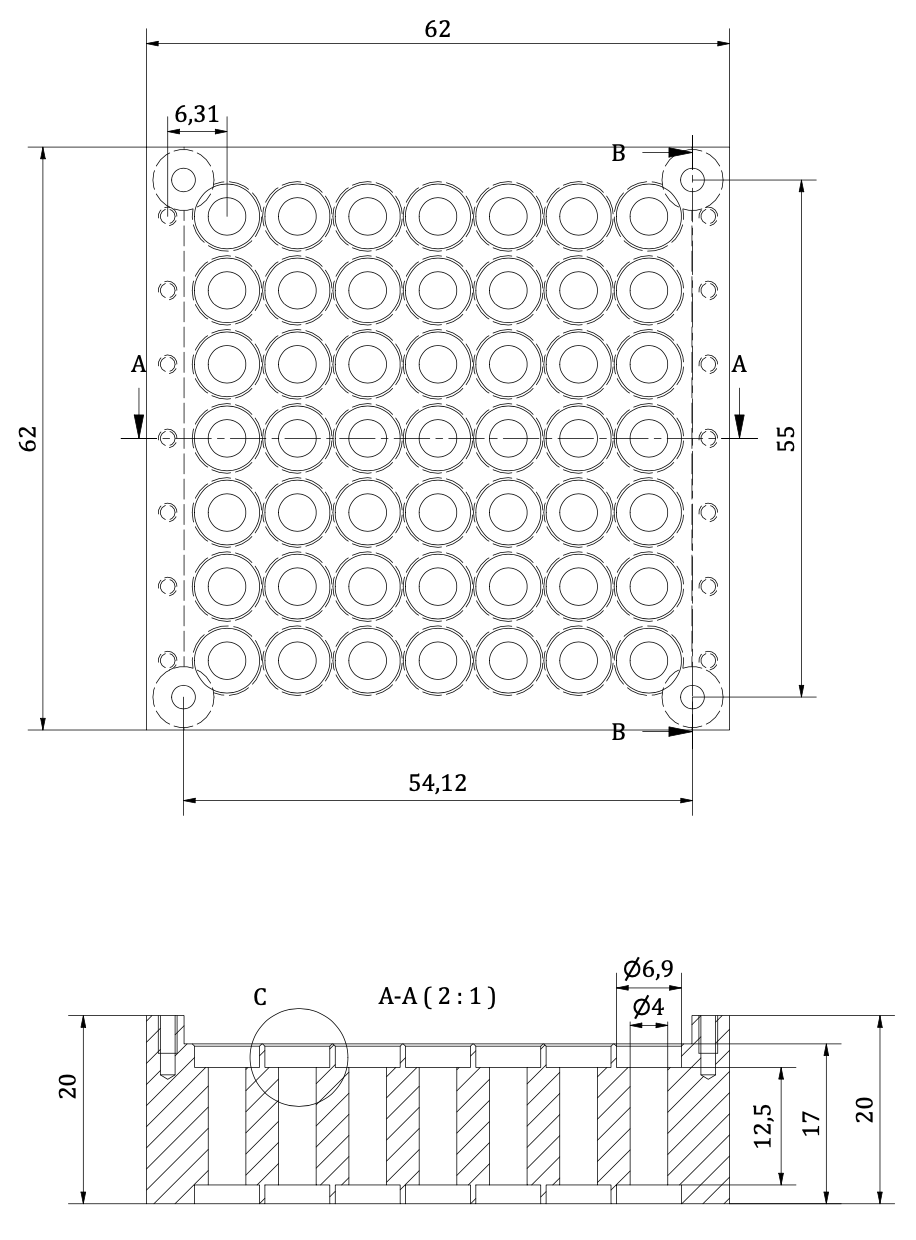}
\centering
\caption{\label{fig:P2_Collector_Design_1} Illustration~(left) and drawing~(right) of the 7$\times$7 \acs{SiPM} collector plate.}
\end{figure}
\begin{figure}[thb!]
\includegraphics[width=0.33\textwidth]{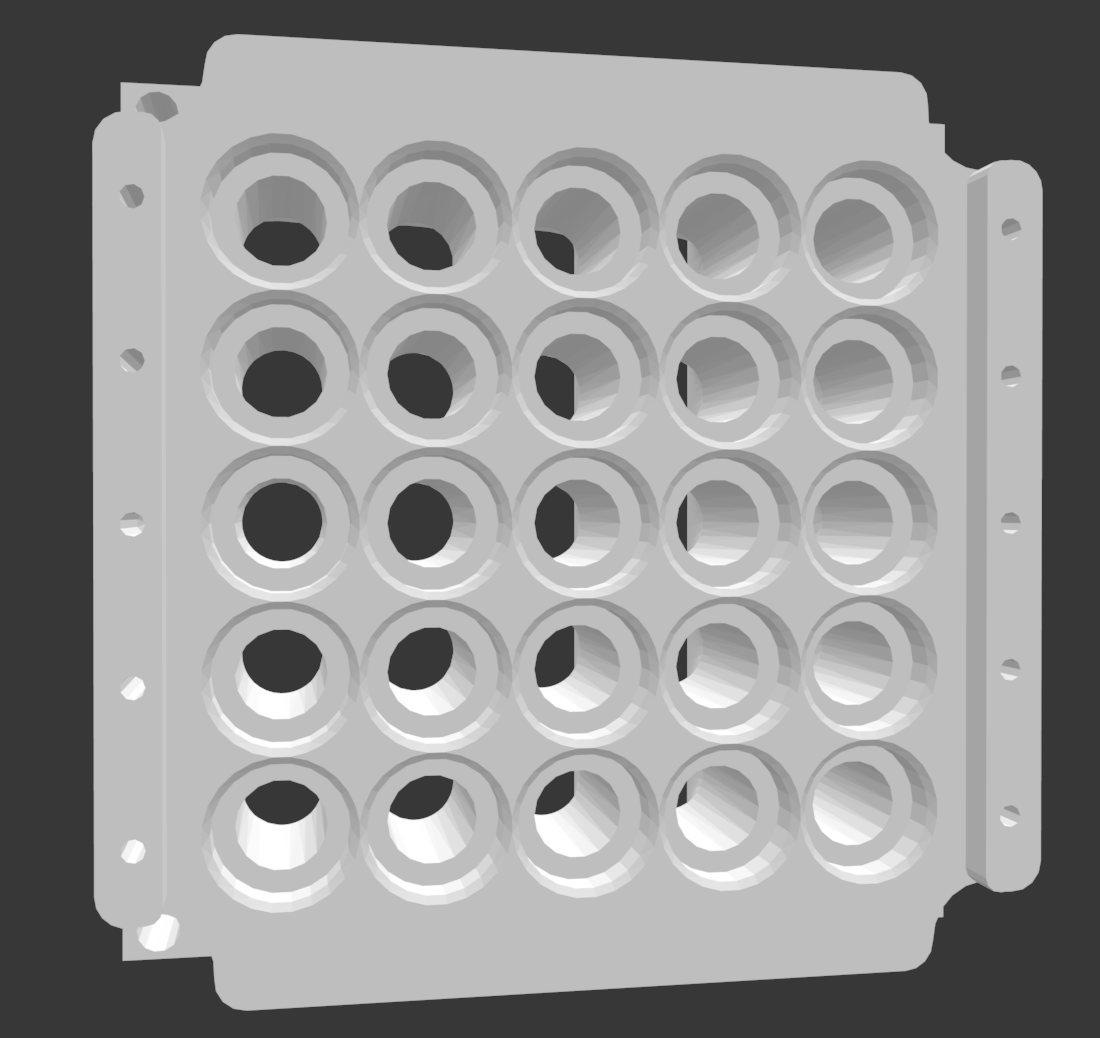}
\hspace{0.75cm}
\includegraphics[width=0.36\textwidth]{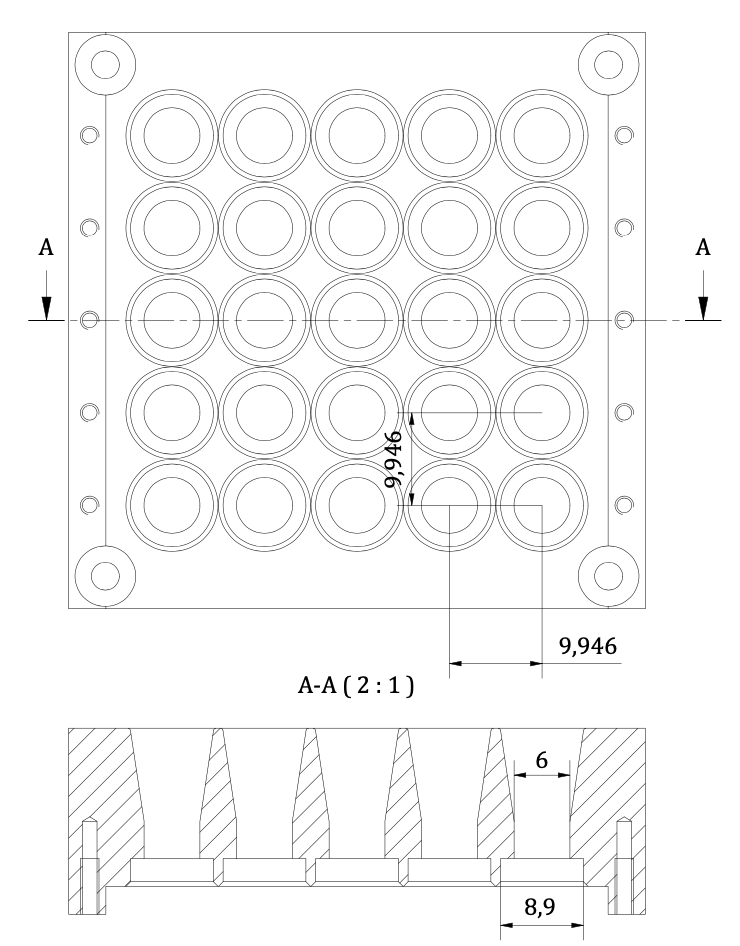}
\centering
\caption{\label{fig:P2_Collector_Design_2} Illustration~(left) and drawing~(right) of the 5$\times$5 \acs{SiPM} collector plate.}
\end{figure}


\newpage
Experience with the first prototype's electronics taught us that saturation of the \acp{ADC} would occur in the channels near the center of hadron showers.  
To mitigate this, a higher granularity for the center module of the second prototype was chosen.
Thus, different collector plates were produced for the center and peripheral modules. 
In the center, the plate collects 49 bundles of 13 or 14 fibers for higher granularity, while in the periphery the fibers are bundled in groups of 26 or 27, and read by 25 \acp{SiPM}. 
The design of the respective collector plate is shown in Figs.~\ref{fig:P2_Collector_Design_1} and~\ref{fig:P2_Collector_Design_2}.
The \acp{SiPM} \acp{PCB} were statically screwed to the collector plates. 
A tight o-ring was placed around fibers and was pushed into the groove in the collector plate to hold the fibers position. 
The light from the fiber ends traverses a short air gap (below 1~mm) 
before reaching the \acp{SiPM}. 

CAEN DT5202~\cite{CAEN-DT5202-manual} boards designed for \ac{SiPM} bias and signal processing have been used as the primary data acquition system for the second prototype.  
Each DT5202 has a pair of Citiroc-1A chips \cite{Citiroc-1A}, for a total of 64 available channels per DT5202 board. 
The operation of multiple boards was synchronized implementing a common external trigger signal which was processed on-board by selecting one of the available trigger modes. 
Instrumenting the full prototype requires four CAEN DT5202 boards. 
Although some early testing was done using fewer boards, these results are not presented here. 
The communication with the readout electronics was via an Ethernet connection to a dedicated server, which performed the control and monitor of the \acp{SiPM} through the vendor supplied JANUS software~\cite{JANUS-User-manual}. 

\FloatBarrier

\section{Simulation setup}
\label{sec:simulation}
\begin{figure}[th!]
\begin{center}
\includegraphics[width=0.48\textwidth]{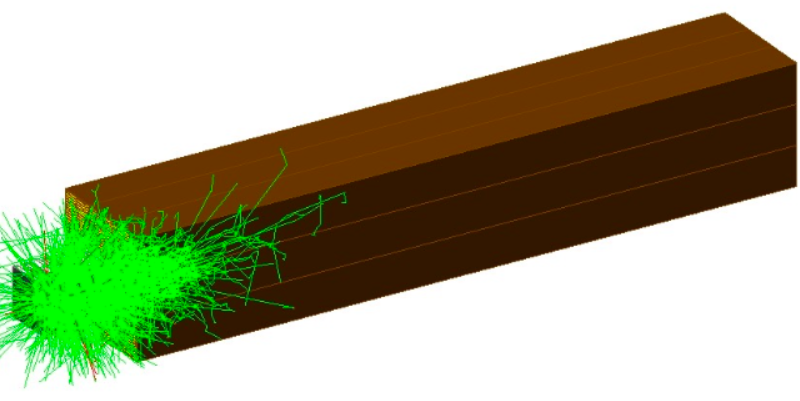}
\hspace{0.2cm}
\includegraphics[width=0.48\textwidth]{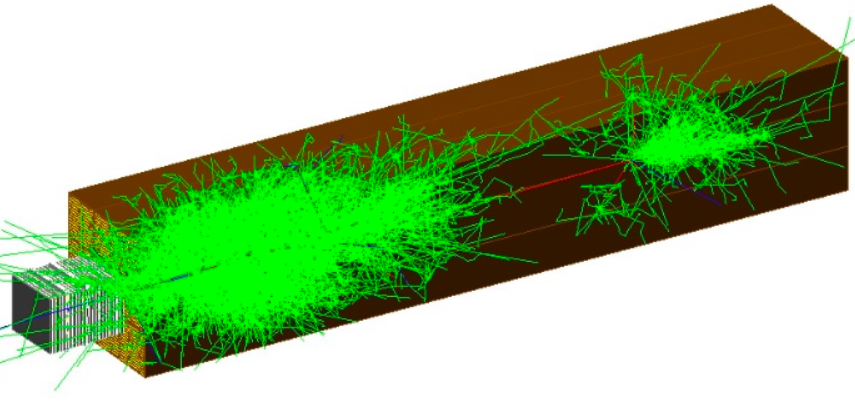}
\caption{\geant simulation of the \ac{FoCal-E} and \ac{FoCal-H}~(second version) prototype detectors for a shower created by a 100\GeV\ electron~(left) and pion~(right). 
The incident particle, which enters the detector from the left, is not shown.}
\label{fig:TBsimSetup}
\end{center}
\end{figure}

To model and understand the performance of the \ac{FoCal} prototypes, a dedicated \ac{MC} simulation was developed using the \geant.10.7 software package~\cite{GEANT4:2002zbu}.
The realistic material budget of the prototype detectors was implemented.
\Figure{fig:TBsimSetup} displays the simulated prototypes.

The \ac{FoCal-E} simulation setup consists of 20 layers with a $1.2~{\rm mm}$ air gap between the layers. 
Each layer comprises a $3.5~{\rm mm}$ thick tungsten alloy absorber and a silicon sensor, where the tungsten alloy consists of 94\% W, 4\% Ni, and 2\% Cu. 
Out of the 20 silicon sensors, 18 are pad sensors, while layers 5 and 10 are pixel sensors. 

The pad sensors have a thickness of $320~{\rm \mu m}$, with an active thickness of $300~{\rm \mu m}$. 
The glue between the tungsten and the sensor and between the sensor and the \ac{FPC} is simulated as a $1.38~{\rm g/cm^{3}}$ density epoxy material, with a thickness of $110~{\rm \mu m}$ and $130~{\rm \mu m}$, respectively. 
The \ac{FPC} is approximated by a $140~{\rm \mu m}$ thick uniform Cu layer to account for the electronic components. 
The corresponding glass-reinforced epoxy laminate~(\acs{FR4}), which has a thickness of $1.595~{\rm mm}$, is simulated as a $1.86 ~{\rm g/cm^{3}}$ density material consisting of 47.2\% of the same material as used in the epoxy glue and 52.8\% of SiO$_{2}$.
An overview of the materials and their relevant properties can be found in \Tab{tab:simulation-pad-materials-part1} and \Tab{tab:simulation-pad-materials-part2}. 
The individual sensors of the pad layers are set as $9\times8$ pads with an individual pad size of $10\times 10 ~{\rm mm^{2}}$.

The pixel layers contain two complementary aluminum carriers hosting three adjacent strings containing three pixel sensors each, and three Kapton flex cables interleaved between the pixel strings. 
In total, they contribute only about 6\% of a radiation length.
The flex cables are approximated as a $1.42~{\rm g/cm^{3}}$ density Kapton film, with a thickness of $50~{\rm \mu m}$. 
The carriers are connected back-to-back to enclose the sensors and provide a uniform active area. 
The pixel sensors have a thickness of $25~{\rm \mu m}$, and the aluminum carriers have a thickness of $1.5~{\rm mm}$.
With an air gap of $2.0~{\rm mm}$ between the aluminum plates, the total thickness of the pixel layers becomes $5.1~{\rm mm}$. 
A complete pixel layer contains 18 sensors, with nine per carrier. 
Each sensor has a matrix of $1024\times 512$ pixels with a pixel size of $29.24 \times 26.88\ \mu\text{m}^{2}$. 
\begin{table}[th!]
\begin{center}
\caption{\label{tab:simulation-pad-materials-part1}
Overview of the materials used in the pad layers of and their relevant properties (part 1/2).\\}
\fontsize{10pt}{10pt}\selectfont
\begin{tabular}{l l c c c}
\textbf{Stack   layer} & \textbf{Material}                         & \textbf{Density (g/cm$^\mathbf{3}$)} & \textbf{Thickness (cm)} & \textbf{Weight (g/cm$^\mathbf{2}$)} \\ \toprule
Sensor active          & Silicon                                   & 2.33                     & 0.030                   & 0.070                  \\
Sensor inactive        & Silicon                                   & 2.33                     & 0.002                   & 0.005                 \\ \midrule
Tungsten plate         & \begin{tabular}[c]{@{}l@{}}94\% tungsten, \\ 4\% nickel, \\ 2\% copper\end{tabular} & 19    & 0.350     & 6.650           \\ \midrule
Glue between W-Si      & PET (epoxy)                               & 1.38                     & 0.011                   & 0.015                 \\ 
Glue between Si-FPC    & PET (epoxy)                               & 1.38                     & 0.013                   & 0.018                 \\ \midrule
Pad PCB material       & FR4                                       & 1.85                     & 0.160                   & 0.295               \\ \midrule
Pad Cu GND layers      & Cu                                        & 8.96                     & 0.014                   & 0.125                 \\ \midrule
Air                    & Air                                       & $1.3 \cdot 10^{-4}$                  & 0.270                   & $3.51 \cdot 10^{-5}$               \\ \bottomrule
\textbf{Sum}           & \textbf{all}                              & \textbf{---}                & \textbf{0.850}          & \textbf{7.178}           
\end{tabular}
\end{center}
\end{table}
\begin{table}[h!]
\begin{center}
\caption{\label{tab:simulation-pad-materials-part2}
Overview of the materials used in the pad layers of and their relevant properties (part 2/2).\\}
\fontsize{10pt}{10pt}\selectfont
\begin{tabular}{l c c c c c}
\textbf{Stack   layer} & \textbf{$X$ (cm)}  & \textbf{$X$ (g/cm$^\mathbf{2}$)} & \textbf{X/X0 per layer (\%)} & \textbf{$E_c$ (MeV)} & \textbf{R$_\textbf{M}$ (mm)} 
\\ \toprule
Sensor active          & 9.37             & 21.80              & 0.32                           & 40                & 4.90      
\\
Sensor inactive        & 9.37             & 21.80              & 0.02                           & 40                & 4.90      
\\ \midrule
Tungsten plate         & 0.37             & 6.80               & 95.44                          & 8                 & 0.93      
\\ \midrule
Glue between W-Si      & 50.31            & 44.80              & 0.02                           & 101               & 10.50     
\\
Glue between Si-FPC    & 50.31            & 44.80              & 0.03                           & 101               & 10.50    
\\ \midrule
Pad PCB material       & 16.00            & 32.20              & 1.00                           & 63                & 6.10     
\\ \midrule
Pad Cu GND layers      & 1.60             & 12.90              & 0.88                           & 19              & 1.60      
\\ \midrule
Air                    & $3.04\cdot 10^4$ & 36.62              & 0.00                           & 88                & 7300.0   
\\ \bottomrule
\textbf{Sum}           & \textbf{---}     & \textbf{---}       & \textbf{97.70}                 & \textbf{---}         & \textbf{---}   
\end{tabular}
\end{center}
\end{table}

The \ac{FoCal-H} prototype simulation setup consists of nine square calorimeter modules organized in a $3\times 3$ matrix, placed $44~{\rm mm}$ behind \ac{FoCal-E}. 
Each calorimeter module has outer dimensions of $64.75\times 64.75\times 1100 ~{\rm mm^{3}}$ and inner dimensions of $61.75\times 61.75\times 1100 ~{\rm mm^{3}}$, and contains $24 \times 28$ copper capillary tubes filled with polystyrene scintillating fibers. 
Both the copper tubes and the scintillating fibers are described in detail to account for the shower development and the energy deposition in the different materials of the detector, with only the fibers considered as active volume. 
The copper tubes have an outer diameter of $2.5~{\rm mm}$ and an inner diameter of $1.1~{\rm mm}$. 
The diameter of the scintillating fibers is $1.0~{\rm mm}$. 

The \textit{Guide For Physics Lists} by the \geant Collaboration~\cite{geant4_guidePhysicsLists} recommends the \acs{FTFP}\_\acs{BERT} physics list for collider physics application and high energy calorimetry. 
The guide also mentions \acs{QGSP}\_\acs{BERT}, the previous default \geant physics list before the default changed to \acs{FTFP}\_\acs{BERT} in the release of \geant~v10.0, as an alternative. 
Tests were performed using both of the recommended lists.
For \ac{FoCal-E}, there was no difference between the two, while for \ac{FoCal-H} 
\acs{QGSP}\_\acs{BERT} gave slightly more consistent results comparing with the data.
Hence, \acs{QGSP}\_\acs{BERT} was chosen for all of the performance simulations presented in this paper.

For \ac{FoCal-H} the following corrections are implemented.
The energy deposition in the plastic scintillators is corrected according to Birks' law using \geant's energy model for electromagnetic saturation. 
The effect was expected to be small due to the impinging particles' high energy and the relatively small \dedx. 
However, there was a decrease of about 5\% in the visible energy with respect to the deposited energy. 

A factor of 8000 photons/MeV is implemented to simulate a more realistic light yield for the BCF-12 fibers. 
Furthermore, to account for the light attenuation, the number of photons from each energy deposit event is corrected with a value of $\exp(-\Delta l / \lambda)$, where $\Delta l$ is the distance from the position of the energy deposition event to the silicon photomultipliers and $\lambda=2.7$~m is used, as indicated for BCF-12 fibers. 
A constant factor of 2.5\% is also assumed for the trapping efficiency. 
This factor considers the solid angle for which total internal reflection occurs and assumes a 100\% loss at the non-equipped fiber end. 
Finally, the light output from each individual hit in a fiber is combined to create a single output quantity from that fiber.

The simulated data is stored in a \acs{ROOT} format on an event-by-event basis and is further processed by the dedicated analysis and reconstruction module. 
At this stage, several other effects are considered. 
The light from individual fibers in the simulation is combined since the fibers in the real prototype are bundled into channels in the readout system. 
The total light directed towards each of the silicon photomultipliers in the readout is, hence, calculated by adding the light output from each bundled fiber in the channel.

Losses in the number of photons will also occur in the propagation from the fiber ends to the silicon photomultipliers since neither optical silicon nor glue is used between the fiber ends and the front surface of the \acp{SiPM}. 
Here, a factor of 0.5 is implemented to account for the light losses. 
The photodetection efficiency of the \ac{SiPM} is assumed to be 0.4. 
This efficiency is used as a conversion factor translating the number of arriving photons to the mean number of produced photoelectrons. 
The mean number of photoelectrons is then smeared according to a Poissonian distribution to obtain the number of fired \ac{SiPM} channels.
In addition, the light loss factor and the photodetection efficiency factor were varied between 0.1 and 0.5 each.
Due to the high photostatistics at the SiPM, this results has a negligible effect in the energy resolution, which is dominated by the shower fluctuations.

As described, the simulation module assumes a uniform detector response and uniform properties of all the fibers and optical contacts in the constructed prototype. 
Furthermore, equal gain and photodetection efficiency are used for each silicon photomultiplier. 
The readout and digitization electronics are also considered to behave linearly and uniformly. 

\section{Test beam setup}
\label{sec:setup}
Extensive test beam campaigns have produced the data to compare the prototypes performance with the simulation results and to guide the R\&D for the final design. 
Beams of electrons and hadrons with energies ranging from 1 to 350\GeV\ provided by various beam lines at the \ac{PS} and \ac{SPS} test beam facilities were used to cover most of the possible shower topologies for the electromagnetic and hadronic components of the calorimeter. 

\begin{figure}[t!]
\begin{center}
\includegraphics[width=\textwidth]{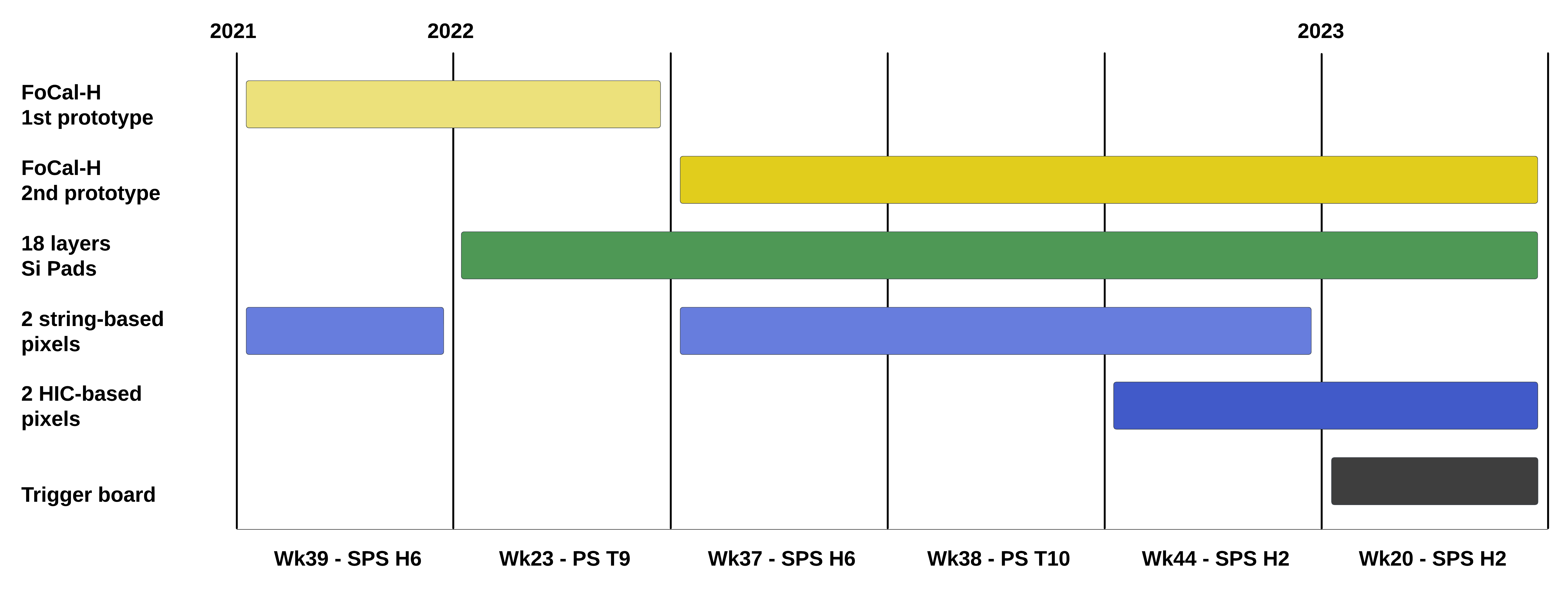} 
\caption{
Timeline of the \ac{FoCal} subsystems and related electronics in 2021, 2022 and 2023. 
In 2021, additionally a single pad layer was tested to validate the detector performance of the electronics readout.}
\label{fig:timeline}
\end{center}
\end{figure}

\Figure{fig:timeline} summarizes the development of the \ac{FoCal} prototype systems and instrumentation employed during the test beam campaigns of 2021, 2022 and 2023. 
Over the years, the tested prototype detectors increased in size and complexity, resembling more and more the prototype of the envisioned final \ac{FoCal} detector at the end of 2022.

In 2021, two \ac{pCT}~(string)-based pixel layers, and the first prototype of \ac{FoCal-H} were tested, as well as a single pad layer.
The trigger and readout for all systems was independent, making the combination of data between the subsystems impossible. 

The situation did not change for the PS campaign of 2022, where the trigger logic of the systems were independent and without possibility of a collective combined data taking. 
However, for the first time, we operated 18 pad layers in the \ac{FoCal-E} prototype stack, without the pixel layers. 
The \ac{DAQ} of the pad layers was operated in the \ac{UDP} mode, in which commands and data are received and sent through Ethernet connections to a dedicated server. 
This infrastructure was built to perform the in-beam calibrations and start, stop, and monitor physics acquisitions. 
A custom-made offline analyzer performed a very preliminary reconstruction of the data to provide a rapid feedback of basic quantities. 

\begin{figure}[t!]
\begin{center}
\includegraphics[width=1\textwidth]{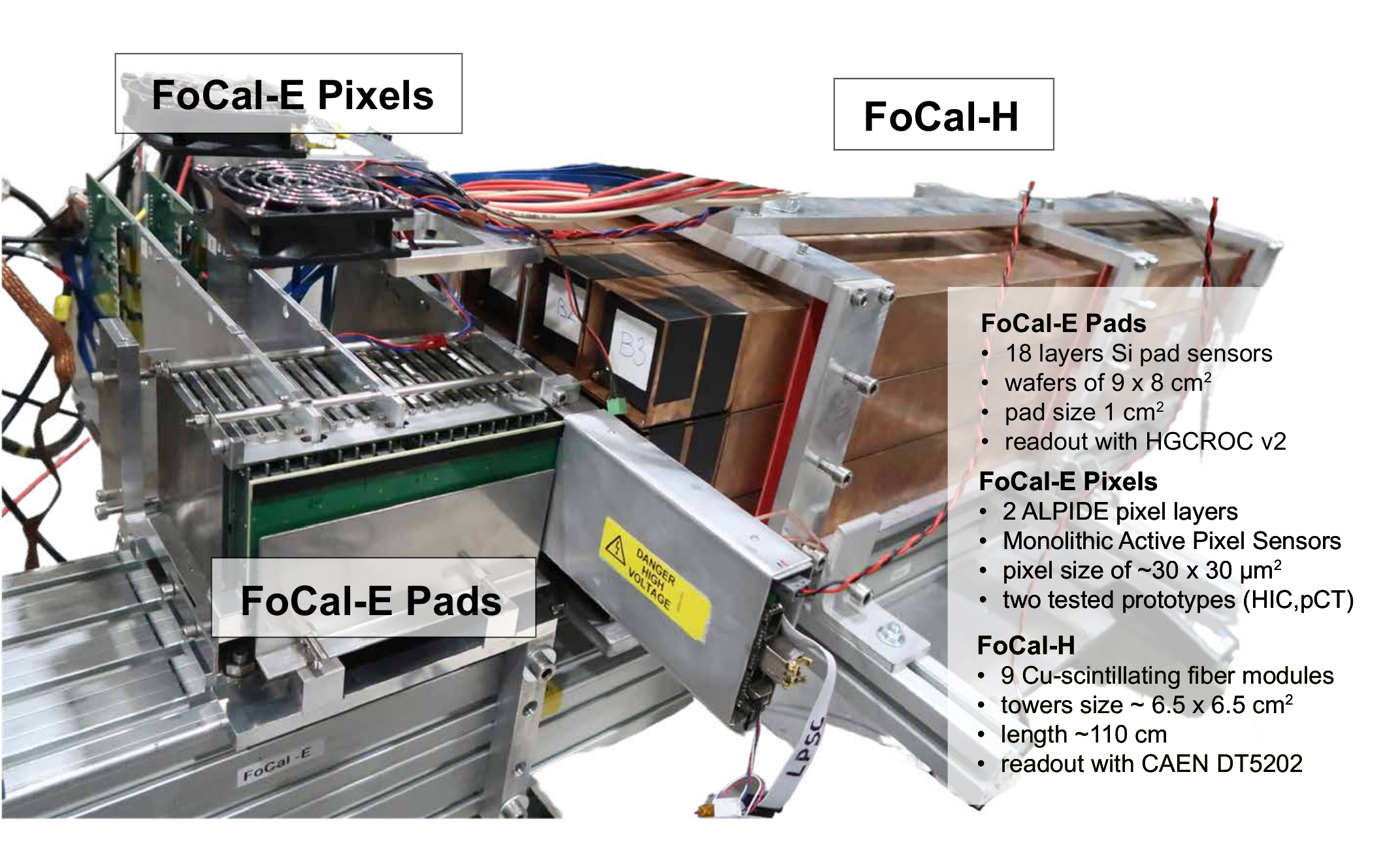}
\caption{Picture of the full \ac{FoCal} test beam setup used during the fall 2022 and 2023 campaigns.
The \ac{FoCal-E} consisted out of 18 pad layers, and 2 pixel layers~(shown are \ac{pCT}-based layers but also \ac{HIC}-based layers were used).
The second prototype for \ac{FoCal-H} was used.
}
\label{fig:setup_pic}
\end{center}
\end{figure}


In September 2022, the second prototype of \ac{FoCal-H} as well as the pixel layers were included, as shown in \Fig{fig:setup_pic}.
The \ac{FoCal-E} \ac{DAQ} was based on the \acs{ALICE} \ac{O2}~\cite{Buncic:2011297} system~(explained below), providing coherent treatment of the pixel and pad layers.
As described in \Sec{subsubsec:Prototype_Hcal}, the readout of the \ac{FoCal-H} was based on the JANUS software~(v2.0.0) provided by CAEN.
At that time, rudimentary event matching between the \ac{FoCal-E} and \ac{FoCal-H} prototype systems was achieved using timestamps of the recorded events and externally vetoing the data collection while the systems are still busy with the signal acquisition and processing. 
From the point of the hardware, this completed the \ac{FoCal} prototype setup.

\begin{figure}[t!]
\begin{center}
\includegraphics[width=1\textwidth]{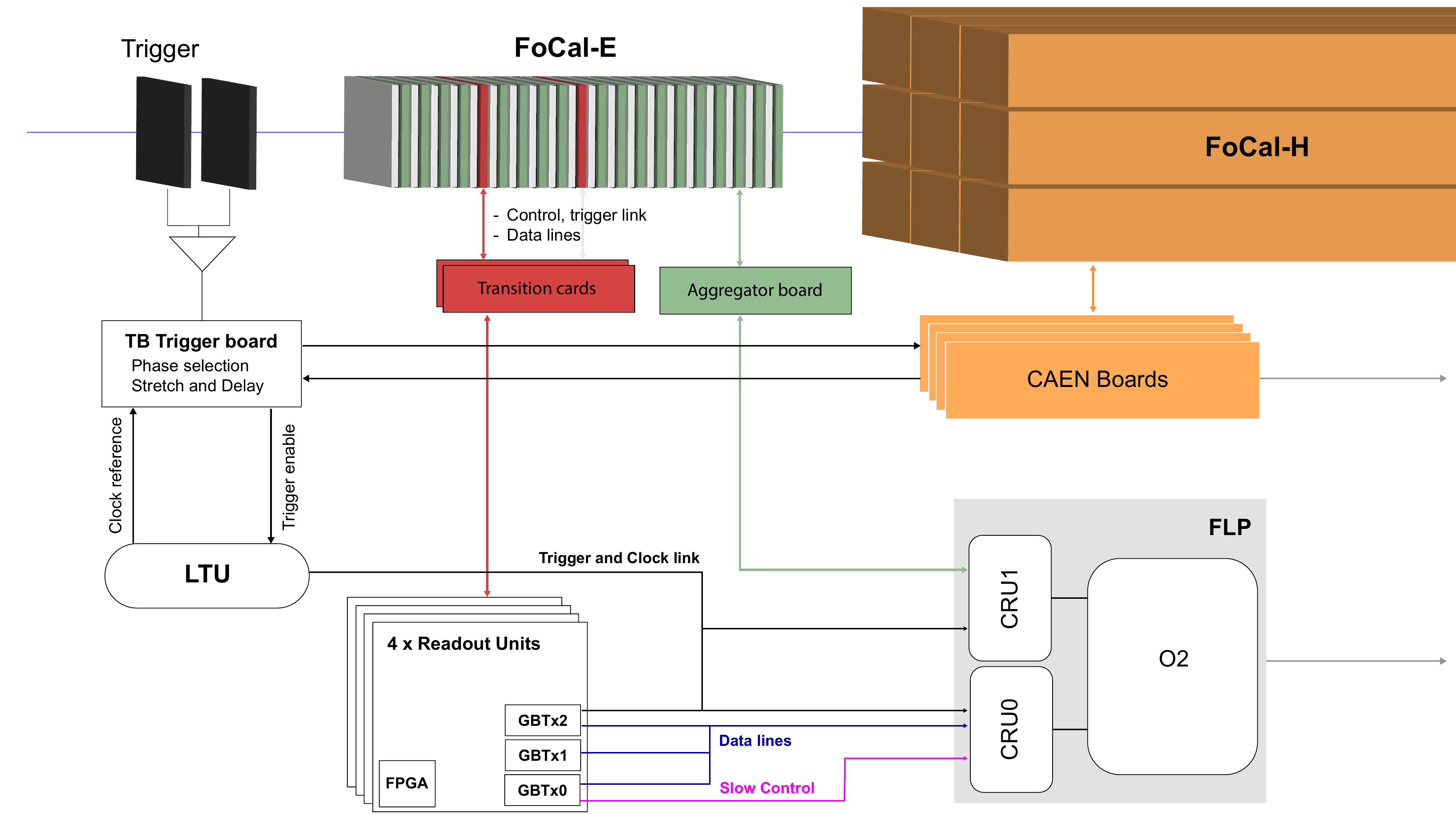}
\caption{Sketch of the setup at the H2 beam line of the \acs{SPS} in May 2023. 
The particles enter the detector from the left to right, having to pass a set of two scintillators used as trigger.
}
\label{fig:setup_new}
\end{center}
\end{figure}

Refinements and improvements in the \ac{DAQ} and trigger were done for the data taking periods at the H2 beam line of the \ac{SPS} in November, 2022 and May, 2023.
In May, 2023, an additional custom-made component, the trigger board (described below), was put in place to provide a common acquisition dead time to the \ac{FoCal-E} and \ac{FoCal-H} subsystems, and enable a more advanced trigger operation.
\Figure{fig:setup_new} provides an overview of the May, 2023 setup, displaying the prototype and the associated readout and trigger systems.
With the occurrence of a coincidence between the two scintillators placed in front of the setup (displayed in \Fig{fig:setup_new}), the trigger board simultaneously sends out busy signals for the \ac{FoCal-H} CAEN boards, and provides an external trigger for the \ac{LTU}. 
The \ac{LTU} emulates the \ac{CTP} functions of \acs{ALICE}, providing in turn the trigger at the \ac{CRU} end-points and enabling the recording of the data for the two systems. 
The \acp{CRU} used to collect the data from the two \ac{FoCal-E} subsystems are hosted inside the \ac{FLP}. 
These machines are equipped with the \ac{O2} software, which handles the detectors readout via the \ac{CRU} interface and prepares the data for online and offline processing, and ultimately also storage. 
In the \ac{FLP} the collected data can be stored locally in the native \ac{O2} format, and organized in sequentially recorded time intervals referred to as \acp{TF}. 
At the same time, locally storing the raw data ensures a reliable backup which can prevent data corruption arising from the online reconstruction procedure. 
The online data-quality monitoring is provided by a detector-specific component based on the \ac{QC} workflows  which allows to inspect a fraction of the produced data stream. 
The monitoring objects created by the \ac{QC} were periodically pushed to a database from which they can be queried and monitored through a web interface.

\begin{figure}[t!]
\begin{center}
\includegraphics[width=0.8\textwidth]{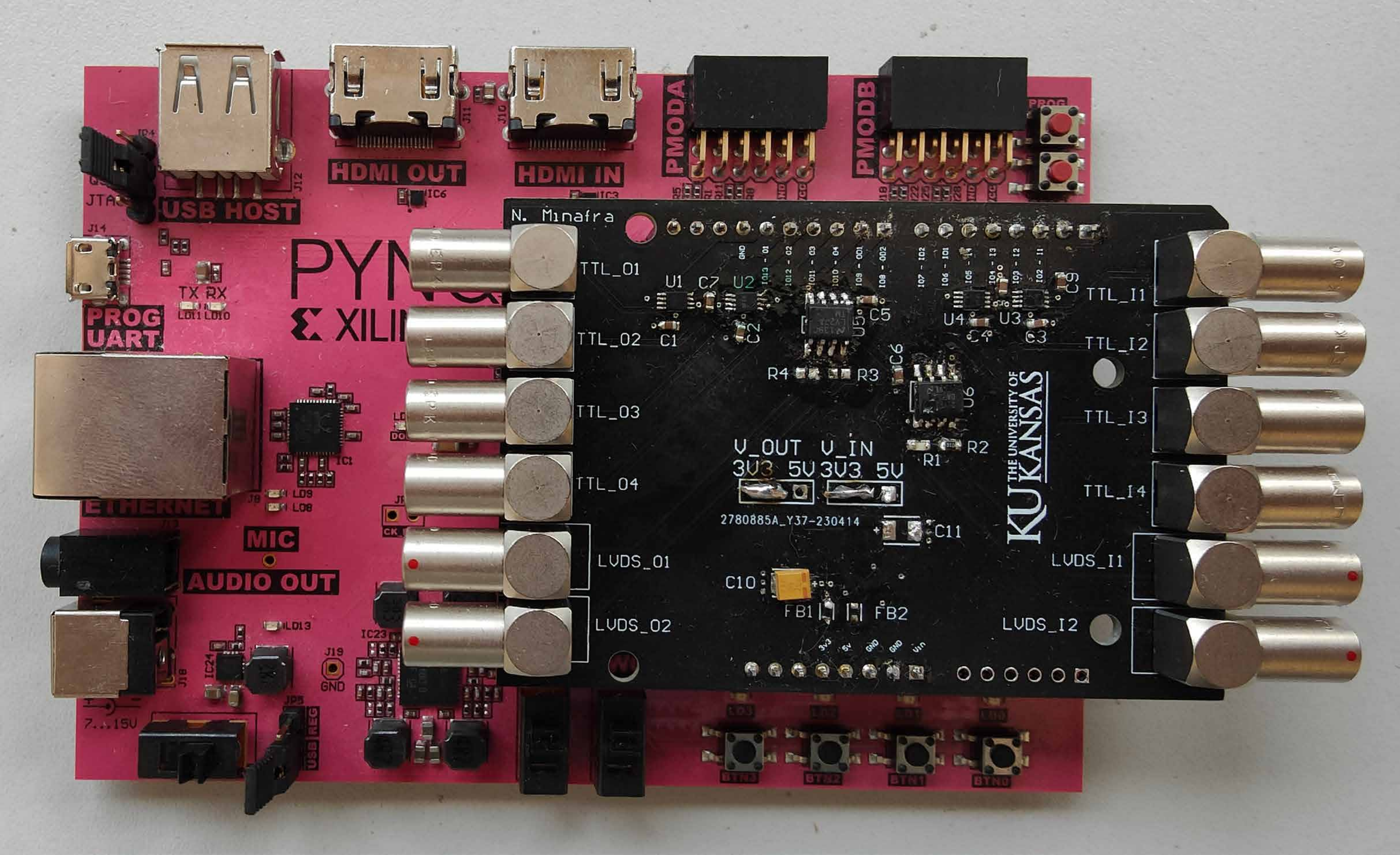} 
\caption{Test beam trigger board assembled on top of the PYNQ Z2 board.}
\label{fig:tb2}
\end{center}
\end{figure}

The trigger board consists of a custom \ac{PCB} with \ac{COTS} components that convert the input and output signals to 3.3V \ac{TTL}, installed on a PYNQ Z2 board. 
A picture of the  assembled the trigger board can be found in \Fig{fig:tb2}. 
The board hosts a Xilinx ZYNQ \ac{SoC} that uses both a microcontroller and an \ac{FPGA}.
The \ac{FPGA} was used to implement all the real-time processing, i.e.\ coincidences, signal synchronization and dead time.
The microcontroller was used to implement the control interface using Ethernet and \acs{TCP}/\acs{IP}.
The board collects signals from 4 different \ac{TTL} inputs, and can produce up to 4 independent output triggers.
Every input can be positive or negative and they can be stretched and delayed digitally to allow an easy synchronization with the other signals. 
Each signal can also be set as normal input or as a veto.
Each output signal can be configured to trigger with a determined pattern of coincidences presented on the inputs.
For example, the typical case is with 2 or more scintillators in coincidence and 1 or more busy signals in veto.
At the same time, it is possible to limit the trigger rate introducing a dead-time.
This feature was particularly useful to guarantee proper synchronization between \ac{FoCal-E} and \ac{FoCal-H} during combined acquisitions.

One of the more advanced features is the synchronization of the output triggers with an external clock. 
For example, it is possible to accept trigger only when in phase with a determined phase of the clock.
This feature is particularly useful when testing a system that is designed to work on a beam with bunch structure~(like at the \ac{LHC}), which is tested on a debunched beam~(like for most of the test beam facilities).
The most obvious application is for the \ac{HGCROC} \ac{ADC}, that is sampling at a fixed phase of the 40~MHz clock. 
When working with a 40~MHz bunch structure, the \ac{ADC} is sampling always the same part of the signal, ideally the maximum, producing a digitized value that is correlated with the deposited charge. 
If the particles are not synchronized with the clock, the \ac{ADC} would sample at a random point of the signal (within the clock period), producing randomly distributed signals.
Without the trigger board one needs to record all events with triggering particles and attempt to filter offline all signals that are out-of-time, hence with an unreliable \ac{ADC}. 



\begin{figure}[t!]
\begin{center}
\includegraphics[width=0.8\textwidth]{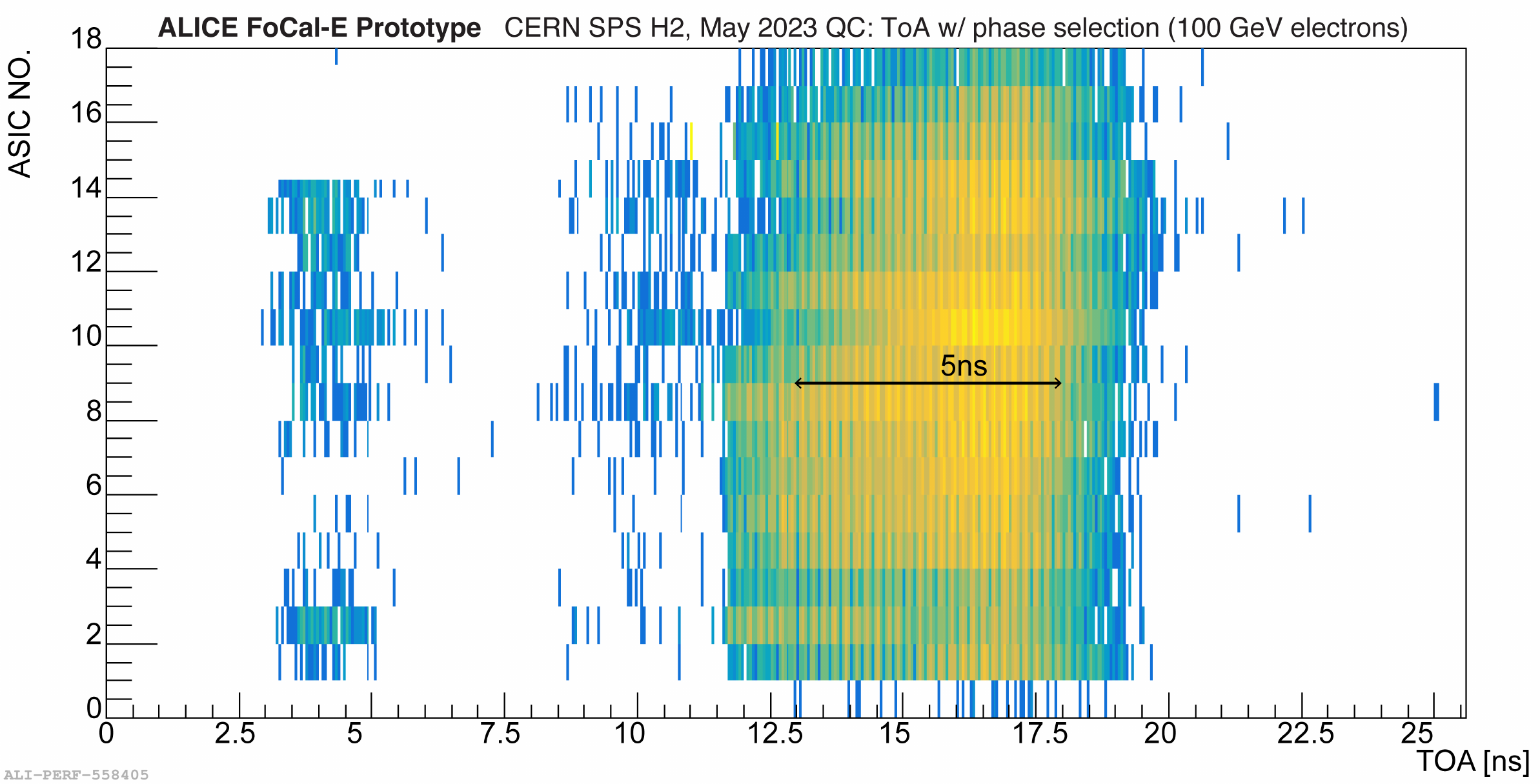} 
\caption{Per-layer \ac{ToA} distributions for all \ac{FoCal-E} pad layers after phase alignment measured during a 100~GeV electron run. 
The plot, extracted online from \ac{QC}, demonstrates the correct relative time alignment of the 18 \ac{HGCROC} chips,  indicating a healthy working point of the detector.}
\label{fig:TOA_layers}
\end{center}
\end{figure}

This was the approach of all test beams before May, 2023, but it made it difficult and rather impossible to have all the 18 \acp{HGCROC} in sync.
Using the trigger board, the phase selection was operated using the online monitoring provided by \ac{QC}.
Events where particles arrive in phase with the \ac{HGCROC} clocks were selected, providing a definite procedure to correctly tune the latency between the event and the readout trigger. 
As the timing can be independently set on a layer-by-layer basis, the information provided by the online monitoring allowed the adjustment of the relative phases between all \acp{HGCROC}. 
\Figure{fig:TOA_layers} shows the \ac{ToA} distributions for all \ac{FoCal-E} pads layers  after the phase alignment.
The resulting \ac{ToA} distributions have a spread that is contained within about 5\,ns window, achieving a stable working point for the subsequent data analysis.
The 5\,ns window is due to the fact that the trigger board is selecting 1/10 phase of the clock with a spread of 2.5\,ns and a spread from the time-walk of the \ac{ToA} for signals with no or low \ac{ToT} value.

\section{FoCal-E results}
\label{sec:focaleres}
The data for the \ac{FoCal-E} prototype systems presented in the following were obtained at the T9 beam line of the \ac{PS} in June, 2022 and at H2 beam line of the \ac{SPS} in November, 2022 and May, 2023.

The data recorded with the \acp{HGCROC} contain the digitized \ac{ADC}, \ac{ToT}, and \ac{ToA} information for every channel of the 18 low granularity layers.
In addition, for a recorded event the data payloads contain the timestamp of the event, as well as the trigger information provided by the \ac{HGCROC} for 20 time intervals~(each 25~ns) around the recorded event for every channel.

\subsection{Pad-layer \acs{MIP} response}
\label{subsec:MIP}

Standard signal processing techniques are applied to separate the charge deposited by the shower particles from contributions of the noise.
The two main contributions to the noise are addressed as follows: 
the signal pedestal, which results from the combination of the electronics noise in the amplification circuit and the intrinsic fluctuation of the silicon sensors, is treated for each channel by fitting the average response in the \ac{ADC} with a Gaussian.
The width of the Gaussian gives the spread of the noise around the average noise value.
The mean of the Gaussian fits are then subtracted to center the reconstructed \ac{ADC} signal distributions around zero.
A second, smaller contribution to the noise results from phenomena equally affecting all the channels of a layer, namely the \ac{CMN}.
This is often enhanced by slower acting effects, such as temperature and bias voltage drifts, however a more impacting source can be identified in the high frequency components coming from the surrounding environment and picked up by the silicon sensors.
The \ac{CMN} is measured event-by-event by sorting the pulse heights of the pedestal-subtracted data in increasing order and selecting the median value.
This median value is then subtracted from all amplitudes in the given event. 
For the \ac{MIP} analysis, the magnitude of the noise-subtracted signals is used to select the seed channel, by determining the pad with the largest energy deposited for each of the pad layers. 
Around the seed channel, a cluster of $N\times N$~(with $N=3$) adjacent cells is used to determine the total energy collected by a given layer, expressed in \ac{ADC} counts.

%
\begin{figure}[t!]
\begin{center}
\includegraphics[width=0.6\textwidth]{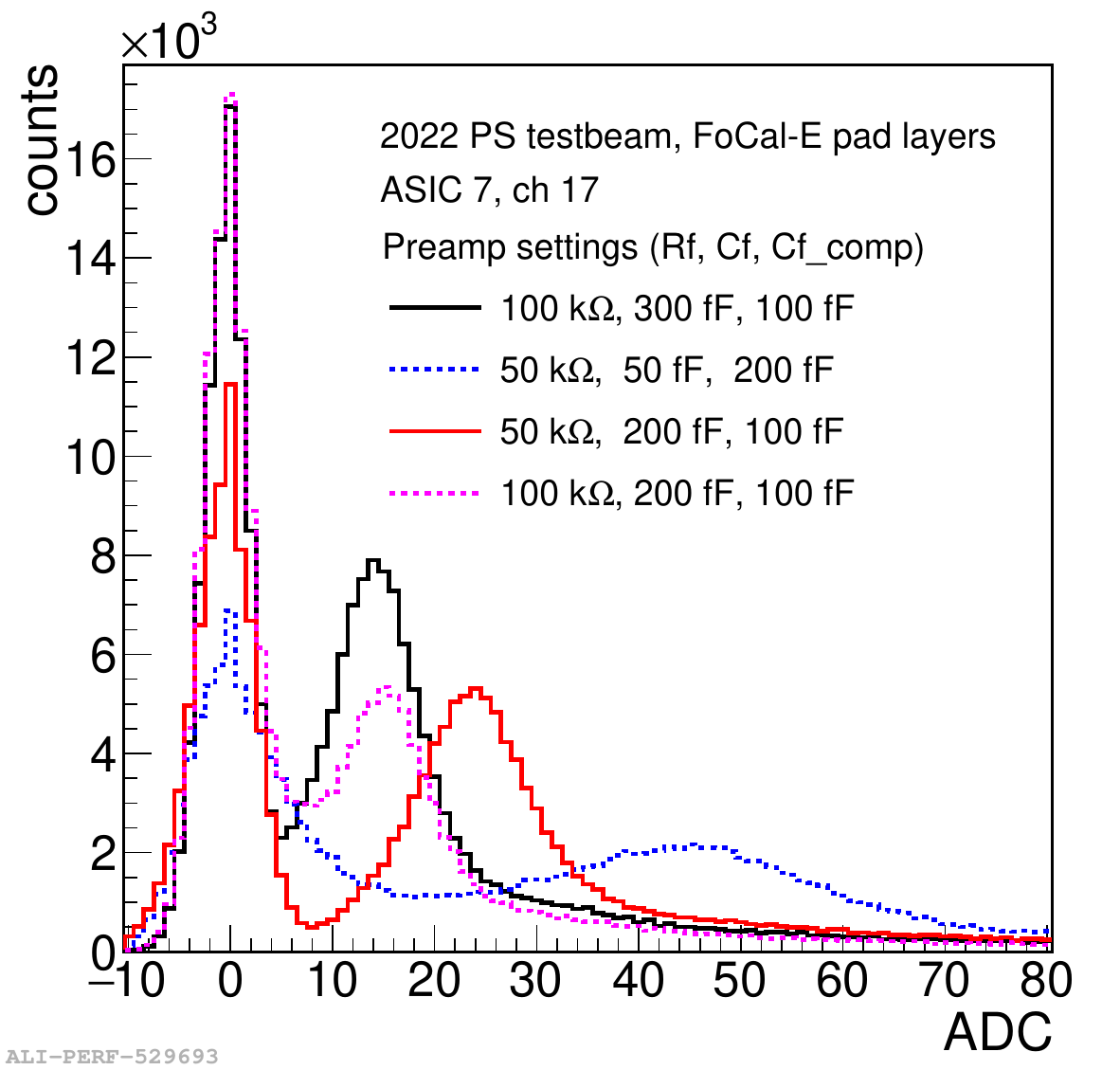} 
\caption{Response from the \ac{FoCal-E} pads \ac{HGCROC} gain scan performed during the June, 2022 \ac{PS} test beam.
Shown are 4 out of the 24 combinations of the preamplifier parameter settings.
The modified parameters include the feedback resistors and two parallel feedback capacitors.}
\label{fig:gain_scan}
\end{center}
\end{figure}

\begin{figure}[t!]
\begin{center}
\includegraphics[width=0.58\textwidth]{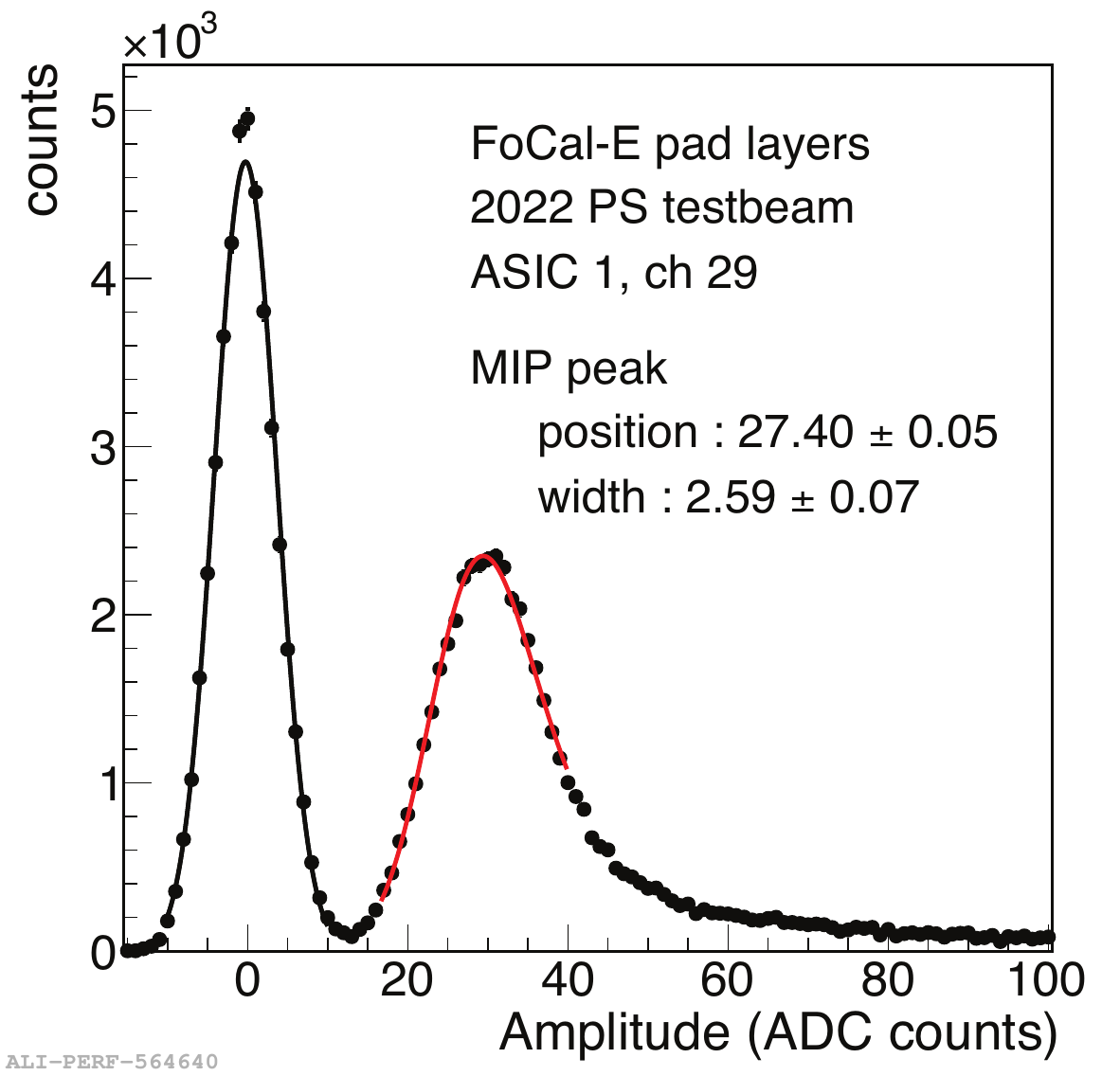}
\caption{\ac{MIP} signal in the seed channel~(29) the second pad layer~(\ac{ASIC} 1) collected during June, 2022 \ac{PS} test beam.
The data are fit with a Gaussian~(black line) to describe the pedestal, and with a \acs{langaus} distribution~(red line) to describe the \ac{MIP} signal.}
\label{fig:MIP_example}
\end{center}
\end{figure}

A detailed characterization of the \ac{FoCal-E} pad-layer response was done using low energy~(5--15~GeV) hadron beams at the PS in June, 2022~\cite{rytkonen2023simulated}. 
This included a comprehensive study of the \ac{HGCROC} preamplification settings and their comparison with the nominal value ranges provided by the \ac{ASIC} developers. 
In particular, the feedback passive components of the \acrshort{HGCROC} preamplification stage can be modified to attempt the optimization of the detectors \ac{SNR}. 
A number of combination of the parameters were tested to find the optimal working point of the readout \acp{ASIC}.
A selection of the results for 4 settings~(out the 24 reasonable combinations of $\rm{R_{f}}$ in range of 25\,-\,100~k$\Omega$ and $\rm{C_{f}}$, $\rm{C_{fcomp}}$ in range of 50\,-\,400~fF) are shown in \Fig{fig:gain_scan}. 
The goal was to choose settings which lead to good separation between the \ac{MIP} signal and noise. 
As a result of the study, in particular, the settings ``100~k$\Omega$, 300~fF and 100~fF''~(black line in the figure) and ``50~k$\Omega$, 200~fF and 100~fF''~(red line in the figure) were used in the following.
The resulting shape of the distribution can be described with a Landau distribution that models the energy deposition in the detector material convoluted with a Gaussian distribution describing the detector's resolution~(called \acs{langaus} distribution in the following).
The \ac{MPV} of the distribution is usually quoted as the \ac{MIP} peak~(or position) and the width of the Gaussian as the \ac{MIP} width~(or resolution).
An example of \ac{MIP} distribution and fit results for a specific chip and seed channel is shown in \Fig{fig:MIP_example}.
Results for all channels can be found in Ref.~\cite{rytkonen2023simulated}.

\begin{figure}[th!]
\begin{center}
\includegraphics[width=0.58\textwidth]{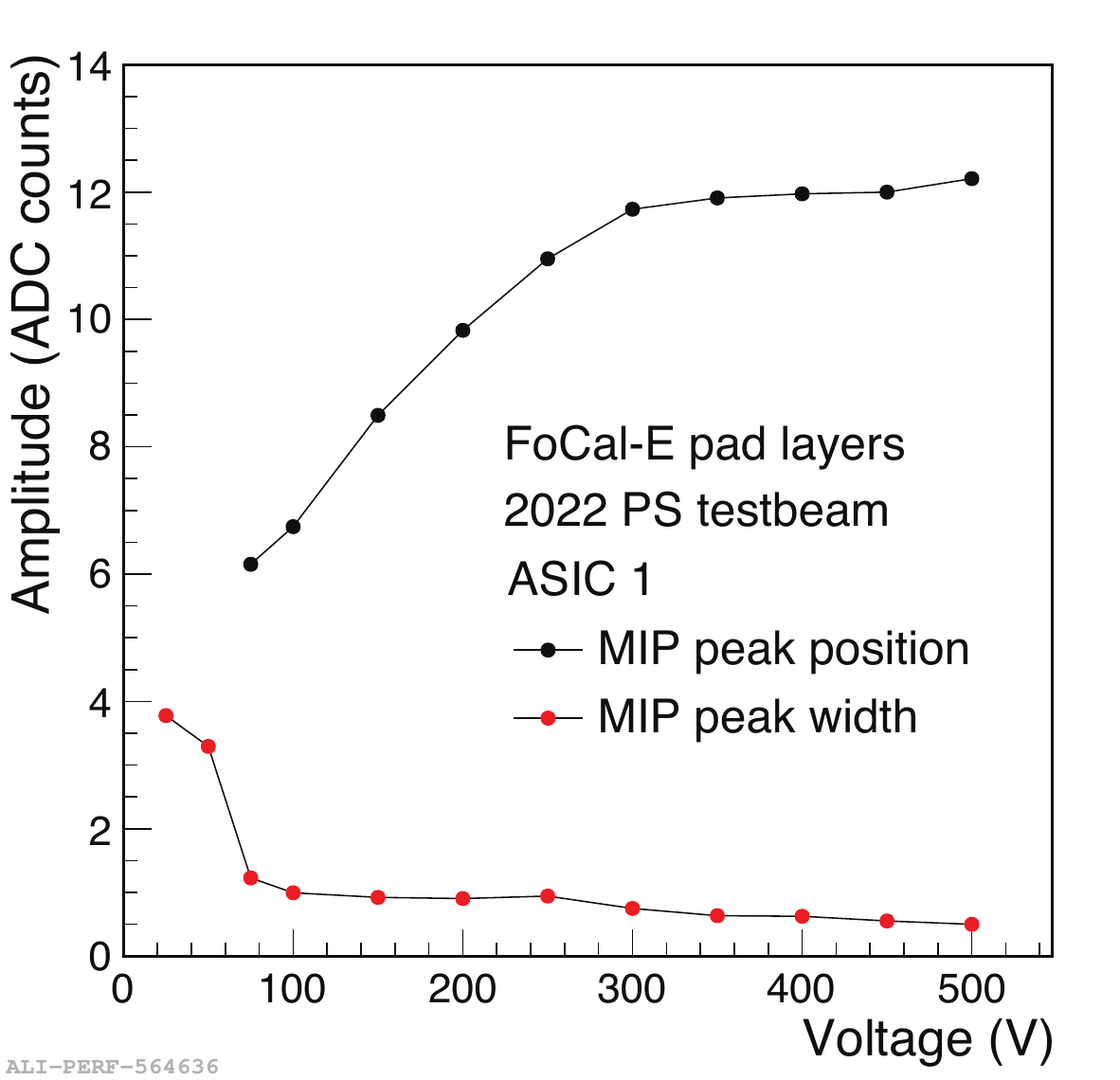} 
\caption{\ac{MIP} position and width as a function of the bias voltage applied to the silicon sensors for a specific seed channel on the second pad layer~(\ac{ASIC} 1). 
The data points were extracted from \acs{langaus} fits of the individual \ac{ADC} distributions obtained at the June, 2022 \ac{PS} test beam.}
\label{fig:voltage_scan_mip}
\end{center}
\end{figure}

The most probable value and the width of the \ac{MIP} distributions were studied as a function of the reverse bias voltage applied to the sensors~\cite{rytkonen2023simulated}. 
As can be seen in \Fig{fig:voltage_scan_mip}, the \ac{MIP} position~(given in \ac{ADC}) increases with increasing bias voltage, reaching a plateau for bias voltages larger than about 300\,V. 
For all of the reported results in the following, the \ac{FoCal-E} pad sensors were depleted with an applied bias of 350\,V. 

Systematically measuring the \ac{MIP} response provides a reliable feedback for monitoring the stability of the \ac{FoCal-E} prototype detector under varying conditions. 
In particular, the \ac{MIP} studies were repeated with data collected at the \ac{SPS} beam lines in November, 2022 and May, 2023 to investigate the detector behavior at higher energies. 
Initially, the pedestal and \ac{CMN} removal procedures described above were performed.
Typically, the width of the \ac{CMN} distribution was about 4.5~\ac{ADC} counts, and the width of the pedestal after \ac{CMN} subtraction about 1.5~\ac{ADC} counts. 
The \ac{MIP} position varied by less than 3\% in runs with a duration of up to 5h.

\begin{figure}[thb!]
\begin{center}
\includegraphics[width=0.49\textwidth]{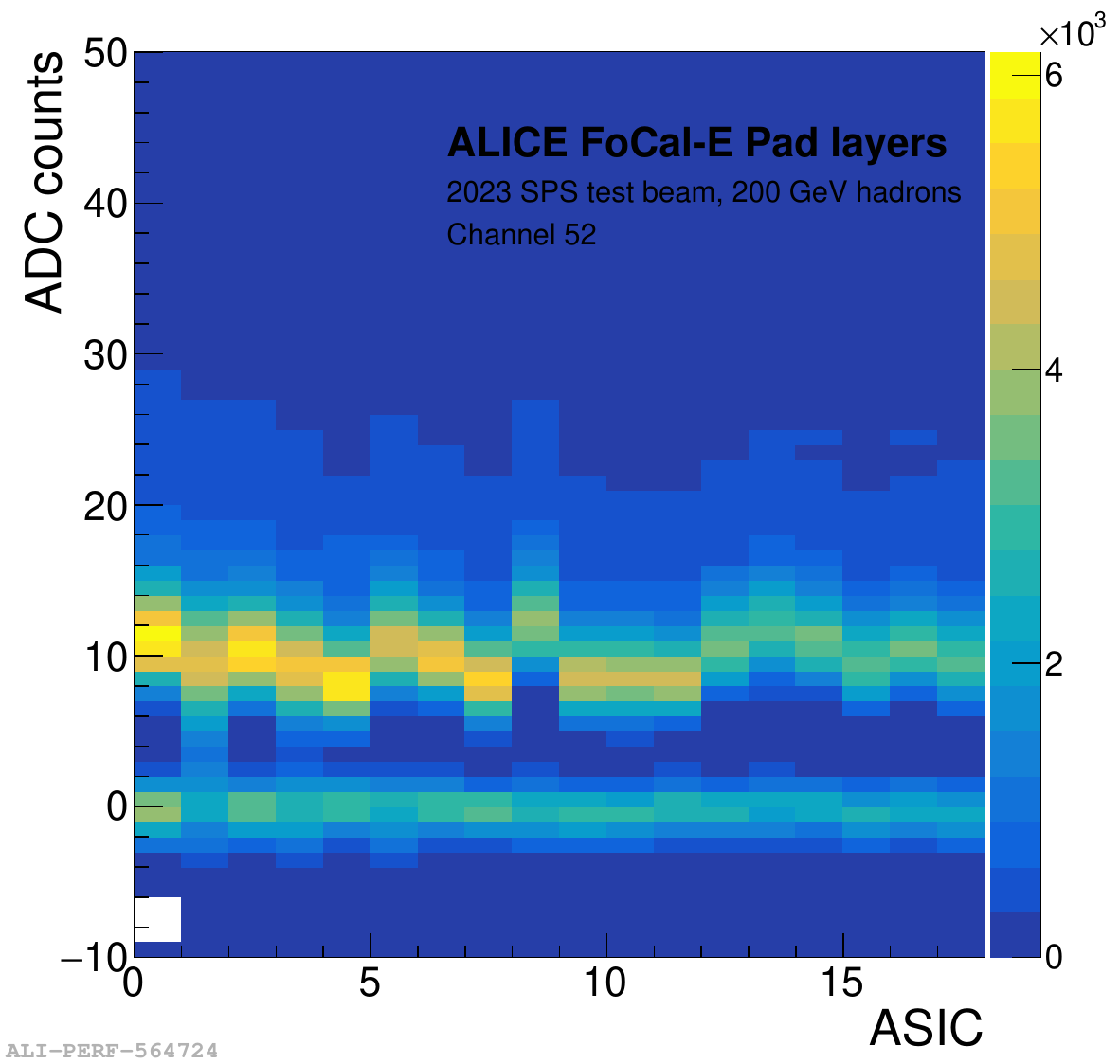} 
\includegraphics[width=0.49\textwidth]{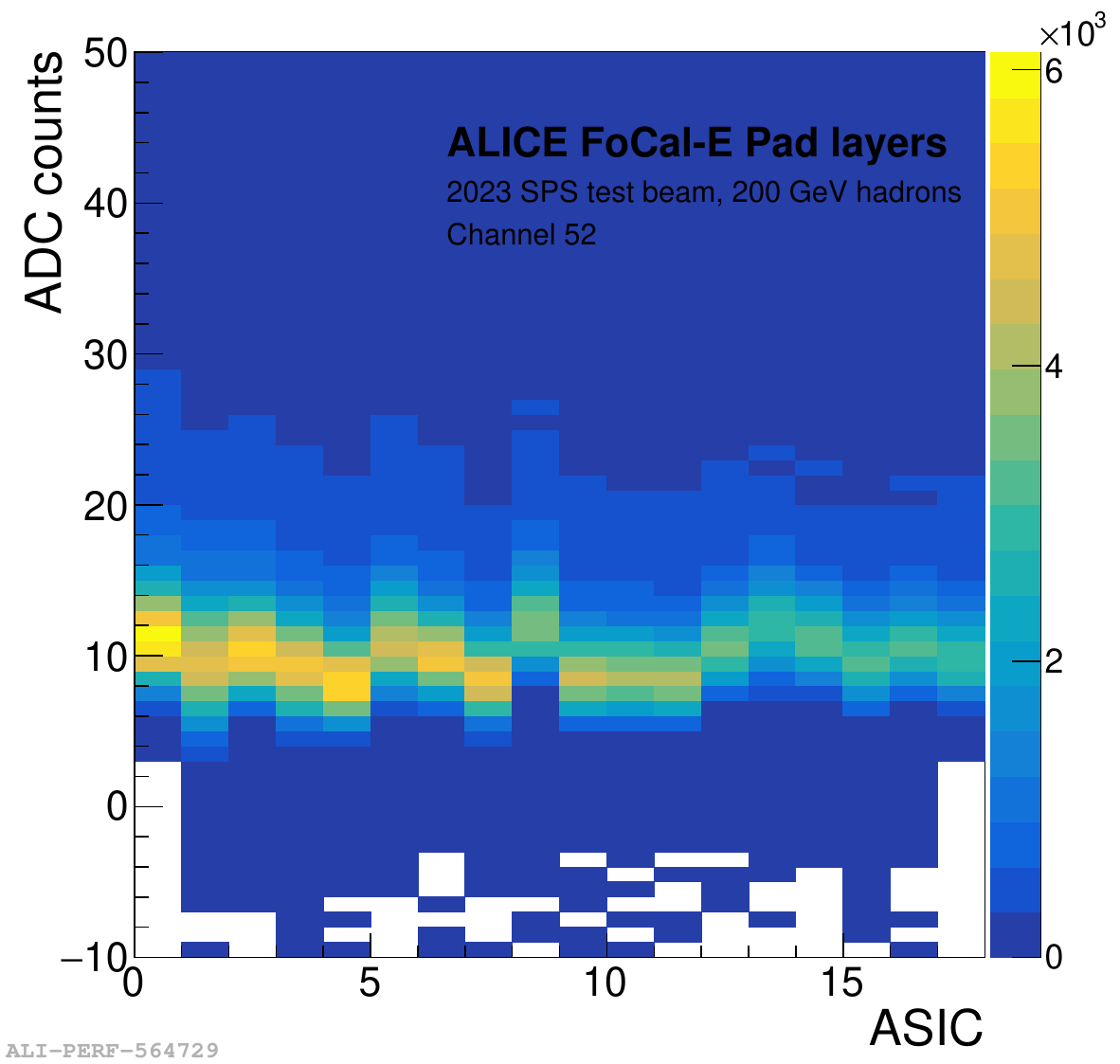} 
\caption{
Left panel: \ac{ADC} distributions of the 18 FoCal-E Pad layers for a given channel~(52) for 200~GeV hadrons at the May, 2023 \ac{SPS} test beam.
Right panel: Same distributions after requiring the presence of a signal in the first and last layer as explained in the text.
}
\label{fig:mip-coincidence-cut}
\end{center}
\end{figure}

\begin{figure}[thb!]
\begin{center}
\includegraphics[width=0.49\textwidth]{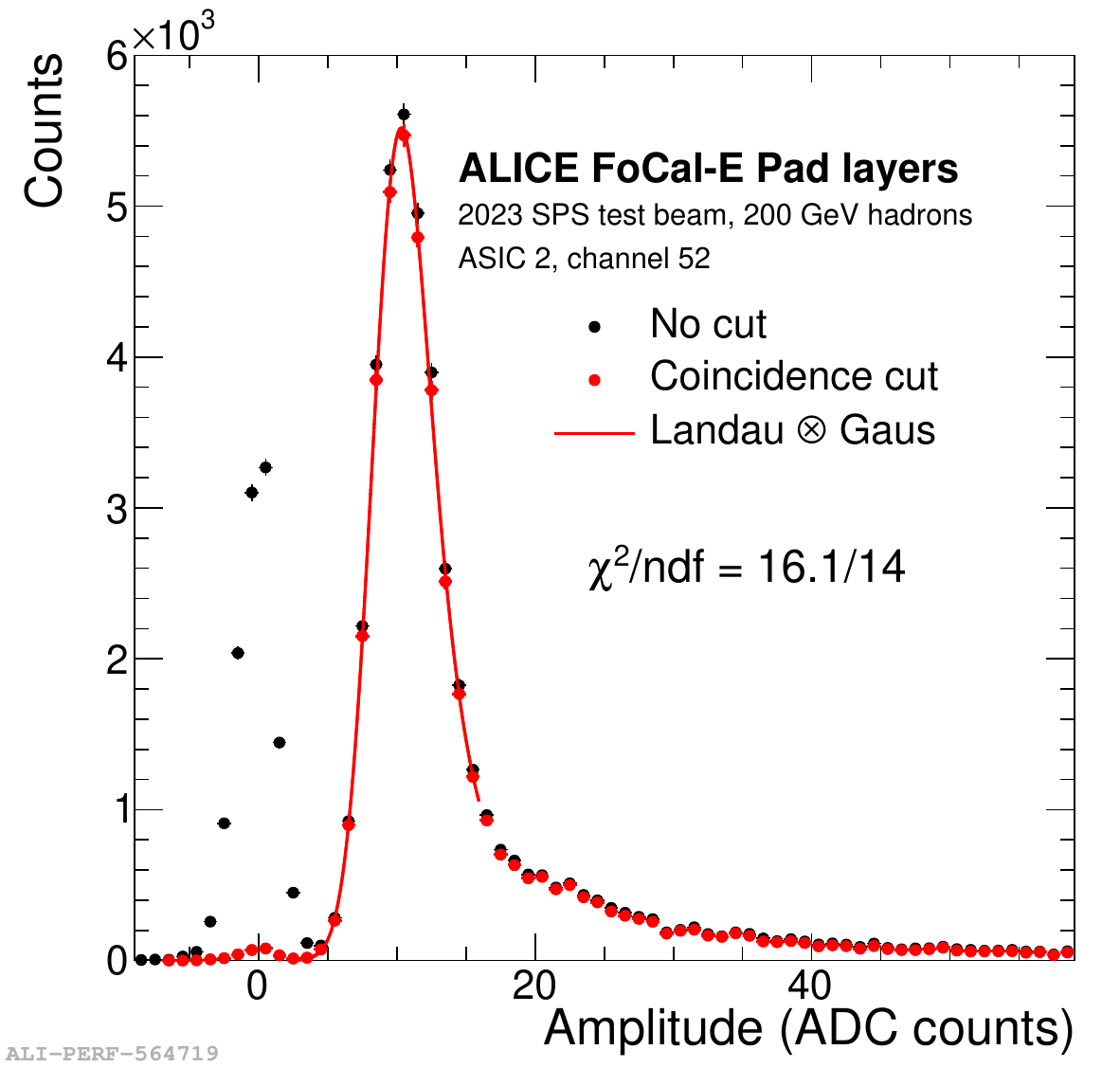} 
\includegraphics[width=0.49\textwidth]{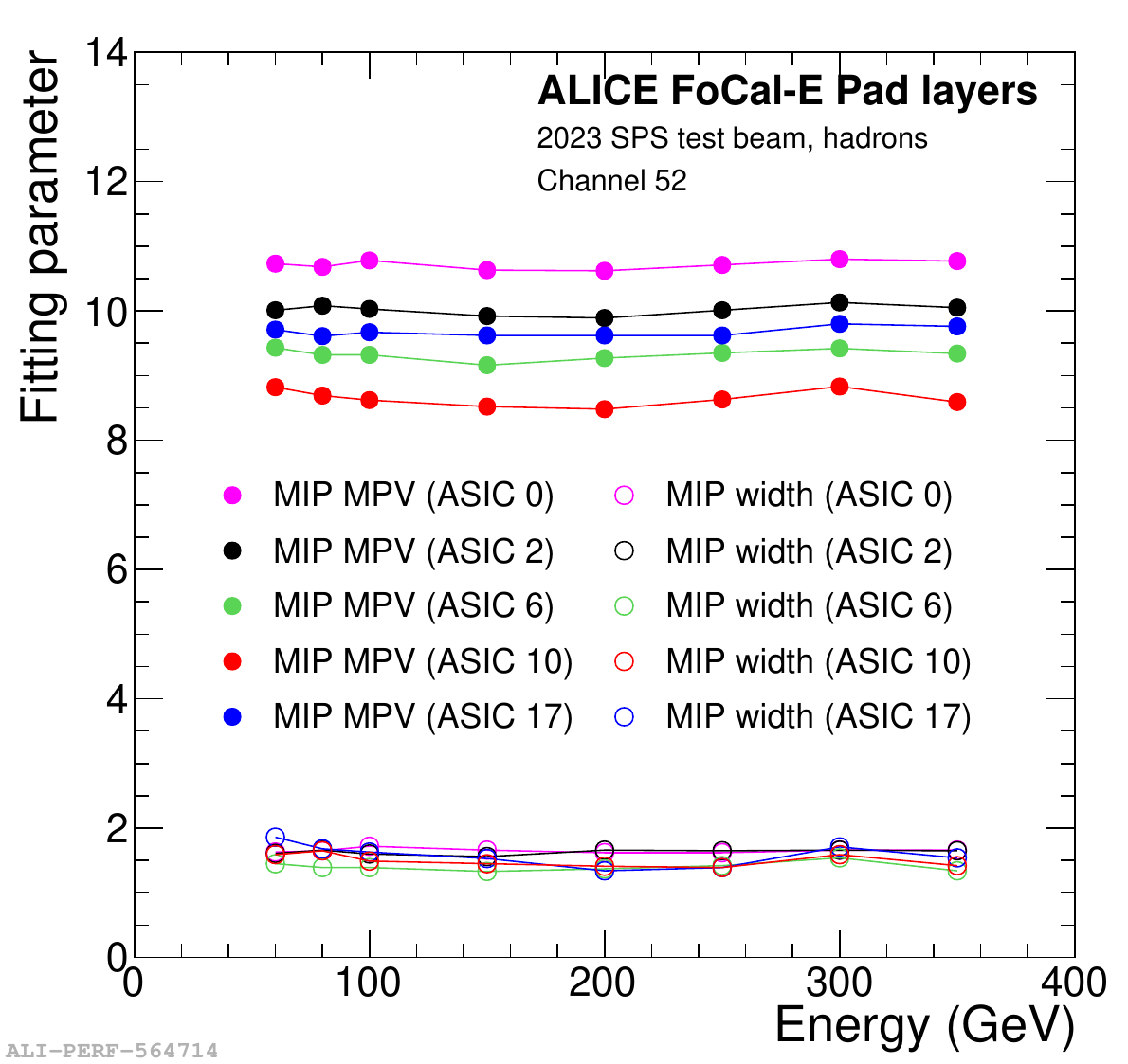} 
\caption{
Left panel: \ac{ADC} distribution for a specific seed channel~(52) on the 3rd layer~(\ac{ASIC} 2) for a 200~GeV hadron beam.
Black markers are before and red markers~(fitted with a \acs{langaus} distribution) after the coincidence cut~(on first and last layer as described in the text). 
Right panel: \ac{MIP} position and width for hadrons between 20 and 350\GeV\, for a specific seed channel~(52) for a few pad layers~(\ac{ASIC} 0, \ac{ASIC} 2 \ac{ASIC} 6, \ac{ASIC}10, \ac{ASIC} 17) after applying the coincidence cut.
Data are from the May, 2023 \ac{SPS} test beam.
}
\label{fig:mip-pos-coincidence-cut}
\end{center}
\end{figure}

The left panel of \Fig{fig:mip-coincidence-cut} shows the measured \ac{ADC} distributions of a specific channel for every pad layer~(\ac{ASIC}) obtained for 200~GeV hadrons at the May, 2023 \ac{SPS} test beam.
Contribution of noise~(seen around zero \ac{ADC} value) can also originate from the response to particles which arrived not well-aligned with the \ac{HGCROC} phase.
To suppress contributions of such events the following additional selection was applied.
Provided the position of a seed channel, the pedestals of the first and last layer~(\ac{ASIC}~0 and 17) were fitted by Gaussian functions. 
The \ac{ADC} values on both layers are then requested to exceed the average pedestal magnitude by 3$\sigma$. 
The effect of this  coincidence cut is shown in the right panel of the \Fig{fig:mip-coincidence-cut}, essentially removing the entries around an \ac{ADC} signal value of 0.
This is also illustrated in the left panel of \Fig{fig:mip-pos-coincidence-cut},
which shows the projection of the \ac{ADC} distribution for \ac{ASIC}~2.
The pedestal around zero is essentially removed by requiring the coincidence cut.
The resulting distribution is fit with a \acs{langaus} distribution, obtaining an excellent description of the data.
The right panel of \Fig{fig:mip-pos-coincidence-cut} shows the \ac{MIP} position and width from the fit of \acs{langaus} distribution obtained after the coincidence cut as a function of hadron beam energy for numerous pad layers~(\acp{ASIC}), demonstrating very little variation over the range from 20 to 350~GeV.
The \ac{MIP} positions are distributed around 10~\ac{ADC} counts with a variation of about 10\%, corresponding to 1~\ac{ADC} count. 
This can be attributed to marginal differences in the \ac{HGCROC} phase settings and tolerances of the components in the detectors readout boards. 
The \ac{MIP} width, which is about factor 5--6 smaller than the \ac{MPV},  does exhibit very little variation between the different \acp{ASIC}.
The corresponding \ac{SNR}, which can be approximated by the \ac{MIP} position divided by the width of the pedestal, is $7\pm1$. 


\FloatBarrier

\subsection{Electron dataset}
\label{subssec:electrondataset}
The electron data presented in the following were recorded in two test beam measurement series at the \ac{SPS} H2 beam line, in November, 2022 and May, 2023.
In November, 2022, electron beams with momenta between 20~GeV and 300~GeV with high purity~(electron fraction $> 90\,\%$ for all energies) were available.
In May, 2023, the electron beams were only available up to momenta of 150\GeV, with a significant drop of electron purity at the higher momenta.
In November 2022, both the \ac{FoCal-E} pixel and the \ac{FoCal-E} pad layers were suffering from few minor technical problems, leading to loss in acceptance. 
However, the detector was operational, with successful data acquisition.
For the pixel layers, the top two horizontal rows of \ac{ALPIDE} chips did not take data, since no control and data taking could be established to one \ac{HIC} board via the \ac{DAQ} system.
This lead to an acceptance loss of 2/3 in layer~5, while layer~10 was fully operational.
For the pad layers, the \ac{ToT} channels of one half of layer 7 were found to provide an unstable response during data taking and calibration.
It was excluded from all November 2022 analyses, in order to have a clean data sample.
The issues were resolved before the May 2023 test beam.
For May, 2023, pad layer 7 was replaced with a fully working spare layer, providing a fully functional \ac{FoCal-E} stack.
The full 18 pad layer performance was thus analyzed with the data from May, 2023 up to electron momenta of 150~GeV.
In order to study the detector performance over the full energy range, the two data-sets were combined, using energies from 20 to 100~GeV from May, 2023 and energies from 150 to 300~GeV from November, 2022.
Here, the layer 7 was excluded in the analyses for both datasets.
Since high-energetic electron beams significantly loose energy by deflections in the beam optics, the real beam energies deviate from the nominal beam energies.
We used the estimates and uncertainties for the real energies from~\cite{Akchurin:2798347}.
In particular, a nominal electron energy of 300~GeV translates to a real energy of 287~GeV, 250~GeV to 243~GeV, 200~GeV to 197~GeV, and 150~GeV to 149~GeV, and the uncertainties on the energies are 0.2--0.3\%.

\begin{figure}[ht!]
\begin{center}
\includegraphics[width=0.6\textwidth]{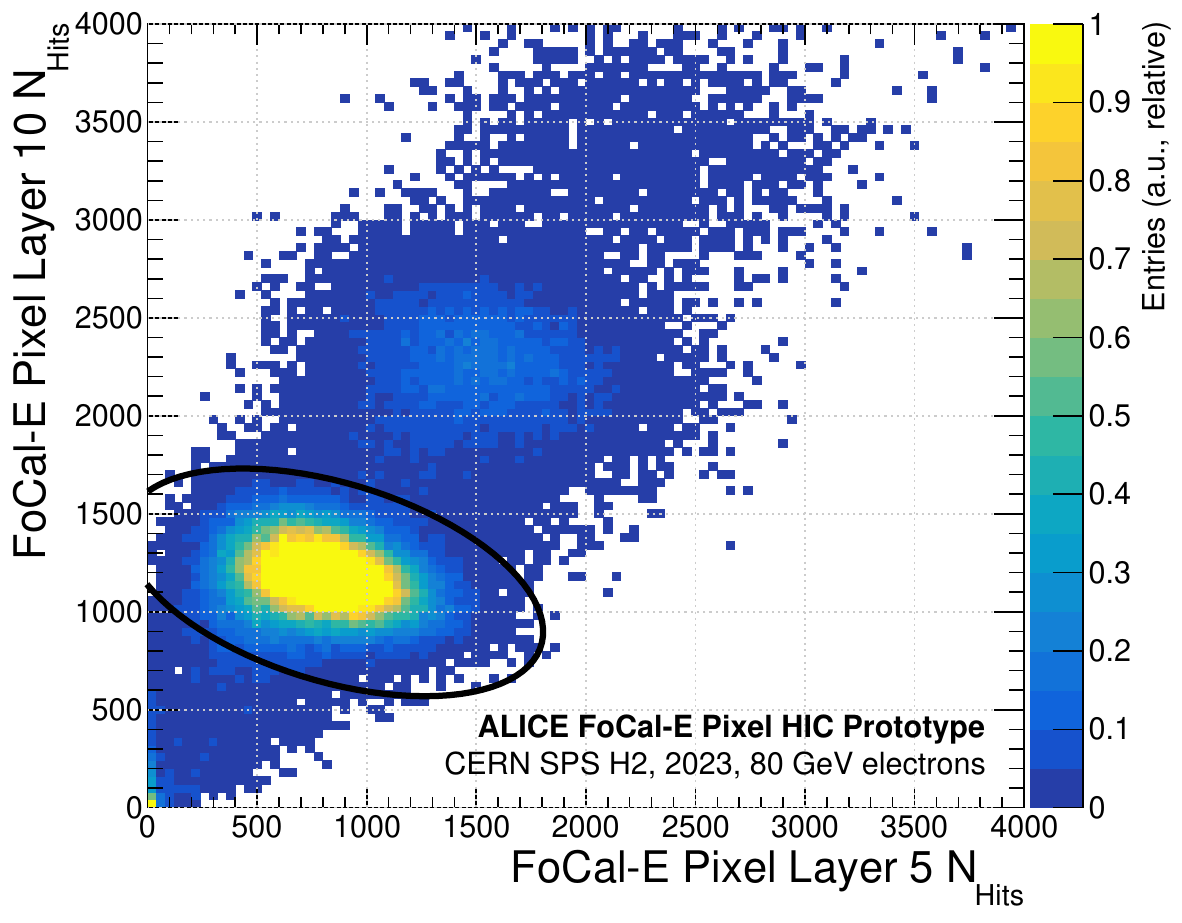} 
\caption{Two-dimensional representation of the number of hits in pixel layer 5 and 10 from the beam test in May, 2023. Single electron events are selected within an ellipse drawn in black.  
}
\label{fig:pixel-event-selection}
\end{center}
\end{figure}

\subsection{One-electron event selection with the pixel layers}
\label{subsec:pixel-event-selection}
While for November, 2022 runs, the data were hardly affected by hadronic contamination, an electron candidate selection was necessary for the May 2023 data.
Because of the relatively long pulse length of the \ac{ALPIDE} pixel front-end~($\approx 5\us$) and the high beam rates, multiple particle events occasionally accumulated in the pixel layers producing pile-up.
We therefore select one-electron events with the pixel layers, and use this selection later also for one-electron candidates for the pad-layer data analysis, since the pad front-end has a significantly shorter integration time of about $25\ns$.
\Figure{fig:pixel-event-selection} shows the correlation of the number of pixel hits~($\Nhits$) in layer~5 and layer~10 for an 80~GeV run from May, 2023.
The two-dimensional representation resolves the various event signatures of one-hadron, one-electron, and multiple-electron events.
Since hadrons traverse \ac{FoCal-E} either as \acp{MIP}, which do not shower, or start showering in the first layers, pure hadronic events are characterized by low values of \Nhits. In the \acs{MIP} case only one pixel cluster with an average size of 3 to 4 occurs, and in the hadron shower case the number of hits roughly follows a falling exponential distribution, up to a few hundreds of hits, well below the number of hits in an electromagnetic shower of the same energy, e.g.\;mostly below 500 hits per layer for a beam energy of 80\,\GeV{ }as shown in \Fig{fig:pixel-event-selection}.
At higher values of \Nhits, electron shower events appear, well separated from the hadronic events.
The regimes of one-electron, two-electron, and three-electron events can be clearly identified.
For the data analysis, we require a signature of exactly one electron in the pixel layers.
In order to optimize the selection of one-electron events, we employed a two-dimensional cut approach based on the correlation of the number of hits in the two pixel layers.
An electron candidate is selected if the number of pixel hits $\{\Nhitsfive, \Nhitsten\}$ lies inside an ellipse~(depicted by the black curve in \Fig{fig:pixel-event-selection}), which we define individually for each energy.
It thus employs the pixel layers to provide a simple and immediate identification of one-electron event candidates for each beam energy.
However, a possible residual contamination by showering hadrons cannot be removed with this method, and remain as a potential background.

\subsection{Pixel-layer hit response to electrons}
\label{subsec:pixel-layer-hits}
\begin{figure}[th!]
\begin{center}
\includegraphics[width=0.32\textwidth]{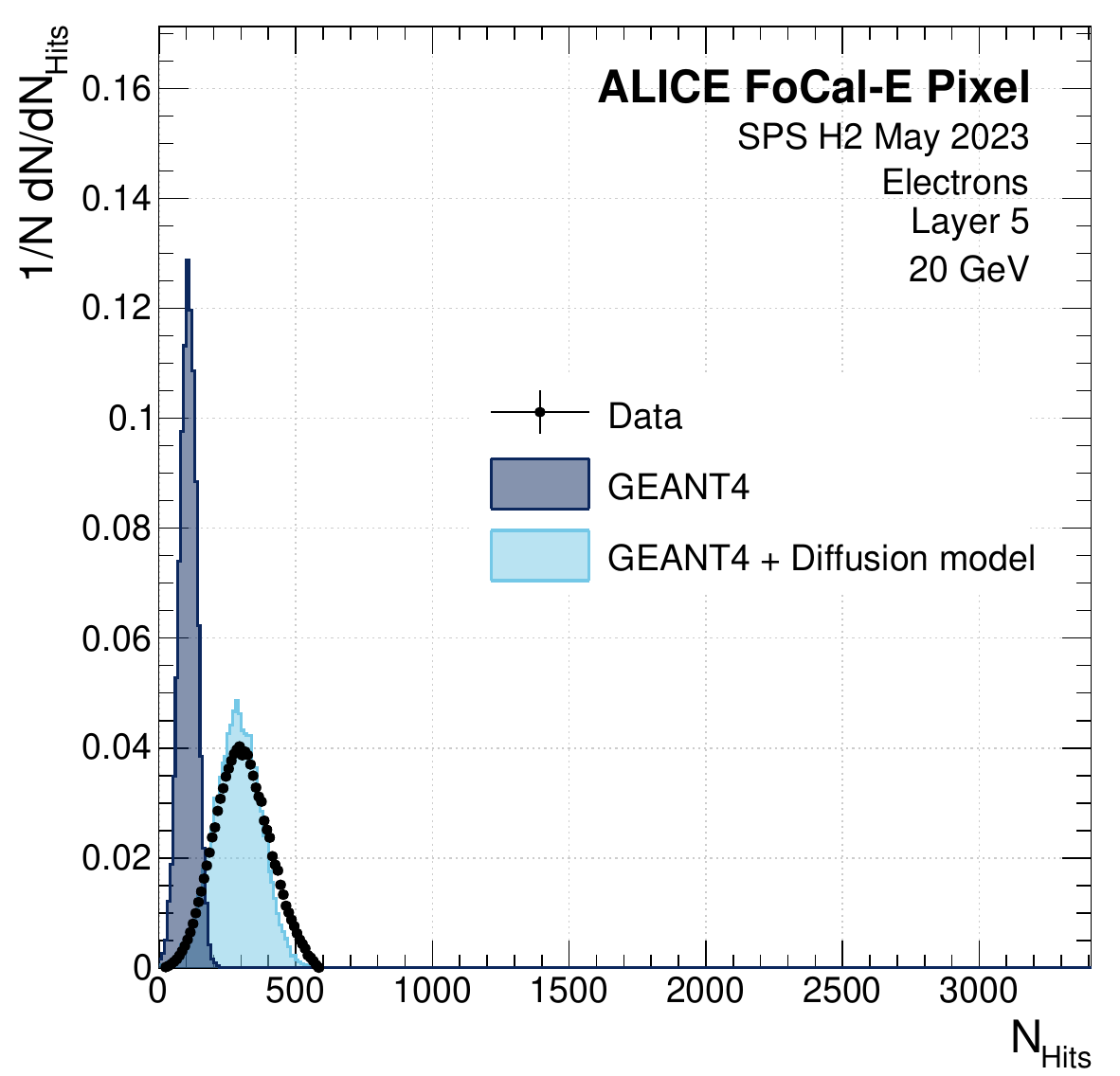}
\includegraphics[width=0.32\textwidth]{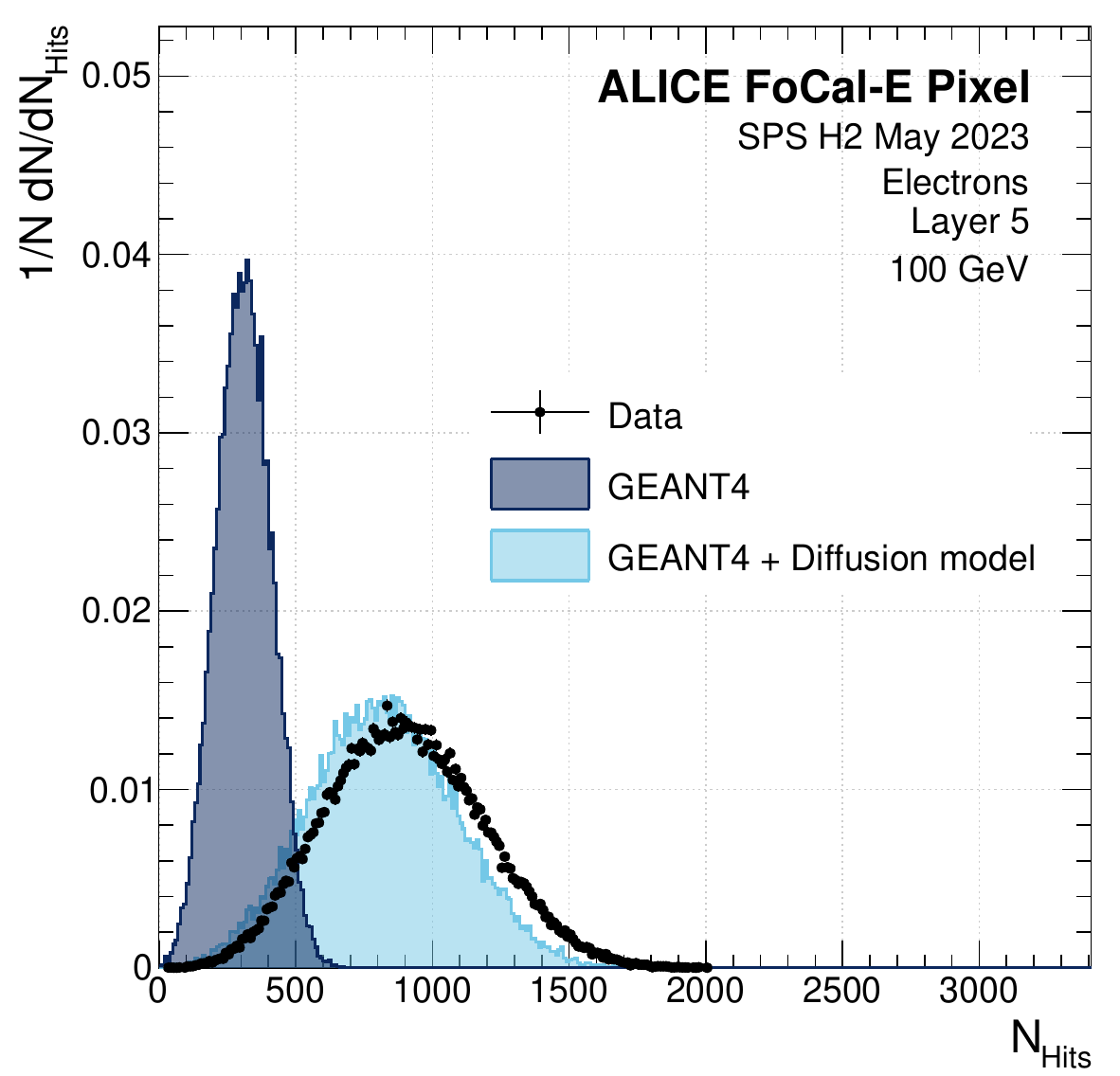}
\includegraphics[width=0.32\textwidth]{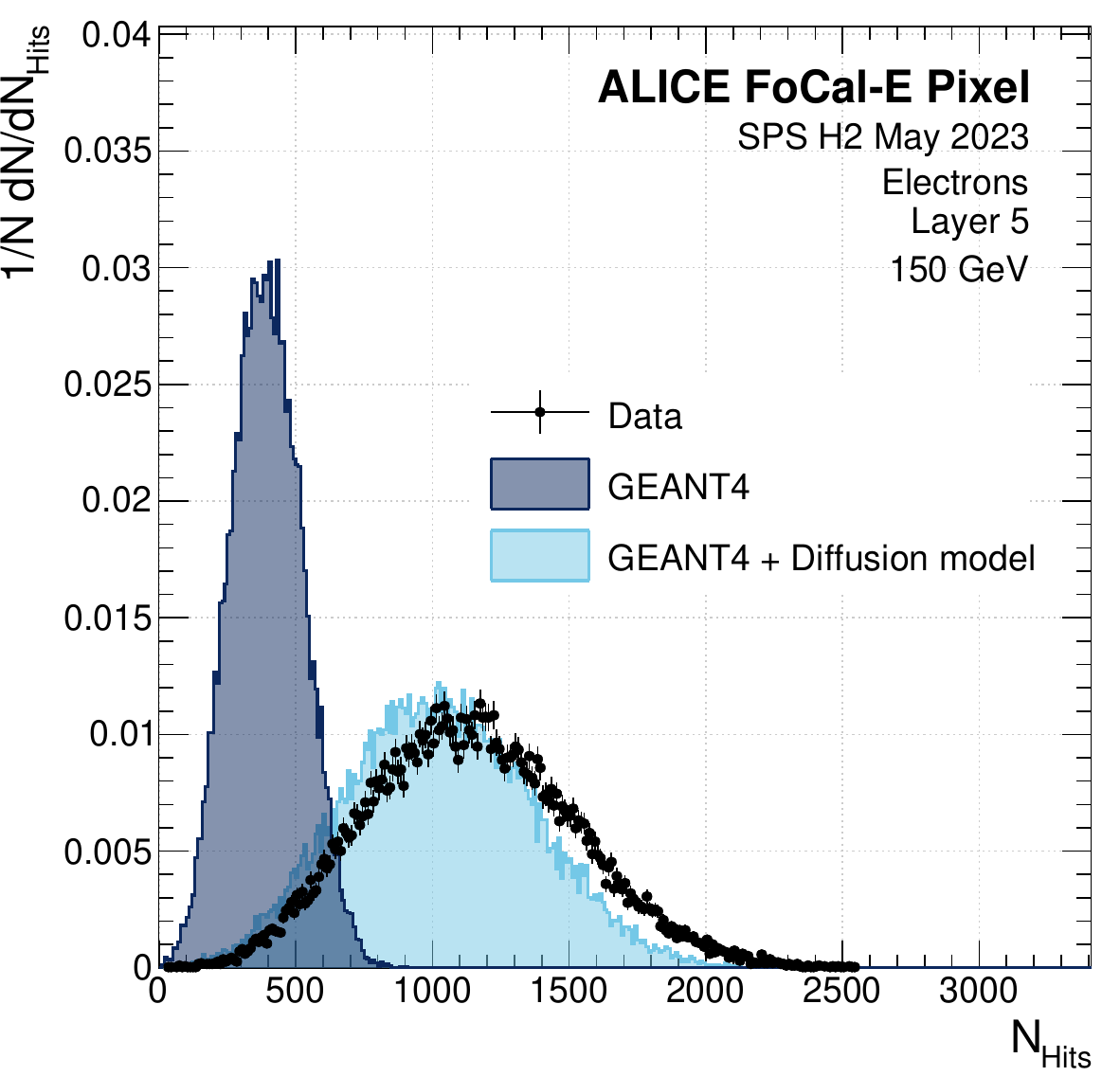}
\includegraphics[width=0.32\textwidth]{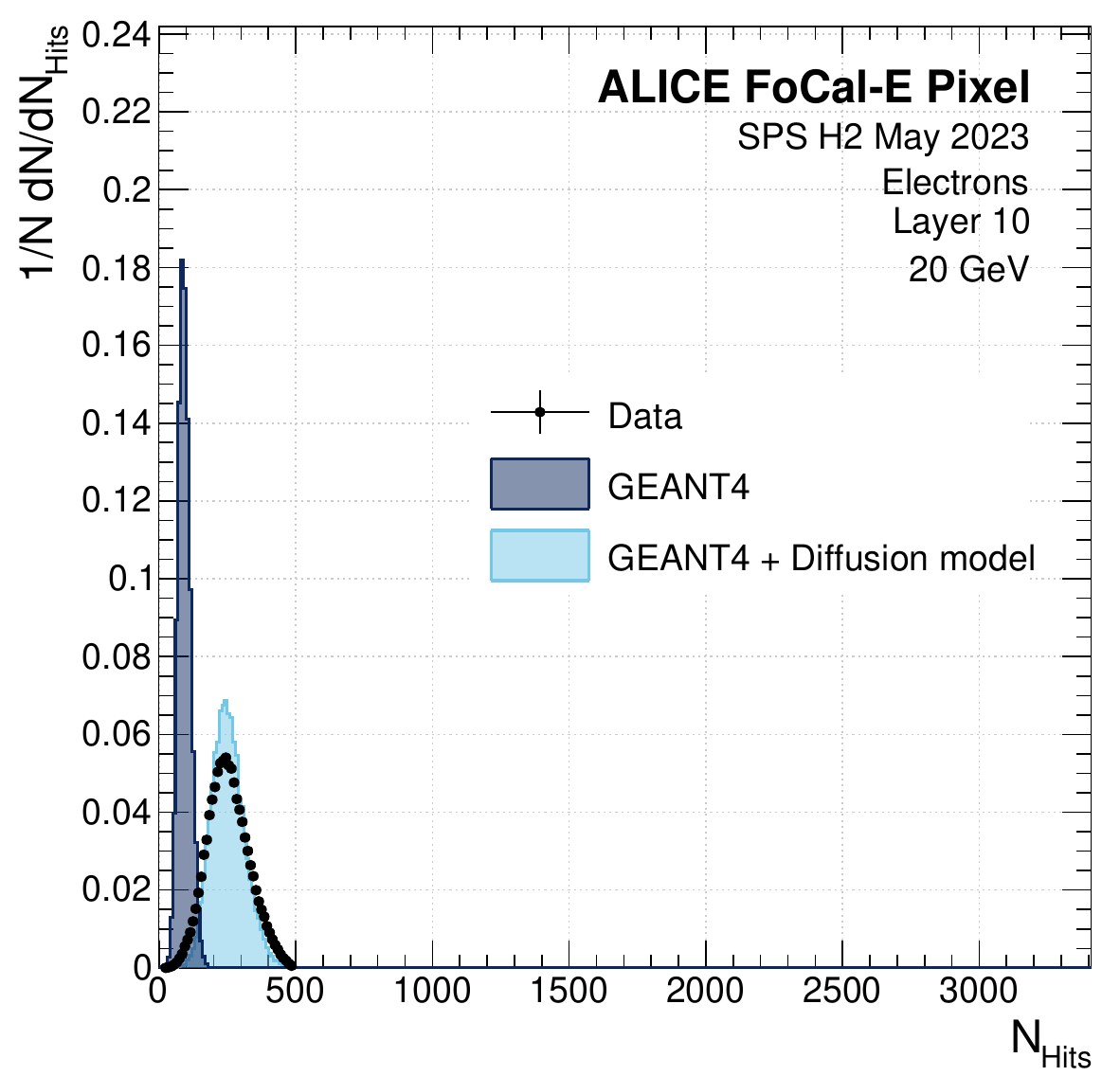}
\includegraphics[width=0.32\textwidth]{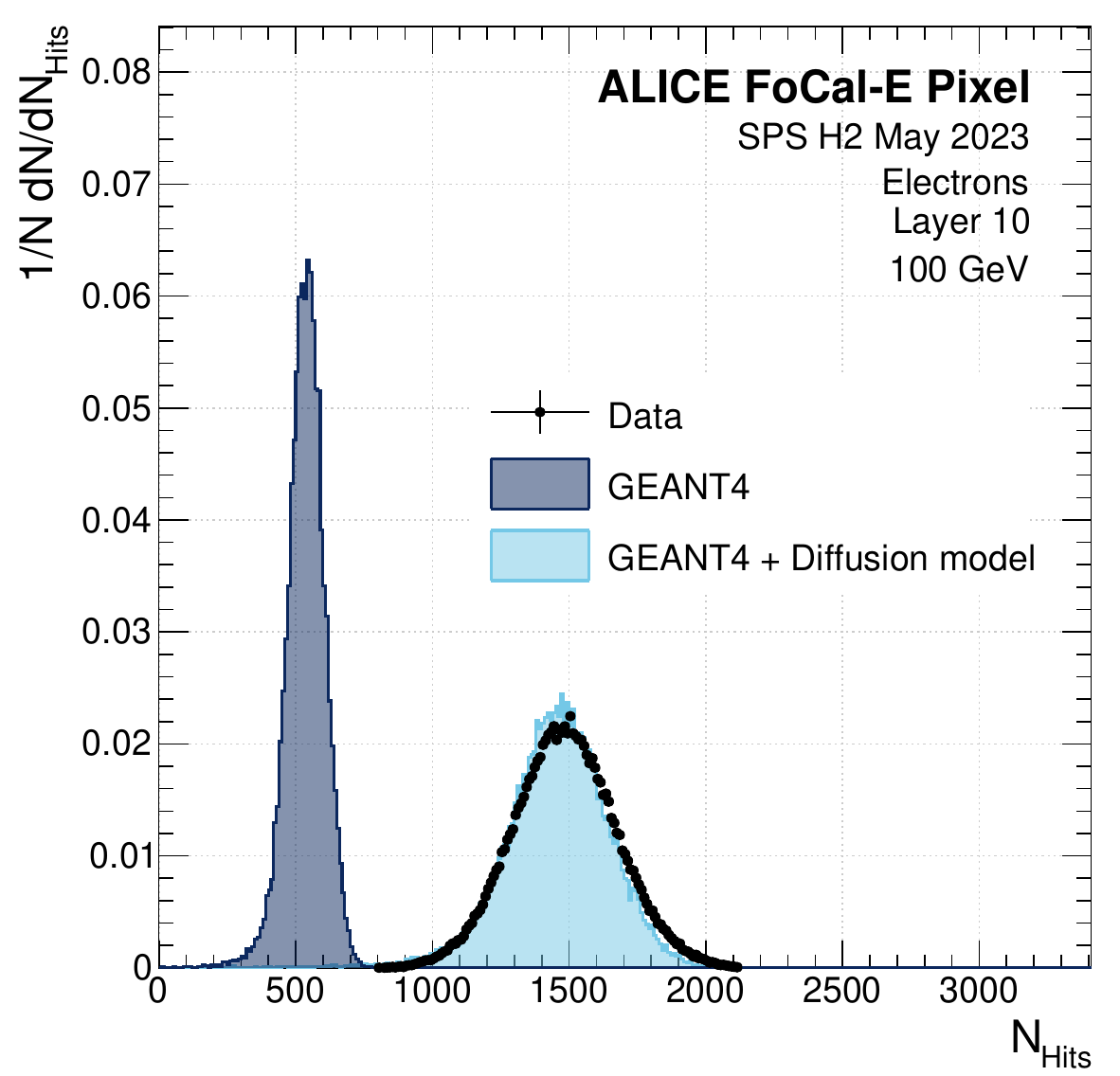}
\includegraphics[width=0.32\textwidth]{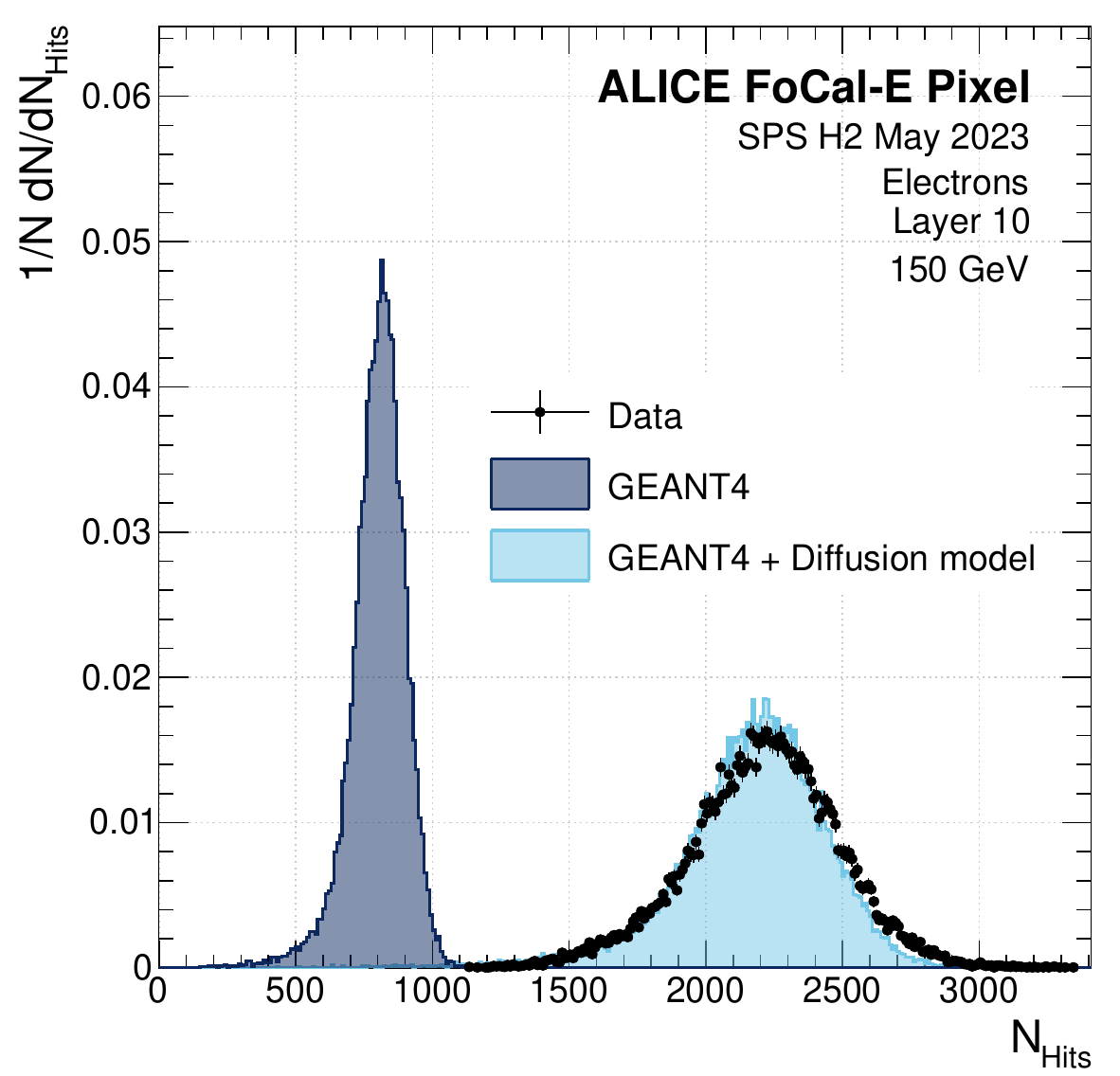}
\caption{\label{fig:nHitsDiffusionL5} Number of hits distributions for data, \geant simulations and \geant simulations with an applied Gaussian diffusion model as a post-processing step for layer~5~(top panels) and layer~10~(bottom panels) for electron beam of 20, 100 and 150~GeV.}
\end{center}
\end{figure}

\Figure{fig:nHitsDiffusionL5} shows the number of pixel hits in data and simulations for pixel layers 5 and 10, respectively, at energies of 20\GeV, 100\GeV, and 150\GeV, after the one-electron candidate selection in the data.
In the current \geant simulation setup, the charge sharing between neighboring pixels is not implemented, and the entire energy deposition process is modeled to occur within a single pixel. 
The \geant simulations will, thus, not generate extended clusters of neighboring pixels since charge diffusion in the epitaxial layer is not simulated. 
This results in a discrepancy between the simulations (dark blue histogram) and the measured data (black points). 

However, creating an accurate description of this process can be extensive and time-consuming.
It requires detailed knowledge of the electric field in the epitaxial layer and around the charge collection node, which is non-disclosure material.
Instead, a simpler model~\cite{pettersen_digital_2018} was implemented as a post-processing step to the \geant simulation, in order to reproduce the experimental pixel hit distributions.
Additional pixels are generated using an empirical model of charge diffusion in the sensor. 
The simulated charge deposited in the sensor is translated into the spatial width of the electron diffusion by using an empiric parametrization.
This width is then used to derive the pixel cluster size from a two-dimensional Gaussian probability density function.

For each simulated pixel hit, a Gaussian distribution is generated in $(x,y)$ surrounding the hit, where the standard deviation is given as $\sigma_x = \sigma_y = \alpha E_\text{dep}^\beta$.
The Gaussian is randomly sampled $N$ times, resulting in $n$ number of unique pixel hits activated around the originally simulated hit~\cite{pettersen_digital_2018, PETTERSEN201751}.
By varying the parameters and comparing the outcome of the model to experimental data, the following parameters were obtained: $\alpha = 0.17 \ \mu\text{m}/(\frac{\text{keV}}{\mu \text{m}})^\beta$, $\beta = 0.41$ and $N= E_\text{dep} \cdot 11 \  \mu\text{m/keV}$, where the energy deposition, $E_\text{dep}$, is given in units of keV/$\mathrm{\mu m}$.
The model is applied to the simulated pixel hits with charge information, which are the output of the \geant model.
The implementation of this post-processing~(light blue histogram) step significantly improved the agreement between data and simulation, as demonstrated in \Fig{fig:nHitsDiffusionL5}.
The data and the resulting \acs{MC} distributions of \Nhits were fitted with a Gaussian distribution function, respectively. Per energy and layer, the mean (width) values of the distributions agree on average within $6\pm5\,\%$ ($9\pm10\,\%$). 

\subsection{Pad channel calibration}
\label{subsec:padchannelcalib}
Each channel of the \ac{HGCROC} is equipped with an internal charge injection circuit used for signal response calibration.
In order to do so, a known charge amount is injected into the chip through a set of  capacitors. 
One branch of the charge injection circuit uses an 0.5\pF\ capacitor for \textit{low-range} injection with charges between 0 and 0.5\,\pC.  
A second branch uses  an 8\pF\ capacitor for \textit{high-range} injection between 0 and 8\,\pC. 
The voltage on the capacitors can be set through an 11-bit DAC-register to a fraction of the chip internal reference voltage, which is 1\,V.

We use the low-range charge injection for the calibration of the \ac{ADC}, and the high-range charge injection for the calibration of the \ac{ToT}.
The calibration measurement is performed by step-wise increment of the voltage at the capacitor, using steps of 5\,mV, and releasing 2000 charge injections at each voltage step. The software records the response of \ac{ADC}, \ac{ToT}, and \ac{ToA} per injection.

For low-range injections, the dependence of the \ac{ADC} on the injection charge is well described by a linear function.
We determine the proportionality factor of the \ac{ADC} with respect to the signal charge from a linear fit whereas the channel-wise pedestal is obtained from dedicated pedestal runs.
For high-range injections above $\approx 2\pC$, we observe a linear response of the \ac{ToT}, and use a linear fit function to calibrate the \ac{ToT}.
For lower charge values (in particular in the turn-on region), the response of the \ac{ToT} does not show the same linear behavior as for the higher charges, and we derive a calibration in this range by linearly interpolating between the \ac{ToT} response at 0 and 2\pC\ (where the former one is always zero). 
The successful channel-by-channel calibrations were then used to convert the response of the \ac{HGCROC} channels to actual charge deposit in the individual sensor pads.

The \ac{ToT} and \ac{ToA} units can be translated directly into the time-over-threshold and time-of-arrival of the charge pulse where 1~\ac{ToT}~unit~=~50\ps, and 1~\ac{ToA}~unit~=~25\ps. For the chosen preamplifier settings, 1\,\ac{ADC} unit is on average equivalent to 0.28\fC.
Further details regarding the calibration procedure can be found in \Sec{appendix:pads_calibration}.

In the \geant simulation model, the energy deposition per pad (i.e.\ per channel) is recorded.
The energy signal is converted to charge by using the average ionization energy per electron-hole-pair in silicon of 3.65\,eV/e-h-pair. A charge collection efficiency of 100\,\% is assumed, and noise (e.g.\;from the sensor or the electronics) is not implemented.

\subsection{Electron selection for pad-layer analysis}
\label{subsec:evtselpads}
For the electron analysis using the \ac{FoCal-E} pad layers, both the November, 2022 and the May, 2023 data required a proper event selection. 
While for November data, events which were recorded with the correct timing with respect to the \ac{HGCROC} clocks had to be enhanced, the event selection for the May, 2023 data needed to focus on a clean electron sample without hadronic contamination.

\begin{figure}[ht!]
\begin{center}
\includegraphics[width=0.75\textwidth]{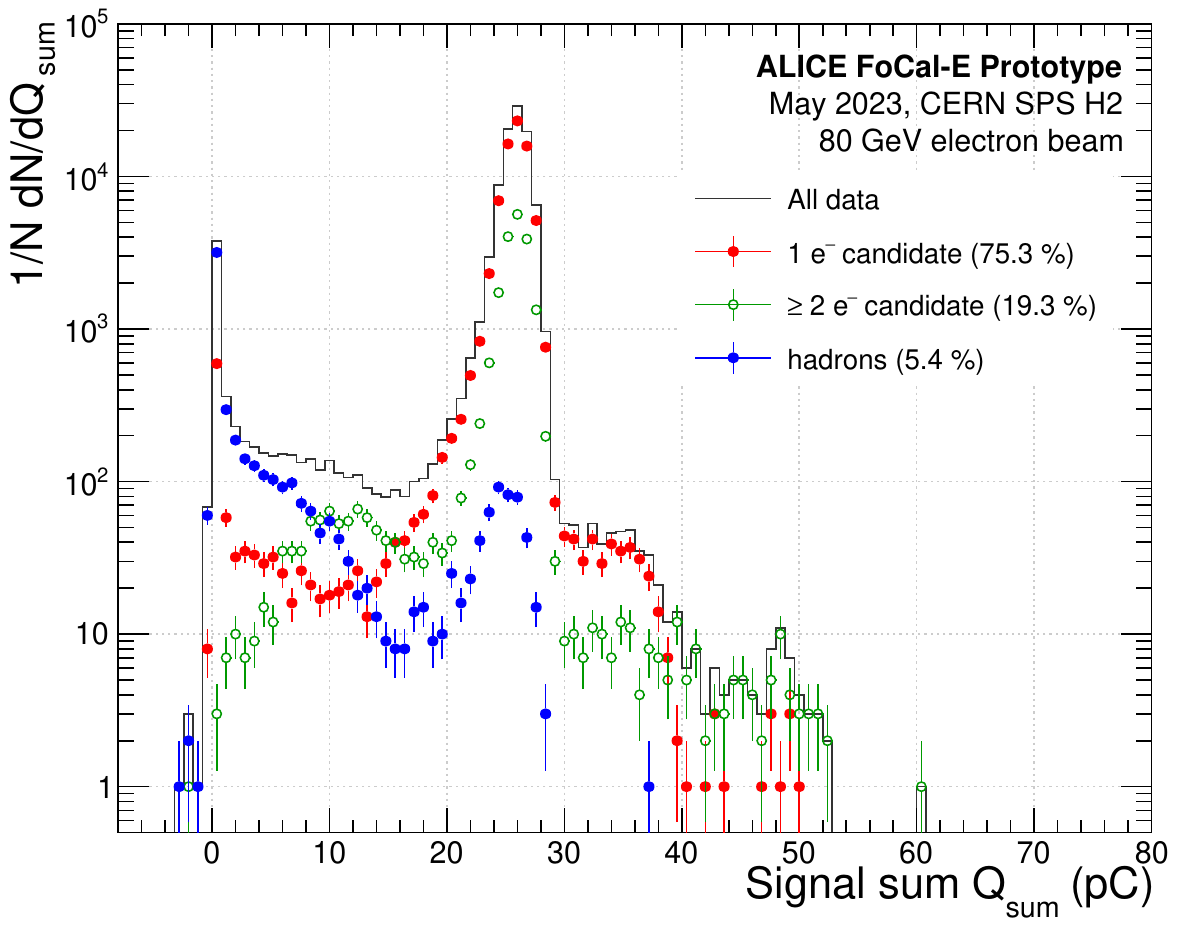} 
\caption{Charge sum~(5x5 around leading cell for all layers) of a) single-electron candidates, b) $\geq2$-electron candidates, and c) hadron candidates measured using the pad layers after applying the selection criteria defined by the ellipse cut, and the shower maximum criterion.}
\label{fig:pixel-pad-selection}
\end{center}
\end{figure}

In both cases, the datasets were pre-processed based on the \ac{HGCROC} trigger sum word, which is measured per bunch crossing~(i.e.\;every 25\ns). 
Each trigger sum word represents the signal sum of nine neighboring pad channels, in a primitive format~(with reduced precision~\cite{Thienpont:2020wau}).
For every recorded event, 8 trigger sums per layer were recorded for offline analysis within 10~bunch-crossing bins around the triggered event.
In order to reject out-of-time events, an event is selected only if no charge deposit before the signal bin was detected in any trigger sum word.
In particular, this rejected events which were recorded one bunch crossing too early.
The decay of such signal charge pulses would still be detected in the signal bin, and would hence distort the dataset.
The black histogram in \Fig{fig:pixel-pad-selection} shows the charge sum distribution of 80~GeV electrons, for all events which have passed the trigger sum requirement.
The charge sum is obtained by summing up the information around the leading cell, in a 5x5 window over all layers. 

While --- due to the high purity of the electron beams --- a specific electron candidate selection was not necessary for the November, 2022 data, hadronic events had to be rejected in the May, 2023 data.
Electron candidate events are obtained from the one-electron dataset in the pixel layers, as described in \Sec{subsec:pixel-event-selection}.
The full red markers in \Fig{fig:pixel-pad-selection} show the charge sum distribution after having applied the one-electron selection criteria.
About 75\,\% of the events are selected with this criterion, showing a prominent electron peak around 26\pC.
The selected one-electron events still show a peak from minimum ionizing particles slightly above zero.
We attribute this to events where a hadron triggered the readout and is measured in the pads, and an additional pile-up electron was measured in the pixel layers.
In order to estimate the hadronic contamination in the beam (but not for further analysis), we requested the signal sums of the central layers~(layer 7--11) to exceed 1.5~times the signal sum detected in later layers~(17--20).
This condition is sensitive to showers which have their maximum in the central layers of the stack, i.e.\ to electron showers.
The hadronic events are depicted with the blue markers in \Fig{fig:pixel-pad-selection}.
Minimum ionizing particles produce the most probable value slightly above zero. 
A small residual electron peak remains, which most likely is related to electrons which do not start showering in the first few layers of the pad layer stack, and therefore pass the selection criteria explained above.
The green open markers in \Fig{fig:pixel-pad-selection} are the remaining events.
They are associated to events where only one electron is recorded in the pad layers, but one or more pile-up electrons are visible in the pixel layers.

Since in the November, 2022 data, triggered events happened asynchronously to the clock of the readout system, an offline phase selection needed to be introduced, making use of the \ac{ToA} measurements of the \acp{HGCROC}. 
This selection was not necessary in the May, 2023 data, because the phase of accepted triggers was selected with the trigger board (described in \Sec{sec:setup}).
Thus they were aligned synchronously to the readout system clock.
The window of the trigger board was measured to have a spread of $2.5\ns$, which makes the time-of-arrival of the particle triggers uniformly distributed with a standard deviation of $5\ns / \sqrt{12} = 0.72\ns$.
In the November, 2022 data, instead, events were selected offline only if the \ac{ToA} in layer~9 was measured to be in a time interval between $1\ns$ and $1.75\ns$ with respect to the 40\MHz\ clock.
This selection condition drastically improved the quality of the data, at the cost of a significant reduction of the sample size.
Since after the \ac{ToA} selection, the events are in a certain phase, such that the sampling of the \ac{ADC} is issued at a constant point in time with respect to the arrival of the particle, the pad-layer response was improved.
We have validated the \ac{ToA}-based phase selection with the $150\,\GeV$ electron run from May, 2023.
Detailed studies confirmed that at 150~GeV the \ac{ToA} criterion was not affected by time-walk effects.
It can therefore be used as an intrinsic time measurement of the particle's phase of arrival with respect to the clock.
We observed that for the lower energies, the shower signal in the leading cell in layer 9 can fluctuate so low that the \ac{ToA} measurement is not fired (i.e.\ $\lesssim 0.2\pC$).
Applying the \ac{ToA} criterion to these energies would therefore enhance the event selection towards high intensity showers, and introduce a bias to the data.
As a result, \ac{FoCal-E} pad data from November, 2022 at electron energies below 150~GeV are not used for analysis, but replaced with the complementary May, 2023 data, resulting in a combined dataset which covers the full energy range from 20 to 300~GeV.

\subsection{Longitudinal shower profiles of electrons}
\label{subsec:shower_profiles}
The longitudinal segmentation of the \ac{FoCal-E} prototype makes the measurement of the longitudinal shower profile of \acs{EM} showers with a granularity of $\approx 1 X_0$ possible.
Here, and in the following this is done by measuring the signal charge per pad layer by summing up the information around the leading cell in a $5 \times 5$ window~(i.e.\ in a $5\cm \times 5\cm$ square), relating to the charge by using the calibration described in \Sec{subsec:padchannelcalib}.

In the following, we present the longitudinal shower profile measured with the \ac{FoCal-E} layers for nine different electron energies~(20 to 300~GeV).
In order to mitigate the effect of transverse shower leakage, only central shower events were used for analysis.
The position of the shower center in transverse direction is defined event-by-event from the channel with highest signal in layer 9.
Since the beam spot fluctuated on a $\cm$-scale between the various beam energies, three central pads were globally determined which were reasonably close to the beam center for all beams, thus providing sufficient statistics for all analyzed runs.
Events are rejected if the shower center is not located in one of the three central channels.

\begin{figure}[H]
\begin{center}
\includegraphics[width=0.28\textwidth]{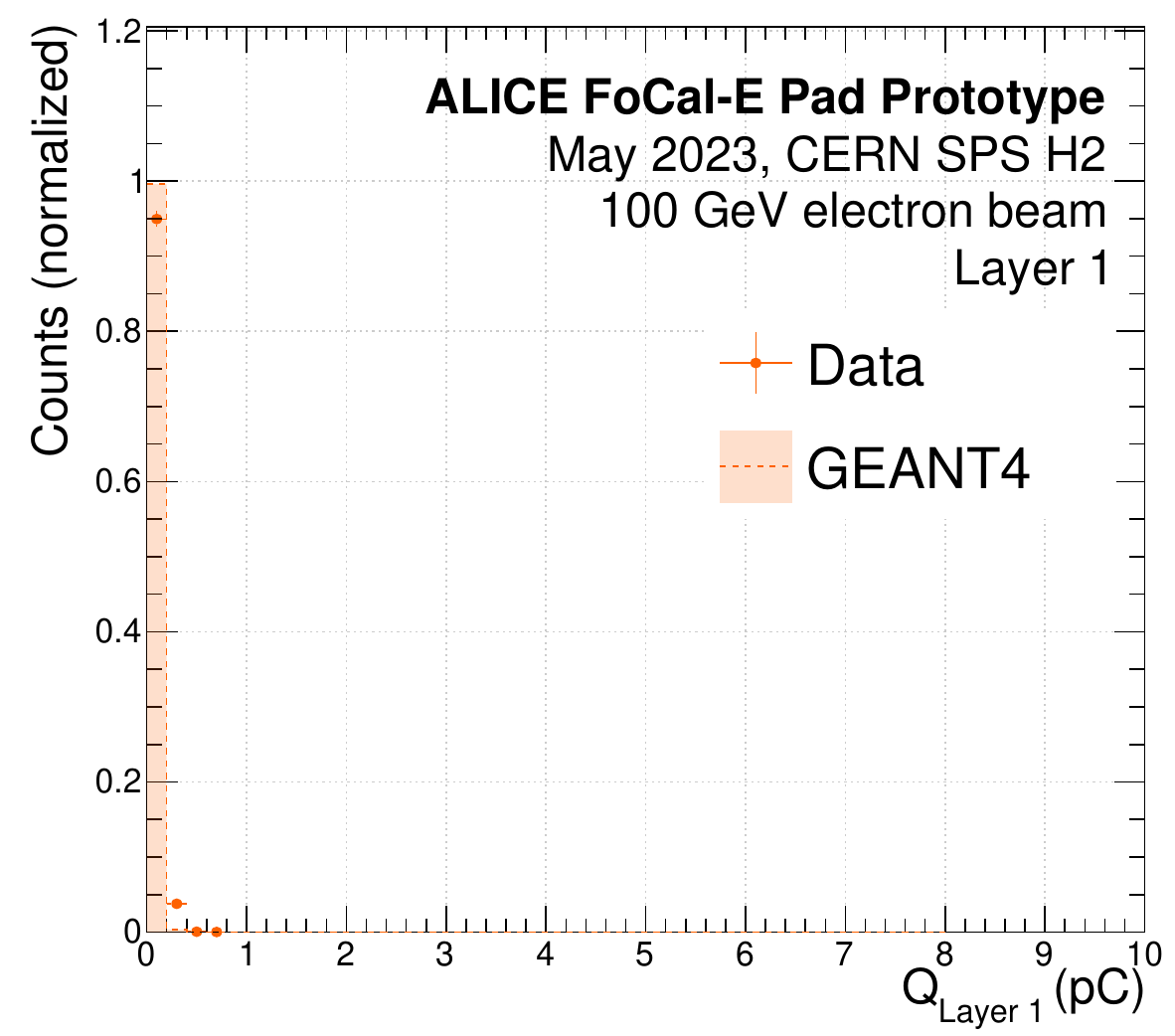}
\includegraphics[width=0.28\textwidth]{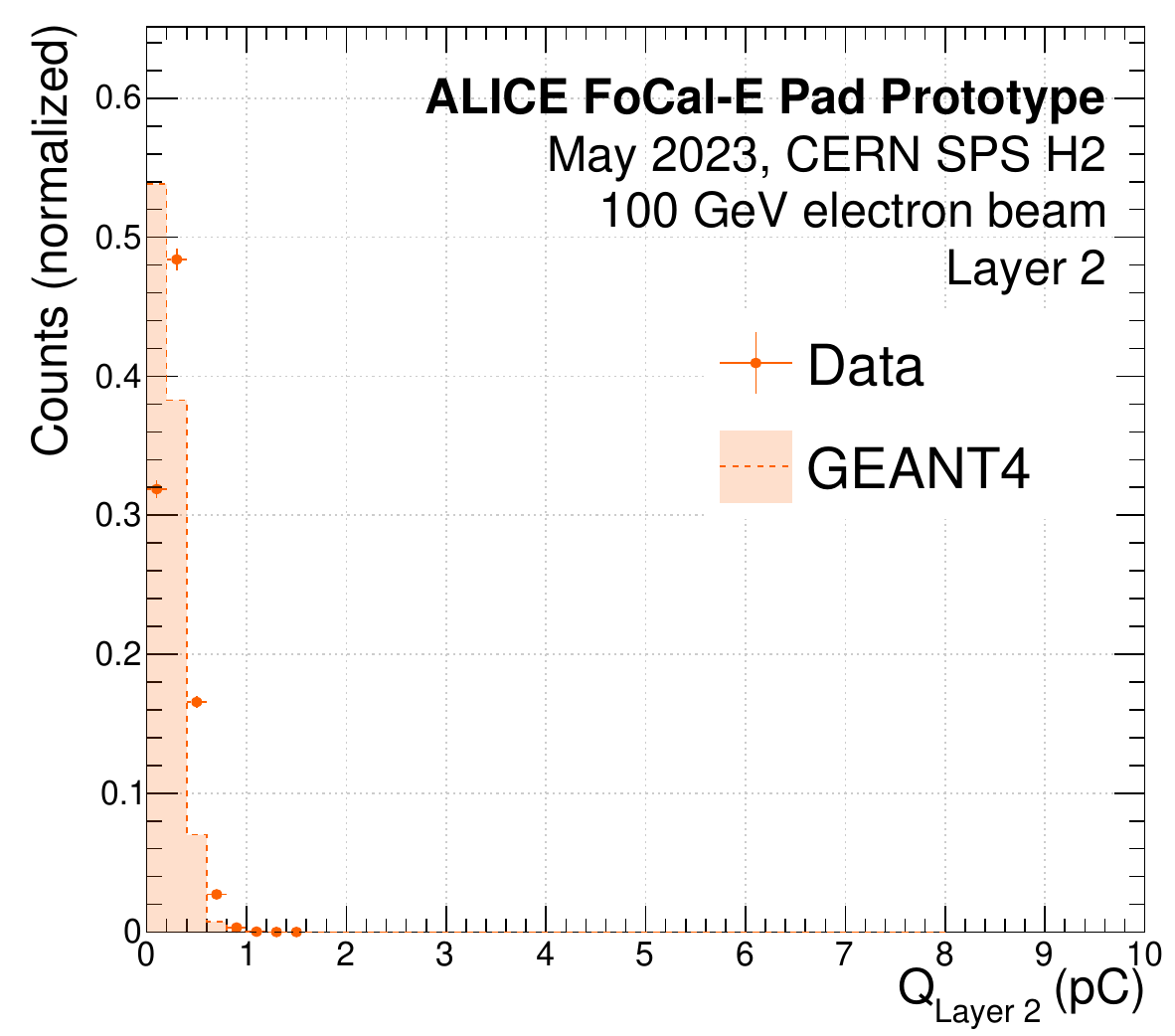}
\includegraphics[width=0.28\textwidth]{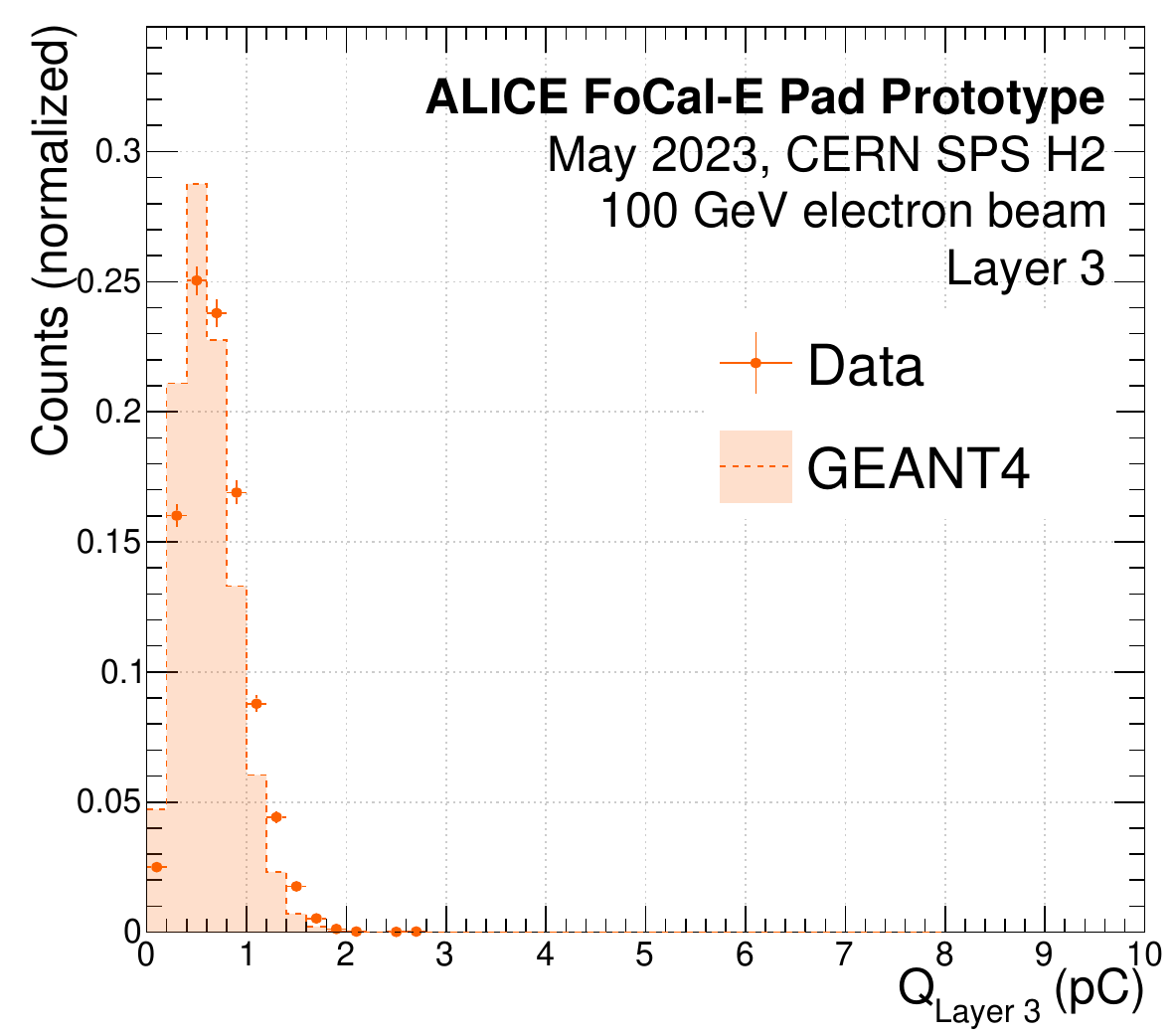}
\includegraphics[width=0.28\textwidth]{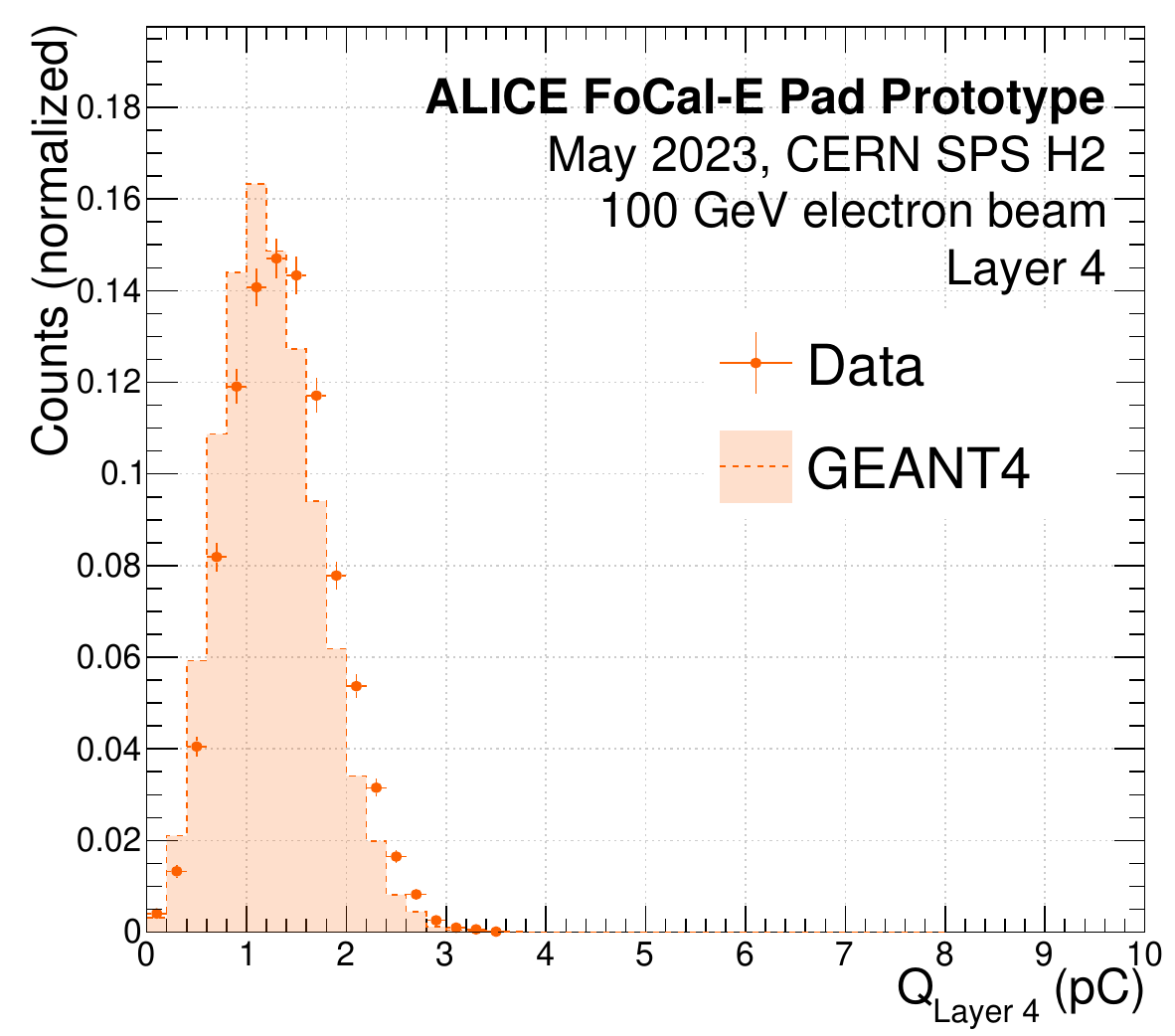}
\includegraphics[width=0.28\textwidth]{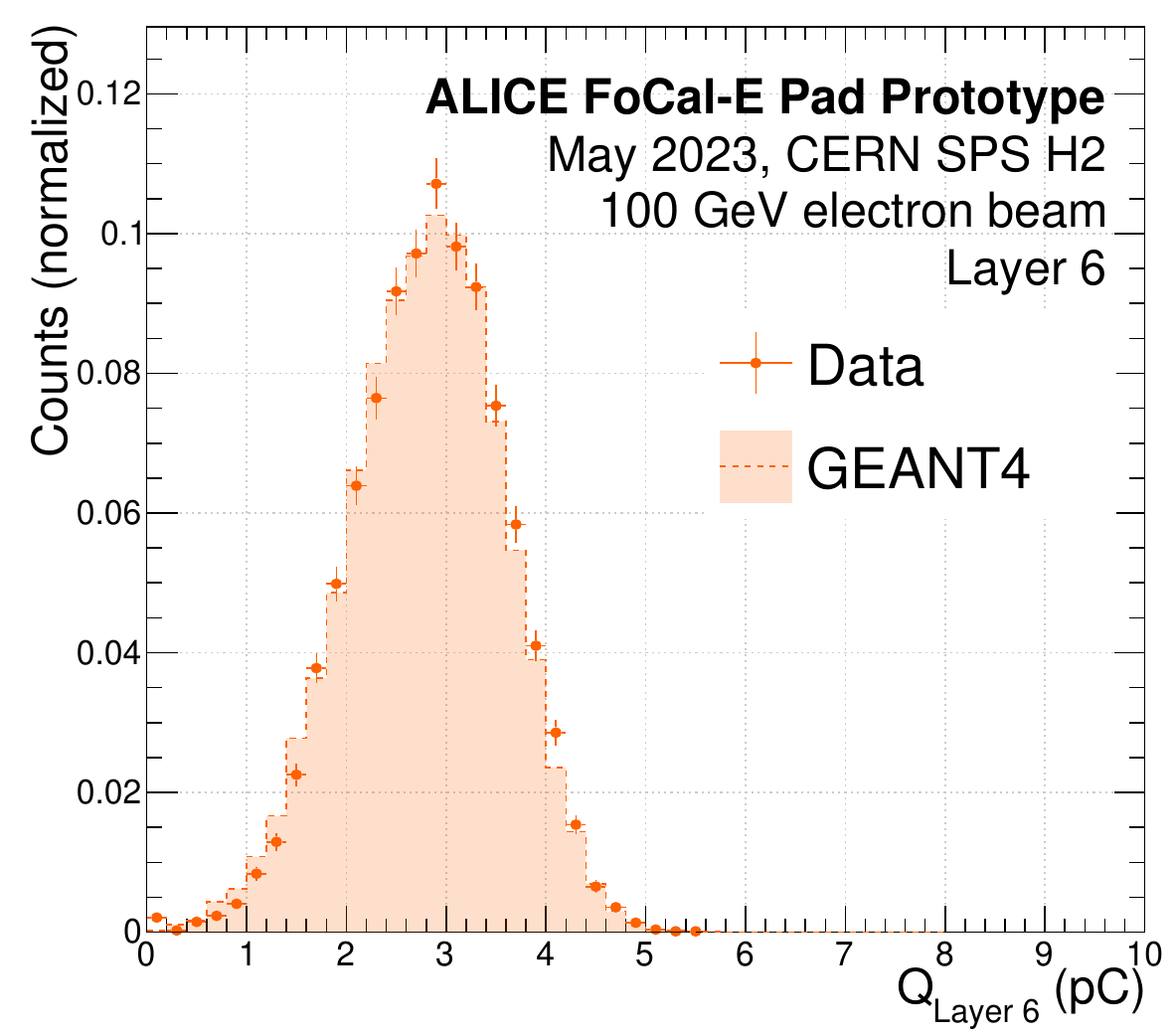}
\includegraphics[width=0.28\textwidth]{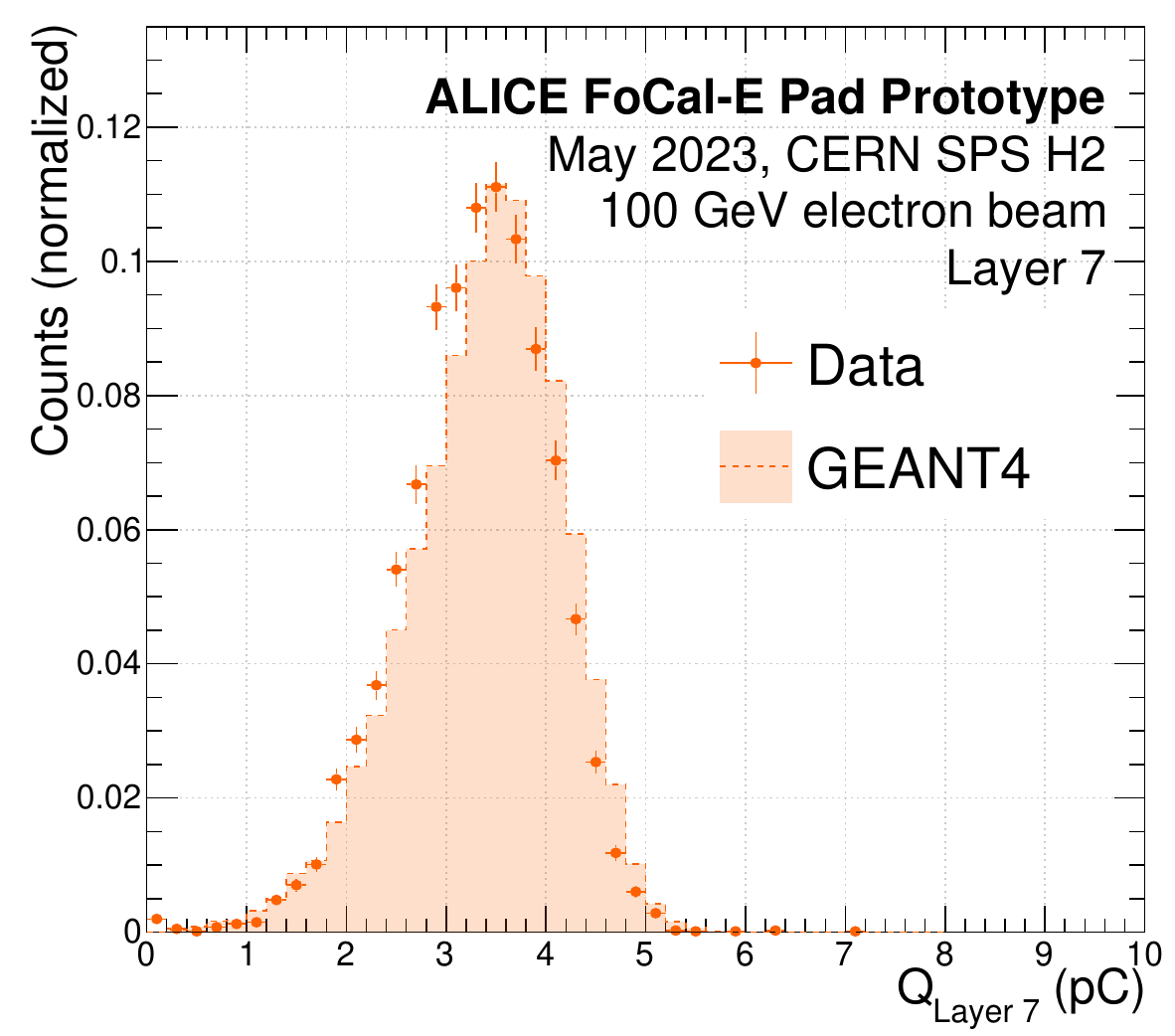}
\includegraphics[width=0.28\textwidth]{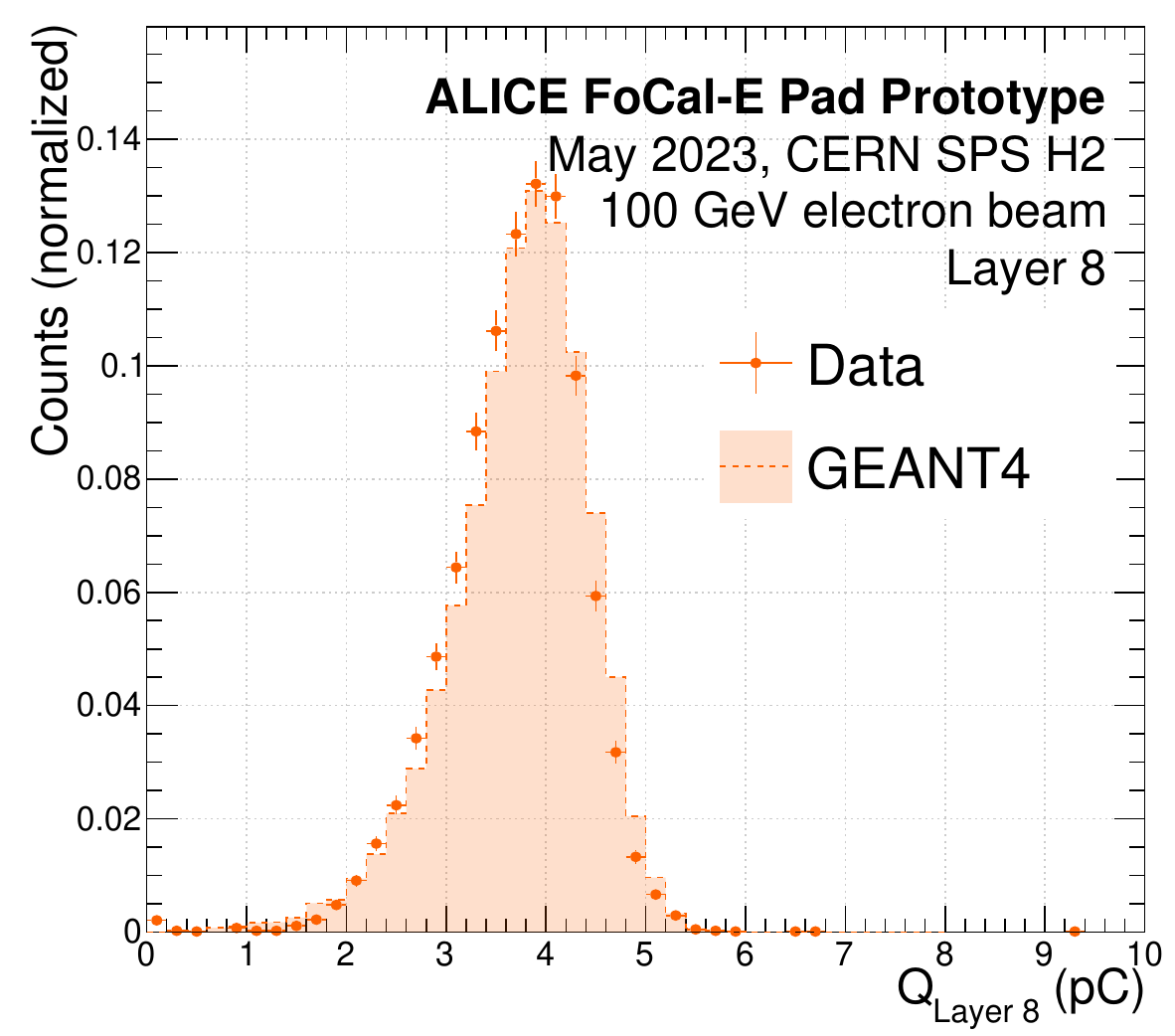}
\includegraphics[width=0.28\textwidth]{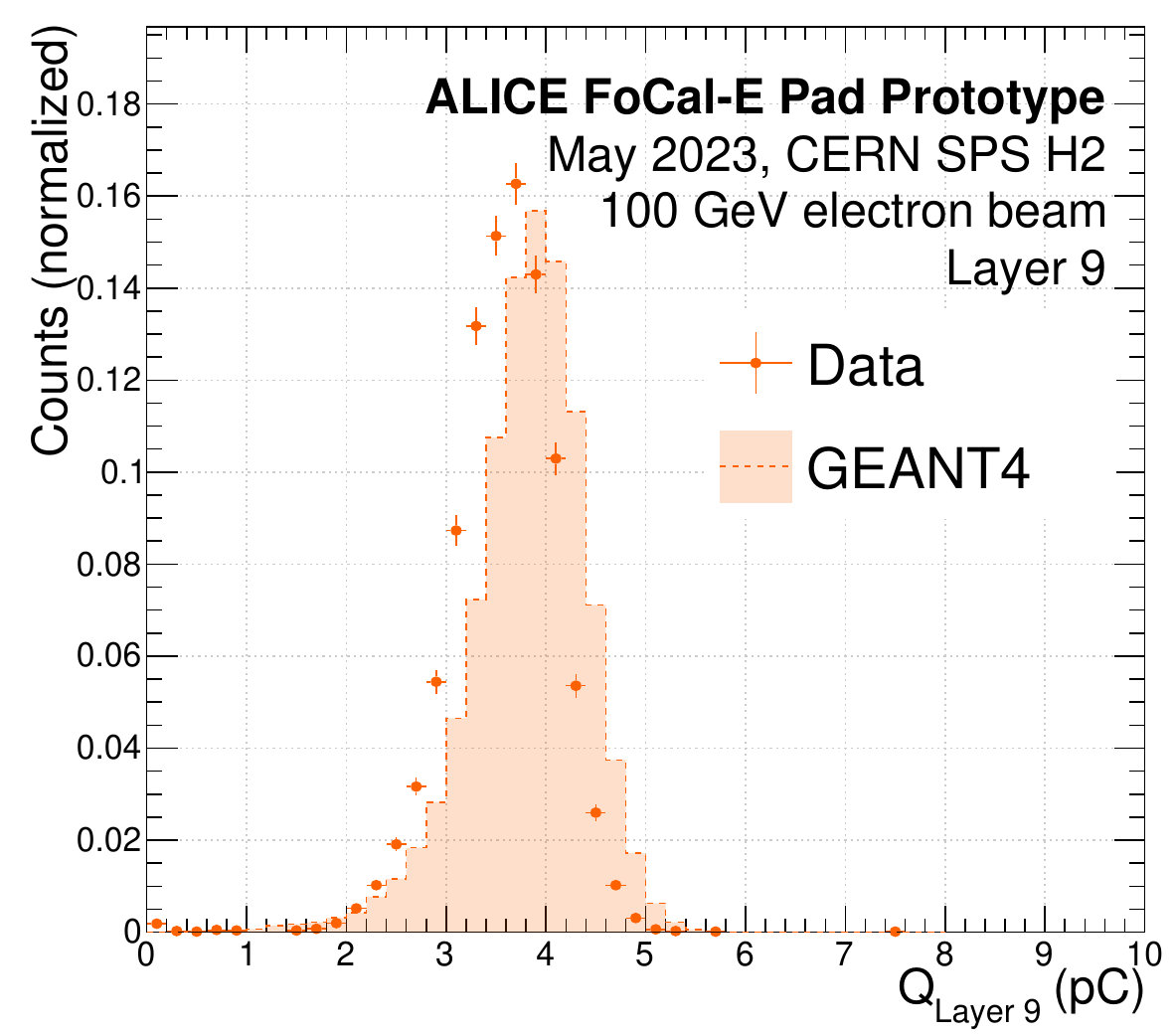}
\includegraphics[width=0.28\textwidth]{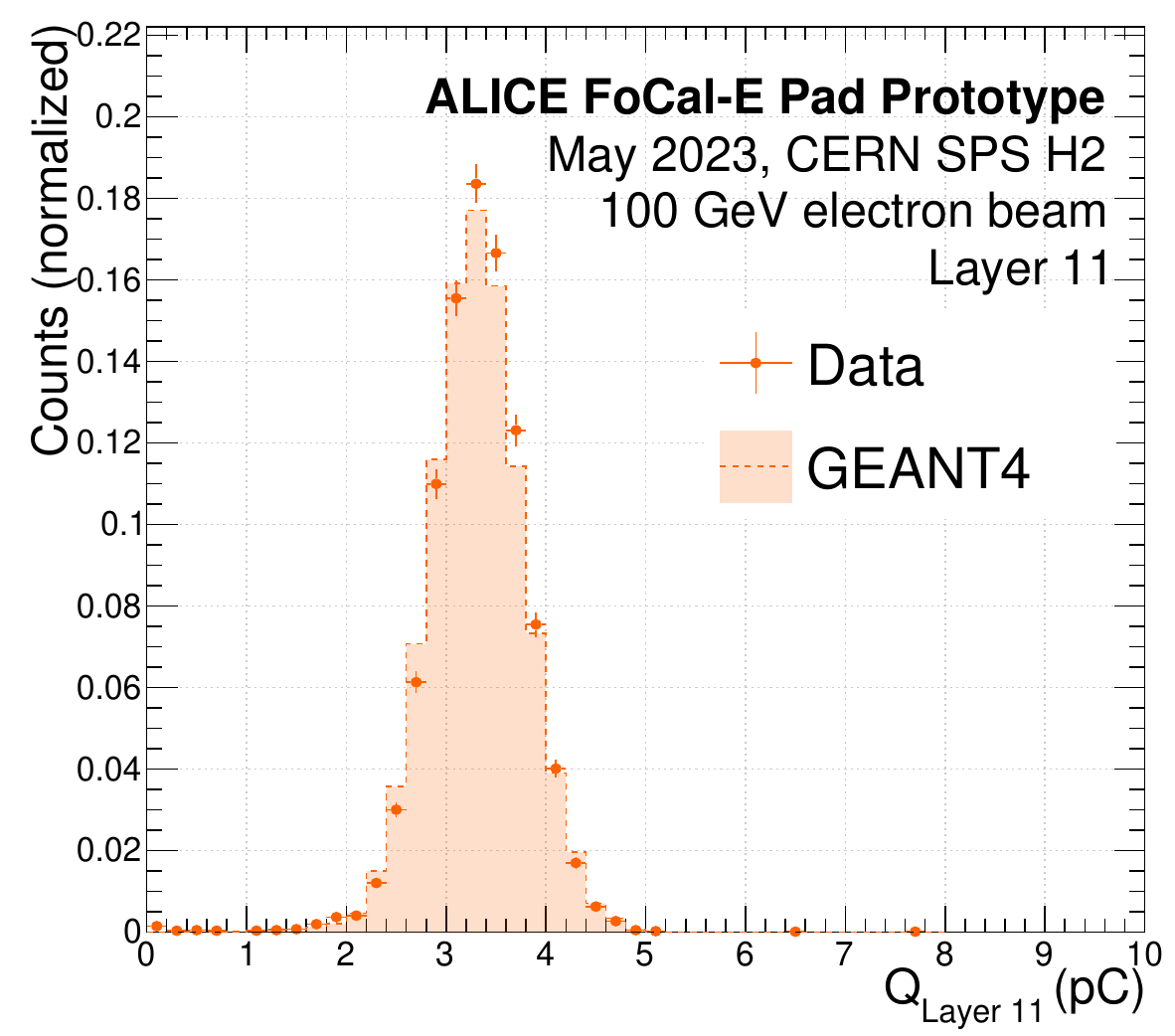}
\includegraphics[width=0.28\textwidth]{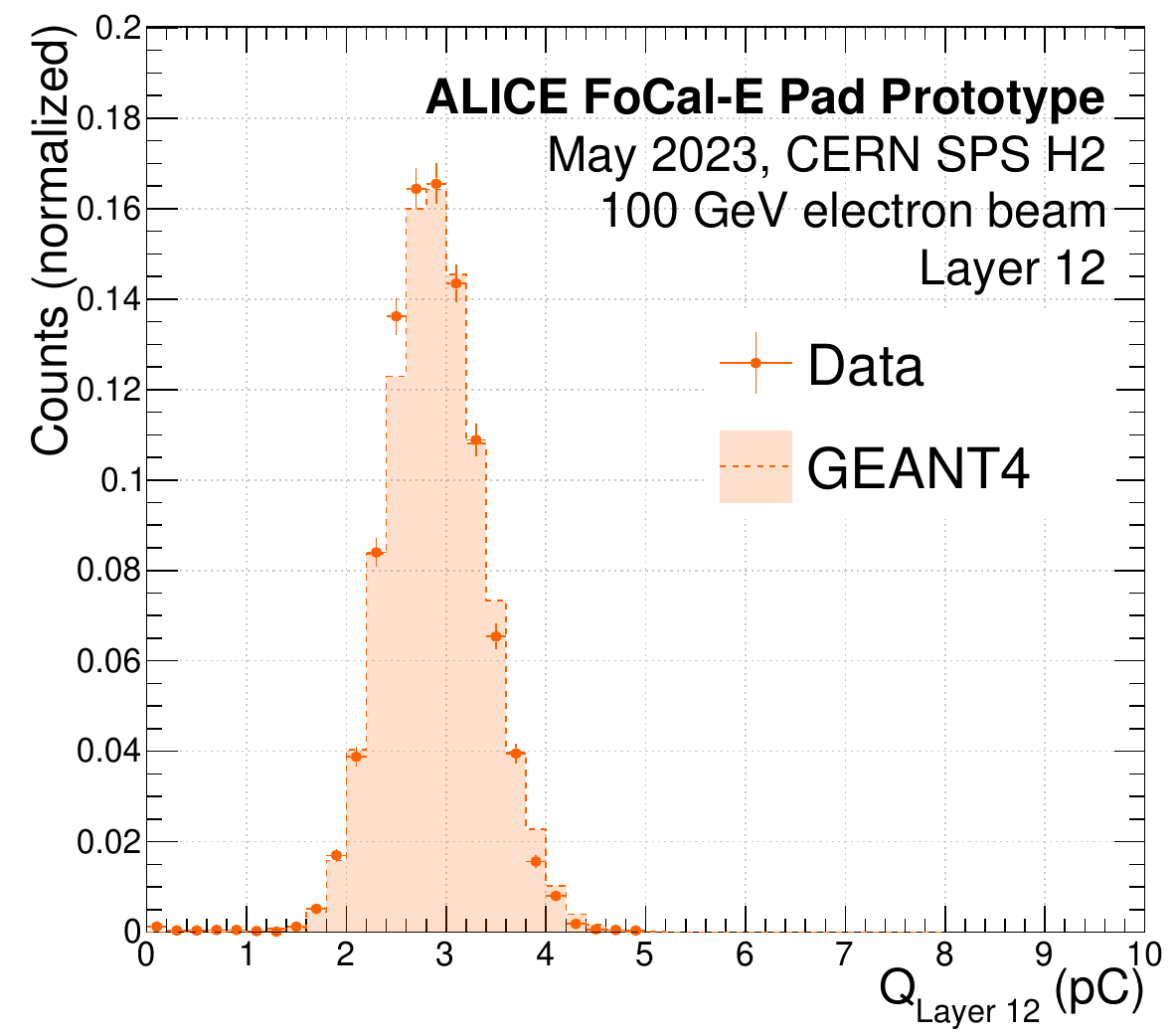}
\includegraphics[width=0.28\textwidth]{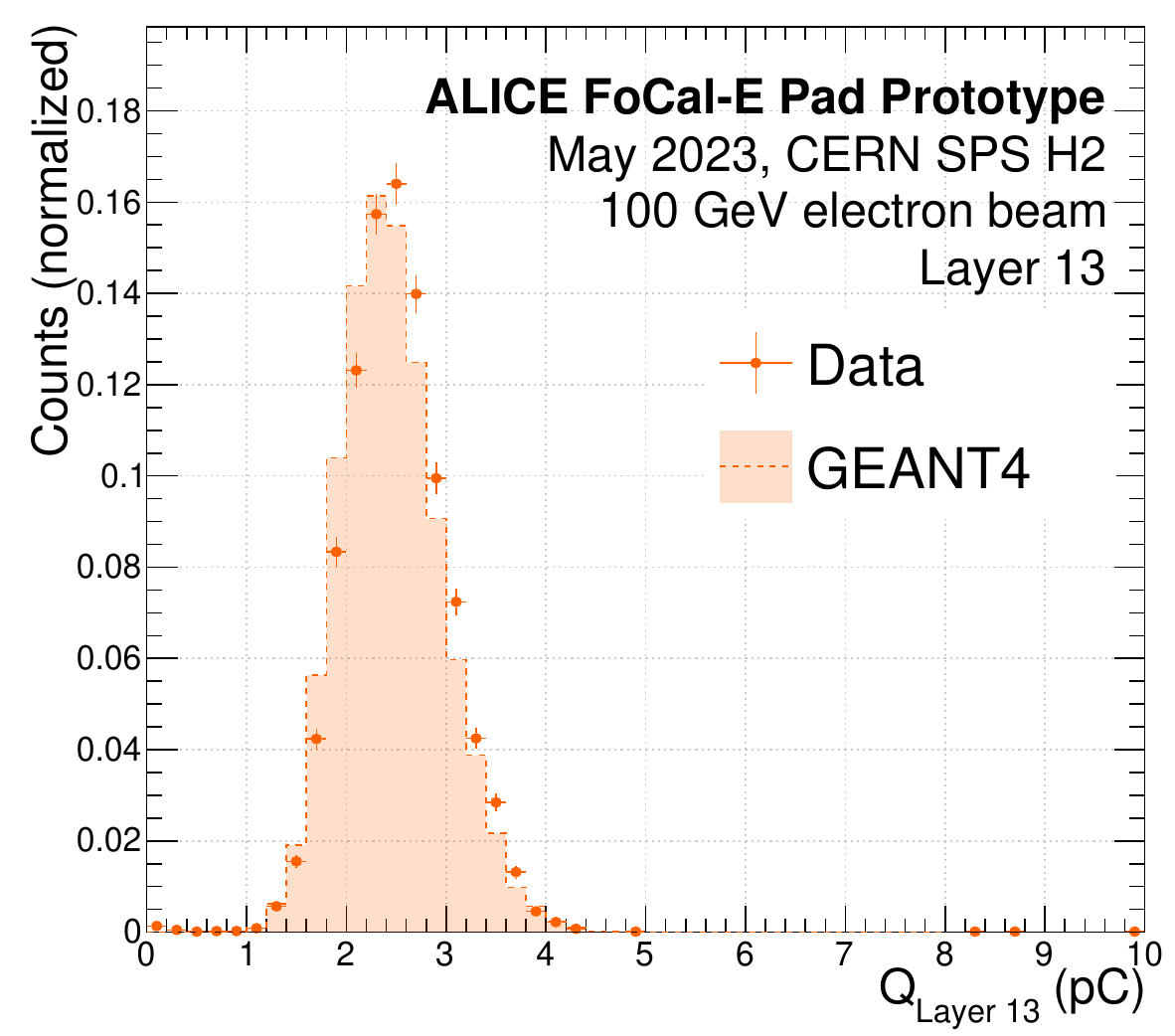}
\includegraphics[width=0.28\textwidth]{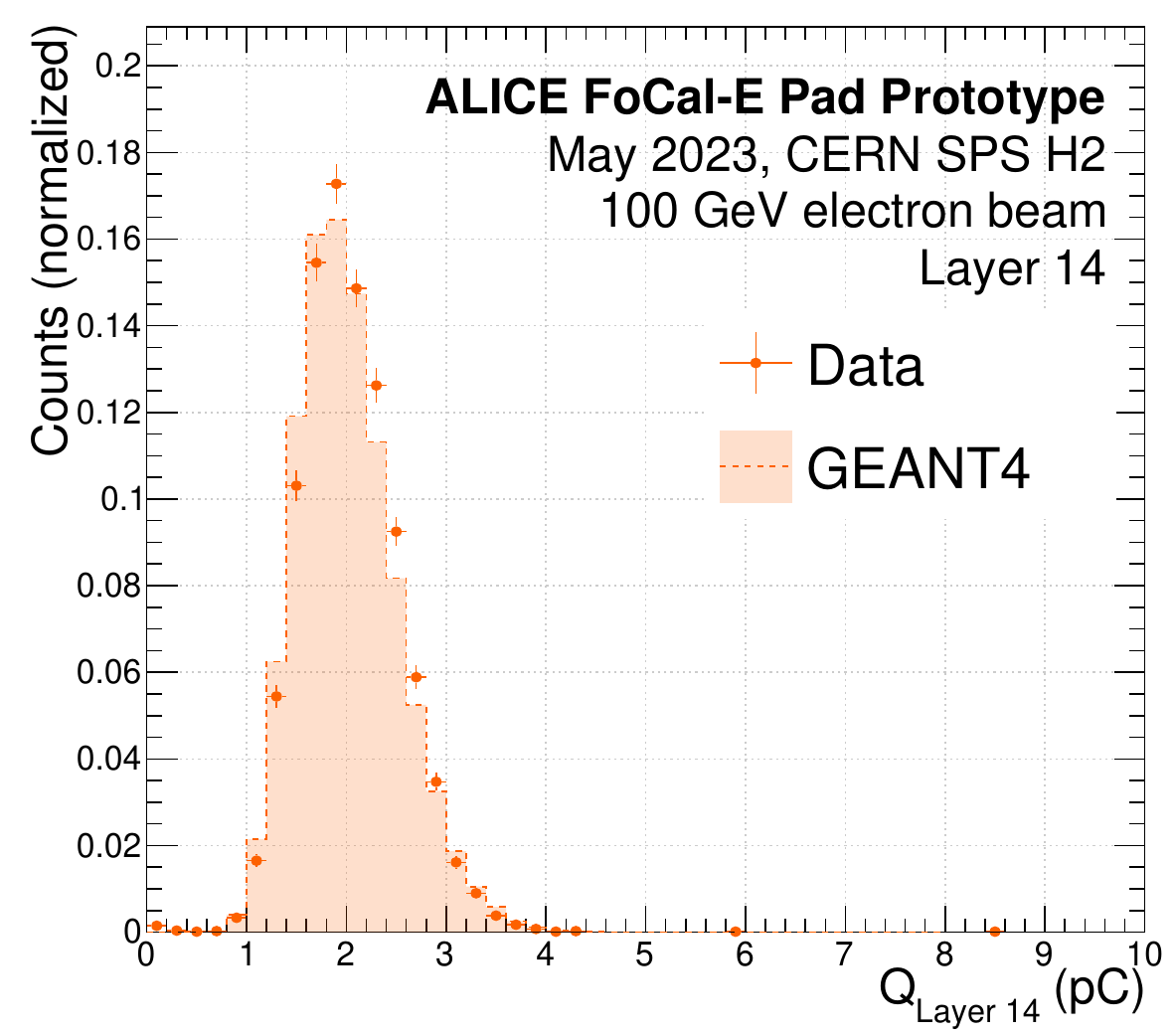}
\includegraphics[width=0.28\textwidth]{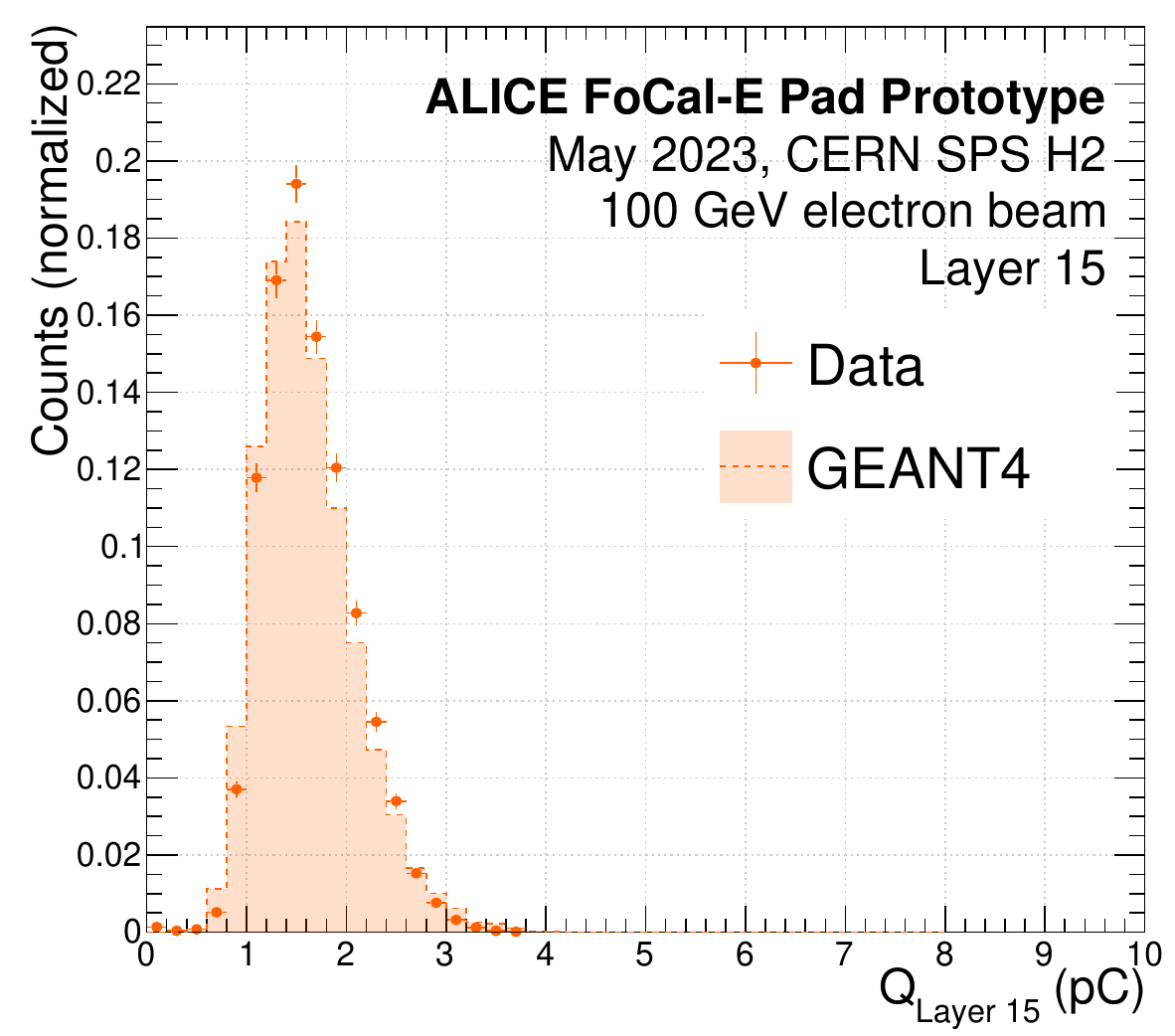}
\includegraphics[width=0.28\textwidth]{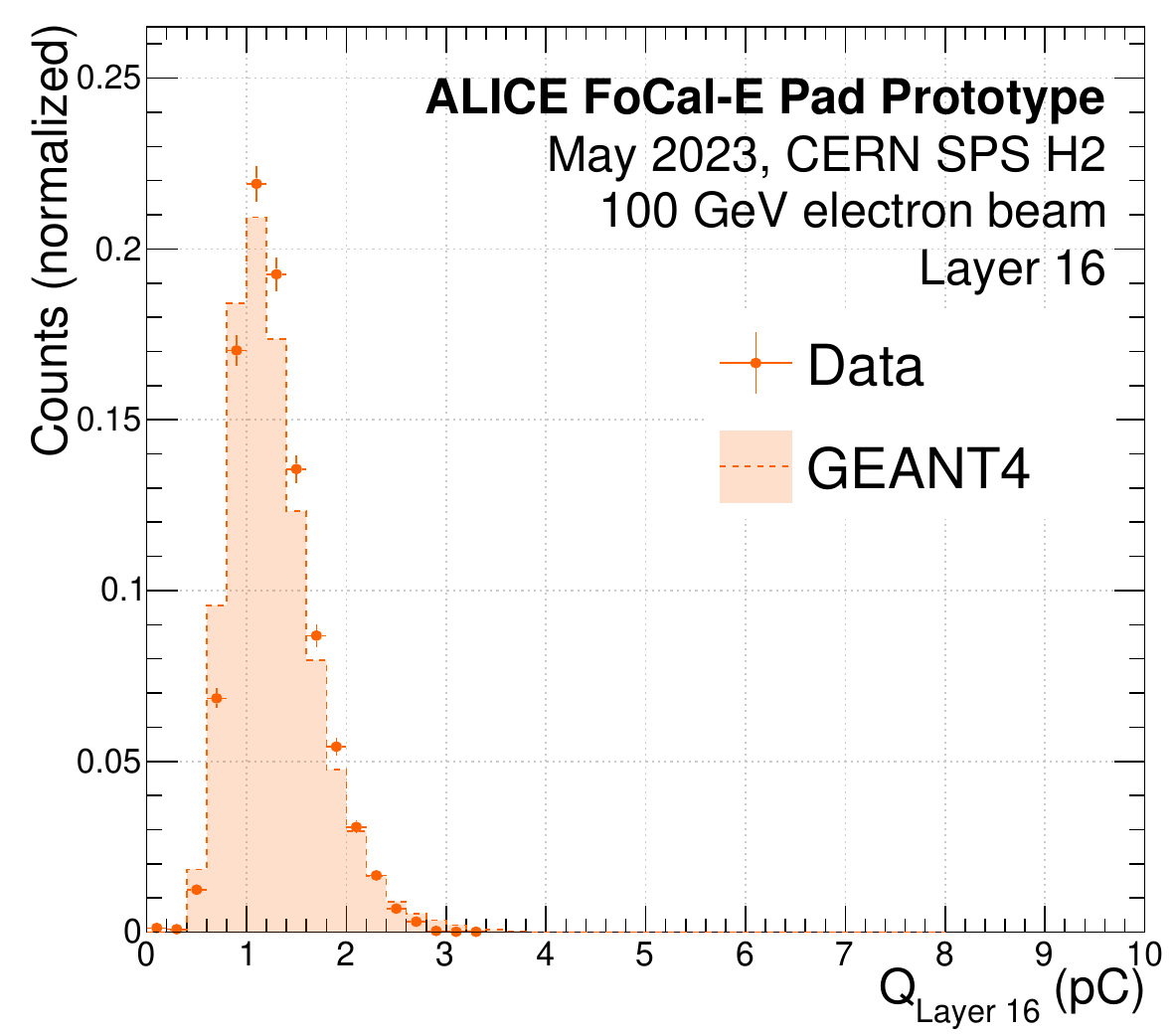}
\includegraphics[width=0.28\textwidth]{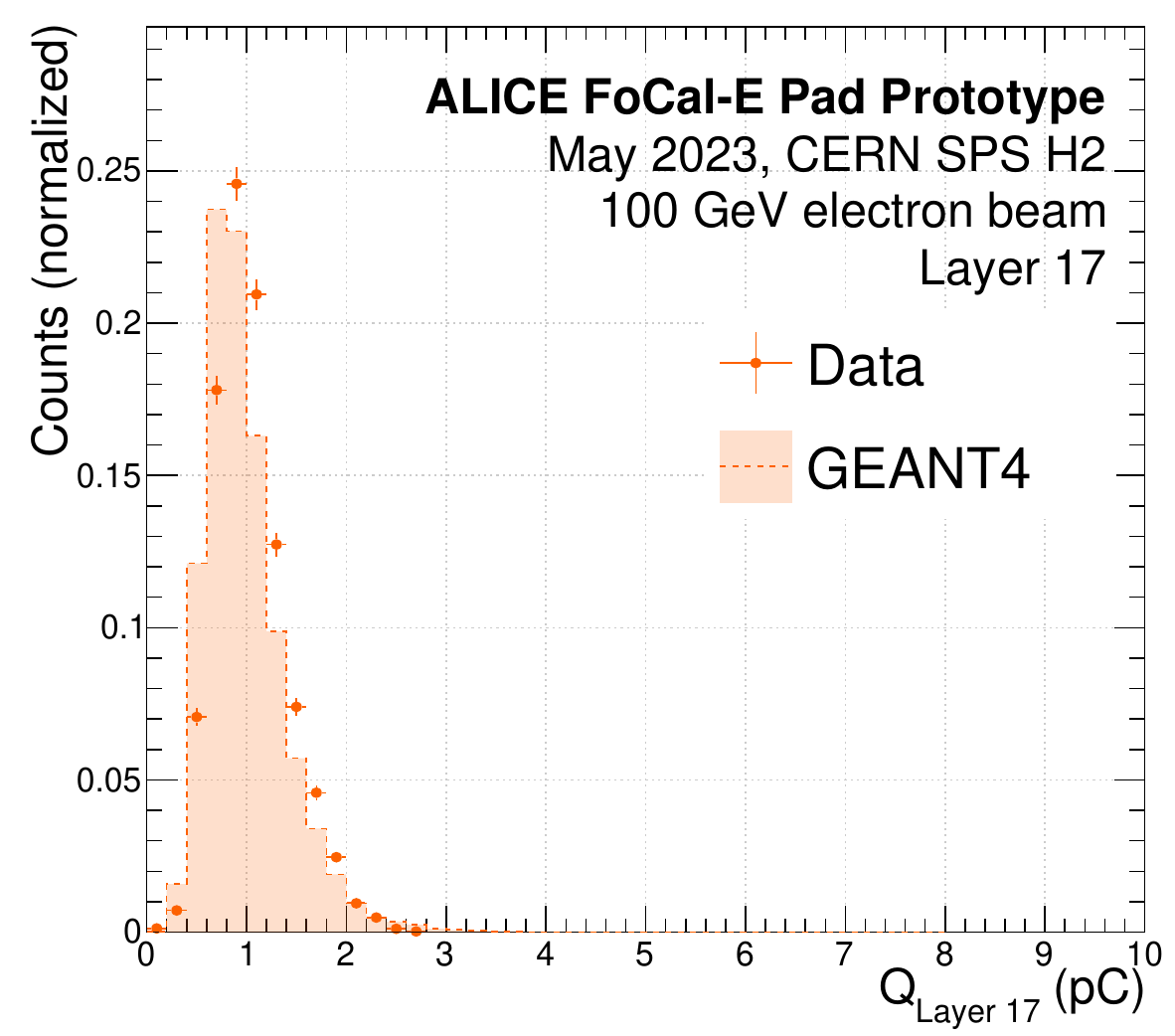}
\includegraphics[width=0.28\textwidth]{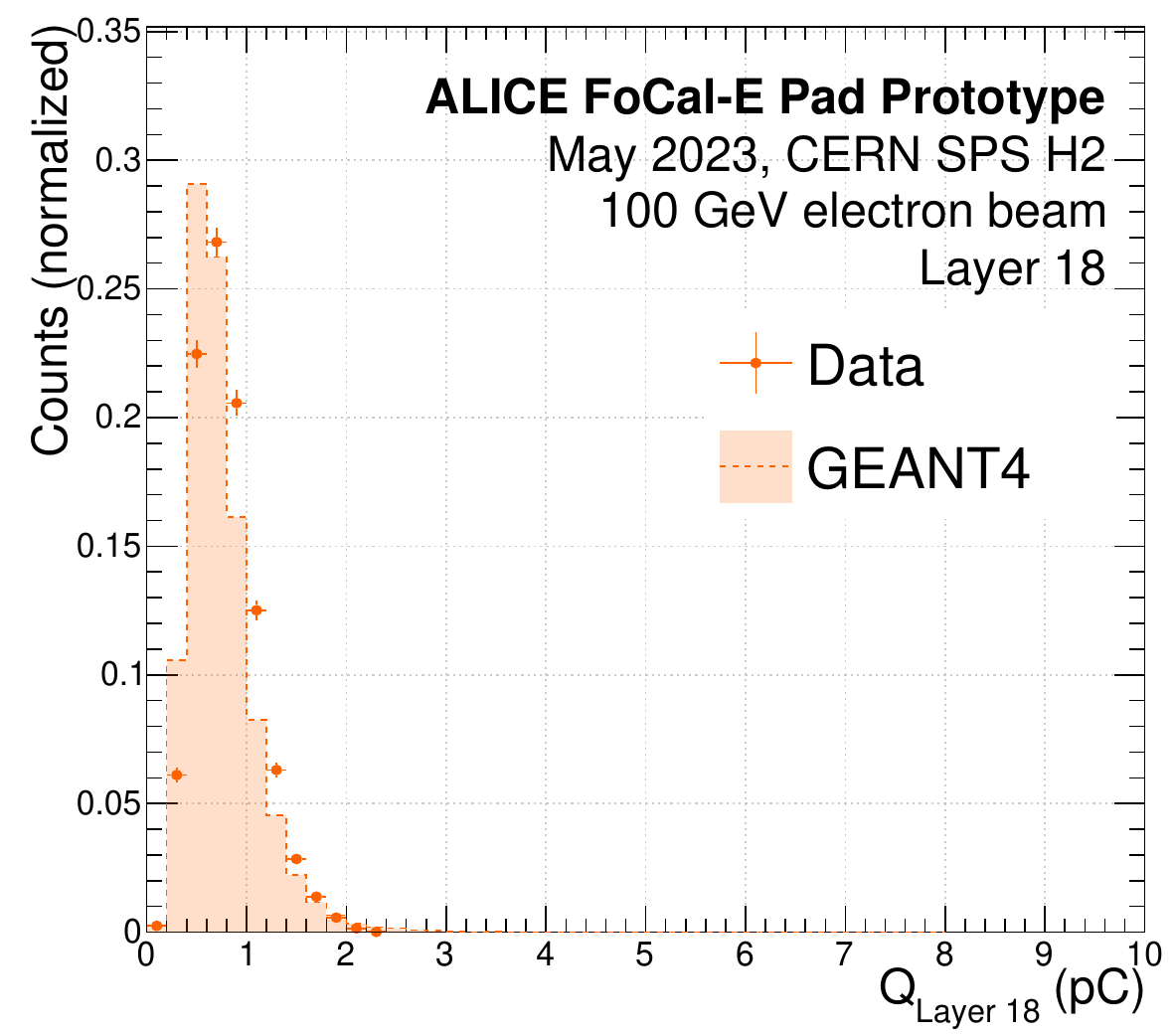}
\includegraphics[width=0.28\textwidth]{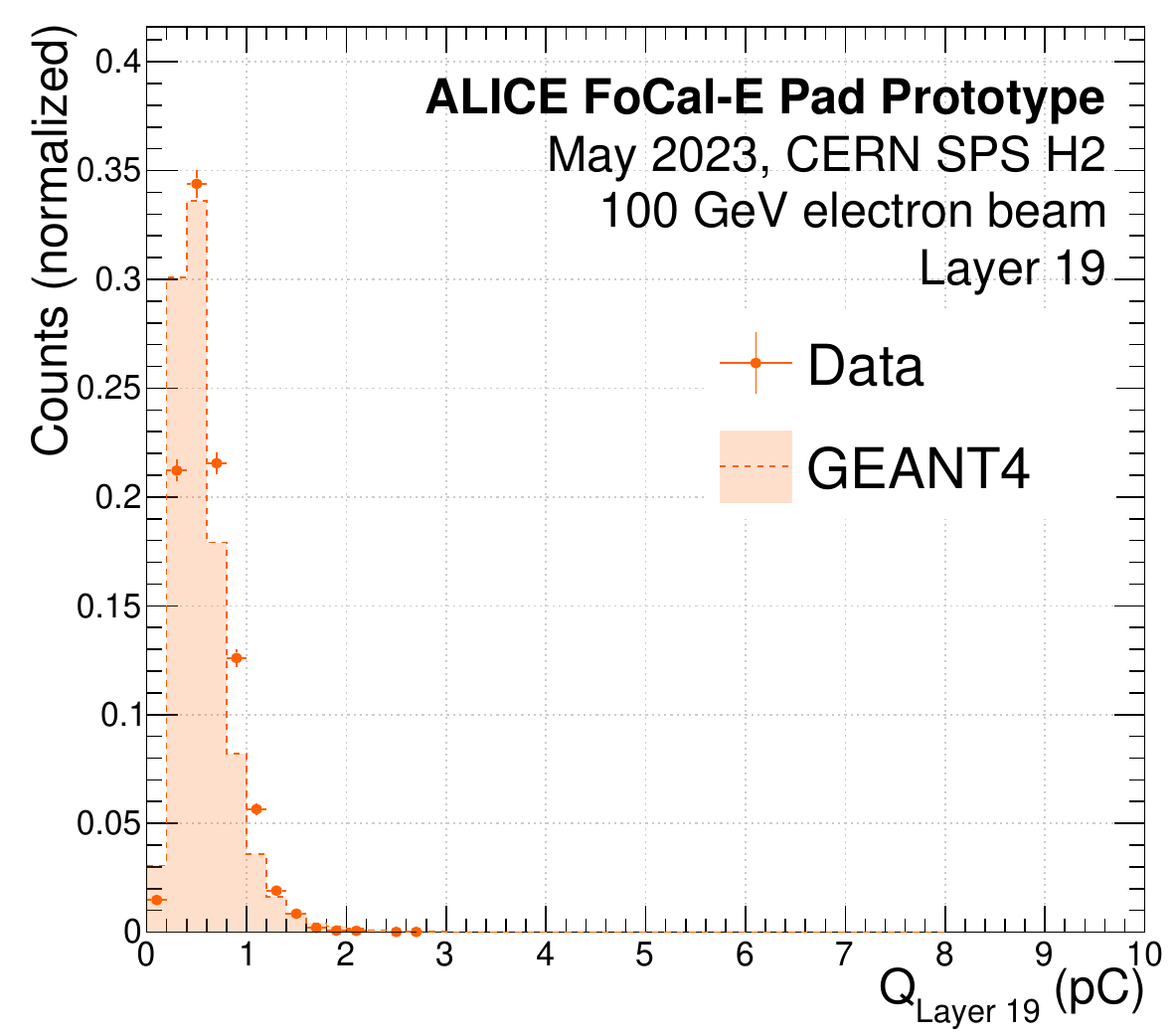}
\includegraphics[width=0.28\textwidth]{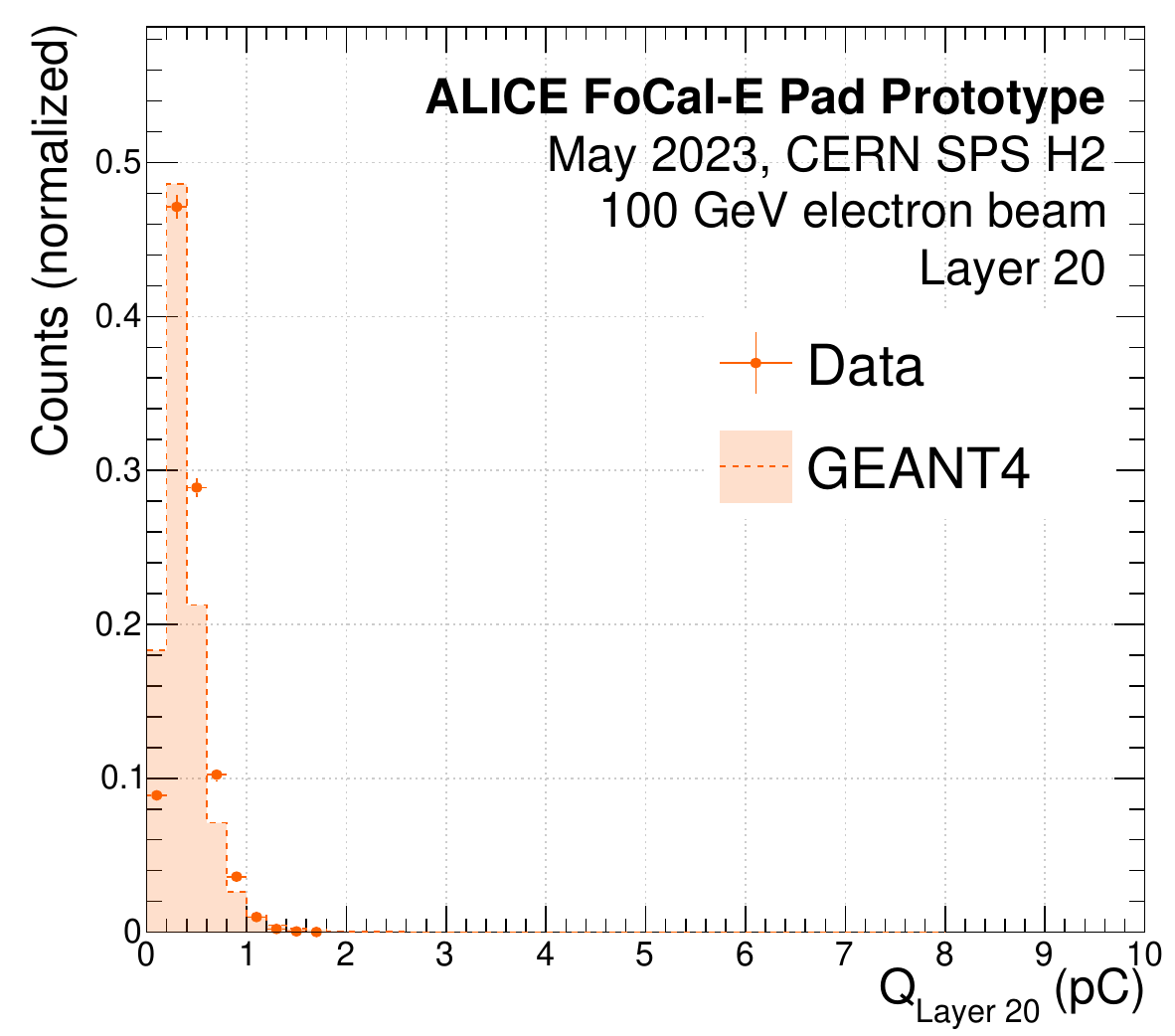}
\vspace{-0.3cm}
\caption{
\label{fig:pad-layer-by-layer-100GeV} 
Per-layer charge signal compared to simulation for 100~GeV electrons for the 18 FoCal-E pad layers, recorded at SPS H2 in May 2022.} 
\end{center}
\end{figure}

\Figure{fig:pad-layer-by-layer-100GeV} shows the signal sum distributions for each of the 18 pad layers for 100~GeV electron events, compared to GEANT4 simulations~(described in \Sec{sec:simulation}).
The line shapes of the distributions, which are in general not Gaussian, are in good agreement between the data and simulation.
The mean value of the signal charge per layer is described with the simulation on a $\pm 10 \,\%$ level for layers 4 to 17.

\begin{figure}[th!]
\begin{center}
\includegraphics[width=0.49\textwidth]{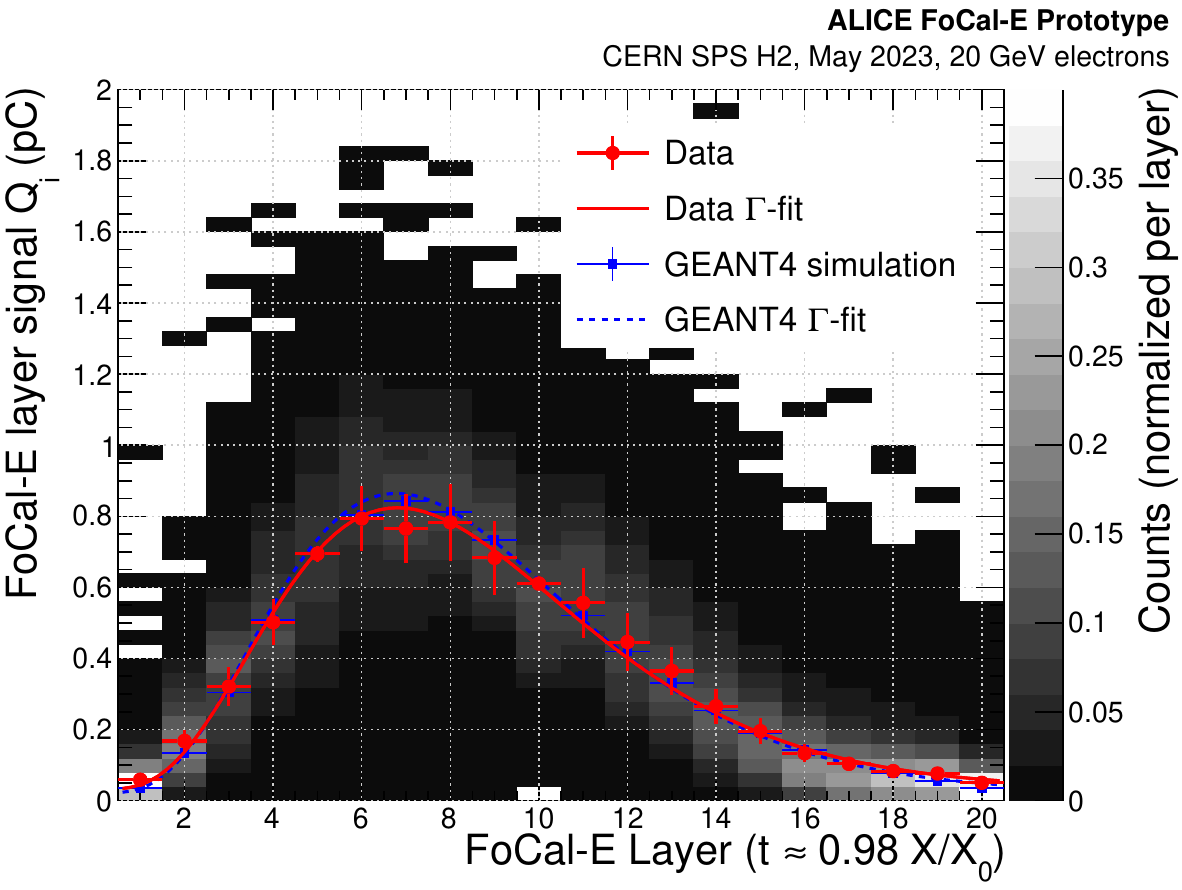}
\hspace{0.1cm}
\includegraphics[width=0.49\textwidth]{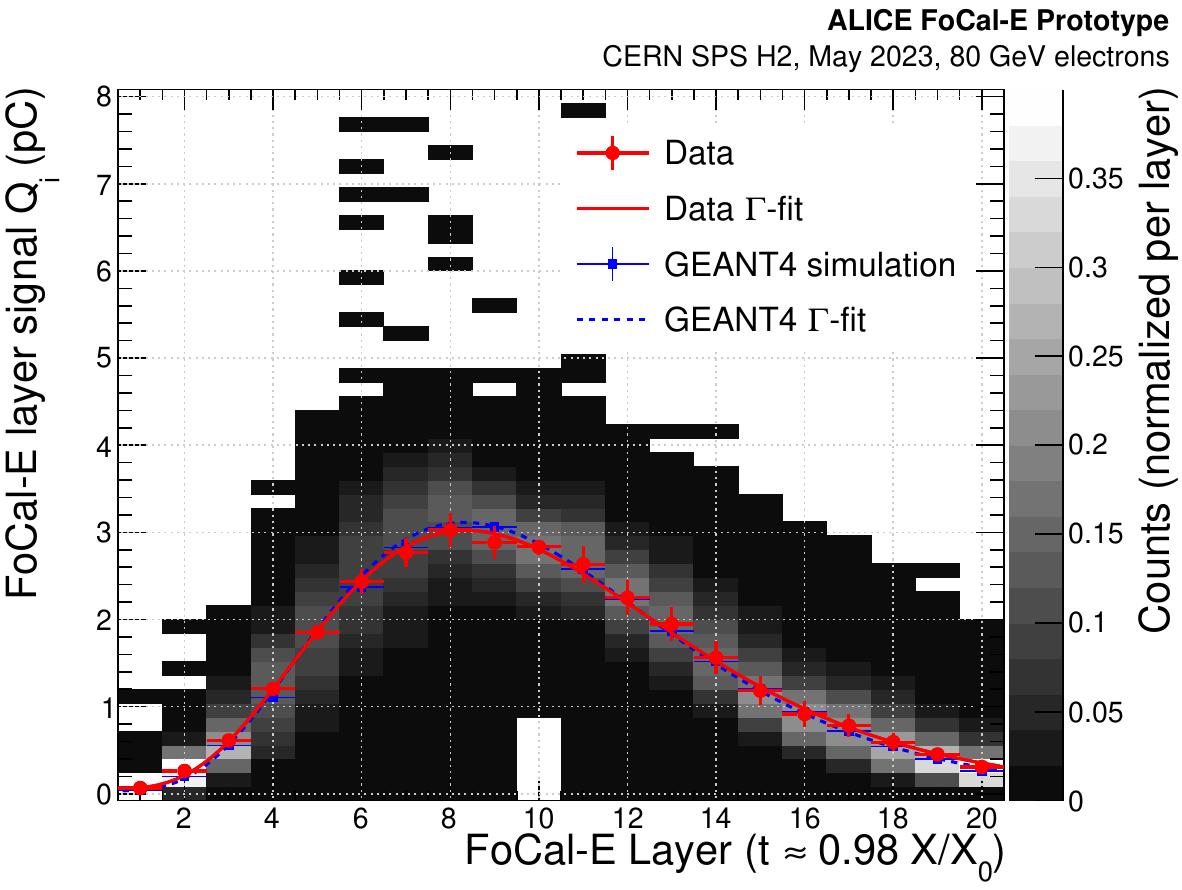}
\includegraphics[width=0.49\textwidth]{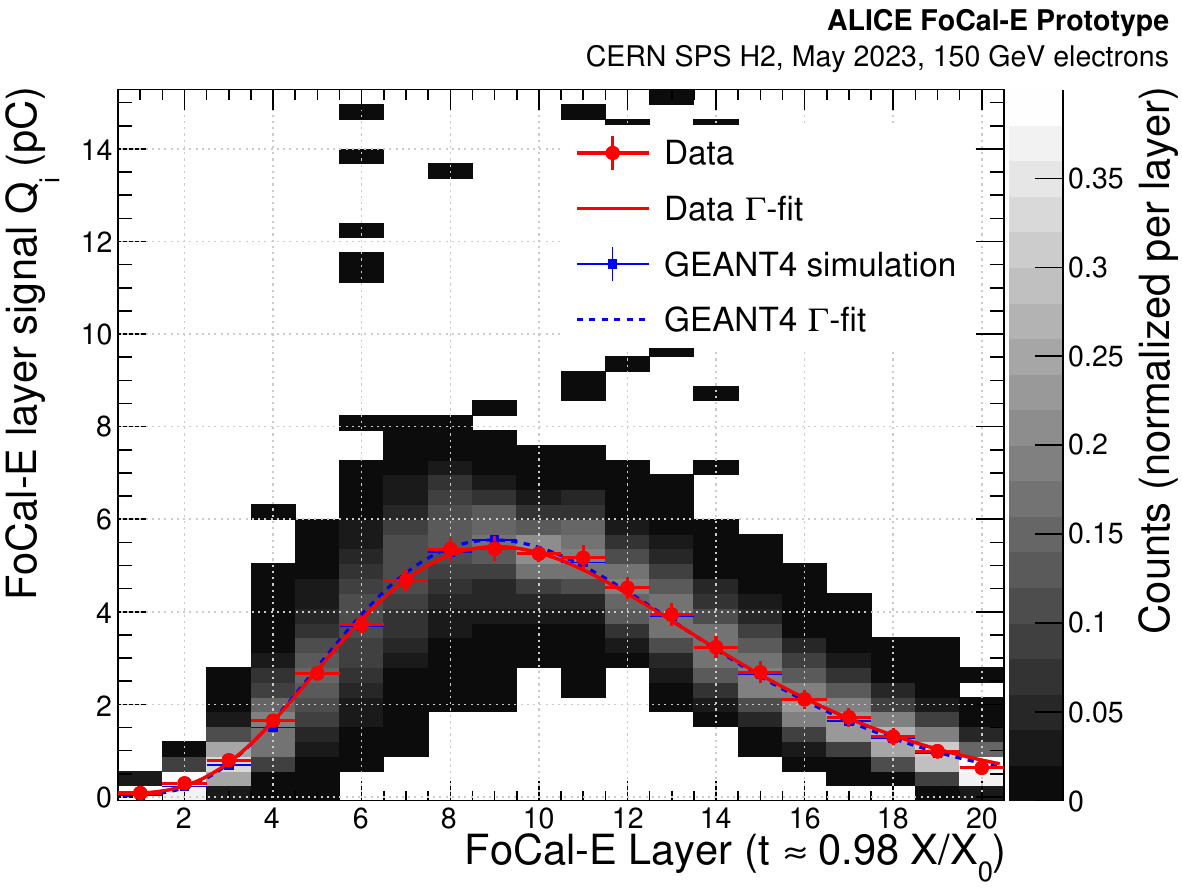}
\hspace{0.1cm}
\includegraphics[width=0.49\textwidth]{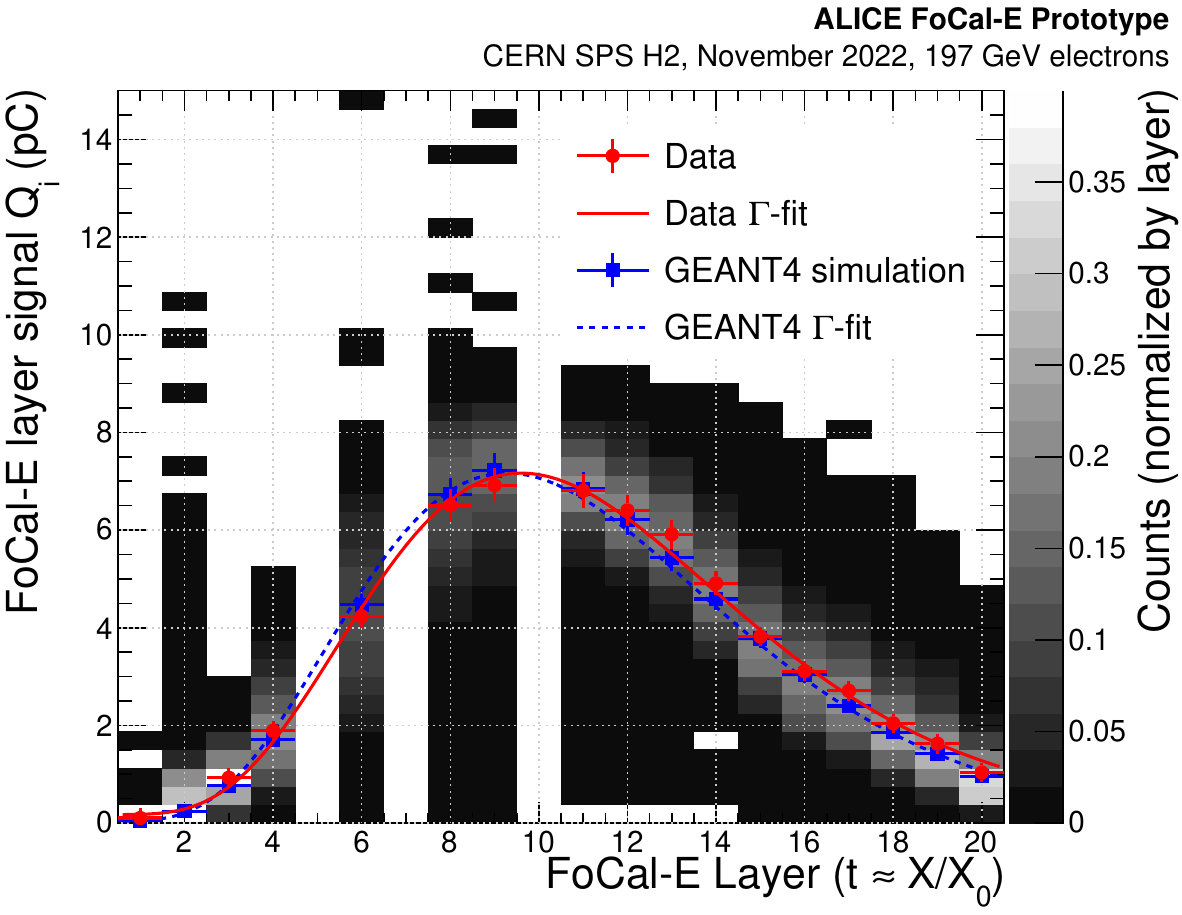}
\includegraphics[width=0.49\textwidth]{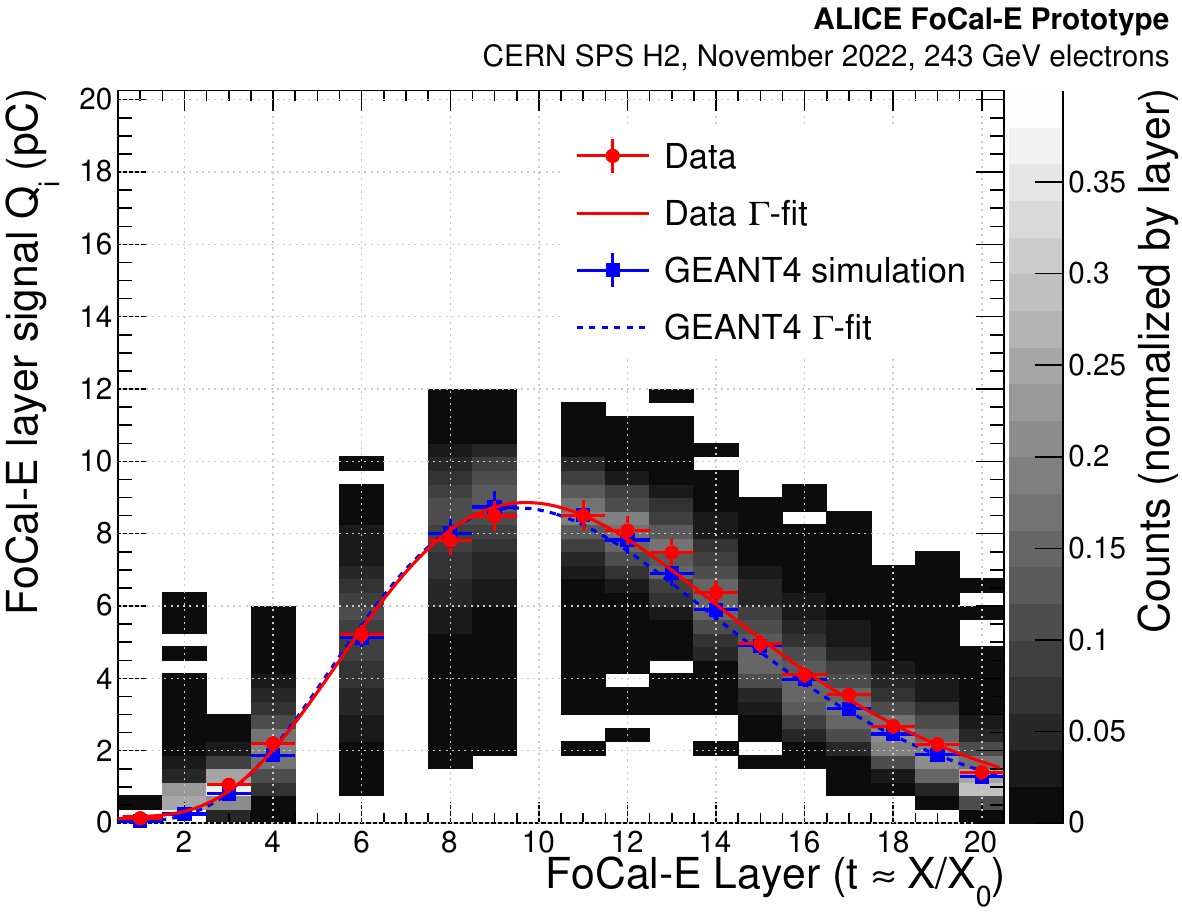}
\hspace{0.1cm}
\includegraphics[width=0.49\textwidth]{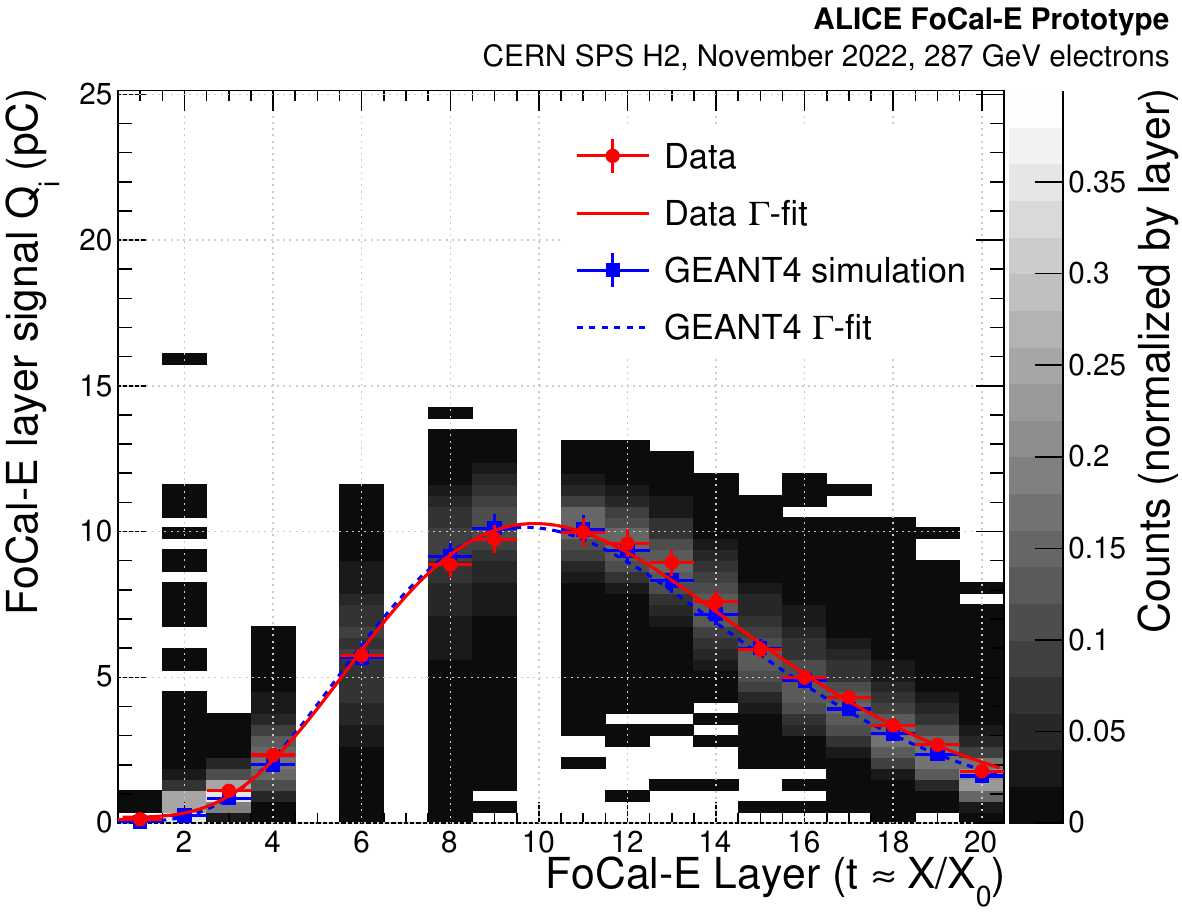}
\caption{
\label{fig:pad-long-shower-profile-all-single} Example longitudinal shower profiles for 20 to 300~GeV electrons. The projections for data and simulations to the vertical axis per layer are identical to the charge distribution measured per layer, as e.g.\ shown in \Fig{fig:pad-layer-by-layer-100GeV} for 100~GeV electrons. 
Fits to the projections are using \Eq{eq:gamma_distribution}.}
\end{center}
\end{figure}

Example distribution of signal sums for 20--300~GeV electrons versus the shower depth~(layer) are shown in \Fig{fig:pad-long-shower-profile-all-single}. 
The longitudinal shower profile is obtained by calculating the average value of the charge sum per layer.
It is shown for data and simulations, with the red and black markers, respectively.
The longitudinal shower profile can be described by a $\Gamma$-distribution~\cite{Workman:2022ynf} plus a small offset correction $Q_{0}$ as
\begin{equation}
\frac{{\rm d}Q}{{\rm d}t}= Q_{E}\,\beta\,\frac{(\beta t)^{\alpha-1}e^{-\beta t}}{\Gamma(\alpha)} + Q_0 .
\label{eq:gamma_distribution}
\end{equation}

In this equation, $Q_E$ is the amplitude which depends on the energy of the primary particles, $t = x/({0.98\,X_0})$ is the depth in the material expressed in units of one layer's radiation length as summarized in Tabs.~\ref{tab:simulation-pad-materials-part1}~and~\ref{tab:simulation-pad-materials-part2}.
The function $\Gamma(\alpha) =  \int_{0}^{\infty} \euler^{-z} z^{\alpha -1}\,dz  $ is the $\Gamma$-function with the parameters $\alpha$ and $\beta$ which can be interpreted as the shape and scale parameter, respectively. 
The constant term $Q_0$ is introduced in order to stabilize the fitting procedure and to account for potential noise contributions in data.
The measured and simulated longitudinal shower profiles are found to be well described by the $\Gamma$-distribution, as also shown in \Fig{fig:pad-long-shower-profile-all-single}.

\begin{figure}[th!]
\begin{center}
\includegraphics[width=0.49\textwidth]{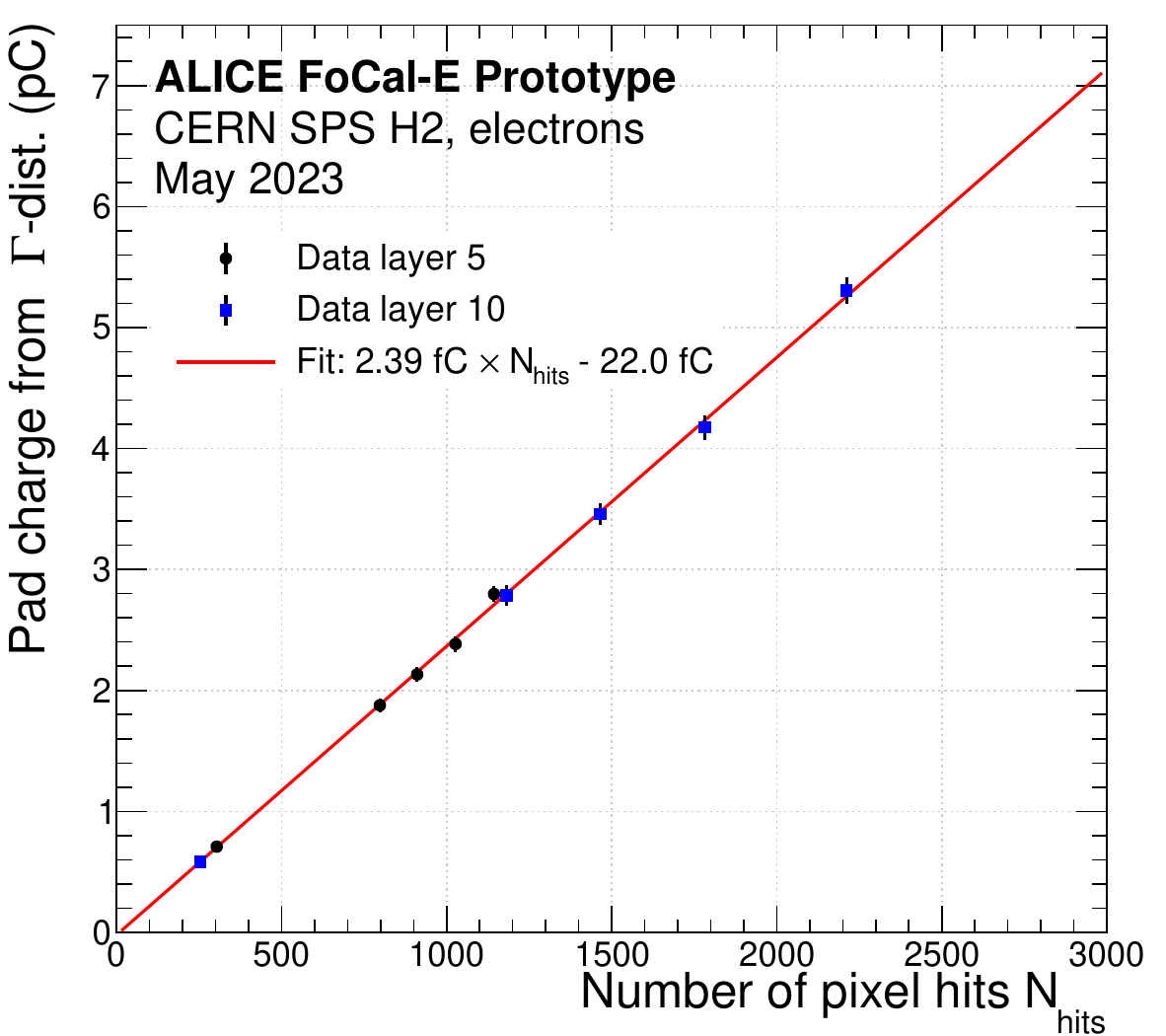} 
\includegraphics[width=0.49\textwidth]{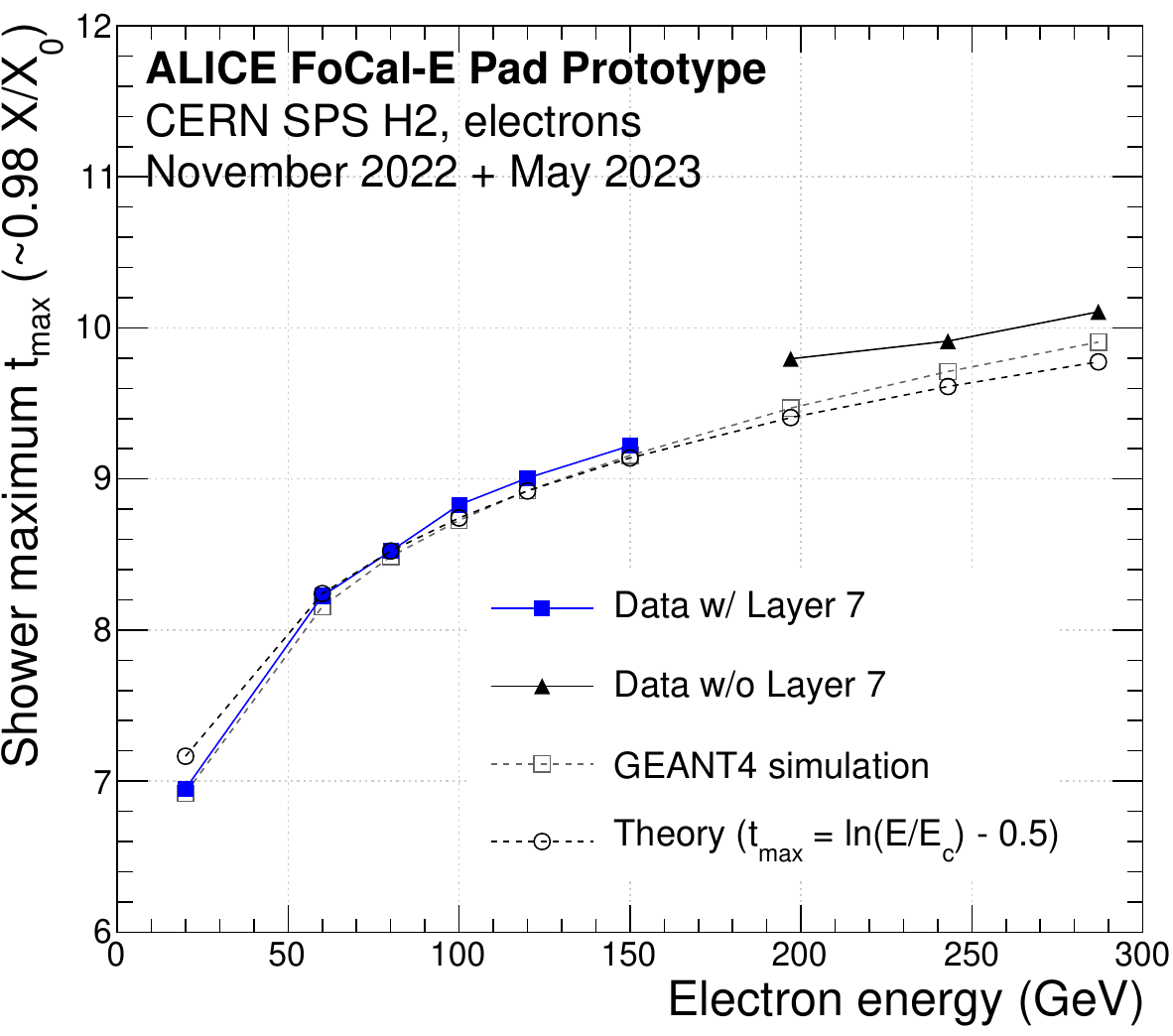}
\caption{Left: Correlation of the \ac{FoCal-E} pad charge evaluated from the $\Gamma$-distribution fit with respect to the number of pixel hits in layer~5 and 10, respectively.
Right:~the position of the shower maximum compared to \geant simulation and the theoretical value of a homogeneous calorimeter.}
\label{fig:pixel-pad-combination-linear-and-shower-max}
\end{center}
\end{figure}

From the measurement of the longitudinal shower profile in the pad layers, a calibration of the pixel layers can be derived, relating the average number of pixel hits to the measured charge in the pads.
The pixel layers are not included for the 60~GeV data because of readout failures for data taken with this specific energy, and neither are they included for the November, 2022 data.
For each energy, the fitted $\Gamma$-distribution is evaluated at the positions of layer 5 and 10 where the pixel layers are located.
The obtained signal charges and their uncertainties derived from the fit function are correlated with the mean number of pixel hits, $\Nhitsfive$ and $\Nhitsten$. 
This correlation, shown in the left panel of \Fig{fig:pixel-pad-combination-linear-and-shower-max}, is found to be linear, following the relation
\begin{equation}
 Q_{\text{pix}} = (2.39 \pm 0.04) \fC \cdot \Nhits + ( 0.02 \pm 0.03) \fC 
\end{equation}
with $Q_{\text{pix}}$ representing the equivalent pad charge, derived from the signal of the pixel layers.
By applying this relation between pixel and pad layers, the pixel signal can be included in the longitudinal shower profile, producing a uniform measurement of the shower profiles with 20 sampling points along the longitudinal shower axis (already done in \Fig{fig:pad-long-shower-profile-all-single}).
After including the pixel layers, the $\Gamma$-distribution fit is repeated, with the fitted parameters listed in \Tab{tab:gamma-parameters-data}.
The parameters of the fitted functions are in good agreement with the expectation from simulation, listed in \Tab{tab:gamma-parameters-simu}. 

\begin{table}[th!]
\caption{\label{tab:gamma-parameters-data} Parameters $Q_0$, $\alpha$, and $\beta$ for the $\Gamma$-distribution fits to data.}
\begin{center}
\begin{tabular}{ c c c c c }
 $E_{\text{nom}}$ (GeV)  & $Q_E \pm \sigma_{Q_E}$ (pC)  & $\alpha \pm \sigma_{\alpha}$ & $\beta \pm \sigma_{\beta}$ & $Q_{0} \pm \sigma_{Q_{0}}$ (pC) \\ \hline
287 & 112.85 $\pm$ 4.29  & 6.28 $\pm$ 0.32  & 0.52 $\pm$ 0.03 & 0.20 $\pm$ 0.18 \\ 
243 & 96.48 $\pm$ 4.24  & 6.17 $\pm$ 0.35  & 0.52 $\pm$ 0.03 & 0.17 $\pm$ 0.18 \\ 
199 & 74.90 $\pm$ 3.95  & 6.32 $\pm$ 0.43  & 0.54 $\pm$ 0.04 & 0.17 $\pm$ 0.18 \\ 
150 & 57.73 $\pm$ 0.79  & 5.75 $\pm$ 0.11  & 0.52 $\pm$ 0.01 & 0.08 $\pm$ 0.00 \\ 
120 & 46.21 $\pm$ 0.63  & 5.56 $\pm$ 0.11  & 0.51 $\pm$ 0.01 & 0.07 $\pm$ 0.01 \\ 
100 & 38.38 $\pm$ 0.57  & 5.42 $\pm$ 0.11  & 0.50 $\pm$ 0.01 & 0.07 $\pm$ 0.01 \\ 
80 & 30.80 $\pm$ 0.62  & 5.39 $\pm$ 0.15  & 0.52 $\pm$ 0.02 & 0.06 $\pm$ 0.02 \\ 
60 & 22.97 $\pm$ 0.62  & 5.23 $\pm$ 0.19  & 0.51 $\pm$ 0.02 & 0.05 $\pm$ 0.02 \\ 
20 & 7.17 $\pm$ 0.26  & 4.78 $\pm$ 0.33  & 0.55 $\pm$ 0.04 & 0.03 $\pm$ 0.01 
\end{tabular}
\end{center}
\end{table}

\begin{table}[th!]
\caption{\label{tab:gamma-parameters-simu} Parameters $Q_0$, $\alpha$, and $\beta$ for the $\Gamma$-distribution fits to simulation.}
\begin{center}
\begin{tabular}{ c c c c c }
 $E_{\text{nom}}$ (GeV)  & $Q_E \pm \sigma_{Q_E}$ (pC)  & $\alpha \pm \sigma_{\alpha}$ & $\beta \pm \sigma_{\beta}$ & $Q_{0} \pm \sigma_{Q_{0}}$ (pC)  \\ \hline
287 & 112.33 $\pm$ 1.62  & 6.15 $\pm$ 0.08  & 0.52 $\pm$ 0.01 & 0.04 $\pm$ 0.00  \\
243 & 95.25 $\pm$ 1.39  & 6.08 $\pm$ 0.08  & 0.52 $\pm$ 0.01 & 0.04 $\pm$ 0.00  \\
197 & 77.09 $\pm$ 1.12  & 5.98 $\pm$ 0.08  & 0.53 $\pm$ 0.01 & 0.04 $\pm$ 0.00  \\
150 & 58.52 $\pm$ 0.28  & 5.85 $\pm$ 0.03  & 0.53 $\pm$ 0.00 & 0.04 $\pm$ 0.00  \\
120 & 46.47 $\pm$ 1.31  & 5.76 $\pm$ 0.16  & 0.53 $\pm$ 0.02 & 0.03 $\pm$ 0.01  \\
100 & 38.99 $\pm$ 0.18  & 5.72 $\pm$ 0.03  & 0.54 $\pm$ 0.00 & 0.03 $\pm$ 0.00  \\
80 & 31.05 $\pm$ 0.15  & 5.62 $\pm$ 0.04  & 0.54 $\pm$ 0.00 & 0.03 $\pm$ 0.00  \\
60 & 23.15 $\pm$ 0.11  & 5.48 $\pm$ 0.04  & 0.55 $\pm$ 0.00 & 0.03 $\pm$ 0.00  \\
20 & 7.50 $\pm$ 0.04  & 4.96 $\pm$ 0.04  & 0.57 $\pm$ 0.00 & 0.02 $\pm$ 0.00  \\
\end{tabular}
\end{center}
\end{table}

\begin{figure}[t!]
\begin{center}
\includegraphics[width=0.95\textwidth]{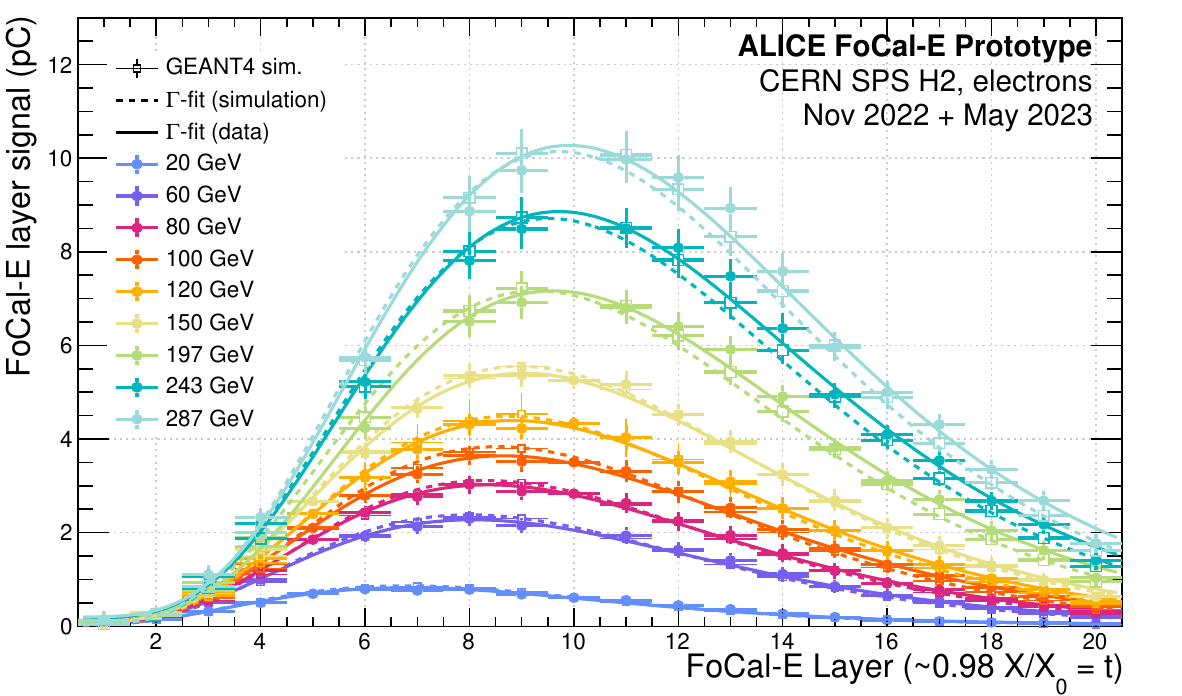}
\caption{
\label{fig:pad-long-shower-profile-all} Longitudinal shower profiles for 20--300~GeV electrons compared to \geant simulations and fitted with a $\Gamma$-distribution.}
\end{center}
\end{figure}

The maximum depth of the showers is given by: 
\begin{equation}
t_{\text{max}} = \frac{\alpha - 1}{\beta} = \ln\bigg(\frac{E}{E_c}\bigg) + C_{e,\gamma} \,.
\end{equation}
The left term is obtained after setting the derivative of \Eq{eq:gamma_distribution} with respect to $t$ to zero and solving for $t$, while the right term is related to the critical energy, $E_c$, at which energy losses by ionization start dominating over those by bremsstrahlung \cite{Workman:2022ynf}.
Here, $C_i$ (with $i \in \{e,\gamma\}$) is a constant which depends on the particle species: it amounts to $-0.5$ for electrons and $+0.5$ for photons.
For $E_c$ we choose a value of $8\MeV$, which is the critical energy of pure tungsten.
The results on the shower maximum studies are reported in the right panel of \Fig{fig:pixel-pad-combination-linear-and-shower-max}, as a function of the electron energy.
The longitudinal position of the shower maximum follows a logarithmic increase with energy, starting from around layer 7 at 20~GeV.
It reaches layer 9 at energies above $120$~GeV and layer 10 at energies above $250$~GeV.
This behavior reflects the theory predictions (for homogeneous calorimeters) and is in very good agreement with the expectation from \ac{MC} up to energies of 150\GeV.
For energies $\geq 200\GeV$ (\woLseven) the shower maximum is measured slightly deeper in the stack than predicted. 
The logarithmic dependence on the energy ensures a sufficient longitudinal shower containment in \ac{FoCal-E} also at higher shower energies.

A comprehensive summary of the measured longitudinal shower profiles is shown in \Fig{fig:pad-long-shower-profile-all}.
The data for electrons of 20--300~GeV are compared to \geant simulations, both fitted with a $\Gamma$-distribution.

\FloatBarrier 

\subsection{Pad-layer linearity and resolution}
\label{subsec:En_linearity}
As mentioned in \Sec{subssec:electrondataset}, two datasets were defined for the analysis of the detector performance:
One covering the full available electron energy range at H2~(20 to 300~GeV) but excluding layer~7~(\textit{\woLseven}), and one with all layers active (\textit{\wLseven}), but only in the energy range from 20 to 150~GeV.
In this section we measure the linearity and resolution for both datasets, hence evaluating the \ac{FoCal-E} performance up to electron energies of 300~GeV.

For the energy response of the pads to electrons the calibrated charge signal in all active pad layers is summed on event-by-event basis, after having applied the electron candidate selection criteria like described above.
\ifextrafigs
\red{Extra:}
\Figure{fig:pad-sum} shows the charge distributions for the 20, 60, 80, and 100~GeV electron datasets, compared to the expectation from \ac{MC}.
\begin{figure}[th!]
\begin{center}
\includegraphics[width=0.49\textwidth]{figures/pad/pad-sum-20GeV.pdf}
\includegraphics[width=0.49\textwidth]{figures/pad/pad-sum-60GeV.pdf}
\includegraphics[width=0.49\textwidth]{figures/pad/pad-sum-80GeV.pdf}
\includegraphics[width=0.49\textwidth]{figures/pad/pad-sum-100GeV.pdf}
\caption{\label{fig:pad-sum} 
\ac{FoCal-E} pad signal total signal sum 20~GeV~(top left), 60~GeV~(top right), 100~GeV~(bottom left), and 100~GeV~(bottom right), fitted with a Gaussian curve.}
\end{center}
\end{figure}
\fi

\begin{figure}[t!]
\begin{center}
\includegraphics[width=0.495\textwidth]{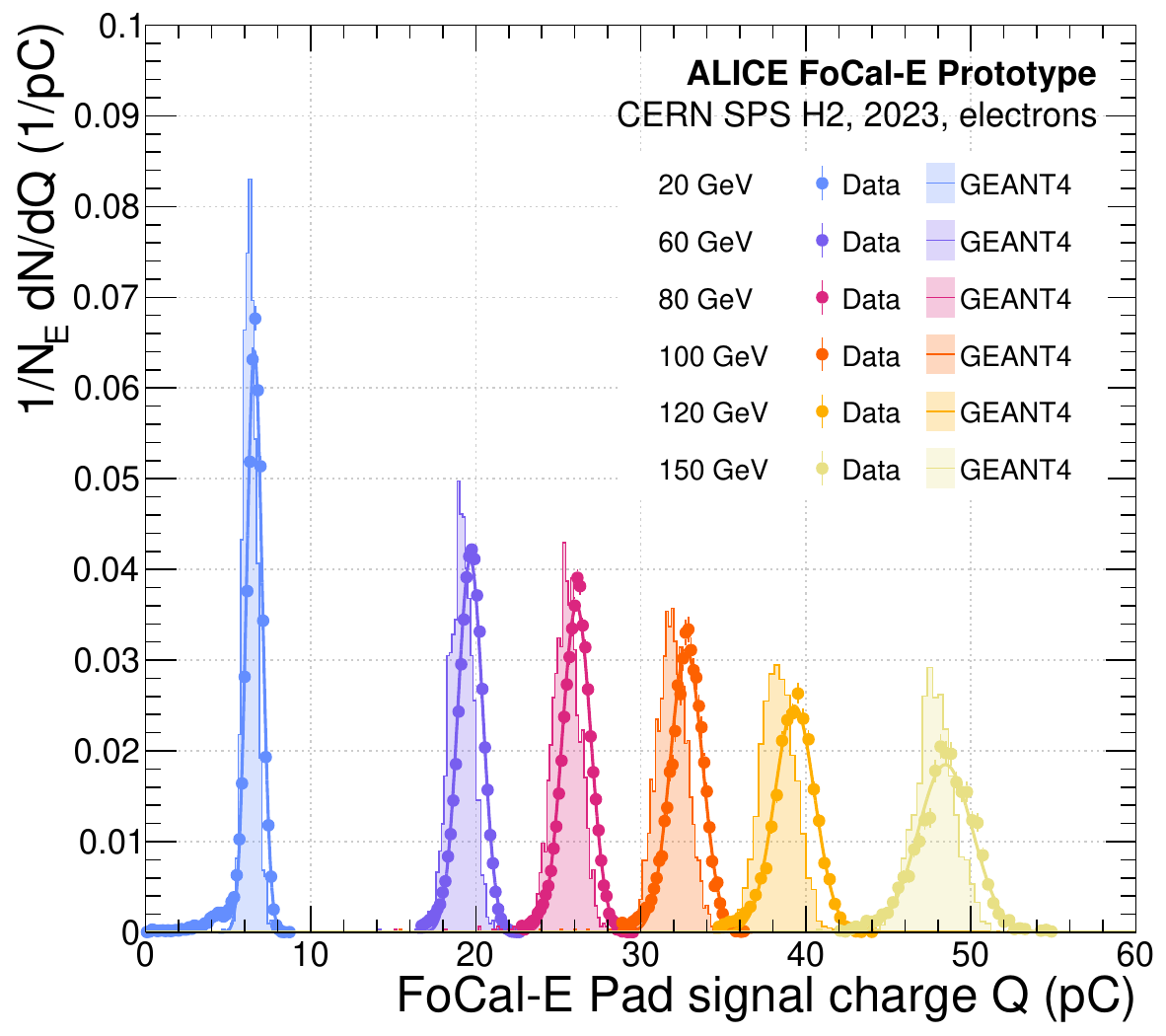}
\includegraphics[width=0.495\textwidth]{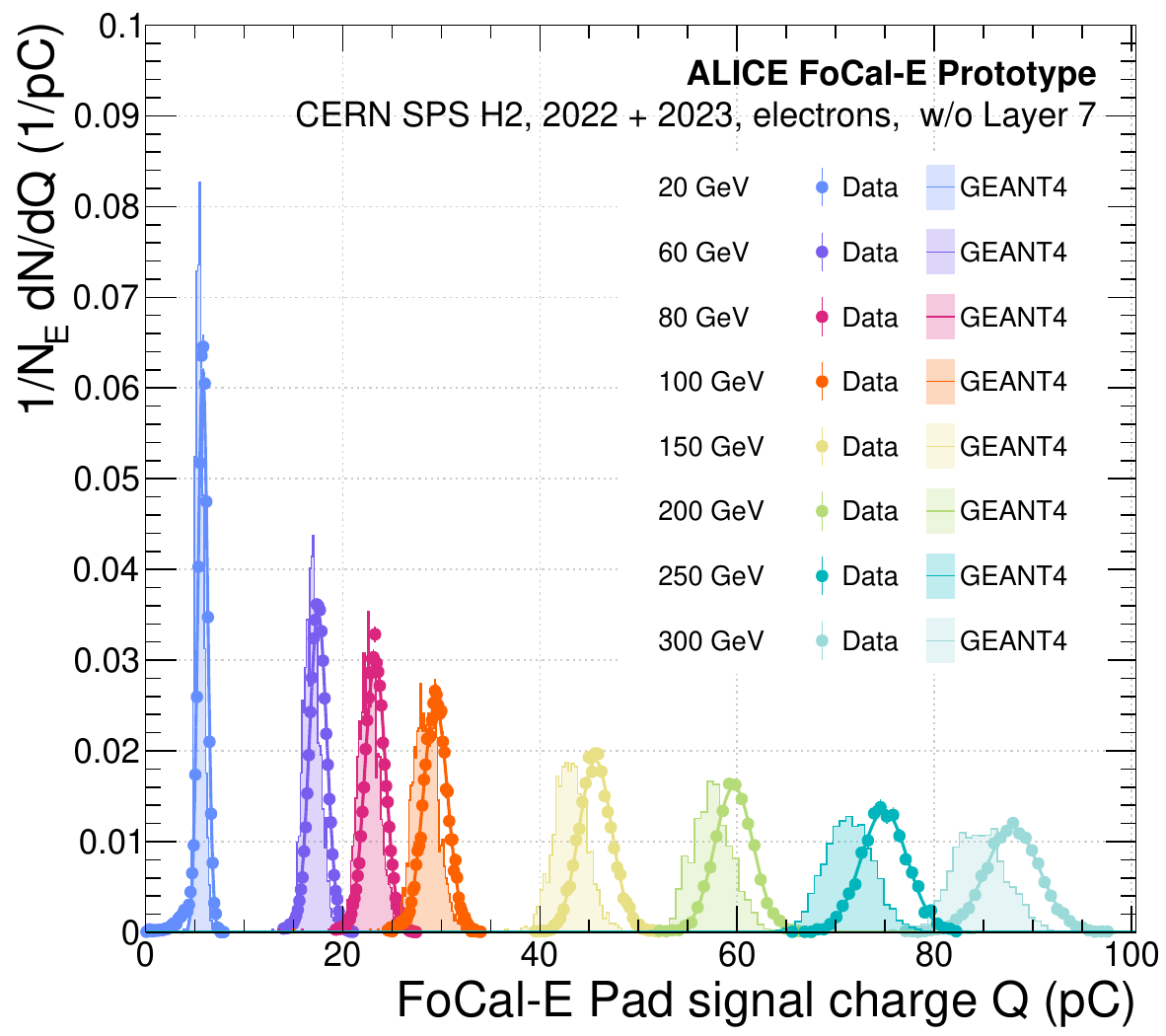}
\caption{
\label{fig:pad-charge-sum-all} 
\ac{FoCal-E} pad signal charge sum distributions for the May, 2023 dataset (left) and the combined dataset from November, 2022 and May, 2023 \woLseven~(right). 
The charge sums are fitted with a Gaussian curve in order to obtain mean and width of the distribution. Outlier points outside the 4-$\sigma$ region of the Gauss fits amount on average to 1--2\,\% and are not drawn for better visualization. The \geant simulated samples are drawn with the filled histograms. The bin width of the histograms increases with higher energies.}
\end{center}
\end{figure}

The pad signal charge sum distributions for the May, 2023 dataset and the combined dataset from November, 2022 and May, 2023 \woLseven\ are shown in \Fig{fig:pad-charge-sum-all}.
The measured charge distributions are well described by Gaussian curves.
For comparison, the distributions obtained from \geant\ simulations are also shown in each case.
The mean, $Q$, and the width, $\sigma_Q$, of the distribution are determined from the fitted Gaussian curves for each energy, both for data and simulation.
For the May, 2023 data, the data mean values and the expectation from Monte Carlo agree on the $2\,\%$ or better level.
For the November, 2022 data, there is a larger deviation~($\approx 5\,\%$).
Potential sources of the deviations and related uncertainties, e.g.\ from detector calibration and  misalignment of the \ac{HGCROC} phases could not be fully identified, and the deviation of $5\,\%$ was considered as acceptable.

\begin{figure}[th!]
\begin{center}
\includegraphics[width=0.65\textwidth]{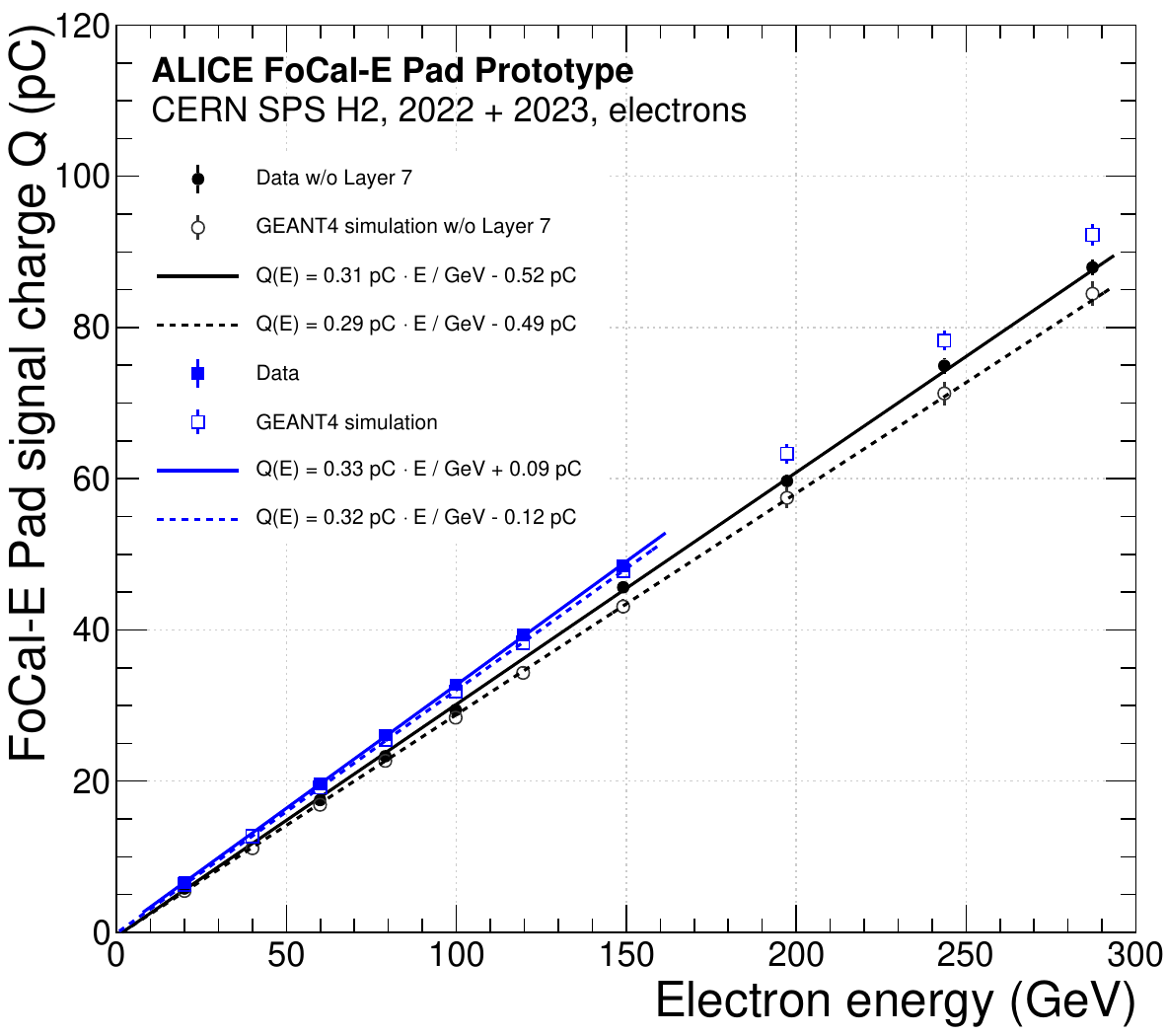}
\includegraphics[width=0.65\textwidth]{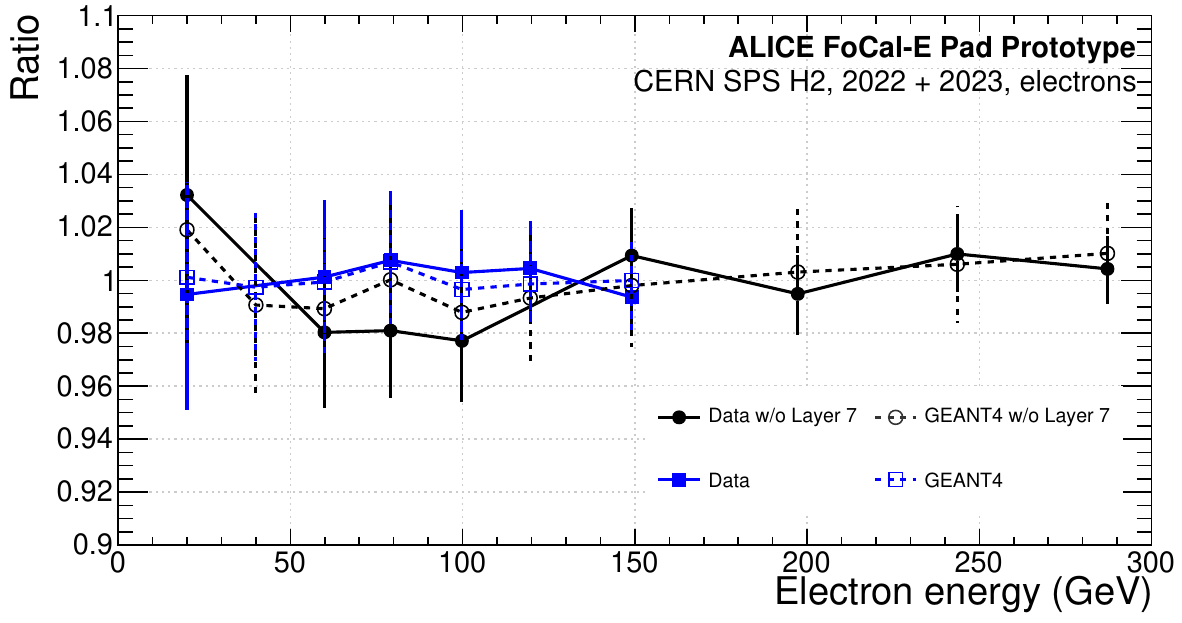}
\caption{\label{fig:pad-linearity} 
Top: Energy response for the \ac{FoCal-E} pad layers measured with the dataset \wLseven~(blue markers) and the combined dataset \woLseven~(black markers), compared to simulation (open markers), and respective linear fits. 
Bottom: Ratio of data and simulations to the respective fits.}
\end{center}
\end{figure}

The top panel of \Fig{fig:pad-linearity} shows the mean signal charge with respect to the electron energy for data and simulations, and in each case with and without layer~7.
Both, data and simulation are well described by a linear fit, $Q(E) = q \times E + Q_0$,
as also demonstrated in the lower panel of the figure.
The values for the parameters and their uncertainties are given in \Tab{tab:linearity}.

For the May, 2023 data, the slope parameter is fitted to $q = 0.33 \pC / \GeV$ with the constant offset of $Q_0 = 0.09 \pC$ compatible with zero, and in agreement with simulations.
For the combined dataset without layer~7, the detector response is still linear, despite differences in the response.
The parameter $q$ is fitted with 0.307\,pC/\GeV, which is $5\,\%$ higher than the expectation from \geant. 
The charge offset $Q_0=-0.52 \pC$, however, compatible with the simulation, deviates by $2\,\sigma_{Q_0}$ from zero.
Since the deviation of $Q_0$ is also observed in the simulation, it can be attributed to the missing layer~7 information.
The missing sampling layer is located in the shower maximum of a 20\GeV{ } shower, but not in the shower maximum of e.g.\ a 300~GeV shower.
Hence, the measured signal is weakened more at lower energies relative to higher energies.

\begin{table}[th!]
\caption{\label{tab:linearity}Parameters and their uncertainties for the fits of the linearity $Q(E) = q \times E + Q_0$ shown in \Fig{fig:pad-linearity}.}
\begin{center}
\begin{tabular}{ l |  c c c c }
 Dataset & $Q_{0} \pm \sigma_{Q_{0}}$ (pC)  & $q \pm \sigma_{q}$ (pC/GeV) & $\chi ^2$ & n.d.f. \\\hline\hline
 Data                      & 0.09 $\pm$ 0.34 	& 0.326  $\pm$ 0.005 	& 0.4 	& 4 \\
 \geant simulation         & -0.12 $\pm$ 0.22 	& 0.322  $\pm$ 0.003 	& 0.2 	& 8 \\\hline
  Data w/o L7              & -0.52 $\pm$ 0.26 	& 0.307  $\pm$ 0.003 	& 3.5 	& 6 \\
 \geant simulation w/o L7  & -0.49 $\pm$ 0.24 	& 0.293  $\pm$ 0.003 	& 1.0 	& 8 
\end{tabular}
\end{center}

\end{table}

\FloatBarrier

\label{subsec:En_resolution}

\begin{figure}[th!]
\begin{center}
\includegraphics[width=0.8\textwidth]{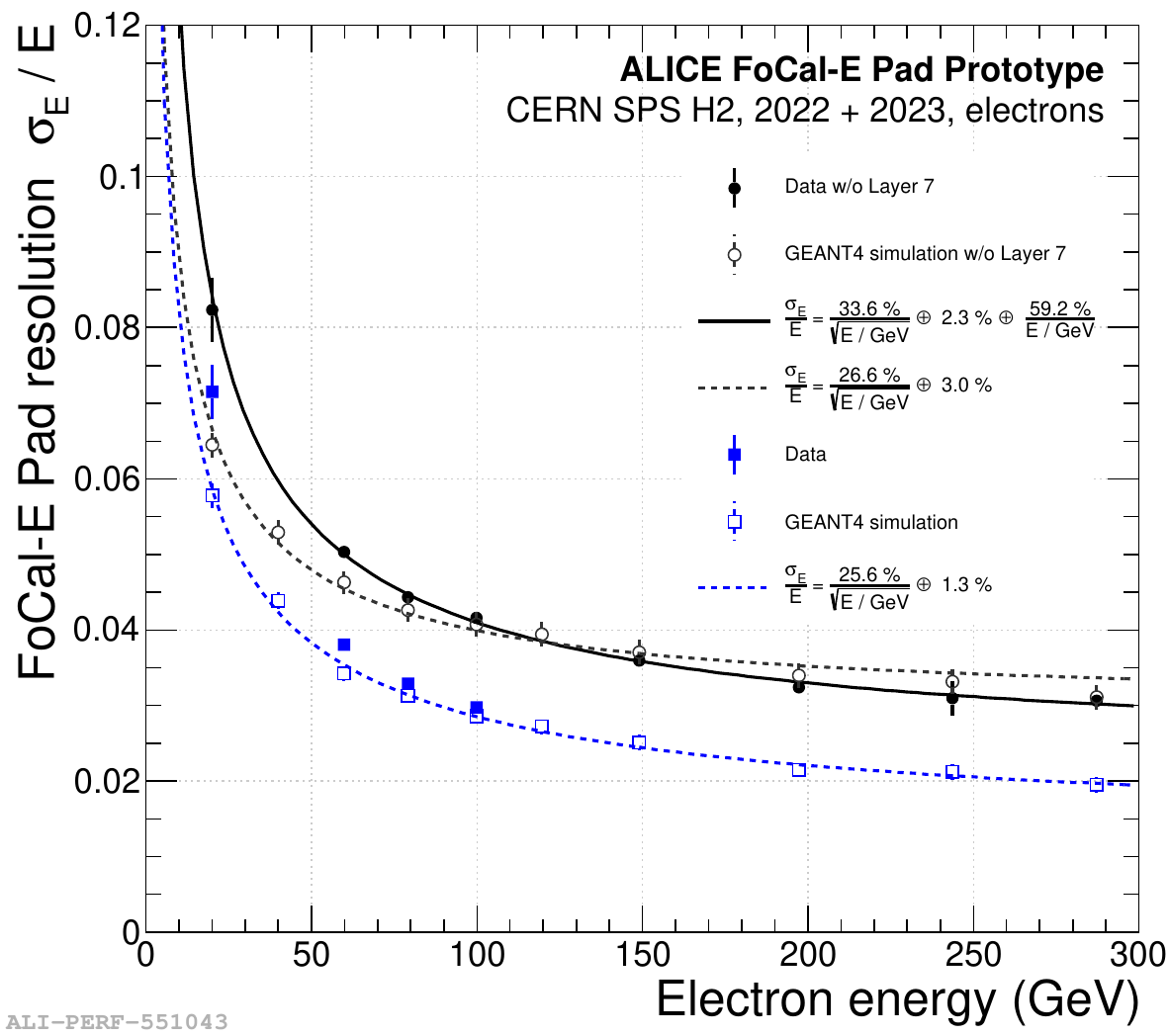}
\caption{\label{fig:pad-linearity-resolution-final} 
Relative energy resolution for the \ac{FoCal-E} pad layers measured with the May, 2023 data \wLseven~(blue markers) and the combined May, 2023 and November, 2022 data \woLseven~(black markers), compared to simulation (open markers), and respective fits.
}
\end{center}
\end{figure}

The relative energy resolution of the \ac{FoCal-E} prototype, $r$, is defined as
\begin{equation}
 r(E) = \frac{\sigma_{E}}{E} = \frac{\sigma_Q(E)}{Q(E)-Q_0} \, ,
\end{equation}
where $Q_0$ is obtained from the linear fits just discussed above.
The slope parameter $q$ cancels out in the derivation of $r(E)$.
\Figure{fig:pad-linearity-resolution-final} shows the relative energy resolution of the \ac{FoCal-E} pad layers for the two datasets and respective simulations using the mean and $\sigma$ of the Gaussian fits presented in \Fig{fig:pad-charge-sum-all}.
The energy resolution, which is commonly denoted as the result of the quadratic sum of three contributions, is expected to follow
\begin{equation}\label{eq:energy_reso}
    \frac{\sigma_E}{E} = \frac{\sigma_{\text{stoch.}}}{\sqrt{E/{\rm GeV}}} \oplus \sigma_{\text{const.}} \oplus \frac{\sigma_{\text{noise}}}{E/{\rm GeV}}
\end{equation}
where $\sigma_E$ and $E$ are given in GeV and $\sigma_i$ are without unit.
The first term takes into account the stochastic shower fluctuations, the second is a constant term representing general limitations of the detector design and calibration, and the last one describes the contribution of the electronic noise.
Numerical results from fits to the combined data, as well as the simulations, shown also in the figure, are listed in \Tab{tab:resolution}.

\begin{table}[th!]
\caption{Stochastic, constant, and noise term parameters for the fits of the energy resolution given in~\Eq{eq:energy_reso} and shown in \Fig{fig:pad-linearity-resolution-final}. The abbreviation ``w L7'' (``w/o L7'') refers to data collected with (without) the presence of a working silicon pad sensor at position 7 of the stack. Since noise was not implemented in the simulation, the noise term, $\sigma_{\text{noise}}$, is not listed for the simulation.}
\begin{center}
\begin{tabular}{ l c c c c c }
Dataset & $\sigma_{\text{stoch.}} (\%)$ & $\sigma_{\text{const.}} (\%)$ & $\sigma_{\text{noise}} (\%)$ & $\chi ^2$ & n.d.f. \\\hline
Data w/o L7 & 33.6 $\pm$ 1.4 	& 2.27  $\pm$ 0.10 	& 59.20  $\pm$ 35.3 	& 4.4 	& 5 \\
\geant simulation w/o L7 & 26.6 $\pm$ 0.9 	& 2.98  $\pm$ 0.13 	& - & 4.6 	& 8 \\
\geant simulation w L7 & 25.6 $\pm$ 0.5 	& 1.27  $\pm$ 0.10 	& - & 5.5 	& 8
\end{tabular}
\end{center}
\label{tab:resolution}
\end{table}

For the combined May, 2023 and November, 2022 datasets~(\woLseven), the resolution values for data and simulation start to agree within the uncertainties from energies of 80~GeV and higher.
For energies $\geq 200\GeV$ we observe the resolution in data and simulations to be nearly constant at $\approx 3\,\%,$ with no obvious trend towards lower values, fulfilling already the required performance of 5\%~\cite{CERN-LHCC-2020-009}.
The points at 20 and 60~GeV deviate significantly from the simulated values, i.e.\
the measured resolution is 1\,\% higher~(in absolute values) than the simulation. 
One possible explanation is that the charge measurement in this energy regime~(especially at 20\,\GeV) is dominated by pad charges in the order of $\approx 0.5\fC$, which lies close to the transition region of the \ac{ADC} and \ac{ToT} range. 
Thus, this charge regime could be affected by additional noise sources, e.g.\ from the turn-on of the \ac{ToT}, which are not so dominant in the higher energy regimes. 
Furthermore, the \ac{ToT} turn-on region is not well described by the calibration procedure because of the non-linearity there.
The fitted curve to the measured resolution data points is given by
\begin{equation}
 \frac{\sigma_E}{E} = \frac{33.6\%}{\sqrt{E/\GeV}}  \oplus 2.3\% \oplus \frac{59.2 \%}{E/ \GeV} \,.
\end{equation}
The fitted parameter for the stochastic (constant) term disagrees with the Monte Carlo by $+26\,\%$~($+31\,\%$), corresponding to a $4.8\,\sigma$ ($4.4\,\sigma$) deviation.
However, this mismatch of the fit parameters likely arises from the worse measured resolution at the low-energy data points.
In the high energy region, which reflects the constant term better than the low energy region, the performance is found to be as expected from simulation.

For the May 2023 dataset (\wLseven), the resolution points are displayed for 20, 60, 80, and 100~GeV with the blue markers using all 18 pad layers.
The resolution of the 18 pad-layer stack is measured to be below 3\,\% at an electron energy of 100\,\GeV.
We excluded the data for 120 and 150~GeV, for which we measured the resolution to be $\approx 3.1\,\%$ and $\approx 3.3\,\%$, respectively.
This unexpected result was investigated carefully, and could be traced back to additional \ac{ToT} noise picked up from the switching of the 40~\MHz\ clock in the \ac{HGCROC}.
This might occur if the detector is operated in an unfavorable timing condition.
The effect may be enhanced by the presence of the test beam trigger board as we did not observe such an effect in November, 2022.
The measured resolution points in the range of 20 to 100~GeV have a similar deviation to simulations as observed for the \woLseven\ case.

With this knowledge, the performance of the detector can be extrapolated to higher energies.
We thus assume that the 18-pad layers prototype may reach an energy resolution performance for electromagnetic showers which follows roughly the fitted resolution curve from simulations at the high tail
\begin{equation}
  \frac{\sigma_E}{E} = \frac{25.6\%}{\sqrt{E/\GeV}} \oplus 1.3 \%  \,,
\end{equation}
plus a potential noise term similar to the \woLseven\ case.
We confirmed this projection by scaling the resolution curve fitted to the \woLseven-data with the ratio of the \wLseven-\ac{MC} to the \woLseven-\ac{MC}.
The resulting curve (not drawn) describes the measured data between 20 and 100\GeV\ well, and is in good agreement with the \wLseven-\ac{MC} at energies above 100\GeV.
We note that for higher electron energies of $\gtrsim 500\,\GeV$, longitudinal shower leakage effects will play a larger role.
This effect would lead to a reduction of linearity and resolution of the \ac{FoCal-E} standalone detector as it is, but can be compensated with the measurement of the leaked shower components by \ac{FoCal-H}.

This study does not incorporate the contribution of the pixel layers to the linearity and resolution of \ac{FoCal-E}.
We have conducted studies on simulation level which indicate that in the lower energy regime ($\approx 20\GeV$) the signal of pixel layers, in terms of number of pixel hits, contributes linearly to the overall \ac{FoCal-E} signal, and the energy resolution can potentially be improved in this energy regime.
At higher energies ($\gtrsim 100\GeV$) effects of pixel hit saturation in the shower center play a non-negligible role, and lead to loss of linearity in the \ac{FoCal-E} signal, thus not improving the resolution.




\FloatBarrier


\subsection{Pixel transverse shower profiles}
\label{subsec:trans_shower_profiles}
The two pixel layers of \ac{FoCal-E} at layer 5 and 10 with a pixel pitch of about $30\um \times 30\um$ make it possible to resolve the structure of particle showers on the sub-millimeter scale.
In particular, this feature will be used to discriminate single photon (electromagnetic) shower events against background two-shower events from \pizero\ decays~\cite{ALICE:2023rol}.

\begin{figure}[ht!]
 \includegraphics[width=0.49\textwidth]{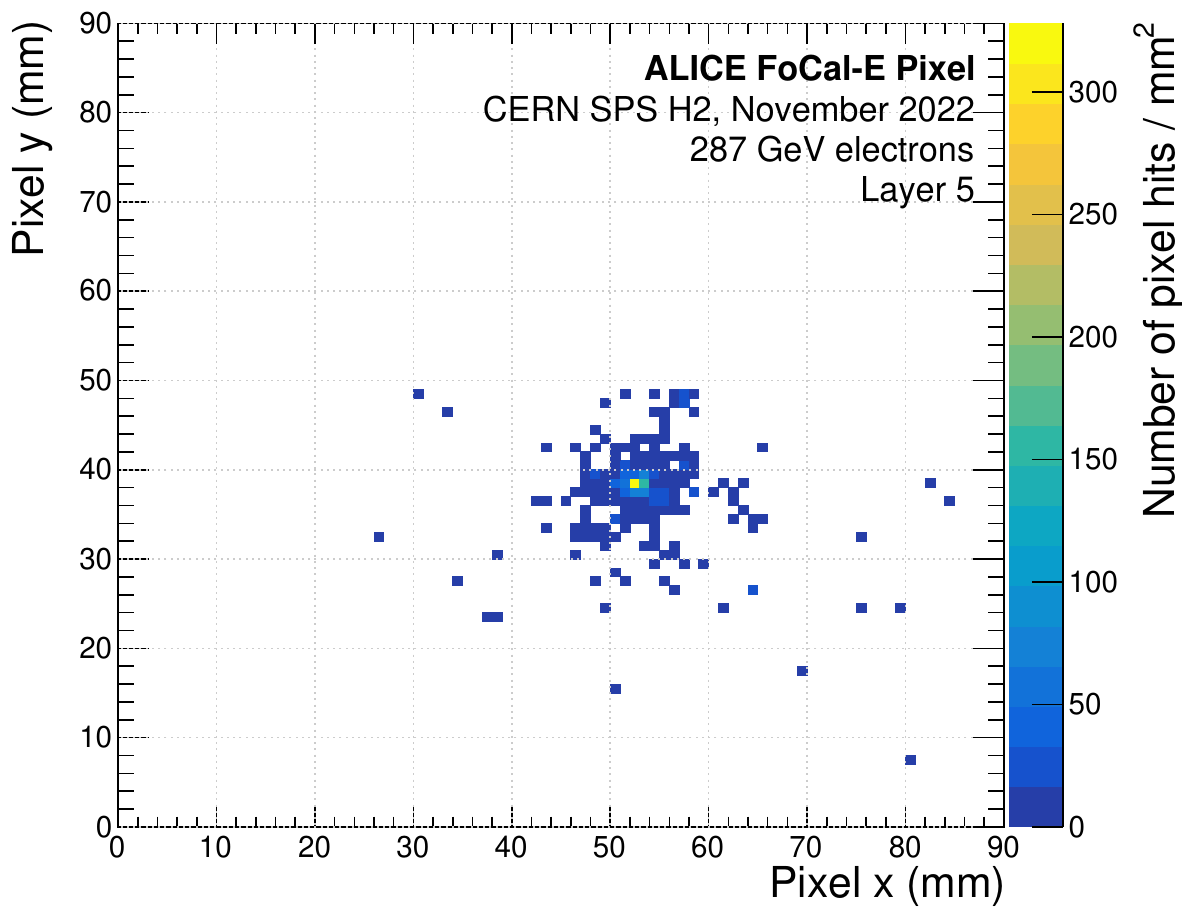}
 \includegraphics[width=0.49\textwidth]{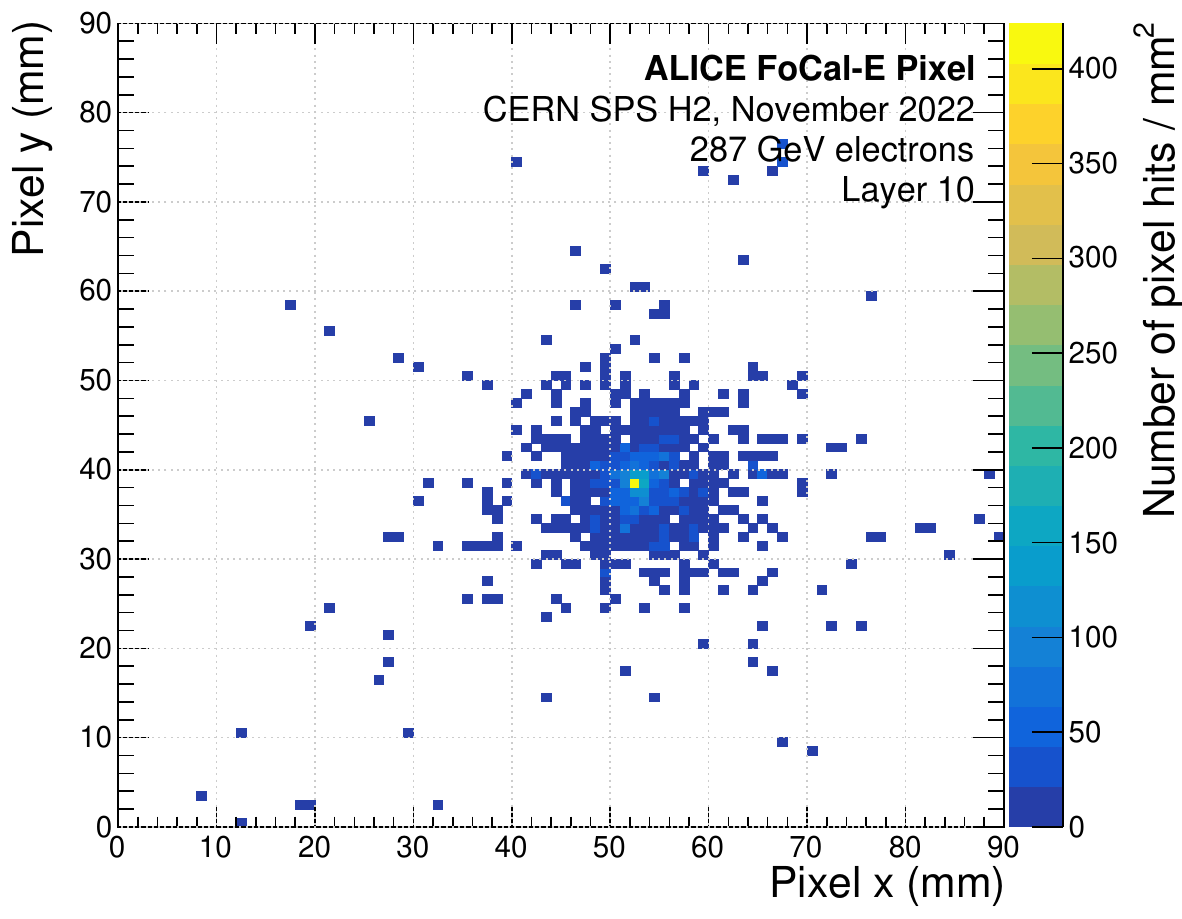}
 \caption{\label{fig:pix-event-display-1e-300GeV}
 \ac{FoCal-E} pixel event display of a 287~GeV one-electron shower in layer 5~(left) and layer 10~(right).}
\end{figure}

As illustration, \Fig{fig:pix-event-display-1e-300GeV} shows an event display of a 287~GeV electron shower event in pixel layer 5~(left) and 10~(right). 
The showers are characterized by pronounced cores in the shower center with hit densities higher than 300~(400) pixel hits/mm$^2$ in layer 5~(10), and are surrounded by tail components with less dense pixel hit occupancy.

\begin{figure}[ht!]
 \includegraphics[width=0.49\textwidth]{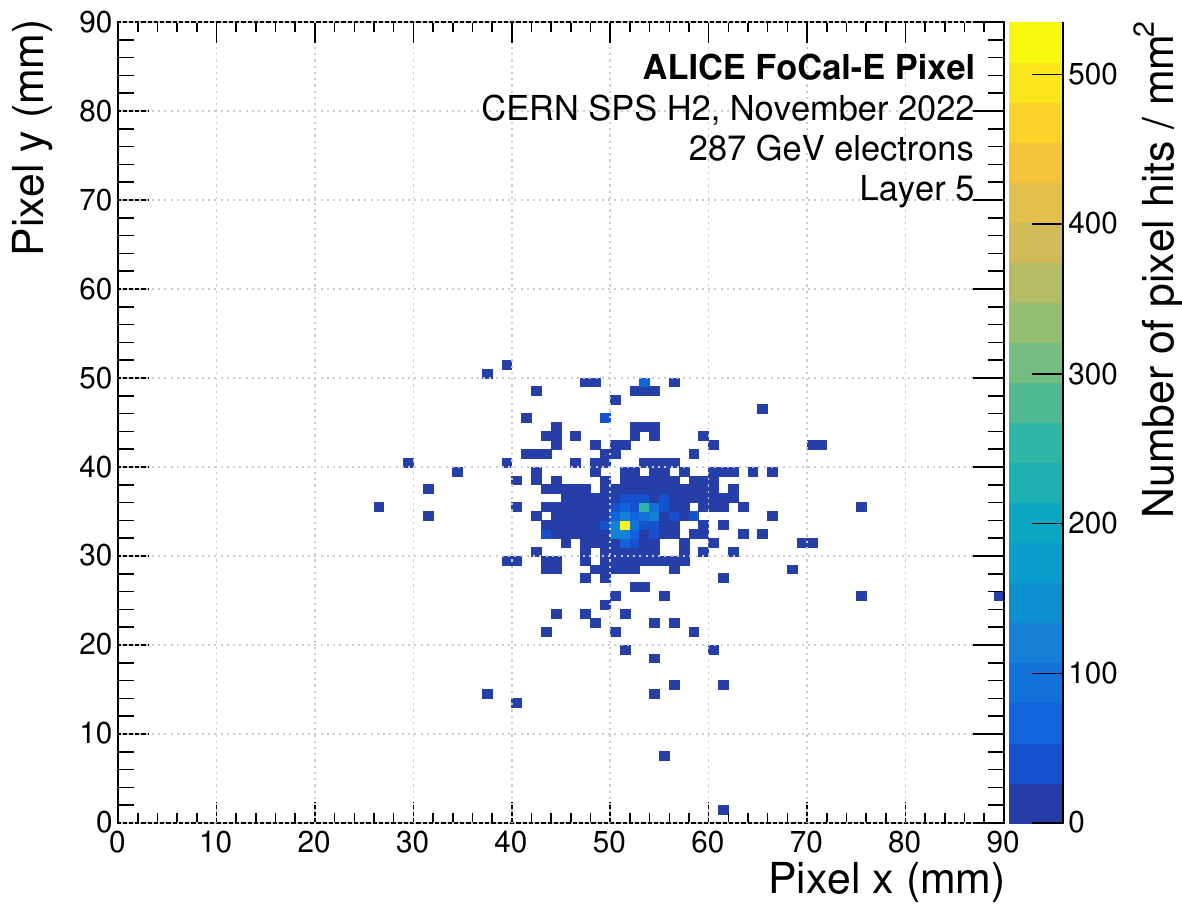}
 \includegraphics[width=0.49\textwidth]{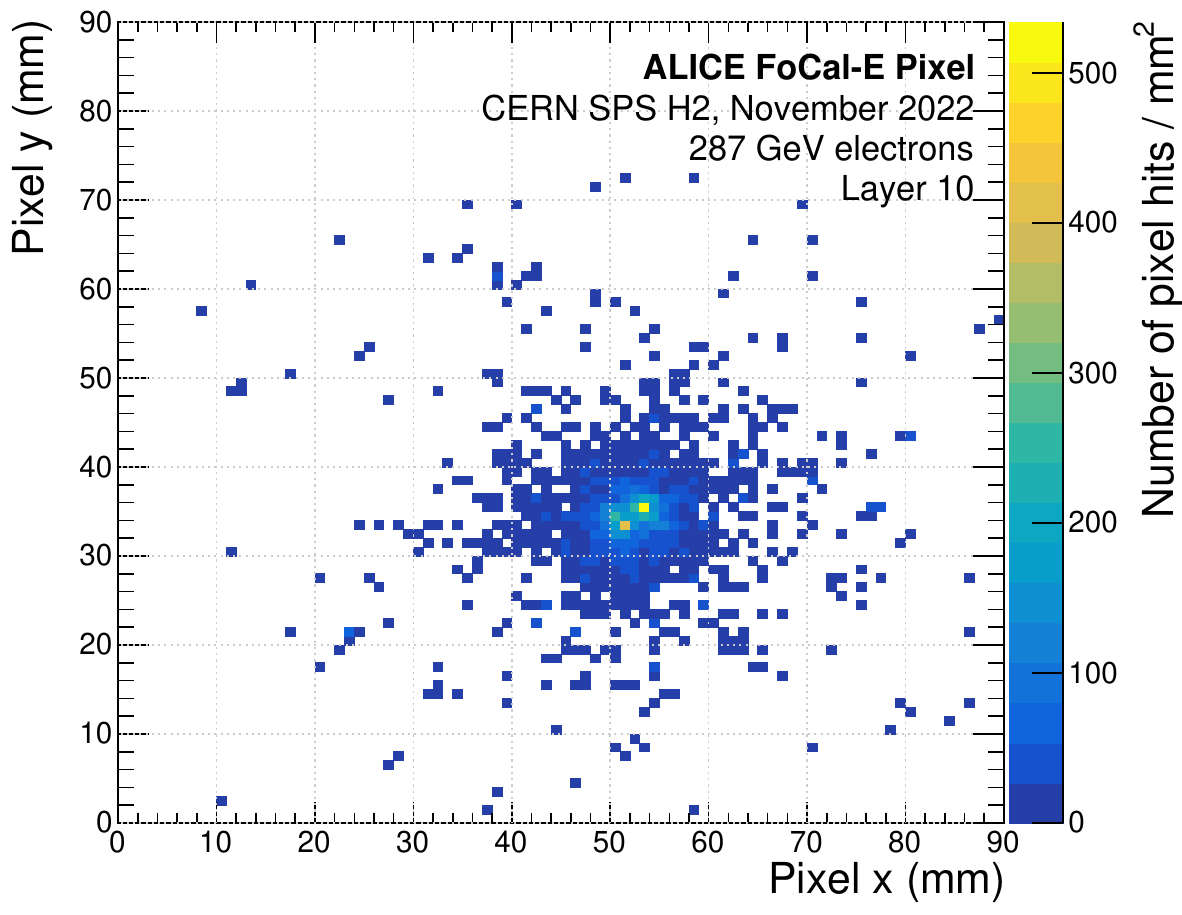}
	\caption{\label{fig:pix-event-display-2e-300GeV}\ac{FoCal-E} pixel event display of a 287\,\GeV{ }two-electron shower in layer 5 (left) and layer 10 (right).}
\end{figure}

As discussed in \Sec{subsec:pixel-event-selection}, due to the high electron rate at the \ac{SPS} H2 beam line and the relatively long integration time of the \ac{ALPIDE} pixel front-end ($\approx 5\us$), we also recorded multiple electron shower events, as shown in \Fig{fig:pix-event-display-2e-300GeV}.
Two electron-showers are clearly visible, and they can be separated by eye on a scale lower than $1\cm$.
The event display also illustrates how varying the longitudinal development of a shower in the calorimeter can be: in layer 5 the left shower produces higher occupancy in the core, and in layer 10 the right shower deposits more hits.
In the final detector, rather complex clustering, particle identification and shower separation algorithms will be implemented, 
to make a decision on a)~whether there are two nearby showers in the event, and b)~how to share the energy fraction between the two showers.
Since for the discrimination of two nearby showers the lateral hit density profiles are one of the key parameters, we present here a study of the transverse electromagnetic shower profile obtained with the pixel layer information.

In order to study the shower profile, we project the hit density distribution to a lateral axis in the pixel layer plane, and measure the functional form $f(\Delta x)$ of the number of pixel hits, \Nhits, in dependence on the lateral distance $\Delta x$ from the shower center $x_0$ as
\begin{equation}
 f(\Delta x) = \frac{1}{\Nhits} \frac{d}{dx} \Nhits (x - x_0) \,,
\end{equation}
where the $x$-axis is identical to the horizontal direction in the setup.
When calculating the distance from a shower-particle hit $i$ from the shower center, $\Delta x_i = x_i - x_0$, the dominating uncertainty originates from the uncertainty on the shower center $x_0$, i.e.{ }$\sigma_{\Delta x} \approx \sigma_{x_{0}}$, whereas each single point $x_i$ is measured with high spatial resolution.

\begin{figure}[t!]
\includegraphics[width=0.495\textwidth]{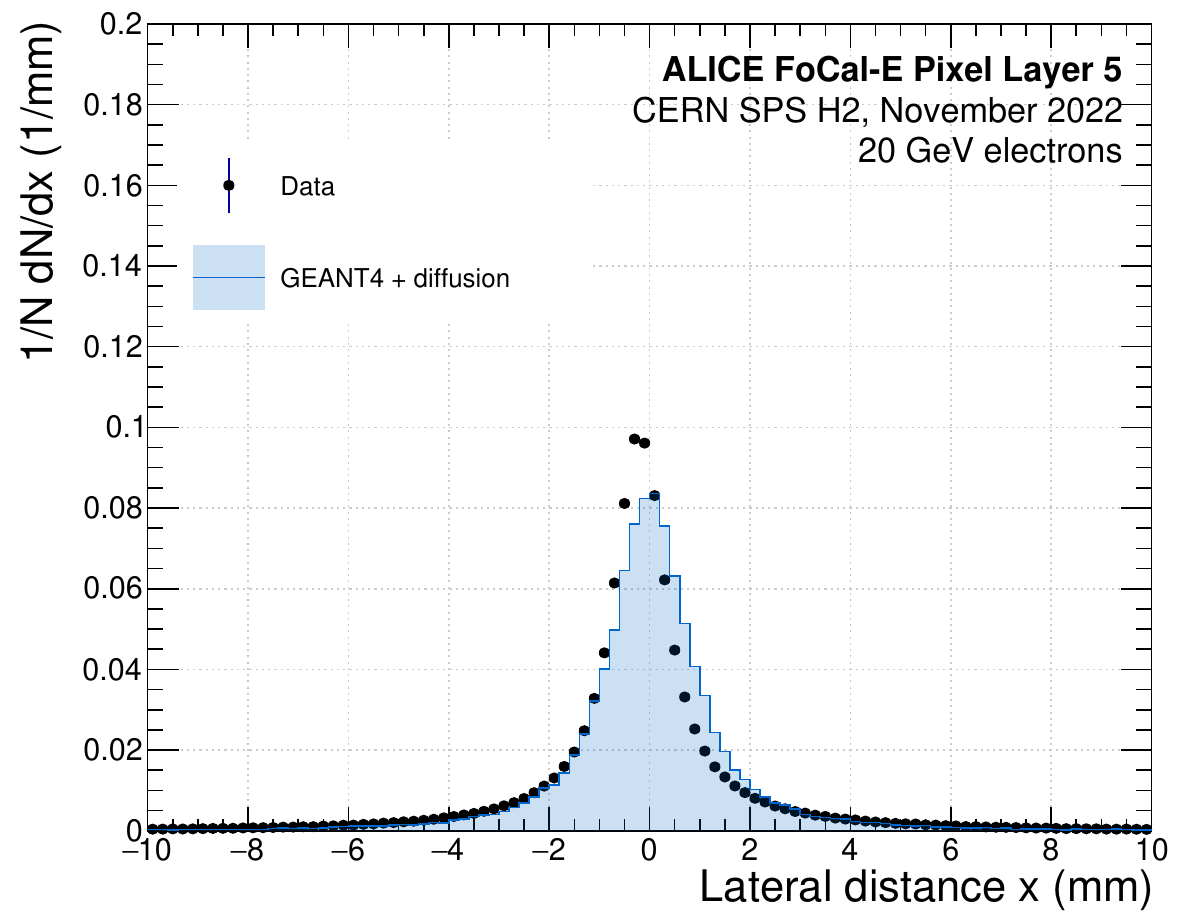}
\includegraphics[width=0.495\textwidth]{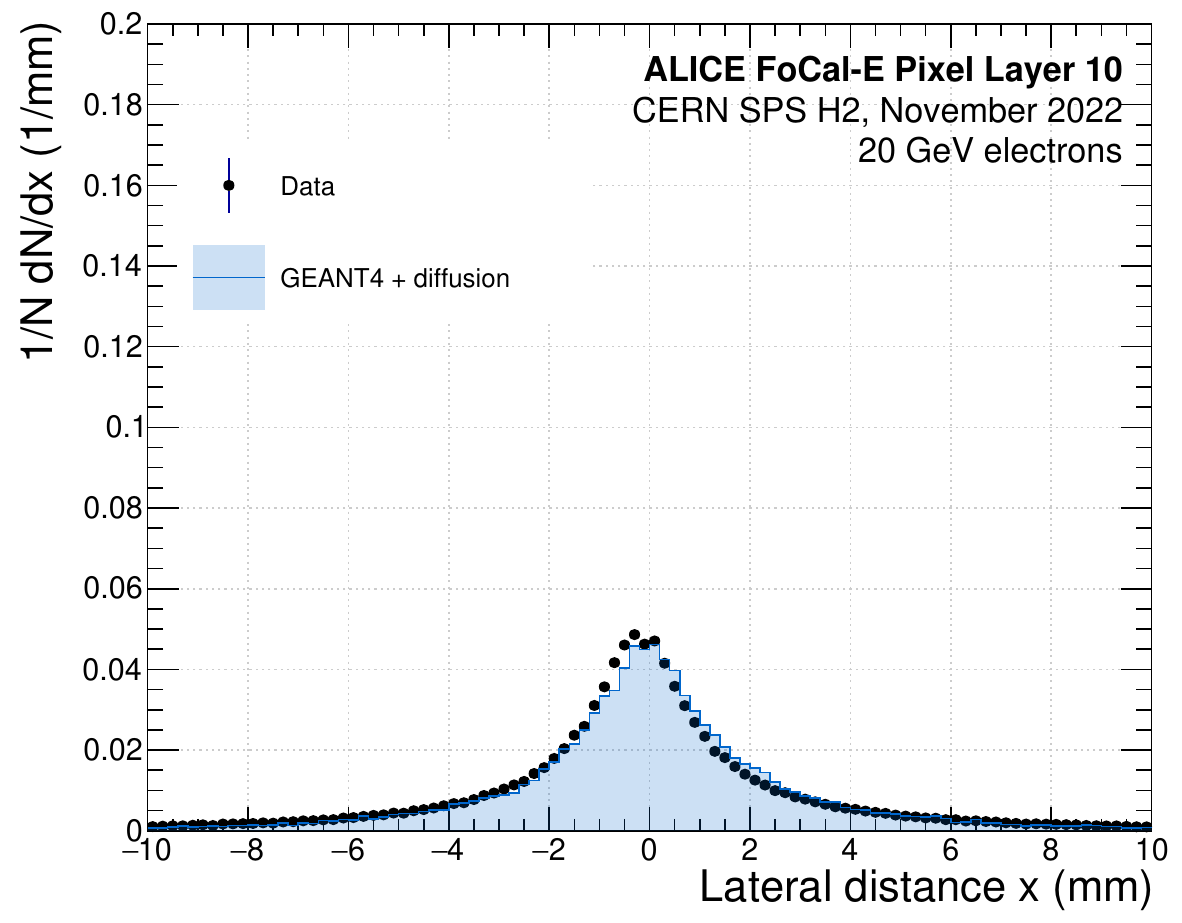}
\includegraphics[width=0.495\textwidth]{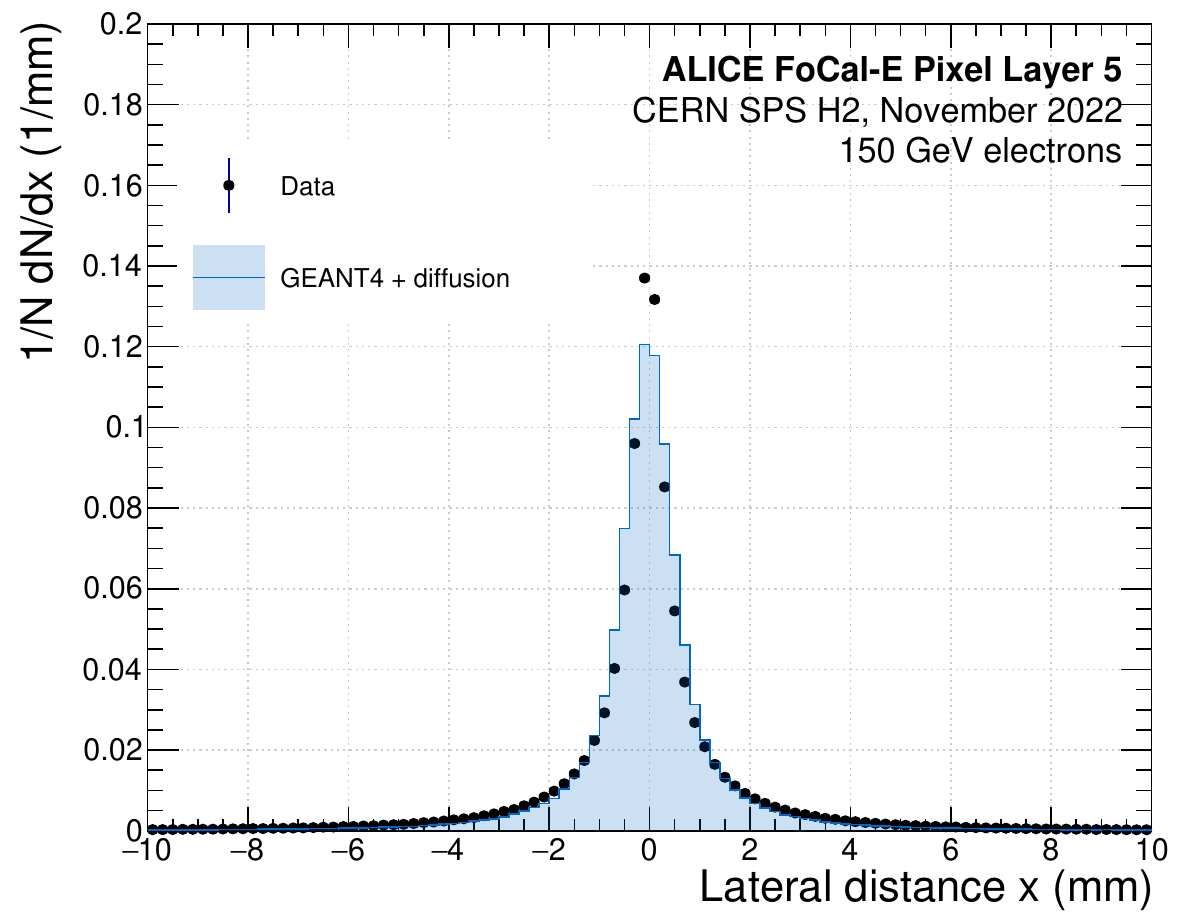}
\includegraphics[width=0.495\textwidth]{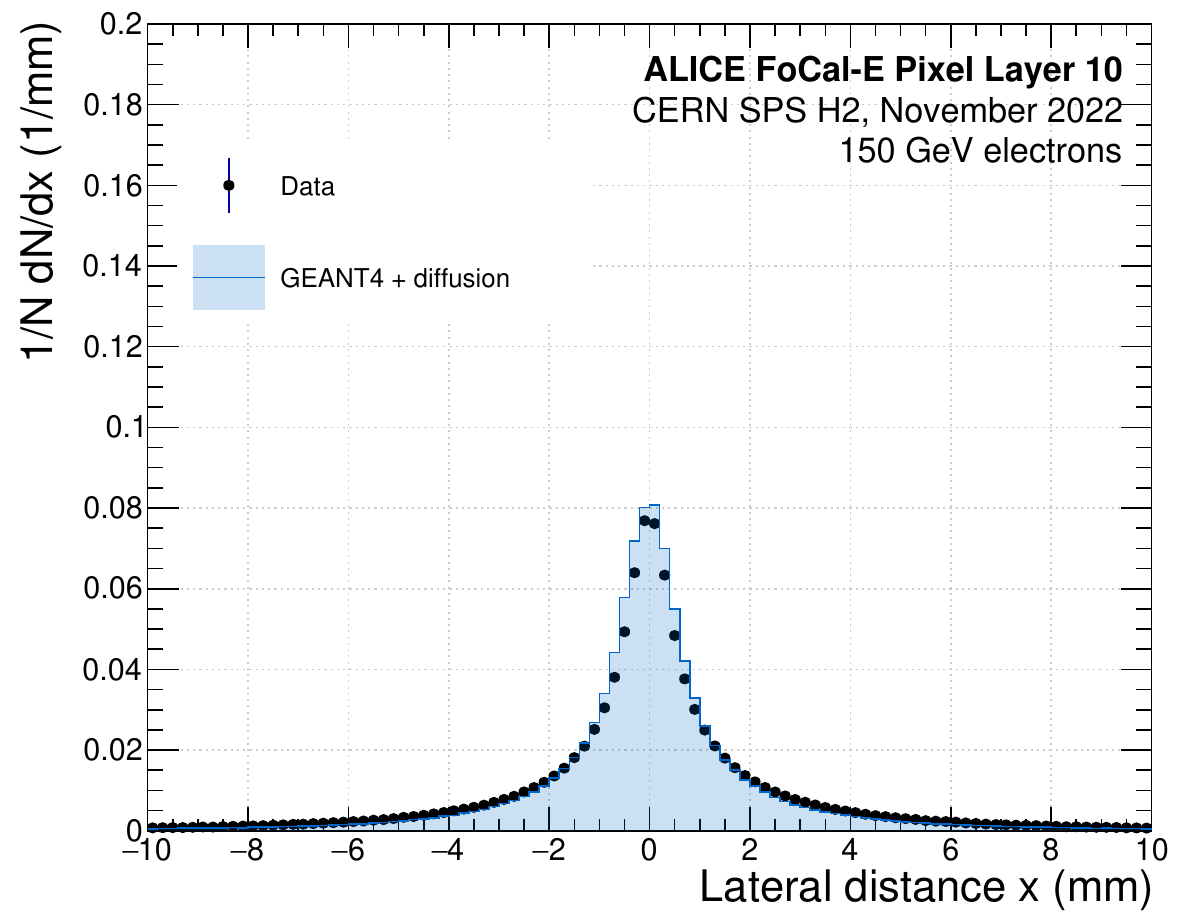}
\includegraphics[width=0.495\textwidth]{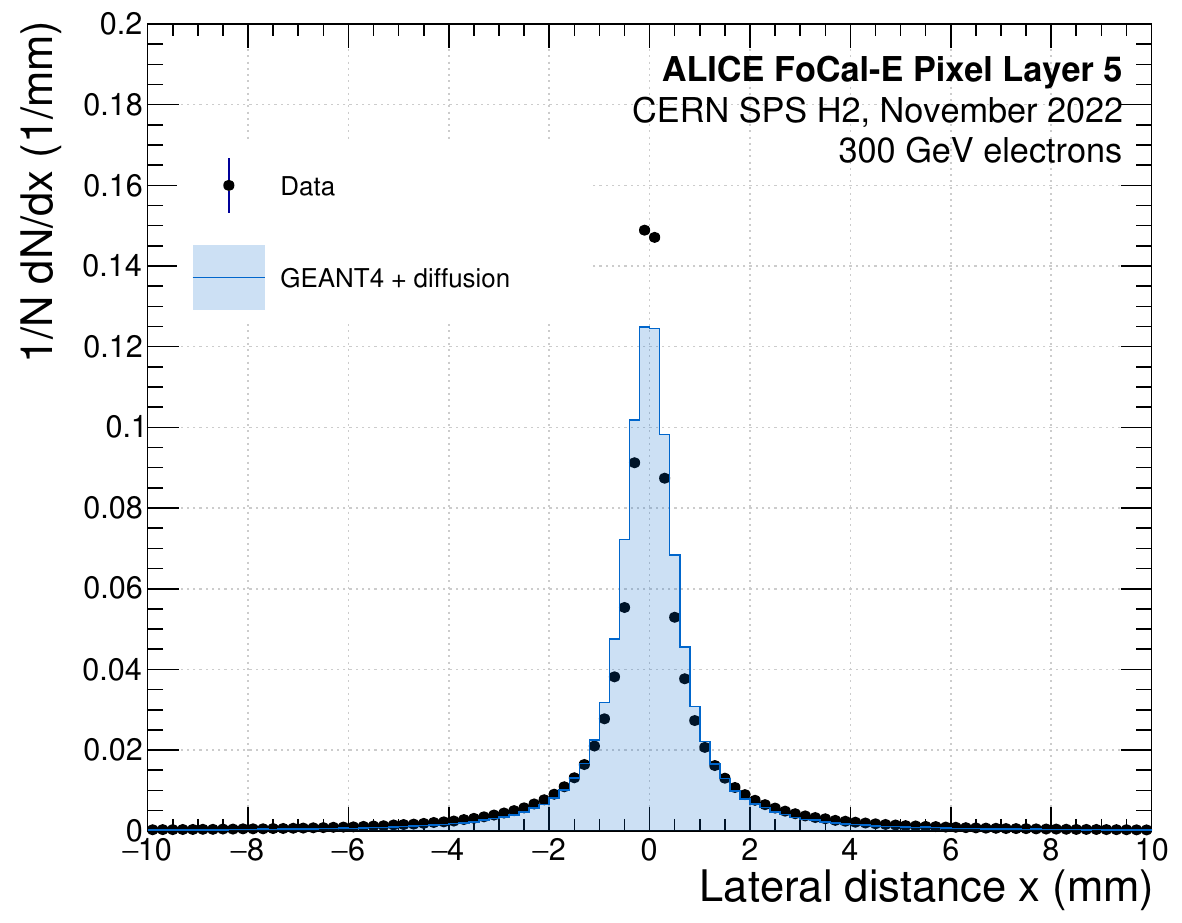}
\includegraphics[width=0.495\textwidth]{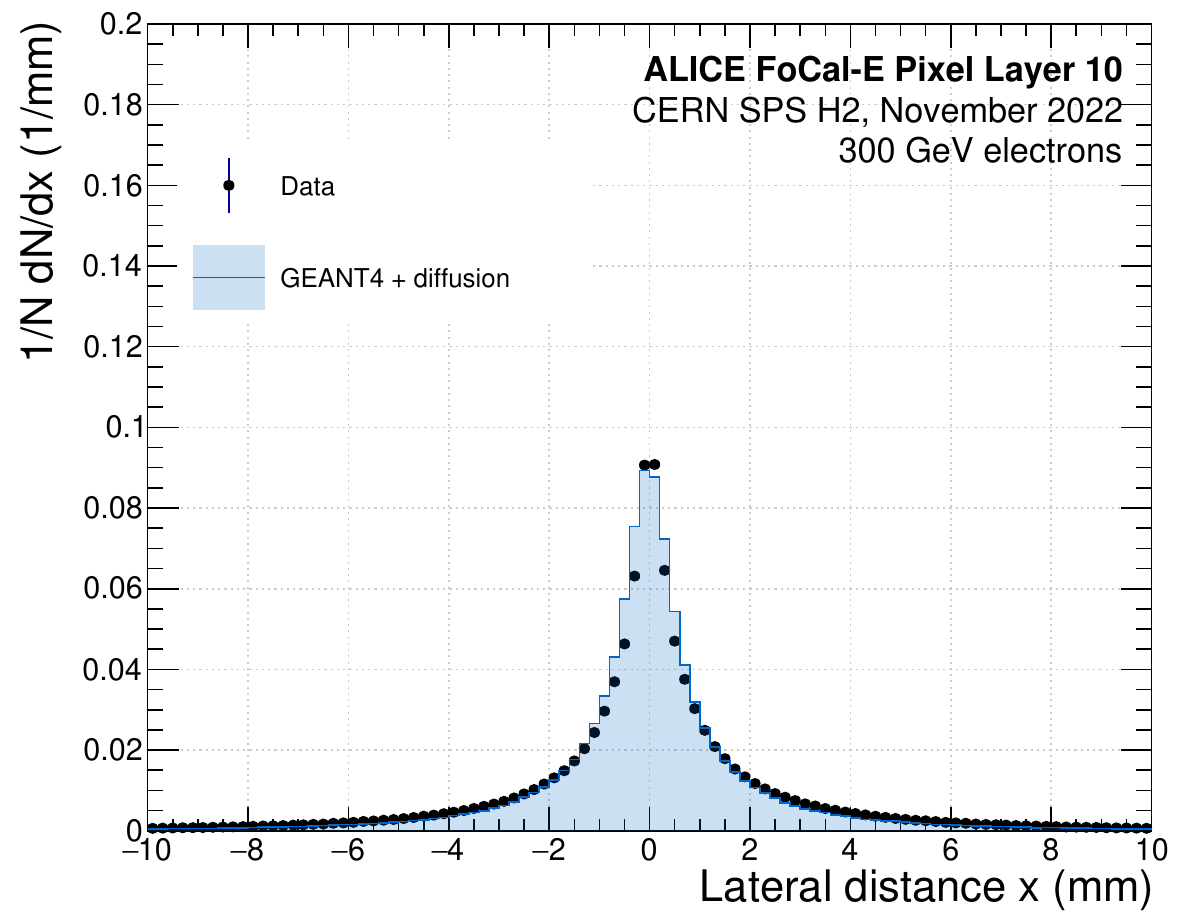}
\caption{\label{fig:pixel-swidth-lin}
Measured and simulated lateral shower profiles for pixel layer 5~(left panels) and 10~(right panels) for 20, 150 and 300~GeV electrons.
The bin width is $200\um$. 
}
\end{figure}

To determine an approximate position for the shower center, pixel hits from a single event are filled into a histogram~(or ``hitmap''), which is binned into macro-pixels of size $40\times40$~pixels~($\approx 1.1\times 1.1\,\mathrm{\mm^2}$).
We calculate the shower center from the weighted mean position of the macro pixels, where we take only macro-pixels into account which contain a minimum number of hits $\Nhitsmin$, and which lie within a distance, $d$, lateral to the bin with the highest number of entries.
We chose $d = 10\;\text{macro-pixels}$ ($\approx 1.1\,\mathrm{cm}$) and \Nhitsmin = 4.
The shower center $(x_0,y_0)$ is calculated event-by-event, and the final distribution, $\mathrm{d}\Nhits / \mathrm{d}x (x)$ is filled with pixel hits integrated in a range $|y_i - y_0 | < 0.5\mm$.
Since this method for the determination of the shower center depends on the number of pixel hits (not particle hits), the more realistic pixel response to particle hits in the \geant simulation is used, which is introduced in \Sec{subsec:pixel-layer-hits}.
From simulation, we derive an uncertainty in the determination of $\sigma_{x_{0}}\approx0.3$~mm with respect to the impact position of the electron, which is in the order of the standard deviation of a uniform probability distribution in the chosen bin width (i.e.~$1.1\mm/\sqrt{12}=0.3\mm$).

\begin{figure}[t!]
\includegraphics[width=0.49\textwidth]{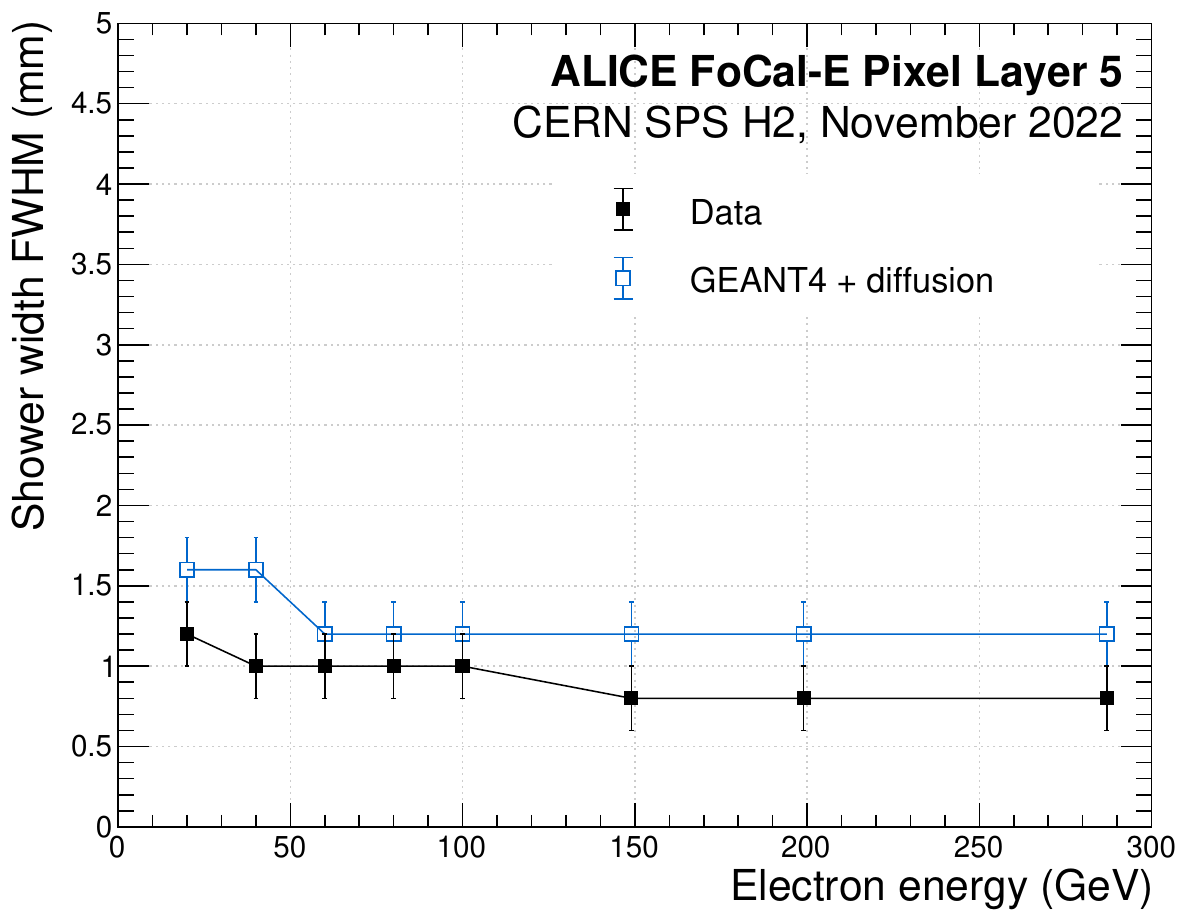}
\includegraphics[width=0.49\textwidth]{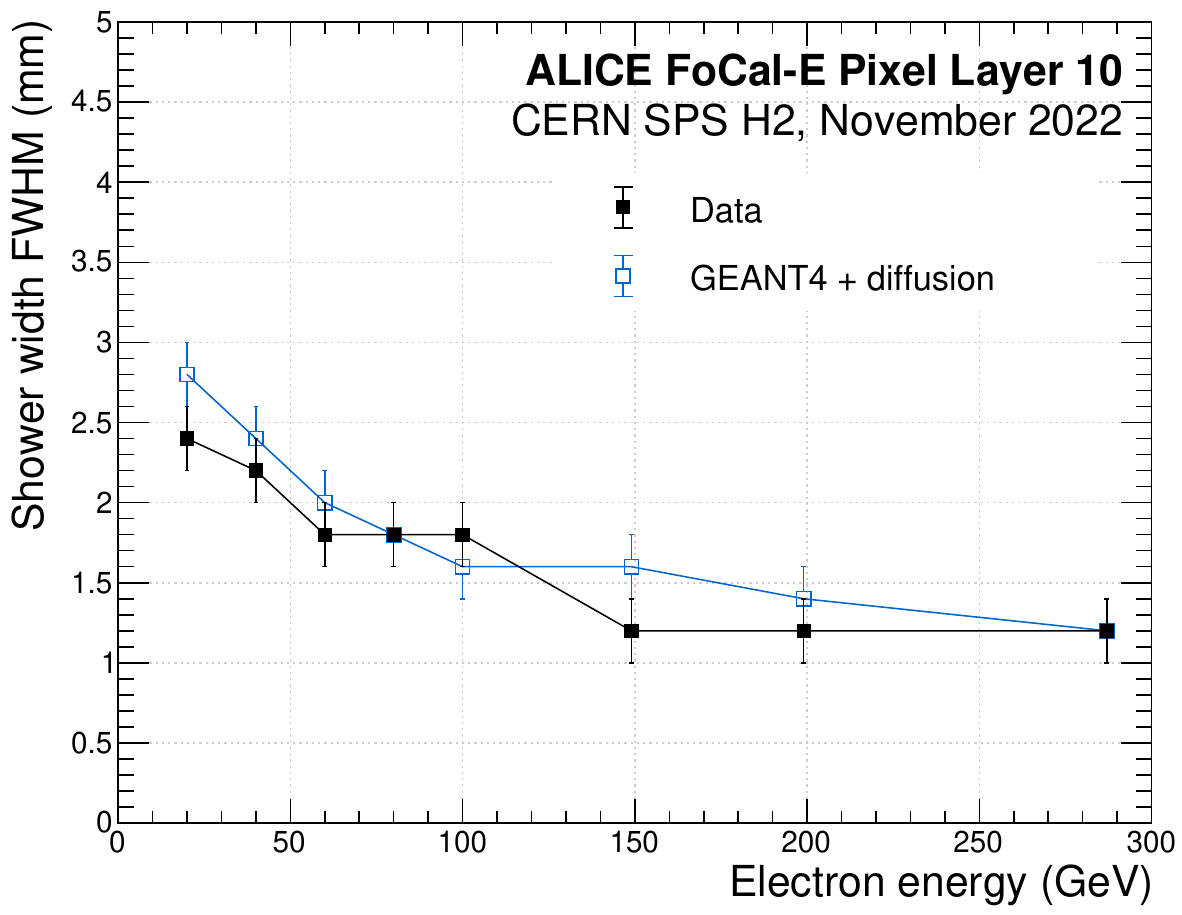}
\caption{\label{fig:pixel-swfwhm}
Measured and simulated \ac{FWHM} for layer 5~(left panel) and 10~(right panel) versus electron energy.
The error bars represent an uncertainty of 0.2\mm.
}
\end{figure}

\Figure{fig:pixel-swidth-lin} shows the measured and simulated transverse shower profiles in layers~5 and 10.
The distributions are characterized by a sharp peak in the center (\textit{core}) and exhibit broader side bands~(\textit{tails}).
As a first-order estimate on which scale a discrimination against another nearby shower should be generally possible, we evaluate the \ac{FWHM} of the measured distributions by finding the maximum bin, and the bin position where the distributions drop below the half of this maximum.
\Figure{fig:pixel-swfwhm} shows the \ac{FWHM} for layer~5 and 10, compared to simulations as a function of electron energy.
From the resulting distributions we obtain an \ac{FWHM} of 1.2\mm\ for electron energies of 20\GeV, and the \ac{FWHM} drop with increasing energies down to 0.8\mm\ for electron energies of 300\GeV.
The \ac{FWHM} values are significantly smaller in layer 5 than in layer 10, which is expected because of the larger transverse spread at higher shower depths.
In layer 10 the \ac{FWHM} values for higher energy electrons are measured to be in the order of $\approx 1.2\mm$, which is up to a factor of 2 higher than in layer 5, and the \ac{FWHM} values increase towards lower electron energies, reaching 2.5\mm\ at 20\GeV.
The simulation dataset --- after having applied the cluster generation model~
--- was analyzed with the same criteria for the determination of the shower center like the data.
We compare the measured values to the simulated \ac{FWHM} values with the diffusion model (introduced in \Sec{subsec:pixel-layer-hits}) applied.
These values in layer 10 are in good agreement with the data in layer 10, but systematically higher than the data in layer 5.
However, the measured and simulated \ac{FWHM} of the transverse shower widths are in general consistent within $\approx 0.5\mm$ or better.
The small discrepancies between data and simulation in layer 5 will be subject to further studies.

The analysis presented above is sensitive to the method of how the shower center is determined.
However, it produces stable results in the full electron energy range tested in beam.
For the higher energies a finer binning in the initial hitmap could be used in order to improve the determination of the shower center and thus the resolution of the shower width.
However at lower energies where the hit density in the shower is lower, our method breaks down if the bin width is decreased to or below the scale of the hit density.
Other experiments (e.g.~\cite{Peitzmann:2022asy}) have circumvented the uncertainty on the determination of the exact shower location by making an independent position measurement of the incident electron before it enters the calorimeter.

\section{FoCal-H results}
\label{sec:focalhres}
To study the \ac{FoCal-H} response to high energy electrons and hadrons, dedicated standalone runs were performed with the same setup as shown in \Fig{fig:setup_new}, but without \ac{FoCal-E} in front of \ac{FoCal-H}. 
In this way we isolate the performance of \ac{FoCal-H} and avoid events in which a shower starts in the volume of \ac{FoCal-E}.
The data used to demonstrate the detector performance were recorded at the \ac{SPS} H2 beam line in May, 2023.
\ifextrafigs
\note{Extra info:}
The energies, exploiting the full range available in the SPS H2 beam line, are listed together with the number of events for each in \Tab{tab:focal-h-data}.

\begin{table}[ht!]
\caption{Data samples of the FoCal-H prototype}
\begin{center}
\begin{tabular}{ l c c c }
Beam energy [GeV] & Total number of triggers & Matched triggers  & Studied events \\  \hline
60 & 111486 & 111486 & 106243\\
80 & 108702 & 108702 & 106798\\ 
100 & 111901 & 111889 & 110734\\ 
150 & 102191 & 102191 & 101952\\ 
200 & 116029 & 116029 & 115296\\ 
250 & 103580 & 103580 & 102579\\ 
300 & 105793 & 105753 & 99358 \\ 
350 & 101678 & 101678 & 99531\\ \hline
\end{tabular}
\end{center}
\label{tab:focal-h-data}
\end{table}
\else
The hadron energies range from 60 to 350~GeV and are known to a relative precision of about 2\%~\cite{deltaEbeam}.
\fi
\ac{FoCal-H} was rotated by 1~degree with respect to the beam direction to avoid particles directly hitting the fibers.
A rotation by the same angle, which corresponds to $\eta\approx5.3$ for particles originating from the nominal interaction point of \acs{ALICE}, was also implemented in the respective simulations.

\subsection{Data processing, pedestal determination and \acs{LG}-\acs{HG} matching
}
\label{subsec:pedestal_h}
The CAEN DT5202 enables the user to set an individual bias voltage for each of the \acp{SiPM} of the detector. 
To equalize the \ac{SiPM} overvoltage and thus the  gain, a dedicated measurement of the breakdown voltage, $\rm{V_{\rm br}}$, was performed for each of the 249 \acp{SiPM} prior to the test beam campaign, as described in \App{appendix:hcal_calibration}.
The nominal operation voltage for each channel was chosen as $ U = {V_{\rm br}}+3$~V. 

Each readout channel of the CAEN DT5202 has independent \ac{HG} and \ac{LG} signal processing chains.
The gains are programmable through variable‐gain pre-amplifiers to provide a broad dynamic range. 
In this study, we use the suitably-scaled \ac{LG} value to replace the \ac{HG} value for events in which the \ac{HG} \ac{ADC} is above a given threshold, as described below.  
There is also the possibility of using time-over-threshold to determine the charge from a \ac{SiPM}, but we have not employed this functionality of the DT5202.  

For the presented results, the measure of the energy of an event is the sum of the \ac{ADC} values of all channels firing in that event.
The position of the FoCal-H prototype was adjusted remotely such that the beam spot was approximately in the geometric center of the front face of the detector. 
The low beam rate ensured that the probability of having more than one particle in the readout window of a given event was small, thus no clustering or other event selection was performed. 

\ifextrafigs
\note{Extra info:}
The CAEN DT5202 output data stream was written in a text format. 
Upon detection and successful processing of a trigger signals, the output file contains the \ac{HG} \ac{ADC} value, \ac{LG} \ac{ADC} value, channel number, and the trigger counter (event ID) with respect to the start of data taking for each event. 
The trigger counter for each CAEN DT5202 readout module is incremented even in the case of unsuccessful processing, allowing for monitoring of the system dead time. 
This is done in case where e.g. the trigger rate is higher than the data throughput, resulting in limitations that causes an event to be lost.
Since the acquisition of the individual boards was asynchronous, a custom event building step was performed by matching the event ID of events from the CAEN DT5202 modules.

The trigger board (see ~\Sec{sec:setup} and \Fig{fig:setup_new}) limited the trigger rate to a maximum of 1--10 kHz, depending on the user settable length of the busy between events, which ensured that most of the events for all acquisition boards responded properly. 
In total, four CAEN DT5202 readout boards were used for \ac{FoCal-H} prototype 2.
The combined events were further processed to determine the energy deposited in the detector. 
\fi




\ifextrafigs
\note{Extra info} 
The CAEN software assumes a pedestal value of 25 \ac{ADC} counts and any channel with an \ac{ADC} value below this threshold is considered to be zero. 
However, this is not the actual pedestal value, which we determine for each of the 249 instrumented channels by summing all events for each channel for a given run, as illustrated by an example channel in \Fig{fig:pedestal_singlechannel}, for both the \ac{HG} and \ac{LG} data. 
In order to estimate the mean position of the pedestal peak, the peak at low \ac{ADC} values is fit with a Gaussian function, as shown in \Fig{fig:pedestal_singlechannel}  and the pedestal is defined to be the mean of this fit. 
This pedestal is then subtracted from the \ac{ADC} value for each channel and event. 
This procedure was carried out for both the \ac{HG} and \ac{LG} distributions.

\begin{figure}[th!]
\begin{center}\includegraphics[width=0.70\textwidth]{figures/hcal/pedestal.png} 
\caption{\label{fig:pedestal_singlechannel} 
Single channel \ac{ADC} response, with the \ac{ADC} range chosen to emphasize the lower end of the spectrum. The narrow peak at an ADC value of approximately 43 is the pedestal and fit via the Gaussian indicated by the full drawn line. The mean of the Gaussian is then used as a pedestal for the channel.}
\end{center}
\end{figure} 
\else
The CAEN software assumes a pedestal value of 25 \ac{ADC} counts and any channel with an \ac{ADC} value below this threshold is considered to be zero. 
However, this is not the actual pedestal value, which we determine for each of the 249 instrumented channels by summing all events for each channel for a given run, for both the \ac{HG} and \ac{LG} data. 
In order to estimate the mean position of the pedestal, the peak at low \ac{ADC} values is fit with a Gaussian, whose mean defines the pedestal value.
The pedestal is then subtracted from the \ac{ADC} value for each channel for each event to give the pedestal subtracted \ac{ADC} values. 
This procedure was carried out for both the \ac{HG} and \ac{LG} distributions.
\fi


To account for the saturation in the \ac{HG} \ac{ADC} values, above a certain \ac{HG} \ac{ADC} threshold the stored \ac{LG} \ac{ADC} values were used and scaled according to the following procedure.
For each individual channel a correlation plot of the \ac{LG} \ac{ADC} versus the \ac{HG} \ac{ADC} was made.

\ifextrafigs
An example of such a plot can be seen in \Fig{fig:chprofilelghg}.
\begin{figure}
\begin{center}\includegraphics[width=0.75\textwidth]{figures/hcal/HG-LG-Profile_not_approved_by_ALICE.pdf} 
\includegraphics[width=0.75\textwidth]{figures/hcal/Slope.pdf} 
\caption{\label{fig:chprofilelghg} Example of a \ac{HG}-\ac{LG} channel profile (blue). The region of the best quality \ac{ADC} signals is illustrated by a linear fit (red). Channel profiles were reconstructed for each channel, and an individual linear fit was estimated for further analysis. 
}
\end{center}
\end{figure} 
\fi 
The \ac{LG}-\ac{HG} correlation \ifextrafigs, seen in \Fig{fig:chprofilelghg},\fi shows a region of linearity which was used to determine the gain ratio between the \ac{HG} and \ac{LG} signals. 
For each channel, the \ac{LG}-\ac{HG}  correlation was fitted using a first order polynomial to estimate this relative gain. 
The average slope for the 249 channels, reflecting the gain ratio between the \ac{HG} and the \ac{LG}, was found to be 9.3 with a spread of 0.2. 
We note that while, in principle, this parameter is derivable from the gain settings of the \ac{HG} and \ac{LG} amplifiers, minor differences in the capacitance values from channel to channel, presumably ascribable to the tolerances on the individual circuit components, cause the channel-to-channel variations. 
In the presented case, the obtained spread of 0.2 translates to about 2\% 
channel-to-channel variations. 
\ifextrafigs
The variation in slopes among all channels  can be seen in the lower panel of \Fig{fig:chprofilelghg}.  
\fi
The resulting fit parameters were used to correct the \ac{LG} gain so that in cases where the \ac{HG} is above a given threshold. 
For the presented results we used a value of 1500 for the \ac{HG}-\ac{LG} gain ratio, but the results were found to be insensitive to the exact value, 
as detailed below.

\ifextrafigs
The effect of the applied \ac{LG}-\ac{HG} method can be seen in \Fig{fig:hcal-hg-lg-comparison}. The points given are the \ac{ADC} sum for each beam energy using the \ac{HG} only. We use a linear fit in the selected range of the first five points, assuming that notable saturation is starting for higher than 200 GeV beam energies. We then apply the described conversion method and we compare the results by fitting the same 5 points. The \ac{ADC} sum for higher than 200 GeV energies are matching the fit range on the level of beam energy uncertainties. This is clear indication that the saturation in the \ac{HG} is reduced. 

\begin{figure}\begin{center}\includegraphics[width=0.7\textwidth]{figures/hcal/HG-linear.pdf} 
\includegraphics[width=0.7\textwidth]{figures/hcal/mixed-linear.pdf} 
\caption{\label{fig:hcal-hg-lg-comparison} The upper figure shows the ADC sum as a function of beam energy.  The line is a fit to the 5 lowest energies (red points).  The blue points are the ADC sum for the 250, 300, and 350 GeV. For these energies, the ADCs of channels near the shower center saturate, leading to the reduction compared to the linear expectation. In the lower panel, the data have been corrected for saturation by using the LG data as described in the text.
}
\end{center}
\end{figure} 
\fi

\subsection{Calibration and linearity}
\label{sec:HCal_calibration}
\begin{figure}[th!]
\begin{center}
\includegraphics[width=0.9\textwidth]{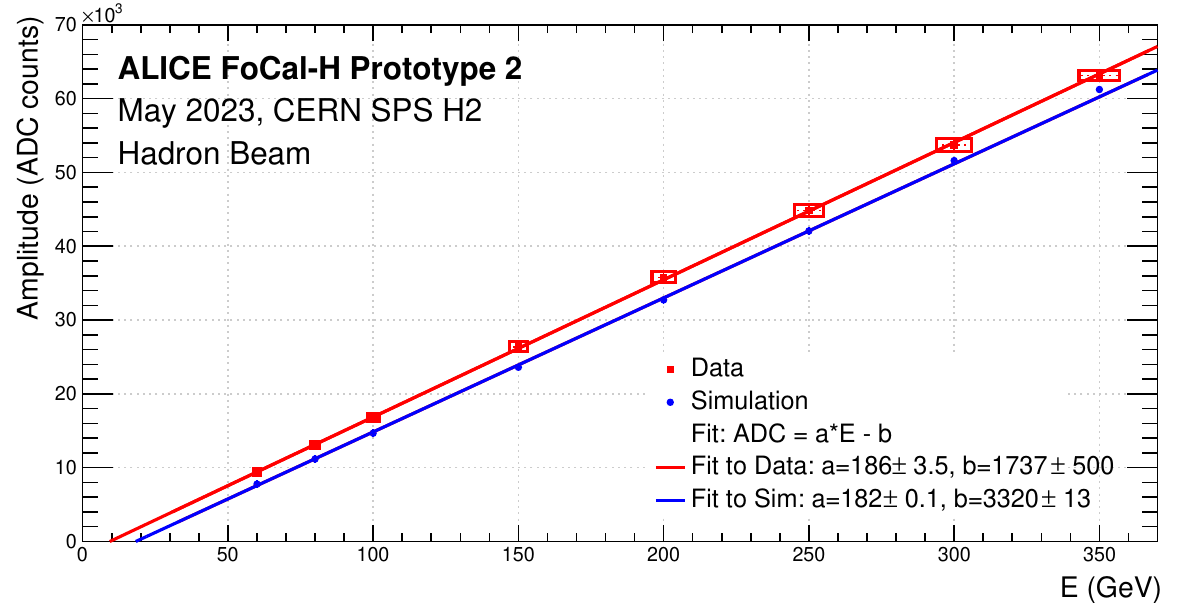} 
\includegraphics[width=0.9\textwidth]{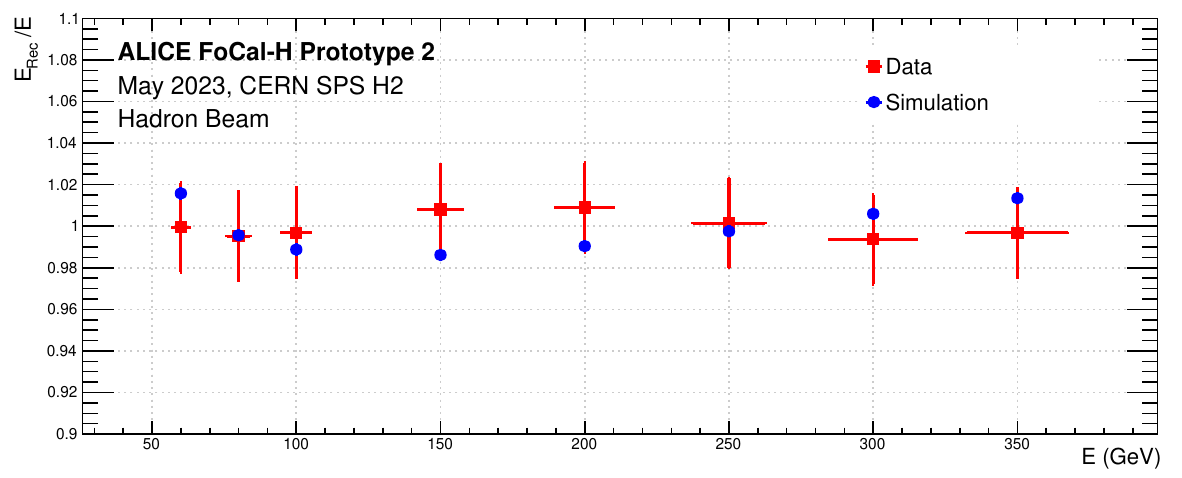} 

\caption{\label{fig:linearity} 
Top: FoCal-H Prototype 2 reconstructed \ac{ADC} sum for 60--350 GeV hadrons obtained at the H2 \ac{SPS} beam line in May, 2023. 
The reconstructed \ac{ADC} sum for the data is compared to simulations~(where the number of photo-electrons was scaled by 1/3 to convert into \ac{ADC} units). 
In each case, the dependence of the detector response is fitted with a linear function.
Bottom: Corresponding ratios between the reconstructed energy in the \ac{FoCal-H} prototype and the beam energy.} 
\end{center}
\end{figure}

For each beam energy, the \ac{ADC} sum distribution was fitted using a Gaussian fit function. 
The mean values of these fits were used to characterize the detector response for each beam energy and the values obtained are plotted as a function of the beam energy in \Fig{fig:linearity}.
For the simulation, the number of photo-electrons was scaled by $1/3$ to convert into \ac{ADC} units.

\ifextrafigs
\begin{figure}[htb]
\begin{center}
\includegraphics[width=0.9\textwidth]{figures/hcal/resp_cor.pdf} 
\caption{\label{fig:linearity-departure} 
Ratio between the reconstructed energy
in FoCal-H Prototype 2 and the beam energy
for 60--350 GeV hadrons obtained at the H2 \ac{SPS} beam line in May, 2023 and compared with the simulation. 
}
\end{center}
\end{figure}
\fi

The points were fitted with a linear function to obtain a relation between the \ac{ADC} value and beam energy, as ${\rm ADC} = a\times E - b$.
The detector response from \ac{ADC} channels to energy is obtained using this calibration function.
As shown in \Fig{fig:linearity}, this was done separately for data and simulations. 
While the slopes of the fits in data and simulations are consistent within the uncertainties, the corresponding intercepts are not.
The ratio between reconstructed energy and the beam energy, assumed to be the impinging particles energy, for both data and MC is shown in the bottom panel of \Fig{fig:linearity}.
The uncertainty of the fit parameters is taken into account. 
The slope term dominates the 2\% uncertainty.
The beam energy uncertainty of 5\%, entering through the denominator,  is not included in the presented uncertainties.
The nonlinearity is within the 2\% range for the entire interval of hadron energy from 60 GeV to 350 GeV. 

\ifextrafigs
The calibrated energy response to the hadrons of the measured energies can be seen in \Fig{fig:hcaldistributiondata} for both data and simulations.
\begin{figure}[ht!]
\begin{center}
\includegraphics[width=0.49\textwidth]{figures/hcal/linearity_plot.pdf} 
\includegraphics[width=0.49\textwidth]{figures/hcal/MC_linearity.pdf} 
\caption{\label{fig:hcaldistributiondata} 
\note{Extra figs}
The reconstructed energy response as a function of the beam energy for the test beam data~(left) and simulations~(right).}
\end{center}
\end{figure}
\fi

\begin{figure}[t!]
\begin{center}
\includegraphics[width=0.49\textwidth]{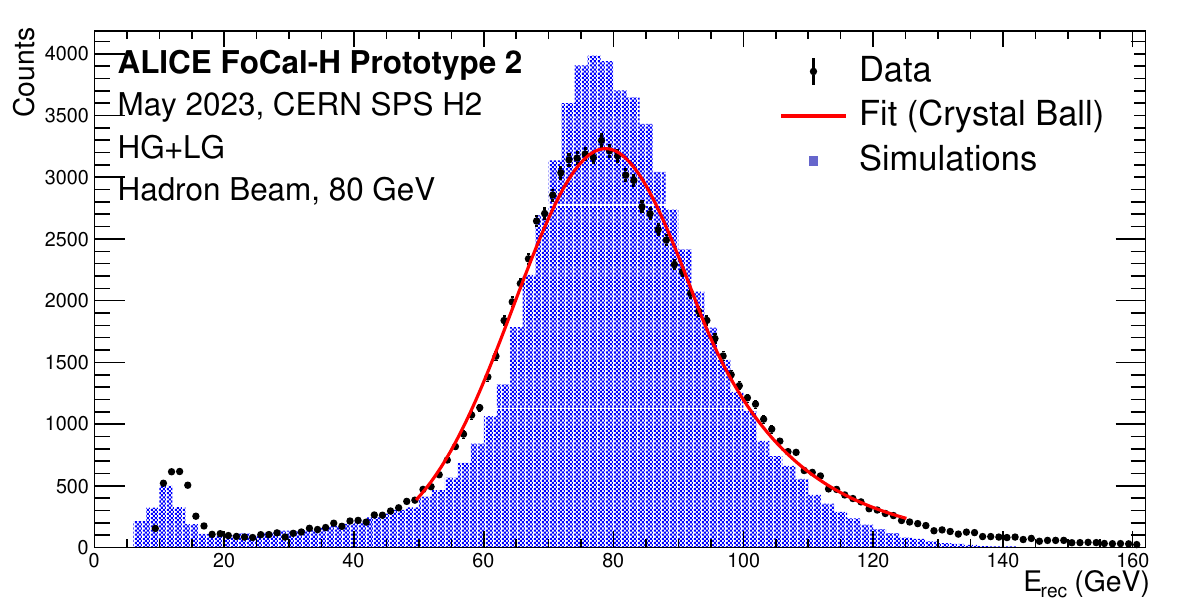} 
\includegraphics[width=0.49\textwidth]{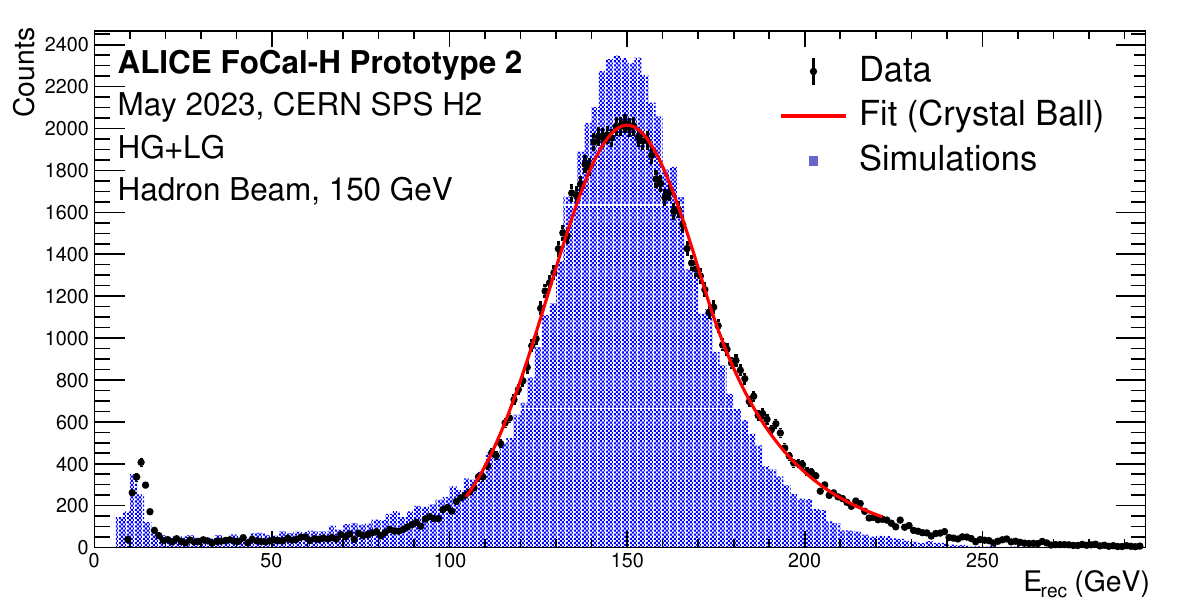} 
\includegraphics[width=0.49\textwidth]{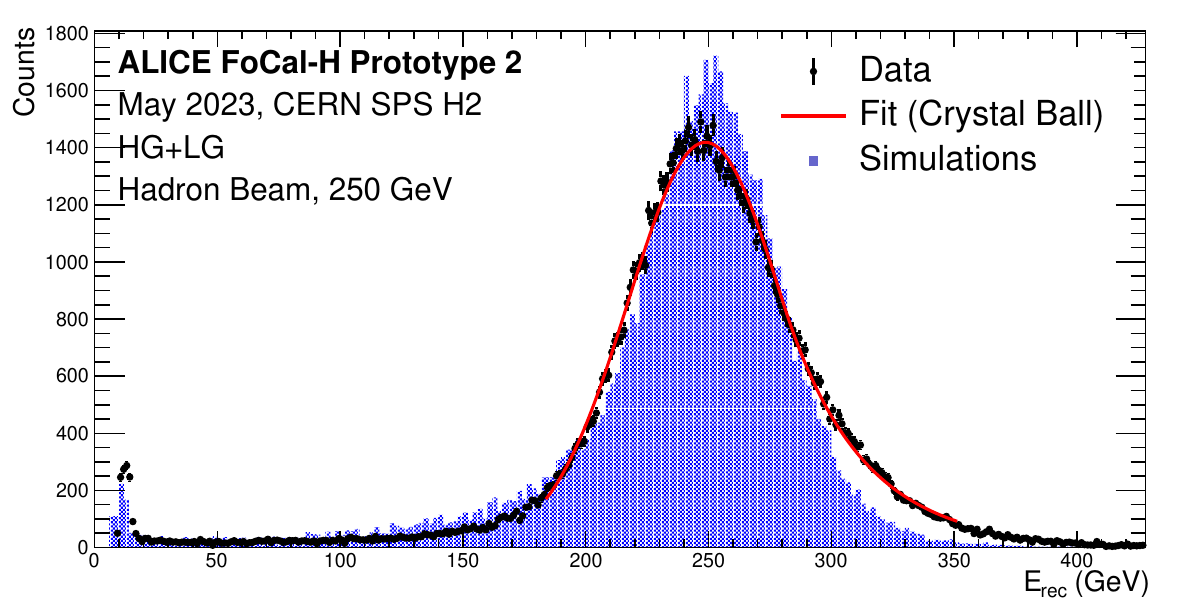} 
\includegraphics[width=0.49\textwidth]{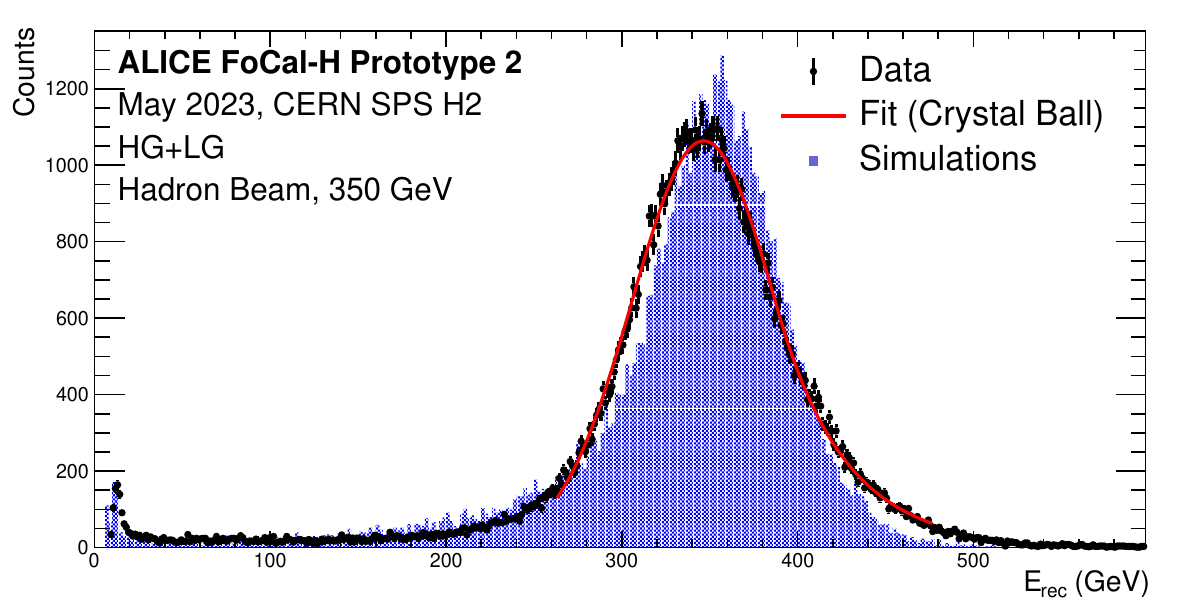} 
\caption{\label{fig:hcaldistribution350} 
FoCal-H Prototype 2 reconstructed energy distributions for data and simulations for hadron energies at 80, 150, 250, and 350 GeV. 
The distributions are normalized to the same total number of events.} 
\end{center}
\end{figure}

The detector simulations were performed as described in \Sec{sec:simulation}. 
Three different hadron types were simulated: $p$, $\pi^+$, $K^+$. 
A weighted sum of the resulting distributions with weights from the Atherton parametrisation~\cite{Atherton:1980vj}
\ifextrafigs
(as seen from \Fig{fig:beam-composition})
\fi
was made to properly reflect the composition of the \ac{SPS} hadron beam at each energy. 
An additional correction for the decay probability of each particle between the production target and the \ac{FoCal-H} prototype, located about 600~m downstream, was applied.
\ifextrafigs
\begin{figure}
\begin{center}
\includegraphics[width=0.49\textwidth]{figures/hcal/atherton-resized-nov.pdf} 
\caption{\label{fig:beam-composition}
\note{Extra fig} 
Relative hadron beam composition as a function of the 
beam energy, computed from the Atherton parametrisation \cite{Atherton:1980vj} for zero 
production angle.}
\end{center}
\end{figure}
\fi
An additional threshold was implemented for each channel in the simulations to mimic the pedestal subtraction in the data. 

After applying the calibration, the measured and simulated response of \ac{FoCal-H} to 80, 150, 250, and 350 GeV hadrons are shown in \Fig{fig:hcaldistribution350}. 
At 60 GeV, pions comprise approximately 75\% of the hadrons in the beam, while at 350 GeV there are almost exclusively protons.  
In all cases, the data exhibit broader tails than the simulation, with the difference decreasing with increasing hadron energy.
An indication of the quality of the calibration is that the left peak in all of the reconstructed signal responses in data and simulations appears at the same energy. 

\subsection{Energy Resolution}  
\label{subsec:energy_res_hcal}
For each beam energy, the \ac{ADC} sum distribution, calibrated as described above, was used.  The distributions were fit using two different fitting functions --- a Gaussian fit function and a Crystal Ball~\cite{ALICE-PUBLIC-2015-006} fit function. 
These fit functions were used individually to compute the energy resolution $\sigma_E$/E. 
A three step fit procedure was employed. 
Initially, a fit with a  Gaussian function was performed in the whole range of the histogram of the distribution.  
The mean and $\sigma$ values were extracted providing the peak position.
The second step was to fit in a selected range around the determined mean position from the first step. 
The final step was to re-fit within the given $n\sigma$ range, where the default range is given by $n=2$, around the mean position from the second step, using either a Crystal Ball fit or a Gaussian fit. 
The results for the parameters of the different fit functions and ranges were used to calculate the systematic uncertainties of the fit.
For each energy, the value of the \ac{FWHM} of the peak was computed from the fit function and divided by 2.355 to extract $\sigma_E$/E, where E is the mean value of the final Gaussian fitting, described above. 


\begin{figure}
\begin{center}
\includegraphics[width=0.9\textwidth]{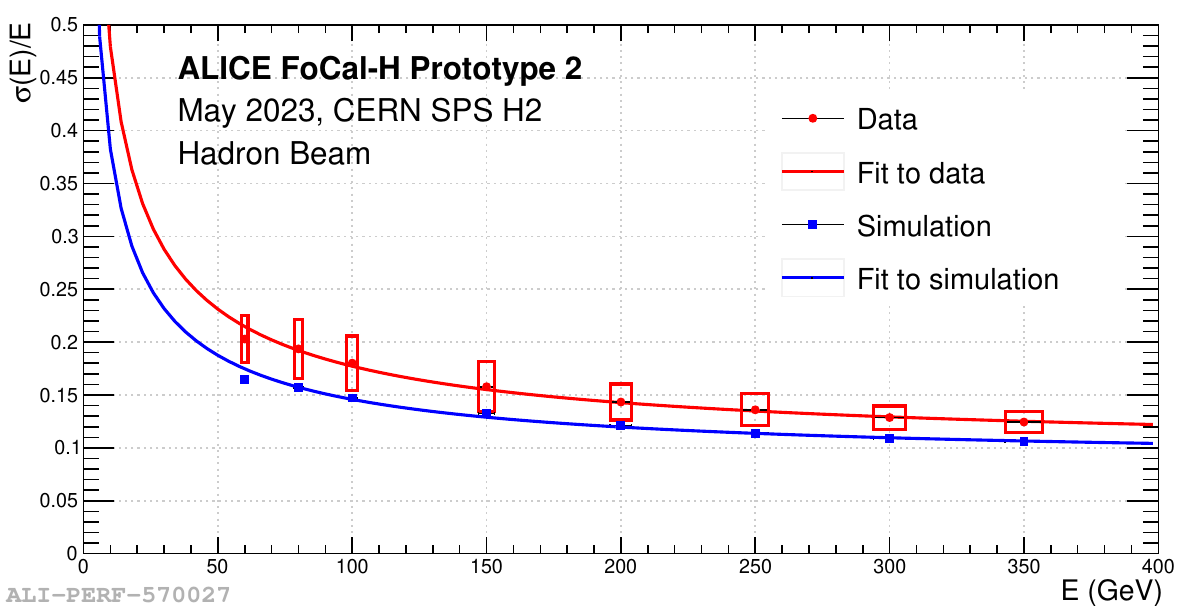} 
\caption{\label{fig:resolutionfwhm} Energy resolution~(obtained from \ac{FWHM}/2.355) of the FoCal-H prototype 2 as a function of the hadron beam energy. The energy dependence was approximated using \Eq{eq:energy_reso}. 
The corresponding fit values are: $\sigma_{\rm stoch.}=1.48$, $\sigma_{\rm noise}=0$, $\sigma_{\rm const.}=0.10$ for data, and $\sigma_{\rm stoch.}=1.18$, $\sigma_{\rm noise.}=0$, $\sigma_{\rm const.}=0.09$ for simulation.
}
\end{center}
\end{figure}

The same procedure was applied both for data and simulations. 
The obtained resolution~($\sigma_E$/E) as a function of the impinging particle energy is shown in \Fig{fig:resolutionfwhm}.
As can be seen, the energy resolution in the simulations is significantly lower than the energy resolution observed in data. 
To account for this difference, a decrease of the number of the photoelectrons by more than factor 100 would have to be implemented. 
However, such large photon loss was considered not to be physical. 
The physics list applied in the simulation is, as mentioned in section \ref{sec:simulation}, the \acs{QGSP}\_\acs{BERT}.
The usage of \acs{FTFP}\_\acs{BERT} physics list in simulation~(not shown) would result in an increase of the discrepancy between data and MC by 20\%. 
Additional source of the discrepancy could be the spread of the channel by channel light collection efficiency, which was not accounted for in the simulation. 

The energy resolution was fitted using \Eq{eq:energy_reso}, where we denote in the following $\sigma_{\rm stoch.}$ as the stochastic, $\sigma_{\rm noise}$ the noise and $\sigma_{\rm const.}$ the constant term. 
To estimate the systematic uncertainty, several effects were considered, most of which were studied for each beam energy and the total systematic uncertainty for each energy is shown with boxes in \Fig{fig:resolutionfwhm}. 
However, being highly correlated, the systematic uncertainties of the individual data points were not used in the fit to the data. 
Instead, fits were performed independently for a given systematic variation~(e.g.\ of the line shape assumptions), and the systematic uncertainty of the resolution parameters was obtained from all the different fits.
In all cases, first the linearity was established~(the resulting variation in the data is also shown in \Fig{fig:linearity}).

The systematic error arising from  different fit ranges was studied by varying the 
limits of the right tail using a Crystal Ball fit for 1$\sigma$, 1.5$\sigma$, 2$\sigma$, 2.5$\sigma$. 
The maximal difference was taken to be the  systematic error and amounts to  $\Delta \sigma_{\rm stoch.} = 0.02$, $\Delta \sigma_{\rm const.} = 0.001$. 
The effect of the line shape was addressed by considering Gaussian and Crystal Ball fits. 
Their contribution to the systematic error is given by $\Delta \sigma_{\rm stoch.} = 0.10$, $\Delta \sigma_{\rm const.} = 0.005$. 
The effect of applying a different \ac{HG} \ac{ADC} threshold, at which the \ac{HG}-\ac{LG}  method is applied was studied. 
The threshold at which the \ac{LG} \ac{ADC} was used rather than the \ac{HG} \ac{ADC} was varied by 500 to 4000 \ac{HG} \ac{ADC} counts, to investigate the impact of this threshold on the final result.  
\ifextrafigs
The obtained distributions of the \ac{ADC} sum are afterwards fit using a Gaussian function to obtain the central value of the \ac{ADC} versus energy correspondence and the energy resolution parameters. 
\Figure{fig:plateau} illustrates the reconstructed \ac{ADC} sum for the \ac{HG} threshold values in the region from 100 to 4095 \ac{ADC}. 
For conversion values close to 4000 \ac{ADC}, the conversion is not able to reconstruct \ac{ADC} signals in a proper way. 
This is due to the fact, that individual channels have different dynamic ranges, 
which means that their saturation buffer is at different \ac{ADC} values. 
For conversion values less than 1000 \ac{ADC}, 
the reconstructed \ac{ADC} sum is steeply decreasing. 
A plateau is obtained for conversion values from 1600 to 2000 \ac{ADC}. 
This correlates with the test beam data from the individual \ac{HG}-\ac{LG} channel profiles. 
\begin{figure}
\begin{center}
\includegraphics[width=0.8\textwidth]{figures/hcal/Plato.pdf} 
\caption{\label{fig:plateau}
\note{Extra fig: }
Reconstructed \acrshort{ADC} sum for a given threshold for using the scaled LG ADC rather than \textbf{HG} \textbf{ADC} value. The obtained plateau is used for determination of the \acrshort{HG}-\acrshort{LG} conversion value for all channels in case of saturation.}
\end{center}
\end{figure}
\fi
The maximal difference was taken as systematic uncertainty and amounts to $\Delta \sigma_{\rm stoch.} = 0.046$, $\Delta \sigma_{\rm const.} = 0.003$. 
The maximal difference between the fit results using \ac{LG} only and the  \ac{HG}-\ac{LG} matching method gives a systematic uncertainty of $\Delta \sigma_{\rm stoch.} = 0.20$, $\Delta \sigma_{\rm const.} = 0.002$. 
The maximum $\Delta p/ p$ of the H2 \ac{SPS} beam line is 2$\%$~\cite{deltaEbeam}, used as the uncertainty on the beam energy~(e.g.\ in \Fig{fig:linearity}), leading to a difference of $\Delta \sigma_{\rm stoch.}=0.04$ and $\Delta \sigma_{\rm const.}=0.003$.
The noise term, denoted as $\sigma_{\rm noise}$, is consistent with zero in all of the above studies. 
The relative systematic uncertainties due to the described effects are summarized in \Tab{tab:focal-h-systematics}.
 
\begin{table}[ht!]
\caption{Estimation of the dominant systematic uncertainties~(given in absolute values) on the stochastic and the constant term in the energy resolution parametrization of the FoCal-H prototype.}
\begin{center}
\begin{tabular}{ l  | c | c } 
\hline
Systematic effect & $\Delta \sigma_{\rm stoch.}$   &   $\Delta \sigma_{\rm const.}$   \\ \hline
Fit range & 0.02 &  0.001 \\
Line shape &  0.10 &  0.005 \\
\ac{HG}-\ac{LG} matching  &  0.05    & 0.003 \\ 
Gain choice & 0.20 & 0.002\\
Global energy scale & 0.04 & 0.003\\
 \hline
Total (added in quadrature) & 0.22     & 0.007 \\ 
\hline
\end{tabular}
\end{center}
\label{tab:focal-h-systematics}
\end{table}

Additional systematic effects such as the pedestal determination and beam composition were found to have negligible influence on the results. 
The final results for the energy resolution parameters are
\begin{eqnarray*}
    \sigma_{\rm stoch.} = (148 \pm 2_{\rm stat} \pm {22}_{\rm syst}) \% \\
    \sigma_{\rm const.} = (10.0 \pm 0.13_{\rm stat} \pm {0.7}_{\rm syst}) \%,
\end{eqnarray*}
with the noise term consistent with zero. 
The uncertainty in these parameters is dominated by the systematic effects.     
The energy resolution of the \ac{FoCal-H} prototype at high energy is dominated by the constant term and is of the order of 10\%. 
This value is consistent with the expected performance of \ac{FoCal-H} at high energies, and it is in line with the physics requirements, thus validating the chosen detector technology.

\section{Summary}
\label{sec:summary}
We constructed a full-length prototype of the \ac{FoCal}, which is being developed for installation at \acs{LHC}, \acs{CERN}, and to take data in Run~4, starting from 2029.
The final detector is designed to cover a pseudorapidity acceptance between 3.2 and 5.8 units and provide unique capabilities in probing non-linear QCD dynamics in unexplored regions at low Bjorken $x$ and $Q^{2}$. 
The detector is composed of a silicon-tungsten electromagnetic sampling calorimeter with 20 layers (\acs{FoCal-E}) and a hadronic calorimeter (\ac{FoCal-H}) in spaghetti design with copper tubes. In \ac{FoCal-E}, 18 sampling layers are equipped with silicon pad sensors of pad sizes of $1 \times 1\cm^2$ and read out with the \ac{HGCROC}, and two layers at position 5 and 10 are instrumented with \ac{ALPIDE} chips with a pixel size of about $30\times 30\um^2$. 
The copper tubes in FoCal-H of an outer (inner) diameter of 2.5~(1.2)\mm\ are filled with scintillating fibers which are connected and read out with \acs{SiPM}s at the rear detector side.
The data were taken in various test beam campaigns between 2021 and 2023 at the \acs{CERN} \acs{PS} and \acs{SPS} beam lines with hadron beams up to energies of 350~GeV, and electron beams up to 300~GeV.

Regarding the \ac{FoCal-E} pad sensors, a comprehensive study of the response to \acp{MIP} across all pad layers was presented, revealing very clean signals and good stability over various hadron energies recorded~(\Fig{fig:mip-pos-coincidence-cut}).
Good agreement between data and simulations for longitudinal shower profiles for 20--300~GeV electrons was demonstrated~(\Fig{fig:pad-long-shower-profile-all}).
Correspondingly, the linearity of the response was found to be in good agreement between data and simulations over the full range of available electron energies~(\Fig{fig:pad-linearity}).
The relative energy resolution was found to be lower than 3\% at energies larger than 100~GeV~(\Fig{fig:pad-linearity-resolution-final}), and projections using data up to 300~GeV with 17 functional pad-layers demonstrate that the energy resolution fulfills the requirements for the physics needs.
Due to the high granularity and good linearity of the pixel layers~(left panel of \Fig{fig:pixel-pad-combination-linear-and-shower-max}), measured transverse shower widths, quantified by \ac{FWHM}, on the scale of mm were obtained, with good agreement to simulations~(\Fig{fig:pixel-swfwhm}). 

The response of \acs{FoCal-H} was found to be linear~(\Fig{fig:linearity}), albeit with a significant intercept that is about factor 2 larger than in simulations.
The shape of the response is non-Gaussian, with wider tails in the data as in simulations~(\Fig{fig:hcaldistribution350}).
The corresponding resolution was quantified using the \acs{FWHM} and decreases from about 16\% at 100~GeV to about 11\% at 350~GeV.
The constant term of the resolution was estimated to be 10\%, satisfying the physics needs.

Initial analyses on combining the \ac{FoCal-E} with \ac{FoCal-H} information show promising results e.g.~for hadronic showers starting in \ac{FoCal-E}, despite the use of the different \ac{DAQ} systems. However, results of those studies are deferred to future publications.

\acknowledgments
We would like to express our most sincere gratitude to the collaborating individuals, groups, and organizations for their invaluable contributions to the success of the test beam campaigns mentioned in this document. 
%
%
We value the contribution of the \acs{ALICE} \ac{FLP} team providing us with the \ac{CRU} and \ac{FLP}.
In particular, we warmly thank F.~Costa, and R.~Divia for their support with the \ac{O2} online infrastructure. 
%
%
We value the contribution of the \acs{ALICE} \ac{CTP} group for providing the \ac{LTU} and their support on the trigger hardware and processing. 
In particular, we warmly thank A.~Jusko, M~.Krivda and R.~Lietava for their kind help in developing and operating the \ac{FoCal} trigger system. 
We are also very grateful for the support of the \acs{ALICE} \acs{ITS} team, and in particular S.~Beole, O.~Grøttvik and F.~Reidt, for providing the necessary hardware and expertise for the design and development of the \ac{HIC}-based high-granularity layers, and their readout infrastructure. 
Finally, we deeply appreciate the invaluable help of C.~De la Taille and the \acs{OMEGA} group in the development and understanding of the \acs{HGCROC} chip based readout employed for the silicon pads sensors of \ac{FoCal}.   
We extend our appreciation to the staff and operators of the \ac{PS} and \ac{SPS} test beam facilities at \acs{CERN} for their assistance in setting up and conducting the experiment operations. 
In particular, we would like to mention D.~Banerjee, N.~Charitonidis, B.~Holzer, M.~Lazzaroni and B.~Rae. 
Their expertise and endless support ensured a smooth and productive data-taking process. 

I.~Bearden, A.~Buhl, L.~Dufke and I.~Pascal acknowledge support from the The Carlsberg Foundation (CF21-0606) and the Danish Council for Independent Research/Natural Sciences. 
M.~Bregant acknowledges financial support by Conselho Nacional de Desenvolvimento Científico e Tecnológico (CNPq) and Fundação de Amparo à Pesquisa do Estado de São Paulo (FAPESP) grant No. 2020/04867-2, Brazil.
T.~Chujo, Y.~Goto, and T.~Sugitate acknowledge the support by JSPS KAKENHI Grant Numbers JP20H05638, JP19H01928, JP21H04484.
A.~Gautam, T.~Isidori and D.~Tapia Takaki are supported by the U.S.\ Department of Energy, Office of Science, Office of Nuclear Physics under award number: DE-SC0020914. 
V.~Kozhuharov and R.~Simeonov acknowledge that partially this study is financed by the European Union-NextGenerationEU, through the National Recovery and Resilience Plan of the Republic of Bulgaria, 
project SUMMIT BG-RRP-2.004-0008-C01.
N.~Minafra and D.~Tapia Takaki acknowledge support by the U.S Department of Energy, Office of Science, Office of Basic Energy Sciences, under Award Number: DE-SC0023510.
H.~Hassan, L.M.Huhta, Y.~Melikyan, S.S.~R\"{a}s\"{a}nen and H.~Rytk\"{o}nen acknowledge financial support from the Helsinki Institute of Physics (HIP), Finland and the Academy of Finland (Center of Excellence in Quark Matter) (grant nos.\ 346327, 346328), Finland.
M.~Fasel, F.~Jonas, C.~Loizides and N.~Novitzky acknowledge financial support by the U.S.\ Department of Energy, Office of Science, Office of Nuclear Physics, under contract number DE-AC05-00OR22725. 
L.~He,  J.~Yi, Z.~Yin and D.~Zhou acknowledge the support by the National Key Research and Development Program of China (2022YFA1602103).
Researchers from Norwegian institutes acknowledge the support by the The Research Council of Norway, Norway.

\bibliographystyle{JHEP}
\bibliography{biblio.bib}
\appendix
\section{Calibration of the \ac{FoCal-E} pads}
\label{appendix:pads_calibration}
\begin{figure}[t!]
\begin{center}
\includegraphics[width=1\textwidth]{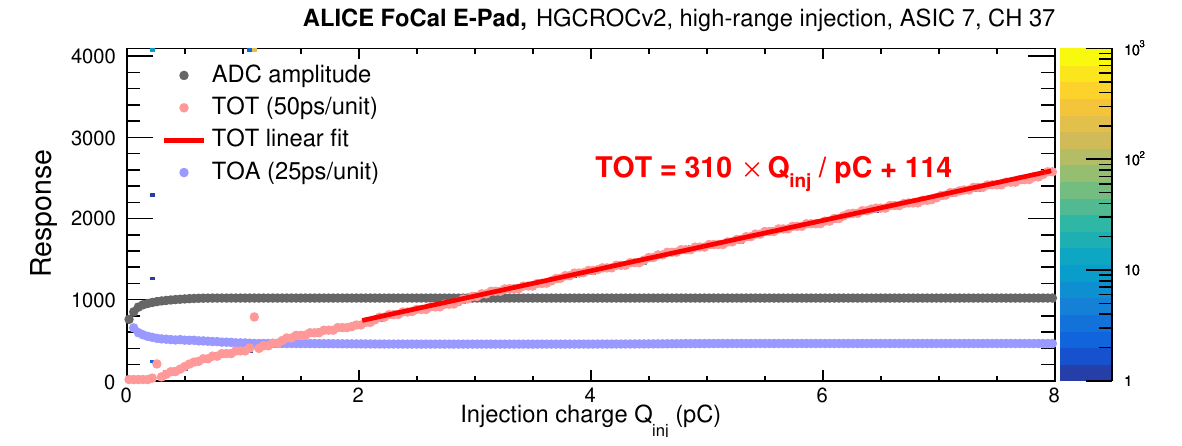}
\includegraphics[width=1\textwidth]{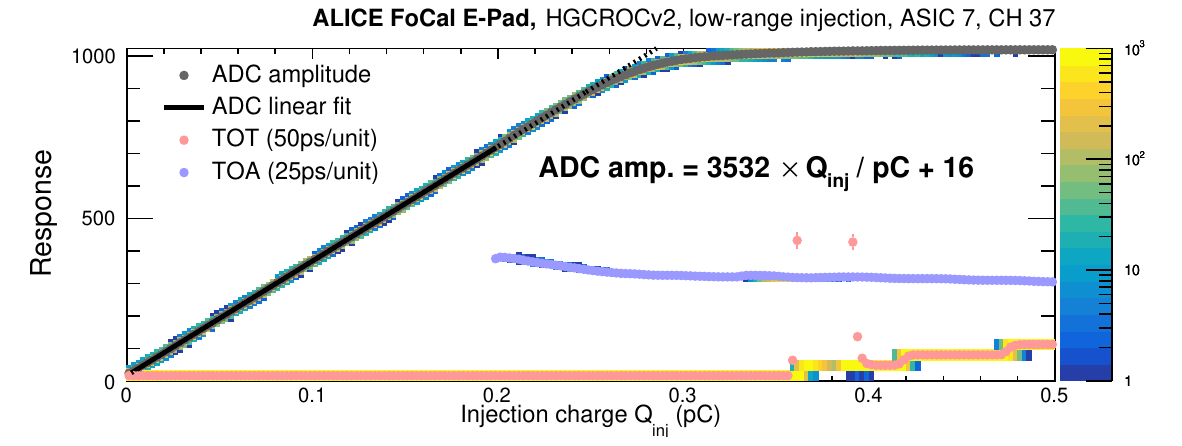}
\caption{Top: Response of the \ac{ADC}, \ac{ToA}, and \ac{ToT} as a function of the injected charge for a single channel of \ac{ASIC}~7. The calibration is operated over the full high-range injection range spanning from 0--8.0\pC. 
Bottom: The response for the low-range injection region going from 0--0.5\pC.
The TOT response is fitted in the range from 2 to 8\pC\ in the high-range injection mode (top), and the ADC amplitude from 0\pC\ to 200\fC\ in the low-range injection mode (bottom).}
\label{fig:calibration_pads}
\end{center}
\end{figure}

This section provides further details to the calibration procedure for the \ac{HGCROC} which was introduced in \Sec{subsec:padchannelcalib} and used for the electron data analysis described in Sections~\ref{subsec:evtselpads}~to~\ref{subsec:En_resolution}.
The calibration pulser supports two charge injection regimes: one for low-range injection with an 0.5\,\pF\ capacitor, $C_{\rm inj, low}$, and one for high-range injection with an 8.0\,\pF\ capacitor, $C_{\rm inj, high}$. 
An integrated 11-bit \ac{DAC}, ${\rm DAC}_{\rm inj}$, allows to set the voltage at the injection capacitors, up to the chip's bandgap voltage of 1\,V.
Thus, two charge injection regimes with a range of 0.5\pC\ and 8.0\pC, respectively, are supported.
The injected charge, $Q_{\rm inj}$, is obtained from the product of the size of the injection capacitor, the voltage fraction set in \ac{DAC} units, and the \ac{HGCROC}'s band-gap voltage reference:
\begin{equation}
\label{eq:q-injection}
Q_{\rm inj} = C_{\rm inj} \times \frac{\rm DAC_{\rm inj}}{2048} \times U_{\rm ref,bandgap} \, .
\end{equation}
The \ac{UDP} operation mode of the aggregator board described in \Sec{subsubsec:Prototype_pads} was used to run and record the calibration of the channel-wise \ac{ADC}, \ac{ToA}, and \ac{ToT} response. 
A dedicated \ac{DAQ} software controls the chip internal injection system through the slow control link. 
For injection, the pulse injection command is fired through the \ac{FCMD} port, and it simulates the signal of the Si pads sensors in each layer~\cite{Thienpont:2020wau}. 
Among the set of options, the calibration script includes the control over the number of injections and the granularity of the injection steps.
For the calibration procedure, the voltage is incremented step-wise in steps of 10 \ac{DAC} units starting from zero, and per step 2000~injections are released.
The custom \ac{DAQ} software records the per-channel response of \ac{ADC}, \ac{ToT}, and \ac{ToA} per charge injection, and the mean response per step is calculated, as displayed in \Fig{fig:calibration_pads}.
The \ac{ToT} and \ac{ToA} units can be translated directly into the time-over-threshold and time-of-arrival of the charge pulse where 1~\ac{ToT}~unit~=~50\ps, and 1~\ac{ToA}~unit~=~25\ps.
We optimized the threshold setting for \ac{ToA} and \ac{ToT} by performing a scan of their response as a function of the applied thresholds.  

For that, a few parameters of the pre-amplifier of the \ac{HGCROC} have to be previously adjusted for characterizing the gain of the chip and optimize the dynamic range of the readout circuit. In particular, a few combinations of feedback resistors and capacitors values can be explored modifying the parameters of the configuration files in the acquisition software. The chip dynamic range and \ac{MIP} response were initially tested at the PS test beam facility with a comprehensive scan of these pre-amplification parameters. Selected results are shown in \Sec{subsec:MIP}; details can be found in~\cite{rytkonen2023simulated}. 

After introducing the use of the \ac{ADC} phase selection through the trigger board (described in~\Sec{sec:setup}), the optimal working point of the detector underwent slight modifications and the pre-amplification  circuit needed additional tuning. The parameters selected for the \ac{FoCal-E} 2023 data acquisitions were chosen to achieve the best compromise between a broad dynamic range, and optimal \ac{MIP} separation in the \ac{ToT} and \ac{ADC} domains. The results relative to the 2023 \ac{MIP} studies are reported in~\Sec{subsec:MIP}, and an example is provided in~\Fig{fig:mip-pos-coincidence-cut}. 
The total \ac{HGCROC} pre-amplifier feedback resistance is defined by the parallel of two resistors. Similarly, the total capacitance is obtained by the parallel of the two feedback capacitors. The values set for the \ac{FoCal-E} setup are: 

\begin{itemize}
    \item $R_F$ = 100\kOhm\, ||\, 66\kOhm\, = 39.8\kOhm
    \item $C_f$ = 100\fF\, ||\, 400\fF\, = 500\fF
\end{itemize}

Once the circuit's gain was characterized, the best \ac{ToT} and \ac{ToA} thresholds were found to amount to, respectively, 300 and 400 units. These settings results in a gap of 100\fC between the dynamic range of the \ac{ADC} and the turn-on region of the \ac{ToT}. However, it significantly reduced the noise of 
\ac{ToT} and \ac{ToA}, leading to an overall more stable detector operation.

\begin{figure}[t!]
\begin{center}
\includegraphics[width=.49\textwidth]{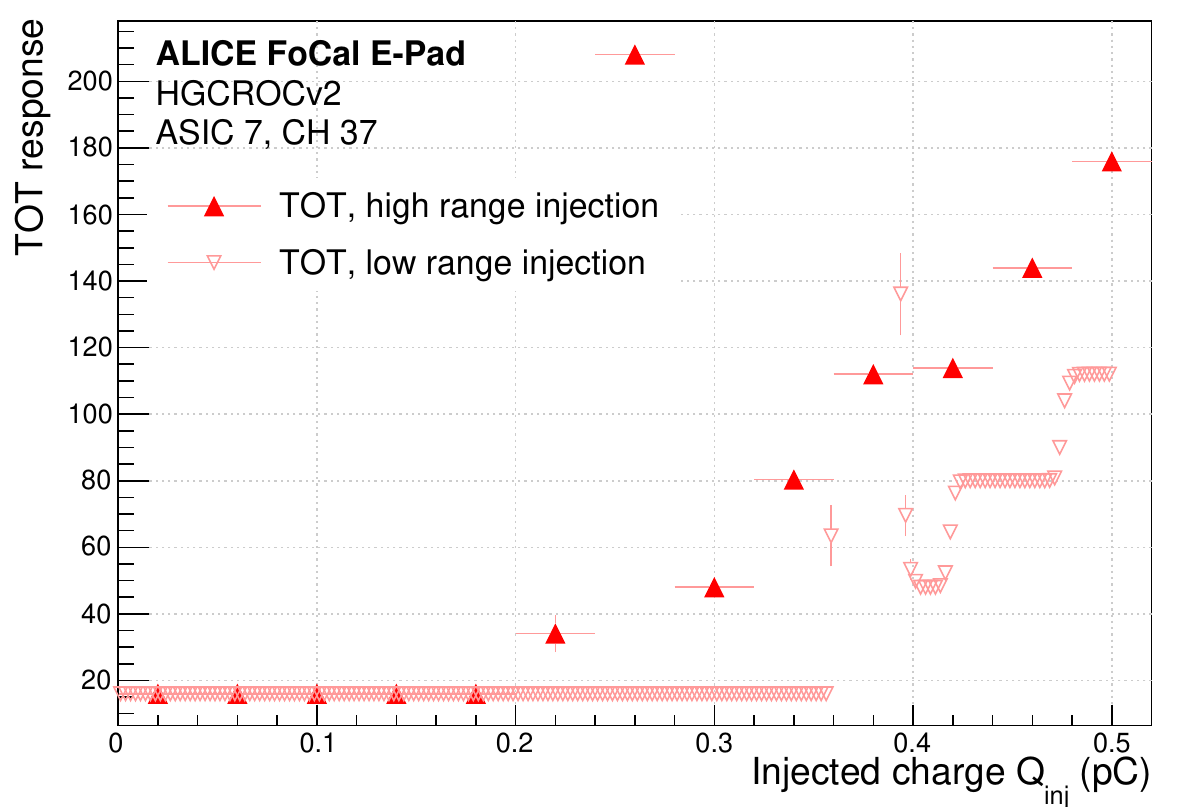}
\includegraphics[width=.49\textwidth]{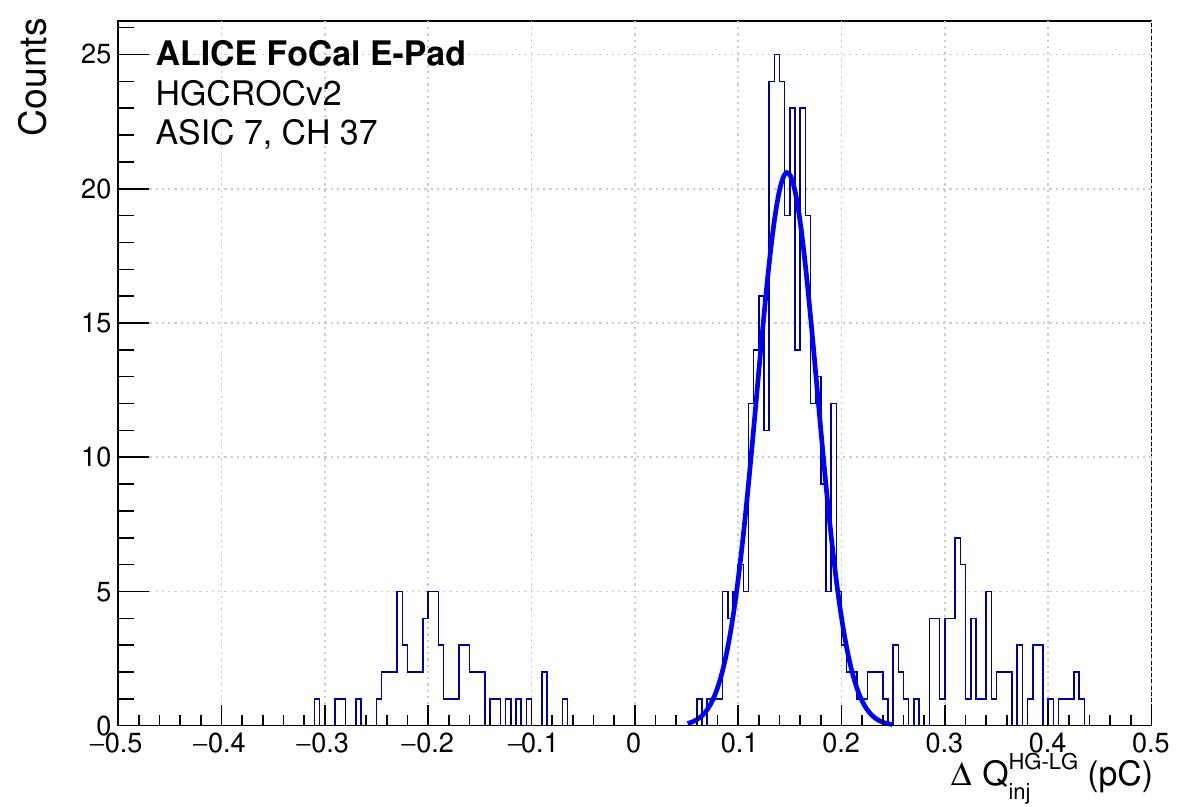}
\caption{
\label{fig:TOT_calibrations}
Left: \ac{ToT} response as a function of the injected charge for high-range and low-range injection regimes.
Right: the difference, $\Delta Q_{\rm inj}^{\rm HG-LG}$, of the \ac{ToT} turn-on position between high-range and low-range injection, fitted with a Gaussian curve in order to extract the mean.}
\end{center}
\end{figure}

\begin{figure}[ht!]
\begin{center}
\includegraphics[width=.49\textwidth]{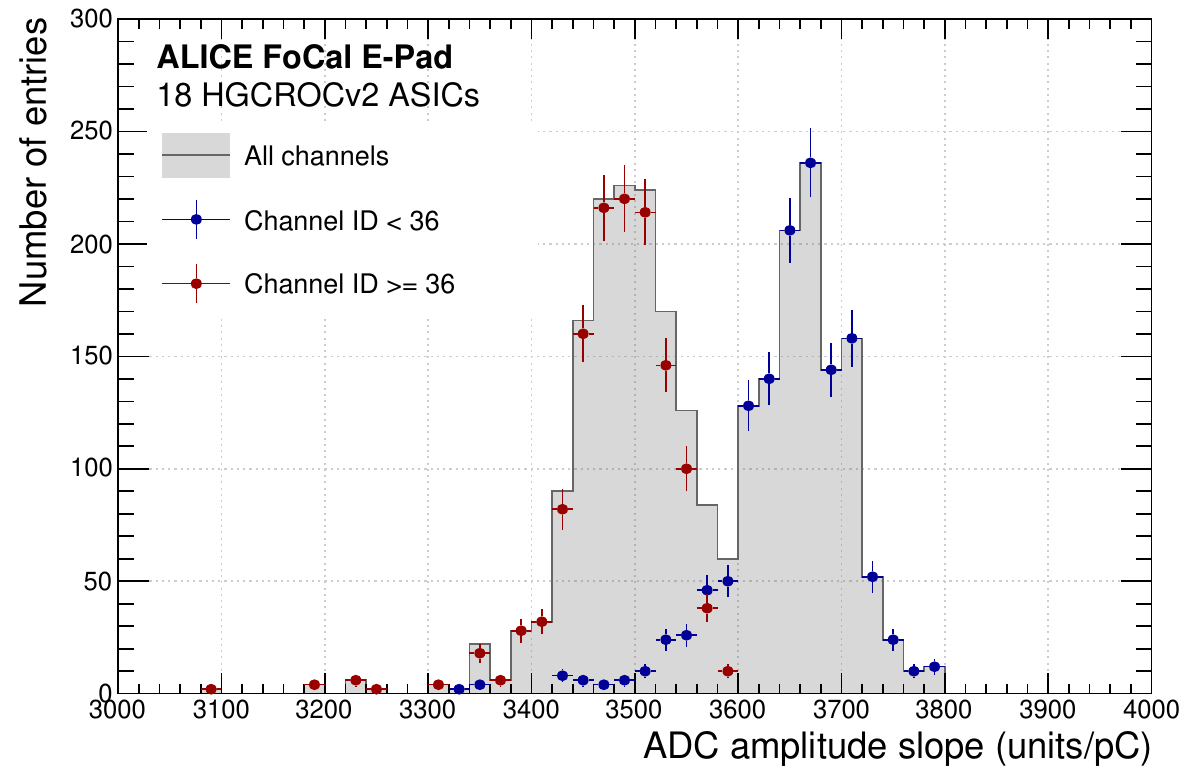}
\includegraphics[width=.49\textwidth]{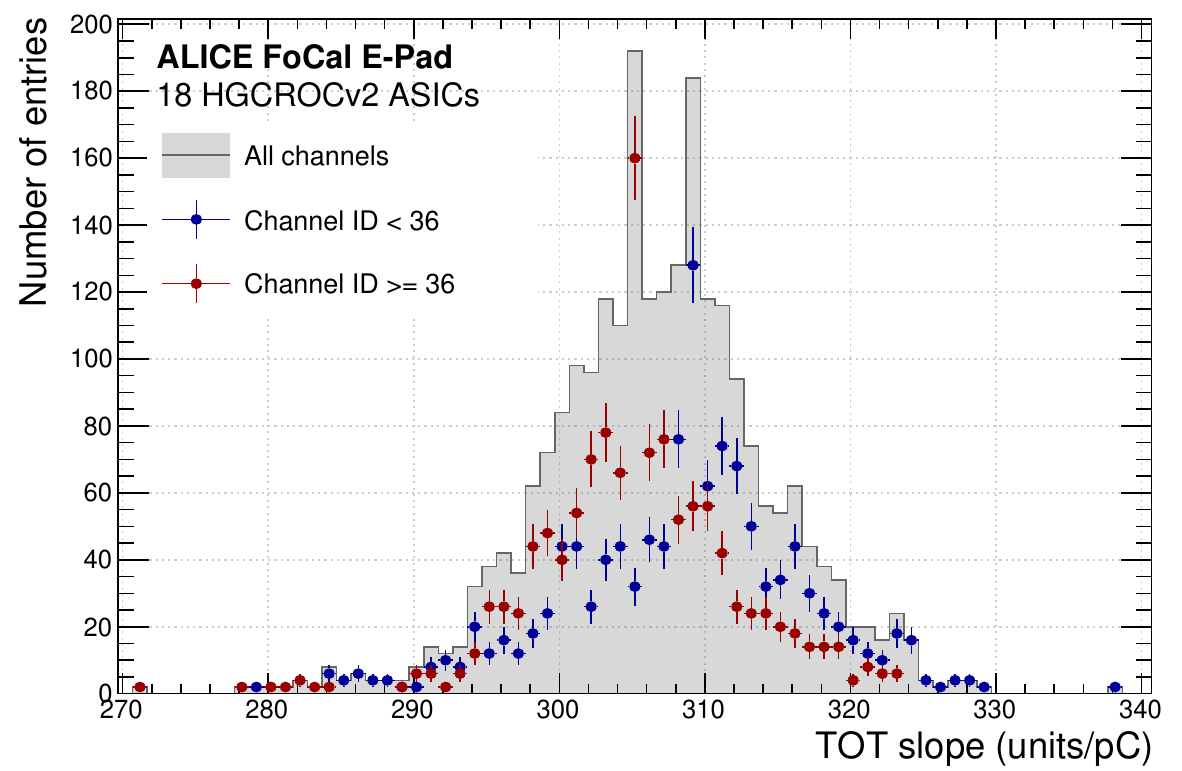}
\caption{\label{fig:calib_slopes} Fitted slope parameter for the calibration of \acs{ADC} and \acs{ToT} response. The two populations in the histograms are related to the slighlty different architectures in the two halves of the \acs{HGCROC}. }
\end{center}
\end{figure}

While the chip's performance shows a good linearity~(about $\pm$ 0.1\% of the full dynamic range) in the tests performed by the development team~\cite{Thienpont:2020wau}, a design error in the grounding routing of the \ac{ASIC} produces an irreducible voltage offset, $\Delta U_0 = 30\mV$, in the charge injection schematic. 
This effectively limits the minimum injection to about $\Delta Q_{\rm min}^{\rm LG} = 0.5\,\pF \times 30\,\mV = 15\fC$. 
Similarly, this contribution results in an offset of $\Delta Q_{\rm max}^{\rm HG} = 8.0\,\pF \times 30\,\mV = 240\fC$ in the high-range regime.
This effect is shown in \Fig{fig:TOT_calibrations}, where the \ac{ToT} is reported as a function of the injected charge. 
For our detector, we determine $\Delta U_0$ from data by evaluating per channel the discrepancy in the turn-on position of the \ac{ToT} between low-range and high-range injection.
A simple computer program was written which coarsely finds the position (expressed in units of charge according to \Eq{eq:q-injection}) where the first \ac{ToT} values $> 0$ are measured, as depicted in \Fig{fig:TOT_calibrations} (left).
The distribution of evaluated offsets between the two ranges is shown in \Fig{fig:TOT_calibrations} (right).
We evaluate an offset of $\Delta Q_{\rm inj}^{\rm HG-LG} = 150\fC$ from a Gaussian fit to the main peak.
\begin{equation}
    \Delta U_0 = \frac{\Delta Q_{\rm inj}^{\rm HG-LG}}{C_{\rm inj,high} - C_{\rm inj,low}} = \frac{150\fC}{8.0\pF - 0.5\pF}=  20\mV \, ,
\end{equation}
which is in reasonable agreement with the expectation of $30\,\mV$.
The channel-wise calibration parameters are obtained from linear fits in the domain $Q_{\rm inj} < 0.2\pC$ for the \ac{ADC}, and $Q_{\rm inj} > 2\pC$ for \ac{ToT}, as shown in \Fig{fig:calibration_pads}.
The calibrated \ac{ToT} values are corrected by the constant charge offset $\Delta Q_{\rm inj}$.
For lower charge values (in particular in the turn-on region), the response of the \ac{ToT} is not linear, and we derive a calibration by linearly interpolating between the \ac{ToT} response at 0 and 2\pC\ (where the former one is always zero).
\Figure{fig:calib_slopes} shows the distribution of the fitted slope parameters for the \ac{ADC} amplitude (left panel) and the \ac{ToT} response (right panel).
The two populations in the histograms are most likely related to the slightly different architectures in the two halves of the \ac{HGCROC}, and therefore to the ID of the front-end channels.
For the \ac{ADC} response we obtain from our calibration procedure a mean response factor of $3650\,\text{units}/\pC = 3.65\,\text{units}/\fC$ for channel IDs $<36$ and $3500\,\text{units}/\pC = 3.50\,\text{units}/\fC$ for channel IDs $\geq 36$.
In particular, the \acs{ADC} response to a minimum ionizing particle is expected to be around $3.6\,\text{units}/\fC \cdot 3.7\fC = 13.3\,\text{units}$, where the \acs{MPV} of 3.7\fC{ }was used to describe the charge generated by a \acs{MIP} in 300\um{ }silicon.
The slope of the \acs{ToT} response was calibrated to be $310\,\text{units}/\pC$ and $305\,\text{units}/\pC$, respectively, and the \acs{ToT} offset was in the order of 120\,units (not shown).
In channels where the high-range \acs{ToT} calibration procedure failed, the mean values of the fitted parameters in the corresponding \ac{ASIC} half were assigned.
For the \ac{ADC} only the proportionality factor was used for the calibration, and the channel-wise pedestal to subtract was determined from dedicated pedestal runs without beam signal.
The successful calibrations were then used to convert the response of the \ac{HGCROC} to actual charge deposit in the pad sensors components.
%

\FloatBarrier

\section{Inter-calibration of the \acsp{SiPM} for \acs{FoCal-H}}
\label{appendix:hcal_calibration}
One of the main ingredients for the calibration of the \ac{FoCal-H} is the gain (inter-)calibration of the \acp{SiPM}.
The second prototype of \acrshort{FoCal-H} uses 249 \ac{HPK} S13360-6025 \acp{SiPM}. 
However, the manufacturer does not guarantee the equality of the breakdown voltage~($\rm{V_{br}}$), declaring it as 53$\pm$5\,V~\cite{S13360-SiPM-datasheet}. 
The CAEN DT5202 used to readout the towers allows one to bias \acp{SiPM} with individually-adjusted voltage values. 
To equalize their overvoltage and thus the \ac{SiPM} gain, a dedicated measurement of $\rm{V_{\rm br}}$ was performed for each photosensor prior to the test beam campaign.

\begin{figure}[th!]
\begin{center}
\includegraphics[width=0.6\textwidth]{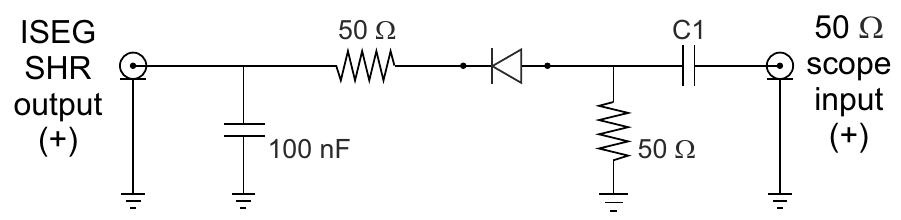}
\caption{\label{fig:SiPM-bias-circuit} Circuitry of the \ac{SiPM} bias and readout board used for the ${V_{\rm br}}$ measurement. Using $C1=1.5$~pF corresponds to the low-gain operation mode of the CAEN DT5202 system, $C1=15$~pF corresponds to the high-gain operation mode.}
\end{center}
\end{figure}

The large number of \acp{SiPM} used in the prototype, and ultimately in the future detector assembly, prioritized the choice of a simplified technique to determine $\rm{V_{\rm br}}$. 
This was done by measuring the individual dependencies of bias current versus bias voltage~(I-V curves). 
For this, each \ac{SiPM} was connected according to the circuitry shown in \Fig{fig:SiPM-bias-circuit}, replicating the one used in the CAEN DT5202 modules~\cite{CAEN-DT5202-manual}.
The I-V curves were measured at a ramp-up rate of 0.1\,V/s, which was low enough to keep the ``background'' current below 5\,\nA. 

\begin{figure}[t!]
\begin{center}
\includegraphics[width=0.6\textwidth]{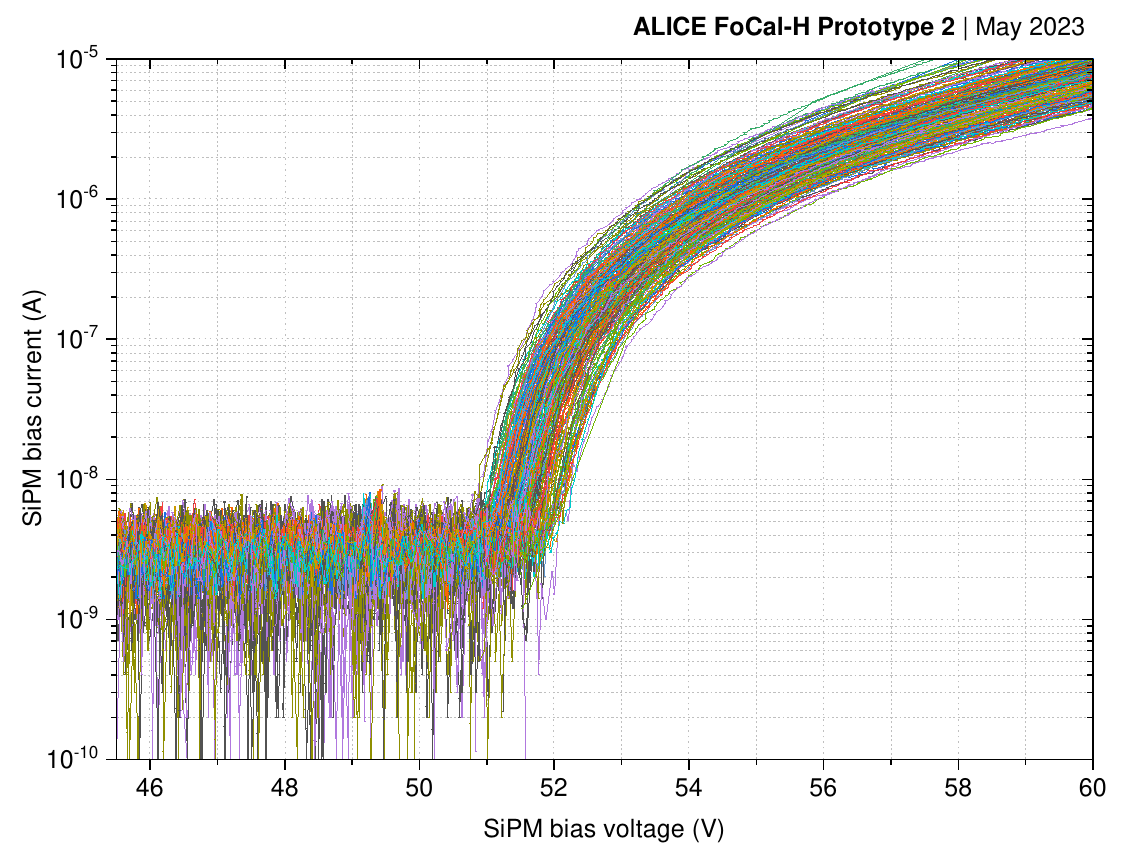}
\caption{\label{fig:249-IV-curves} I-V curves measured for each \ac{SiPM} at 0.1~V/s ramp rate.}
\end{center}
\end{figure}

\Figure{fig:249-IV-curves} shows the 249 I-V curves measured for all \acp{SiPM}. 
To derive the precise point corresponding to the true ${V_{\rm br}}$ value, various techniques could be considered depending on the exact application~\cite{Nagy_2017}. 
For our application, the primary goal was the relative gain-matching of the 249 \acp{SiPM}. 
Therefore, as a simple technique consistent for all devices tested, ${V_{\rm br}}$ was defined as the point at which the current exceeds the 10\,nA threshold.

\begin{figure}[th!]
\begin{center}
\includegraphics[width=0.8\textwidth]{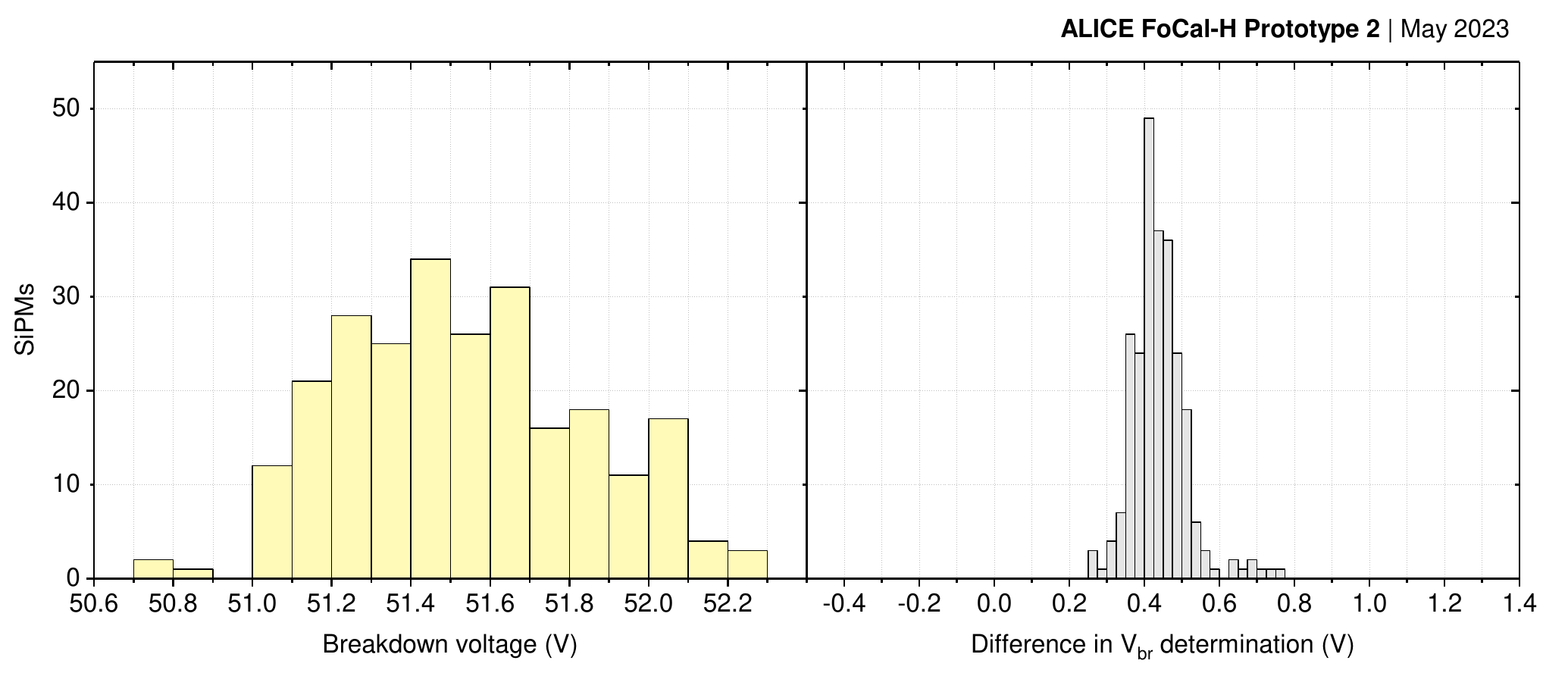}
\caption{\label{fig:Vbr-spread} Left: spread of the $\rm{V_{br}}$ for the 249 \acp{SiPM} defined at the point where the current exceeds the 10\,nA threshold; mean = 51.5~V, standard deviation = 0.3~V. Right: difference in $\rm{V_{br}}$ between the values  determined by a simple threshold or by exponential fitting; mean = 0.43\,V, standard deviation = 0.054\,V.}
\end{center}
\end{figure}

\Figure{fig:Vbr-spread} (left) shows the resulting distribution of ${V_{\rm br}}$, with mean of $51.5$~V and a spread of $0.3$~V.
Another approach is to define the starting point of the exponential rise of the I-V curve above the breakdown voltage with an exponential fit. 
The average difference between this and the former approach is 0.43\,V with a standard deviation of 0.05\,V, as shown in the right panel of \Fig{fig:Vbr-spread}.

\begin{figure}[t!]
\begin{center}
\includegraphics[width=0.6\textwidth]{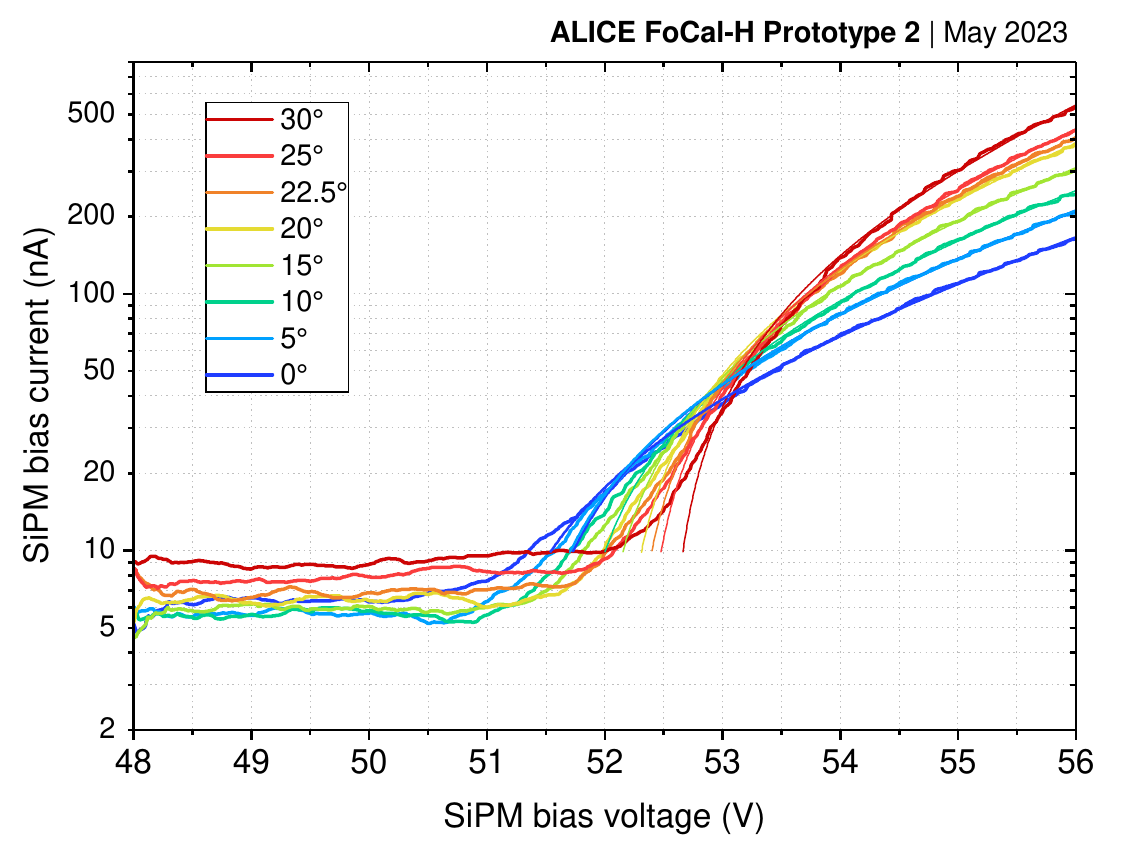}
\caption{\label{fig:4SiPM-0to30deg-Vbr} I-V dependence of one \ac{SiPM} measured for various \ac{SiPM} temperatures.}
\end{center}
\end{figure}

\begin{figure}[t!]
\begin{center}
\includegraphics[width=0.8\textwidth]{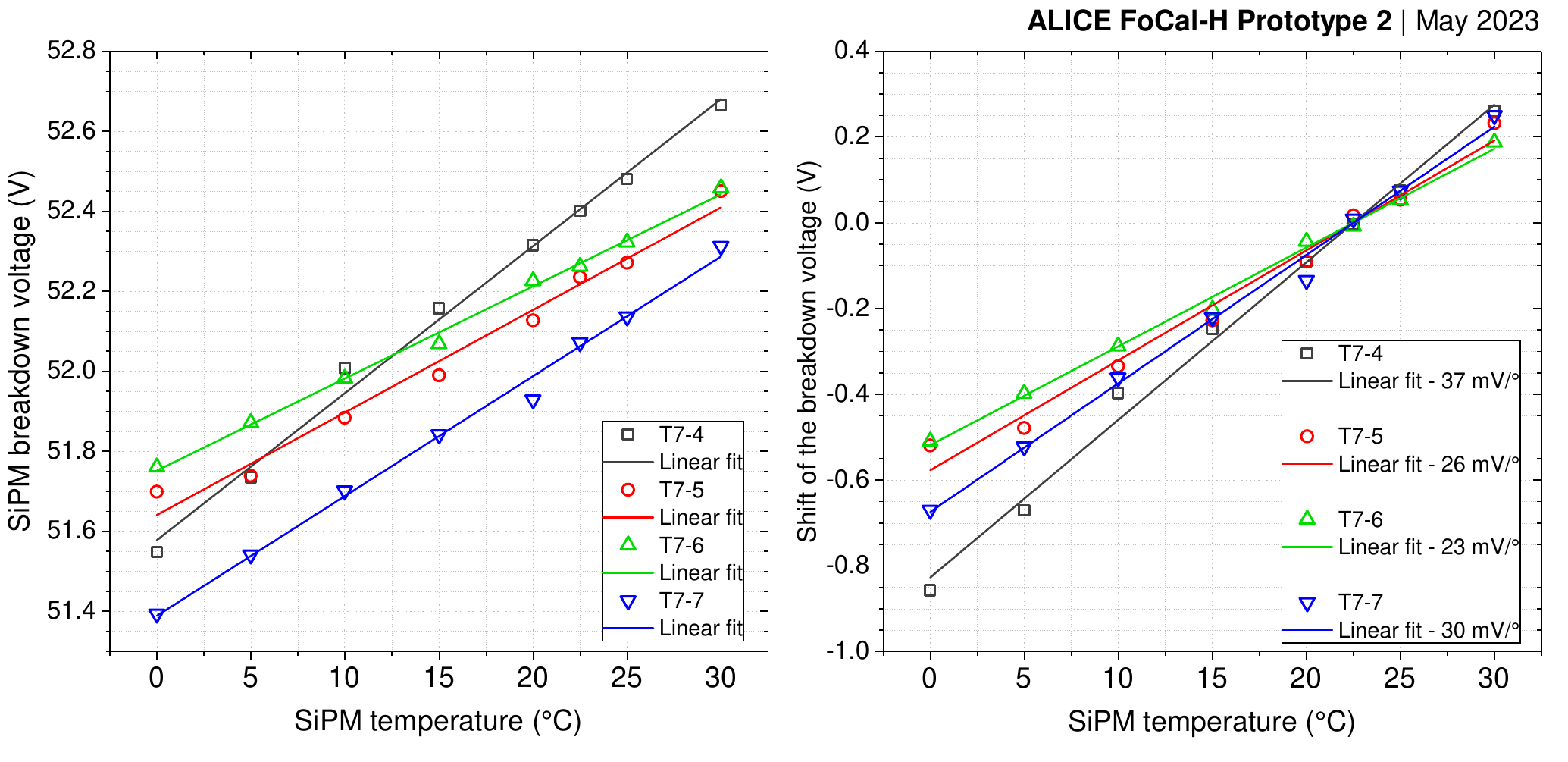}
\caption{\label{fig:4SiPM-temperature} Left: breakdown voltages of the four \acrshort{SiPM}~(named T7-4 to T7-7) on the test board as a function of temperature. Right: shift of the breakdown voltages of the four \acp{SiPM} on the test board as a function of temperature.}
\end{center}
\end{figure}

The light-tight box used as the \ac{SiPM} test bench was not thermostabilized, but the temperature in the \ac{SiPM} vicinity was monitored with a probe. 
The entire characterization process was conducted at $22\pm0.6^\circ$C.
Influence of such temperature spread to the precision of the ${V_{\rm br}}$ was estimated by characterizing the behavior of four \acp{SiPM} on a test board in a climatic chamber. 
\Figure{fig:4SiPM-0to30deg-Vbr} shows the I-V temperature dependence for one of the \ac{SiPM}. 
\Fig{fig:4SiPM-temperature} summarizes the temperature dependence of the breakdown voltage~(left) and the difference to the nominal breakdown voltage~(right) for each of the 4 \acp{SiPM}. 
A typical temperature coefficient of ${V_{\rm br}}$ is 29\,mV/K was found. 
Although there is a significant spread of 25\% (at a fairly small statistics of 4 devices), we estimate minor temperature variations of the \acp{SiPM} in the test board to result in an uncertainty below $\pm10$~mV.

\begin{figure}[t!]
\begin{center}
\includegraphics[width=0.8\textwidth]{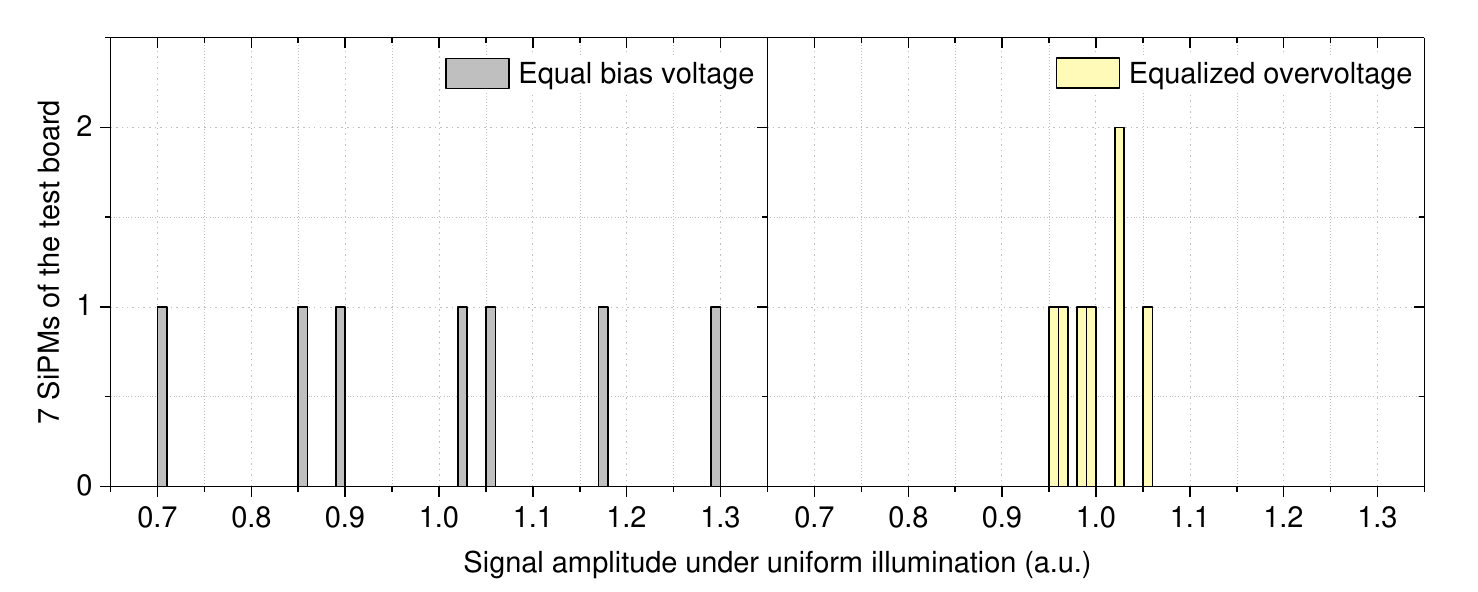}
\caption{\label{fig:7SiPM-equal-bias-OVV} Response function (gain $\times$ \ac{PDE}) of the 7 \acp{SiPM} of the test board under a uniform 440~nm illumination. Left: measured at equal bias voltage with an average overvoltage of 3~V; standard deviation: 20\%. Right: measured at equal overvoltage; standard deviation: 3.7\%.}
\end{center}
\end{figure}

The inherent spread in $\rm{V_{br}}$ seen at the statistics of 249 \acp{SiPM} ($\pm0.6$~V) results in $\pm20$\% spread in gain and $\pm15$\% spread in \ac{PDE} at 3V overvoltage~\cite{S13360-SiPM-datasheet}. 
The outcome of the gain-matching, as measured for the limited set of 7 \acp{SiPM} of a test board under a uniform illumination with 440~nm light, was a $\pm3.7$\% spread in the response function, as shown in \Fig{fig:7SiPM-equal-bias-OVV}.
\ifextrafigs
\newpage
\section{Extra figures: for the transverse shower analysis of pads}

Development in the central \acrshort{FoCal-E} layers for a 100\GeV{} electron beam is shown in \Figure{fig:pad-layer-by-layer-100GeV}. There, the mean of the normalize energy distributions $\langle Q_{layers}\rangle$ increases and retracts after reaching the shower maximum $t_{max}$, with the latter growing as a function of the initial energy. In the plots, the data are overlapped with the simulation results which display a compatible behavior with the experimental data points. 
The particle population of the electromagnetic showers develops with the depth inside the \acrshort{FoCal-E} stack and reaches its maximum within the detector.

The longitudinal profile of the energy deposited per detecting layer is investigated integrating the full charge distribution .  
These measurements are performed as a function of the selected beam energy and they require systematic studies of the shower transversal dimensions, for ensuring the full collection of the secondary products.
Finally, a calibration procedure, described in~\ref{appendix:pads_calibration} is needed for correlating the energy deposited to the response of the \acrshort{HGCROC} readout circuit. 

\begin{figure}[t!]
\begin{center}
\includegraphics[width=0.33\textwidth]{figures/pad/20GeV_LayerTransverseFraction.pdf}
\includegraphics[width=0.33\textwidth]{figures/pad/20GeV_LayerFraction.pdf}
\includegraphics[width=0.33\textwidth]{figures/pad/20GeV_LayerTransverseRMS.pdf}
\caption{Left: Fraction $Q_{N} / Q$ of the 20\,\GeV{ }transverse shower signal contained in a pad window of size $N$ with respect to the FoCal-E layer. Right: RMS of the 20 GeV transverse shower signal with respect to the FoCal-E layer.}
\label{fig:pad-trans-20GeV}
\end{center}
\end{figure}

\begin{figure}[t!]
\begin{center}
\includegraphics[width=0.33\textwidth]{figures/pad/60GeV_LayerTransverseFraction.pdf}
\includegraphics[width=0.33\textwidth]{figures/pad/60GeV_LayerFraction.pdf}
\includegraphics[width=0.33\textwidth]{figures/pad/60GeV_LayerTransverseRMS.pdf}
\caption{Left: Fraction $Q_{N} / Q$ of the 60 GeV transverse shower signal contained in a pad window of size $N$ with respect to the FoCal-E layer. Right: RMS of the 60 GeV transverse shower signal with respect to the FoCal-E layer.}
\label{fig:pad-trans-60GeV}
\end{center}
\end{figure}

\begin{figure}[t!]
\begin{center}
\includegraphics[width=0.33\textwidth]{figures/pad/100GeV_LayerTransverseFraction.pdf}
\includegraphics[width=0.33\textwidth]{figures/pad/100GeV_LayerFraction.pdf}
\includegraphics[width=0.33\textwidth]{figures/pad/100GeV_LayerTransverseRMS.pdf}
\caption{Left: Fraction $Q_{N} / Q$ of the 100 GeV transverse shower signal contained in a pad window of size $N$ with respect to the FoCal-E layer. Right: RMS of the 100 GeV transverse shower signal with respect to the FoCal-E layer.}
\label{fig:pad-trans-100GeV}
\end{center}
\end{figure}

\begin{figure}[t!]
\begin{center}
\includegraphics[width=0.33\textwidth]{figures/pad/150GeV_LayerTransverseFraction.pdf}
\includegraphics[width=0.33\textwidth]{figures/pad/150GeV_LayerFraction.pdf}
\includegraphics[width=0.33\textwidth]{figures/pad/150GeV_LayerTransverseRMS.pdf}
\caption{Left: Fraction $Q_{N} / Q$ of the 150 GeV transverse shower signal contained in a pad window of size $N$ with respect to the FoCal-E layer. Right: RMS of the 150 GeV transverse shower signal with respect to the FoCal-E layer.}
\label{fig:pad-trans-150GeV}
\end{center}
\end{figure}

The transversal development of the showers is investigated along the x and y axis, profiling the layer-by-layer distribution of the charge; this is needed for setting the final transversal acquisition windows. The results for both directions for a 100\GeV{} electron beam are shown in \Figure{fig:pad-trans-layer-by-layer-100GeV}. In the plots, the shaded areas representing the simulation studies consistently envelops the overlapped data points, which give a preliminary idea of the lateral spread of the showers secondary products. 
More detailed representations of these quantities can be found in the panels of \Figure{fig:pad-trans-100GeV}. There, the fraction of the total deposited charge for different collection windows is reported as a function of the shower depth. In particular, the combined studies over the two directions illustrated on the rightmost plot of \Figure{fig:pad-trans-100GeV} provides a guide to the data selection for the following analysis studies. 
For this analysis, the collection window was kept to 3x3 sensors in the initial four layers and increased to a 5$\times$5 matrix in the later layers. This ensure a relative charge collection of about 85\% over the whole detector's length. 

\begin{figure}[t!]
\begin{center}
\includegraphics[width=\textwidth]{figures/pad/layers/20GeV/20GeV_LayerTransverseX.pdf}
\includegraphics[width=\textwidth]{figures/pad/layers/20GeV/20GeV_LayerTransverseY.pdf}
\caption{Transverse shower profile in x direction (top) and y direction (bottom) of \acrshort{FoCal-E}pad Layers 1 to 20 for 20\,\GeV{ }electrons. The data are drawn with the data points, the simulation with the filled area.}
\label{fig:pad-trans-layer-by-layer-20GeV}
\end{center}
\end{figure}

\begin{figure}[t!]
\begin{center}
\includegraphics[width=\textwidth]{figures/pad/layers/60GeV/60GeV_LayerTransverseX.pdf}
\includegraphics[width=\textwidth]{figures/pad/layers/60GeV/60GeV_LayerTransverseY.pdf}
\caption{Transverse shower profile in x direction (top) and y direction (bottom) of \acrshort{FoCal-E}pad Layers 1 to 20 for 60\,\GeV{ }electrons. The data are drawn with the data points, the simulation with the filled area.}
\label{fig:pad-trans-layer-by-layer-60GeV}
\end{center}
\end{figure}

\begin{figure}[t!]
\begin{center}
\includegraphics[width=\textwidth]{figures/pad/layers/100GeV/100GeV_LayerTransverseX.pdf}
\includegraphics[width=\textwidth]{figures/pad/layers/100GeV/100GeV_LayerTransverseY.pdf}
\caption{Transverse shower profile in x direction (top) and y direction (bottom) of \acrshort{FoCal-E}pad Layers 1 to 20 for 100\,\GeV{ }electrons. The data are drawn with the data points, the simulation with the filled area.}
\label{fig:pad-trans-layer-by-layer-100GeV}
\end{center}
\end{figure}

\begin{figure}[t!]
\begin{center}
\includegraphics[width=\textwidth]{figures/pad/layers/150GeV/150GeV_LayerTransverseX.pdf}
\includegraphics[width=\textwidth]{figures/pad/layers/150GeV/150GeV_LayerTransverseY.pdf}
\caption{Transverse shower profile in x direction (top) and y direction (bottom) of \acrshort{FoCal-E}pad Layers 1 to 20 for 150\,\GeV{ }electrons. The data are drawn with the data points, the simulation with the filled area.}
\label{fig:pad-trans-layer-by-layer-150GeV}
\end{center}
\end{figure}

\begin{figure}[t!]
\begin{center}
\includegraphics[width=0.33\textwidth]{figures/pad/20GeV_LayerTransverseFraction.pdf}
\includegraphics[width=0.33\textwidth]{figures/pad/20GeV_LayerFraction.pdf}
\includegraphics[width=0.33\textwidth]{figures/pad/20GeV_LayerTransverseRMS.pdf}
\caption{Right: Fraction $Q_{N} / Q$ of the 20\,\GeV{ }transverse shower signal contained in a pad window of size $N$ with respect to the \acrshort{FoCal-E} layer. Left: \acrshort{RMS} of the 20 GeV transverse shower signal with respect to the \acrshort{FoCal-E} layer.}
\label{fig:pad-trans-20GeV}
\end{center}
\end{figure}

\begin{figure}[t!]
\begin{center}
\includegraphics[width=0.33\textwidth]{figures/pad/60GeV_LayerTransverseFraction.pdf}
\includegraphics[width=0.33\textwidth]{figures/pad/60GeV_LayerFraction.pdf}
\includegraphics[width=0.33\textwidth]{figures/pad/60GeV_LayerTransverseRMS.pdf}
\caption{Right: Fraction $Q_{N} / Q$ of the 60 GeV transverse shower signal contained in a pad window of size $N$ with respect to the \acrshort{FoCal-E} layer. Left: \acrshort{RMS} of the 60 GeV transverse shower signal with respect to the \acrshort{FoCal-E} layer.}
\label{fig:pad-trans-60GeV}
\end{center}
\end{figure}

\begin{figure}[t!]
\begin{center}
 \includegraphics[width=0.45\textwidth]{figures/pad/100GeV_LayerTransverseFraction.pdf}
 \includegraphics[width=0.45\textwidth]{figures/pad/100GeV_LayerFraction.pdf}
 \includegraphics[width=0.45\textwidth]{figures/pad/100GeV_LayerTransverseRMS.pdf}
\caption{Right: Fraction $Q_{N} / Q$ of the 100\GeV{} transverse shower signal contained in a pad window of size $N$ with respect to the \acrshort{FoCal-E} layer. Left: \acrshort{RMS} of the 100\GeV{} transverse shower signal with respect to the \acrshort{FoCal-E} layer.}
\label{fig:pad-trans-100GeV}
\end{center}
\end{figure}

\begin{figure}[t!]
\begin{center}
\includegraphics[width=0.45\textwidth]{figures/pad/150GeV_LayerTransverseFraction.pdf}
\includegraphics[width=0.45\textwidth]{figures/pad/150GeV_LayerFraction.pdf}
\includegraphics[width=0.45\textwidth]{figures/pad/150GeV_LayerTransverseRMS.pdf}
\caption{Right: Fraction $Q_{N} / Q$ of the 150 GeV transverse shower signal contained in a pad window of size $N$ with respect to the \acrshort{FoCal-E} layer. Left: \acrshort{RMS} of the 150 GeV transverse shower signal with respect to the \acrshort{FoCal-E} layer.}
\label{fig:pad-trans-150GeV}
\end{center}
\end{figure}
%
%
%
\begin{figure}[t!]
\begin{center}
\includegraphics[width=0.3\textwidth]{figures/pad/layers/20GeV/20GeV_Layer1.pdf}
\includegraphics[width=0.3\textwidth]{figures/pad/layers/20GeV/20GeV_Layer2.pdf}
\includegraphics[width=0.3\textwidth]{figures/pad/layers/20GeV/20GeV_Layer3.pdf}
\includegraphics[width=0.3\textwidth]{figures/pad/layers/20GeV/20GeV_Layer4.pdf}
\includegraphics[width=0.3\textwidth]{figures/pad/layers/20GeV/20GeV_Layer6.pdf}
\includegraphics[width=0.3\textwidth]{figures/pad/layers/20GeV/20GeV_Layer7.pdf}
\includegraphics[width=0.3\textwidth]{figures/pad/layers/20GeV/20GeV_Layer8.pdf}
\includegraphics[width=0.3\textwidth]{figures/pad/layers/20GeV/20GeV_Layer9.pdf}
\includegraphics[width=0.3\textwidth]{figures/pad/layers/20GeV/20GeV_Layer11.pdf}
\includegraphics[width=0.3\textwidth]{figures/pad/layers/20GeV/20GeV_Layer12.pdf}
\includegraphics[width=0.3\textwidth]{figures/pad/layers/20GeV/20GeV_Layer13.pdf}
\includegraphics[width=0.3\textwidth]{figures/pad/layers/20GeV/20GeV_Layer14.pdf}
\includegraphics[width=0.3\textwidth]{figures/pad/layers/20GeV/20GeV_Layer15.pdf}
\includegraphics[width=0.3\textwidth]{figures/pad/layers/20GeV/20GeV_Layer16.pdf}
\includegraphics[width=0.3\textwidth]{figures/pad/layers/20GeV/20GeV_Layer17.pdf}
\includegraphics[width=0.3\textwidth]{figures/pad/layers/20GeV/20GeV_Layer18.pdf}
\includegraphics[width=0.3\textwidth]{figures/pad/layers/20GeV/20GeV_Layer19.pdf}
\includegraphics[width=0.3\textwidth]{figures/pad/layers/20GeV/20GeV_Layer20.pdf}
\caption{
Layer-by-layer charge signal compared to simulation for 20\GeV{ }electrons. The signal for Layer 4 is top left and the signal for Layer 14 is bottom right.}
\label{fig:pad-layer-by-layer-20GeV}
\end{center}
\end{figure}

\begin{figure}[t!]
\begin{center}
\includegraphics[width=0.3\textwidth]{figures/pad/layers/60GeV/60GeV_Layer1.pdf}
\includegraphics[width=0.3\textwidth]{figures/pad/layers/60GeV/60GeV_Layer2.pdf}
\includegraphics[width=0.3\textwidth]{figures/pad/layers/60GeV/60GeV_Layer3.pdf}
\includegraphics[width=0.3\textwidth]{figures/pad/layers/60GeV/60GeV_Layer4.pdf}
\includegraphics[width=0.3\textwidth]{figures/pad/layers/60GeV/60GeV_Layer6.pdf}
\includegraphics[width=0.3\textwidth]{figures/pad/layers/60GeV/60GeV_Layer7.pdf}
\includegraphics[width=0.3\textwidth]{figures/pad/layers/60GeV/60GeV_Layer8.pdf}
\includegraphics[width=0.3\textwidth]{figures/pad/layers/60GeV/60GeV_Layer9.pdf}
\includegraphics[width=0.3\textwidth]{figures/pad/layers/60GeV/60GeV_Layer11.pdf}
\includegraphics[width=0.3\textwidth]{figures/pad/layers/60GeV/60GeV_Layer12.pdf}
\includegraphics[width=0.3\textwidth]{figures/pad/layers/60GeV/60GeV_Layer13.pdf}
\includegraphics[width=0.3\textwidth]{figures/pad/layers/60GeV/60GeV_Layer14.pdf}
\includegraphics[width=0.3\textwidth]{figures/pad/layers/60GeV/60GeV_Layer15.pdf}
\includegraphics[width=0.3\textwidth]{figures/pad/layers/60GeV/60GeV_Layer16.pdf}
\includegraphics[width=0.3\textwidth]{figures/pad/layers/60GeV/60GeV_Layer17.pdf}
\includegraphics[width=0.3\textwidth]{figures/pad/layers/60GeV/60GeV_Layer18.pdf}
\includegraphics[width=0.3\textwidth]{figures/pad/layers/60GeV/60GeV_Layer19.pdf}
\includegraphics[width=0.3\textwidth]{figures/pad/layers/60GeV/60GeV_Layer20.pdf}
\caption{
\label{fig:pad-layer-by-layer-60GeV} Layer-by-layer charge signal compared to simulation for 60\GeV{ }electrons. The signal for Layer 4 is top left and the signal for Layer 14 is bottom right.}
\end{center}
\end{figure}
\begin{figure}[t!]
\begin{center}
\includegraphics[width=0.3\textwidth]{figures/pad/layers/150GeV/150GeV_Layer1.pdf}
\includegraphics[width=0.3\textwidth]{figures/pad/layers/150GeV/150GeV_Layer2.pdf}
\includegraphics[width=0.3\textwidth]{figures/pad/layers/150GeV/150GeV_Layer3.pdf}
\includegraphics[width=0.3\textwidth]{figures/pad/layers/150GeV/150GeV_Layer4.pdf}
\includegraphics[width=0.3\textwidth]{figures/pad/layers/150GeV/150GeV_Layer6.pdf}
\includegraphics[width=0.3\textwidth]{figures/pad/layers/150GeV/150GeV_Layer7.pdf}
\includegraphics[width=0.3\textwidth]{figures/pad/layers/150GeV/150GeV_Layer8.pdf}
\includegraphics[width=0.3\textwidth]{figures/pad/layers/150GeV/150GeV_Layer9.pdf}
\includegraphics[width=0.3\textwidth]{figures/pad/layers/150GeV/150GeV_Layer11.pdf}
\includegraphics[width=0.3\textwidth]{figures/pad/layers/150GeV/150GeV_Layer12.pdf}
\includegraphics[width=0.3\textwidth]{figures/pad/layers/150GeV/150GeV_Layer13.pdf}
\includegraphics[width=0.3\textwidth]{figures/pad/layers/150GeV/150GeV_Layer14.pdf}
\includegraphics[width=0.3\textwidth]{figures/pad/layers/150GeV/150GeV_Layer15.pdf}
\includegraphics[width=0.3\textwidth]{figures/pad/layers/150GeV/150GeV_Layer16.pdf}
\includegraphics[width=0.3\textwidth]{figures/pad/layers/150GeV/150GeV_Layer17.pdf}
\includegraphics[width=0.3\textwidth]{figures/pad/layers/150GeV/150GeV_Layer18.pdf}
\includegraphics[width=0.3\textwidth]{figures/pad/layers/150GeV/150GeV_Layer19.pdf}
\includegraphics[width=0.3\textwidth]{figures/pad/layers/150GeV/150GeV_Layer20.pdf}
\caption{
\label{fig:pad-layer-by-layer-150GeV} Layer-by-layer charge signal compared to simulation for 150\GeV{ }electrons. The signal for Layer 4 is top left and the signal for Layer 14 is bottom right.}
\end{center}
\end{figure}

Finally, these quantities can be included in a two-dimensional histogram, where the signal amplitude is plotted as a function of the layer number. \Figure{fig:pad-long-shower-profile-all} reports the longitudinal shower profiles measured with 20, 60, 80, 100, 120, and 150\GeV{} beam energies. The figure present profile comprehensive of the information gathered by the \acrshort{FoCal-E} pixels, in layers 5 and 10 of the stack. The charge collected by the digital, high granularity detectors is calculated correlating the number of hits in the pixels arrays to the corresponding energy in the pads layers. These studies are presented in section~\ref{}. \\\textbf{ REF to pixels section} \\ 

As the hadron contamination increases with the increasing of the energy, the data sets acquired with beam energies greater than 100\GeV{} requires selection cuts to sort out the electronic showers. Using the assumption that hadronic showers develop in later layers, the electron-induced events are selected requiring greater signal amplitudes at the center of the detector. In particular, the analysis discriminates the two contributions tagging electron signals only if $Signal_{L7 \div L11} > 1.5\times Signal_{L17 \div L20}.$

\begin{figure}[t!]
\begin{center}
\includegraphics[width=0.495\textwidth]{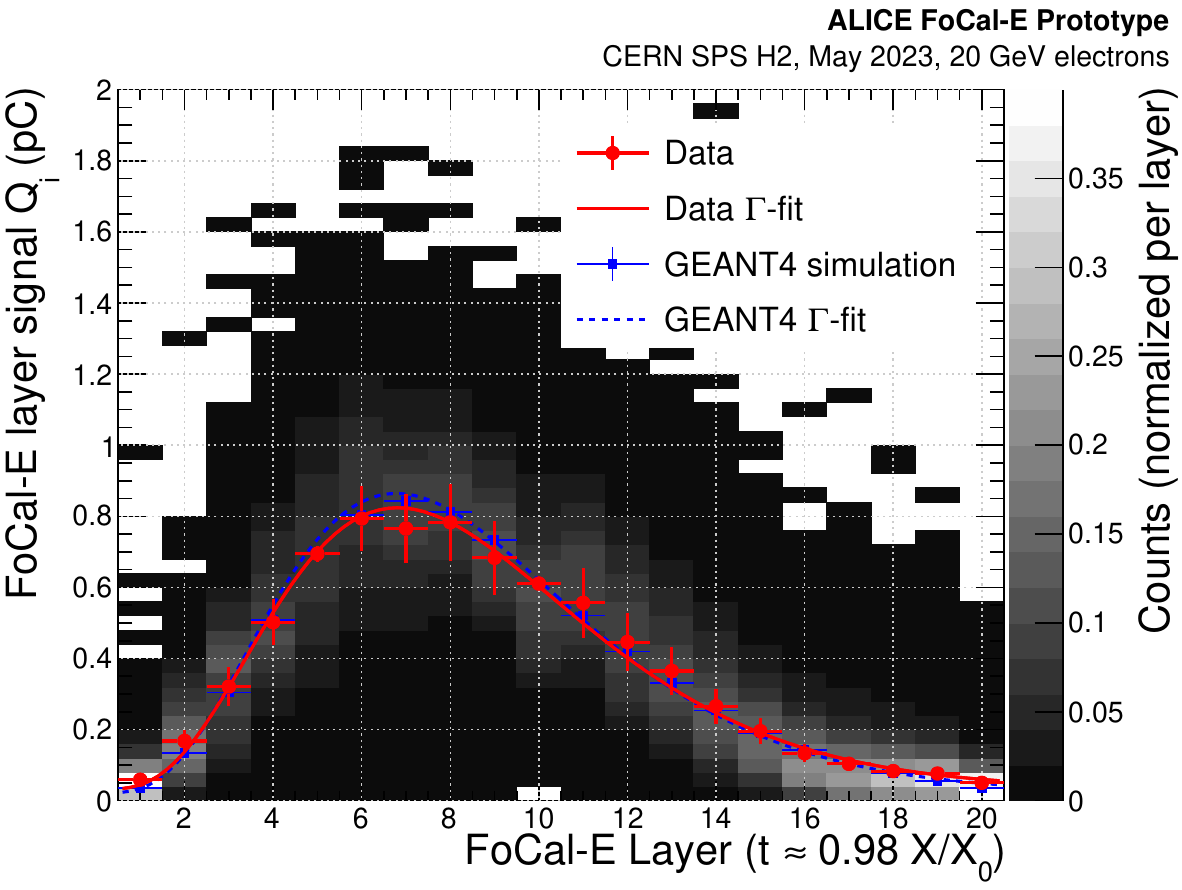}
\includegraphics[width=0.495\textwidth]{figures/pad/pad-long-shower-profile-60GeV.pdf}
\includegraphics[width=0.495\textwidth]{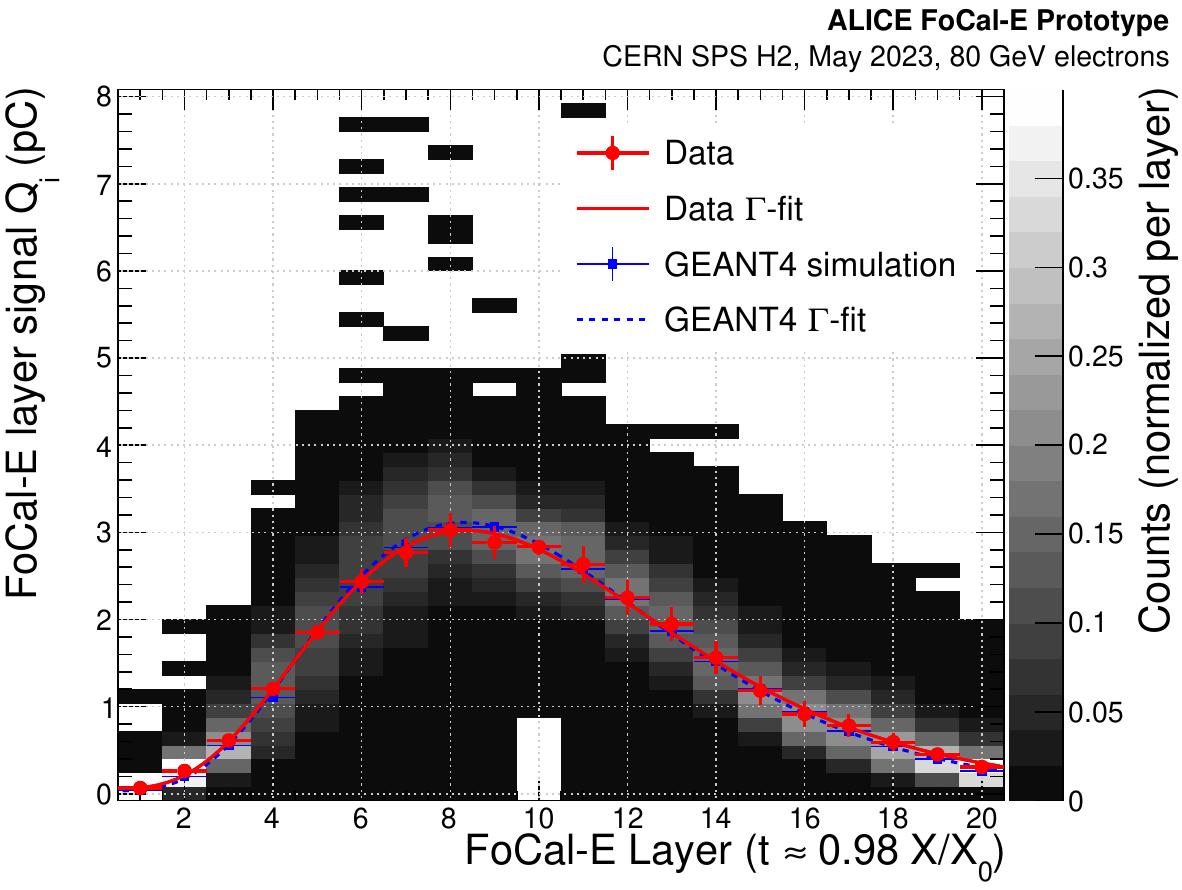}
\includegraphics[width=0.495\textwidth]{figures/pad/pad-long-shower-profile-100GeV.pdf}
\includegraphics[width=0.495\textwidth]{figures/pad/pad-long-shower-profile-120GeV.pdf}
\includegraphics[width=0.495\textwidth]{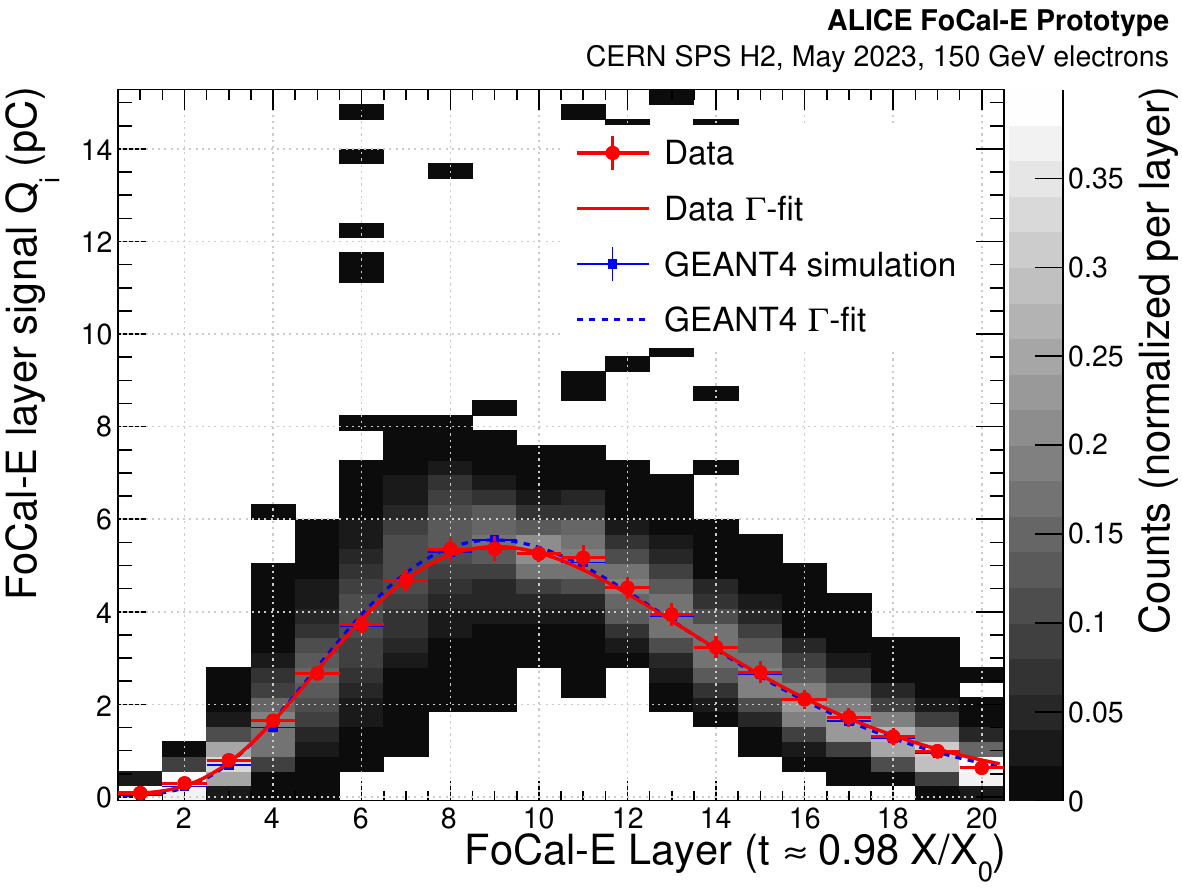}
\caption{
\label{fig:pad-long-shower-profile-all-app-single} Longitudinal shower profiles for 20\GeV{ }, 60\GeV{ }, 80\GeV{ }, 100\GeV{ }, 120\GeV{ }, and 150\GeV{ } electrons.}
\end{center}
\end{figure}
\begin{figure}[t!]
\begin{center}
\includegraphics[width=0.495\textwidth]{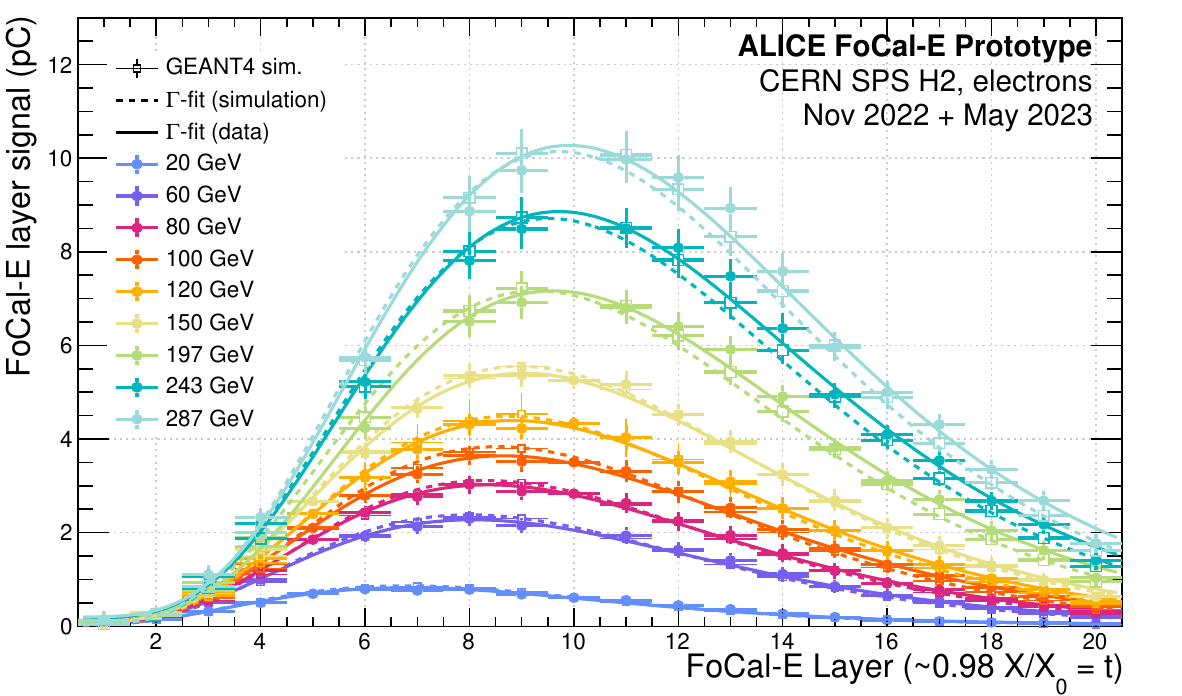}
\includegraphics[width=0.495\textwidth]{figures/pad/pad-long-shower-profile-showermax.pdf}
\caption{
\label{fig:pad-long-shower-profile-all-app} Longitudinal shower profiles for 20, 60, 100, and 150 GeV electrons compared to \geant simulations (left), and the position of the shower maximum compared to \geant simulation and the theoretical value of a homogeneous calorimeter.}
\end{center}
\end{figure}

\section{Extra figures: FoCal-E Pixel Transverse Shower Profiles}
\label{app:extra:transprof}

The two pixel layers of FoCal-E at layer 5 and 10 with a pixel pitch of about $30\,\um \times 30\,\um$ make it possible to resolve the structure of particle showers on the sub-millimeter scale.
In particular, this feature will be used to discriminate single photon (electromagnetic) shower events against background two-shower events from \pizero{ }decays.
\Figure{fig:pix-event-display-one-e-300GeV} shows an event display of a 287\GeV{ }electron shower event in pixel layer layer 5 (left) and 10 (right). The showers are characterized by pronounced cores in the shower center with hit densities higher than 400 pixel hits / $\mm^2$ in layer 10 (300 pixel hits / $\mm^2$ in layer 10), and are surrounded by tail components with less dense pixel hit occupancies.
\begin{figure}
 \includegraphics[width=0.49\textwidth]{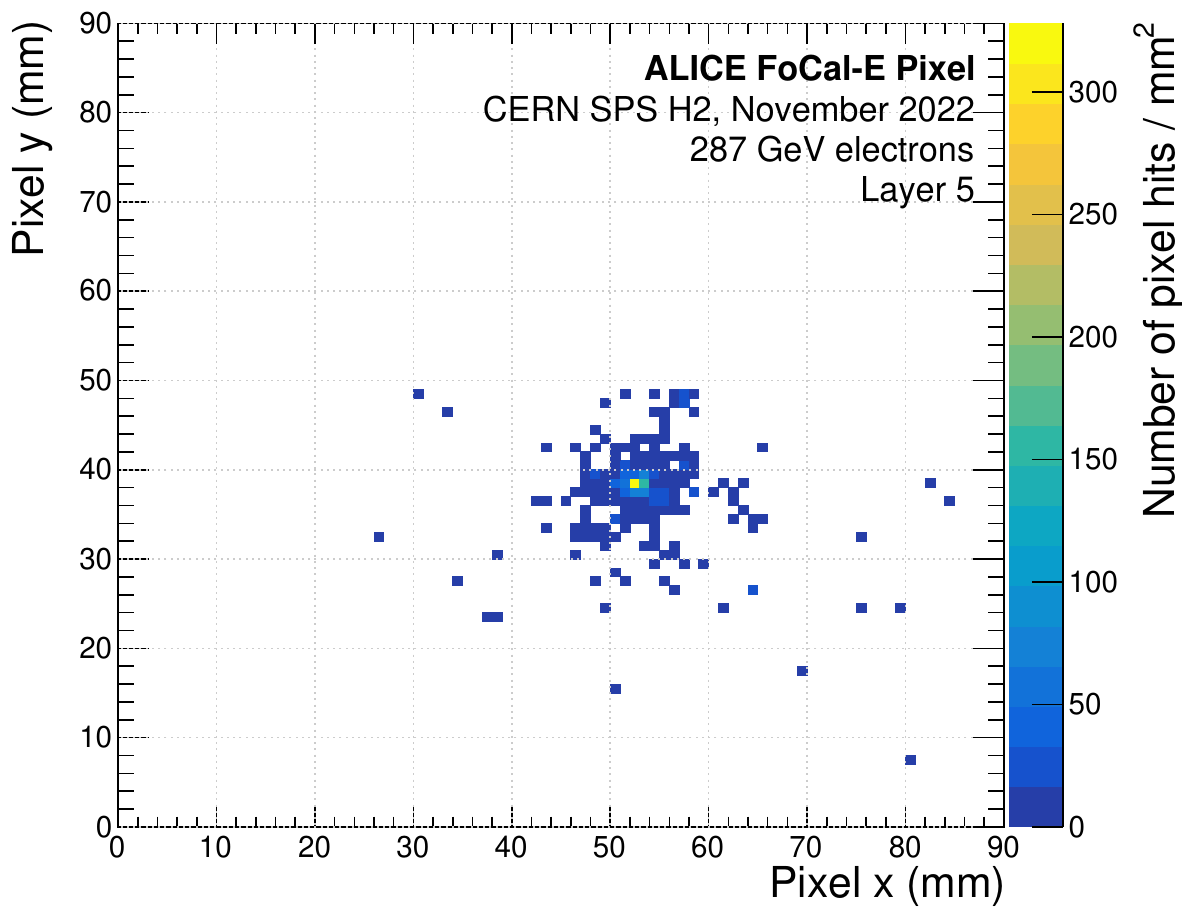}
 \includegraphics[width=0.49\textwidth]{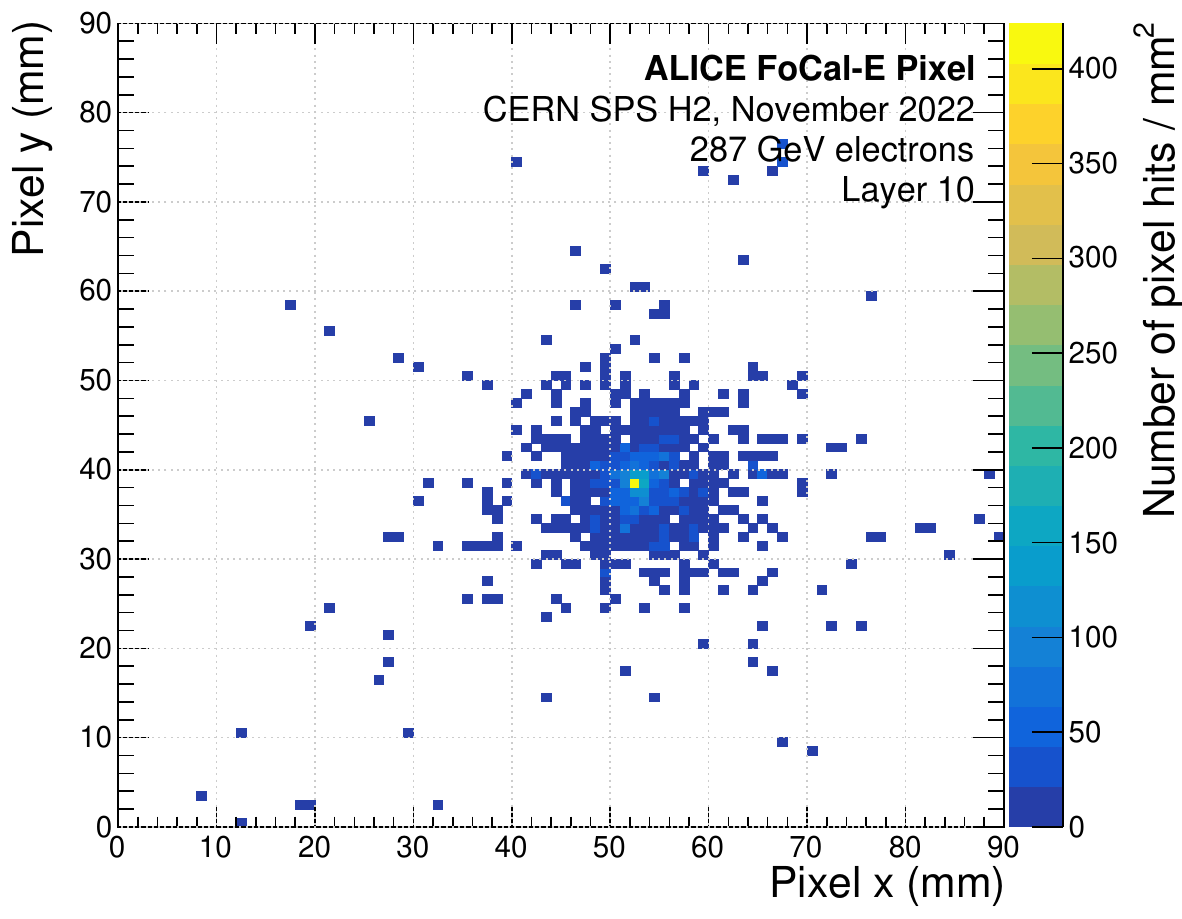}
 \caption{\label{fig:pix-event-display-one-e-300GeV}FoCal-E Pixel event display of a 287\,\GeV one-electron shower in layer 5 (left) and layer 10 (right).}
\end{figure}
Thanks to the high electron rate at the SPS H2 beam line and the relativiley long integration time of the ALPIDE pixel front-end ($\approx 5\us$), the setup was able to record multiple electron shower events, as shown in \Figure{fig:pix-event-display-two-e-300GeV}.
\begin{figure}
 \includegraphics[width=0.49\textwidth]{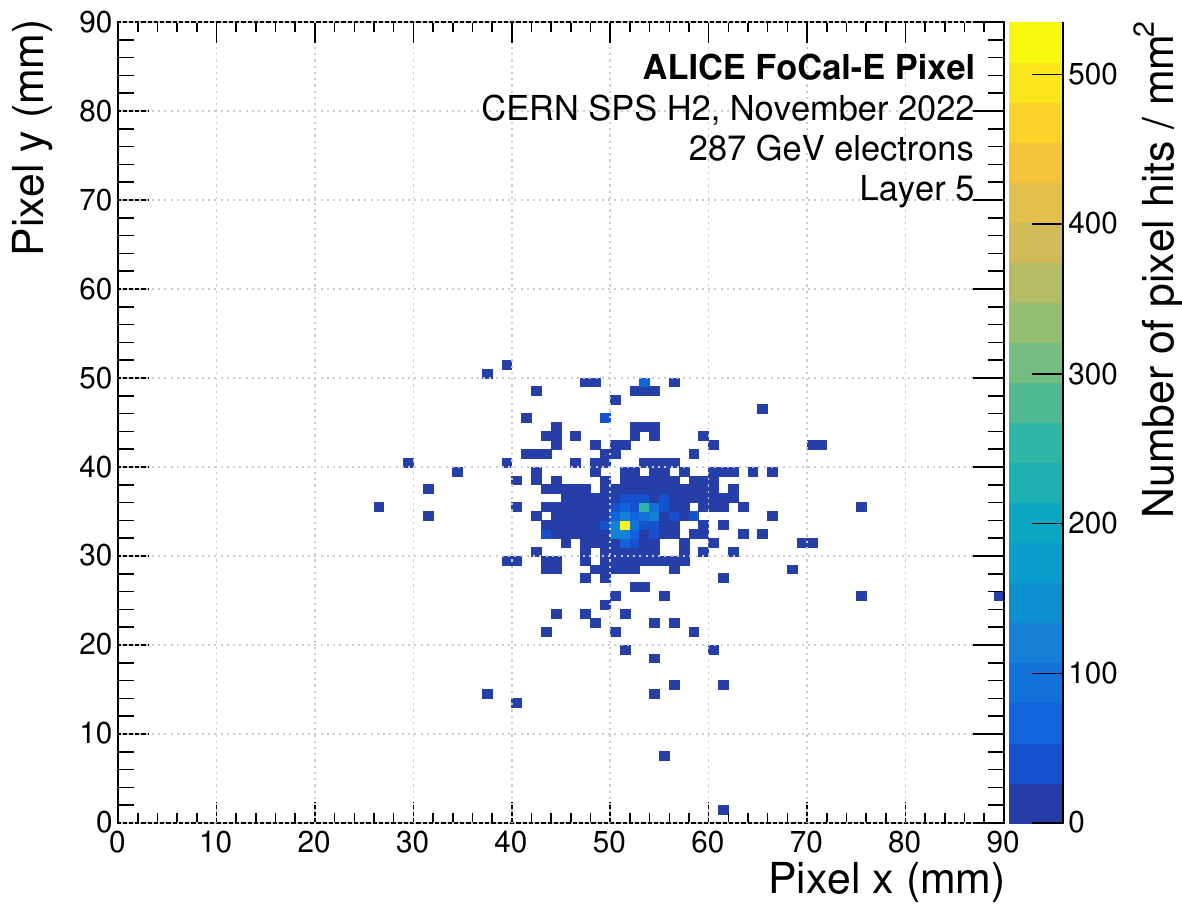}
 \includegraphics[width=0.49\textwidth]{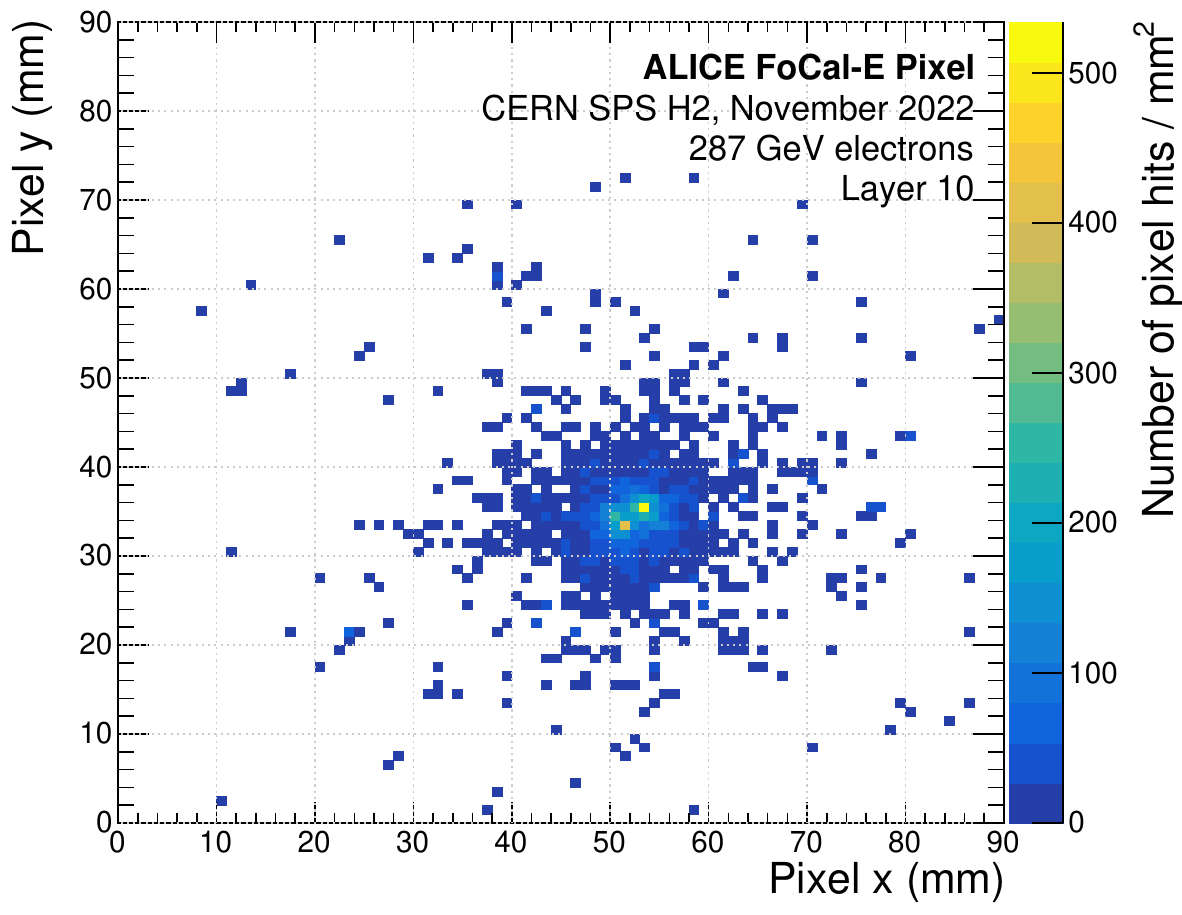}
 \caption{\label{fig:pix-event-display-two-e-300GeV}FoCal-E Pixel event display of a 287\,\GeV two-electron shower in layer 5 (left) and layer 10 (right).}
\end{figure}
Two electron shower events are clearly visible, and they can be separated in the hitmap only by eye on a scale lower than $1\,\cm$.\\
The event display also illustrates how varying the longitudinal development of a shower in the calorimeter can be: in layer 5 the left shower produces higher occupancies in the core, and in layer 10 the right shower is more intense.
In the final experiment rather complex clustering, particle identification and shower separation algorithms will be implemented, which will make use also of the FoCal-E Pads and FoCal-H information, to make a decision a) whether there are two showers in the event, and b) how to share the energy fraction between the two showers.\\
Since for the discrimination of two closeby showers the lateral hit density profiles are one of the key parameters, we present here a study of the transverse electromagnetic shower profile with the FoCal-E Pixel layers.
In order to do so, we project the hit density distribution to a lateral axis in the pixel layer plane, and measure the functional form $f(\Delta x)$ of the number of pixel hits, \Nhits, in dependence on the lateral distance $\Delta x$ from the shower center $x_0$:
\begin{equation}
 f(\Delta x) = \frac{1}{\Nhits} \frac{d}{dx} \Nhits (x - x_0) 
\end{equation}
where the $x$-axis is identical to the horizontal direction in the setup.
When calculating the distance from a particle hit $i$ from the shower center, $\Delta x_i = x_i - x_0$, the dominating uncertainty originates from the uncertainty on the shower center $x_0$, i.e.{ }$\sigma_{\Delta x} \approx \sigma_{x_{0}}$.
We therefor consider two different methods for the determination of the shower center in our analysis, one with \textit{stringent} criteria and one with rather \textit{loose} criteria.
In both methods pixel hits from a single event are filled into a hitmap, which is binned to macro-pixels of size 40 pixels $\times$ 40 pixels ($\approx 1.1\,\mathrm{\mm^2}$).
We calculate the shower center from the weighted mean position of the macro pixels, where we take only macro-pixels into account which contain a minimum number of hits $\Nhitsmin$, and which lie within a distance, $d$, lateral to the bin with the highest number of entries.
For the loose method we choose $d = 30 \text{macro-pixels}$ ($\approx 3.3\,\mathrm{cm}$) and \Nhitsmin = 1, and for the stringent method $d = 10\;\text{macro-pixels}$ ($\approx 1.1\,\mathrm{cm}$) and \Nhitsmin = 4.
Since the various methods for the determinaton of the shower center depend on the number of hits, a more realistic pixel response to particle hits in the GEANT4 simulation is implemented.
The technique is based on a pixel cluster generation model with Gaussian charge diffusion in the sensor, and an empiric parameterization curve deriving for the computation of the pixel cluster size. (\todo{cite Helge's thesis or whatever}).\par
\subsection{Characterization by shower width FWHM}
Figures \ref{fig:pixel-swidth-L5-x-stringent-log} to \ref{fig:pixel-swidth-L10-x-loose} show the measured transverse shower profiles in pixel layers 5 and 10 binned to bin widths of $200\um$.
\begin{figure}
\includegraphics[width=\textwidth]{figures/pixel/showerwidth/pixel-swidth-L5-x-stringent-log.pdf}
\caption{\label{fig:pixel-swidth-L5-x-stringent-log}FoCal-E Pixel Layer 5: measured and simulated lateral shower profiles for the \textit{stringent} shower center method in logarithmic scale.}
\end{figure}
\begin{figure}
\includegraphics[width=\textwidth]{figures/pixel/showerwidth/pixel-swidth-L5-x-loose-log.pdf}
\caption{\label{fig:pixel-swidth-L5-x-loose-log}FoCal-E Pixel Layer 5: measured and simulated lateral shower profiles for the \textit{loose} shower center method in logarithmic scale.}
\end{figure}
\begin{figure}
\includegraphics[width=\textwidth]{figures/pixel/showerwidth/pixel-swidth-L5-x-stringent.pdf}
\caption{\label{fig:pixel-swidth-L5-x-stringent}FoCal-E Pixel Layer 5: measured and simulated lateral shower profiles for the \textit{stringent} shower center method in linear scale.}
\end{figure}
\begin{figure}
\includegraphics[width=\textwidth]{figures/pixel/showerwidth/pixel-swidth-L5-x-loose.pdf}
\caption{\label{fig:pixel-swidth-L5-x-loose}FoCal-E Pixel Layer 5: measured and simulated lateral shower profiles for the \textit{loose} shower center method in linear scale.}
\end{figure}
\begin{figure}
\includegraphics[width=\textwidth]{figures/pixel/showerwidth/pixel-swidth-L10-x-stringent-log.pdf}
\caption{\label{fig:pixel-swidth-L10-x-stringent-log}FoCal-E Pixel Layer 10: measured and simulated lateral shower profiles for the \textit{stringent} shower center method in logarithmic scale.}
\end{figure}
\begin{figure}
\includegraphics[width=\textwidth]{figures/pixel/showerwidth/pixel-swidth-L10-x-loose-log.pdf}
\caption{\label{fig:pixel-swidth-L10-x-loose-log}FoCal-E Pixel Layer 10: measured and simulated lateral shower profiles for the \textit{loose} shower center method in logarithmic scale.}
\end{figure}
\begin{figure}
\includegraphics[width=\textwidth]{figures/pixel/showerwidth/pixel-swidth-L10-x-stringent.pdf}
\caption{\label{fig:pixel-swidth-L10-x-stringent}FoCal-E Pixel Layer 10: measured and simulated lateral shower profiles for the \textit{stringent} shower center method in linear scale.}
\end{figure}
\begin{figure}
\includegraphics[width=\textwidth]{figures/pixel/showerwidth/pixel-swidth-L10-x-loose.pdf}
\caption{\label{fig:pixel-swidth-L10-x-loose}FoCal-E Pixel Layer 10: measured and simulated lateral shower profiles for the \textit{loose} shower center method in linear scale.}
\end{figure}
The figures show the results for the loose and the stringent method, both on a linear and logarithmic ordinate.
The distributions are characterized by a sharp peak in the center (\textit{core}) and non-gaussian sidebands (\textit{tail}).
Independently from an analytical description of the transverse shower shapes, the FWHM can be used to quantify how pronounced the core of the shower is.
We interprete it as a first order estimator on which scale a discrimination against another closeby shower is generally possible.
We evaluate the full-width-half-maximum (FWHM) of the measured distributions by finding the maximum bin, and the bin position where the distributions drop below the half of this maximum.
Figures \ref{fig:pixel-swfwhm-L5-x} and \ref{fig:pixel-swfwhm-L10-x} show the FWHM for the two shower center methods in layer 5 and 10, respectively. 
\begin{figure}
\includegraphics[width=0.49\textwidth]{figures/pixel/showerwidth/pixel-swfwhm-L5-x-loose.pdf}
\includegraphics[width=0.49\textwidth]{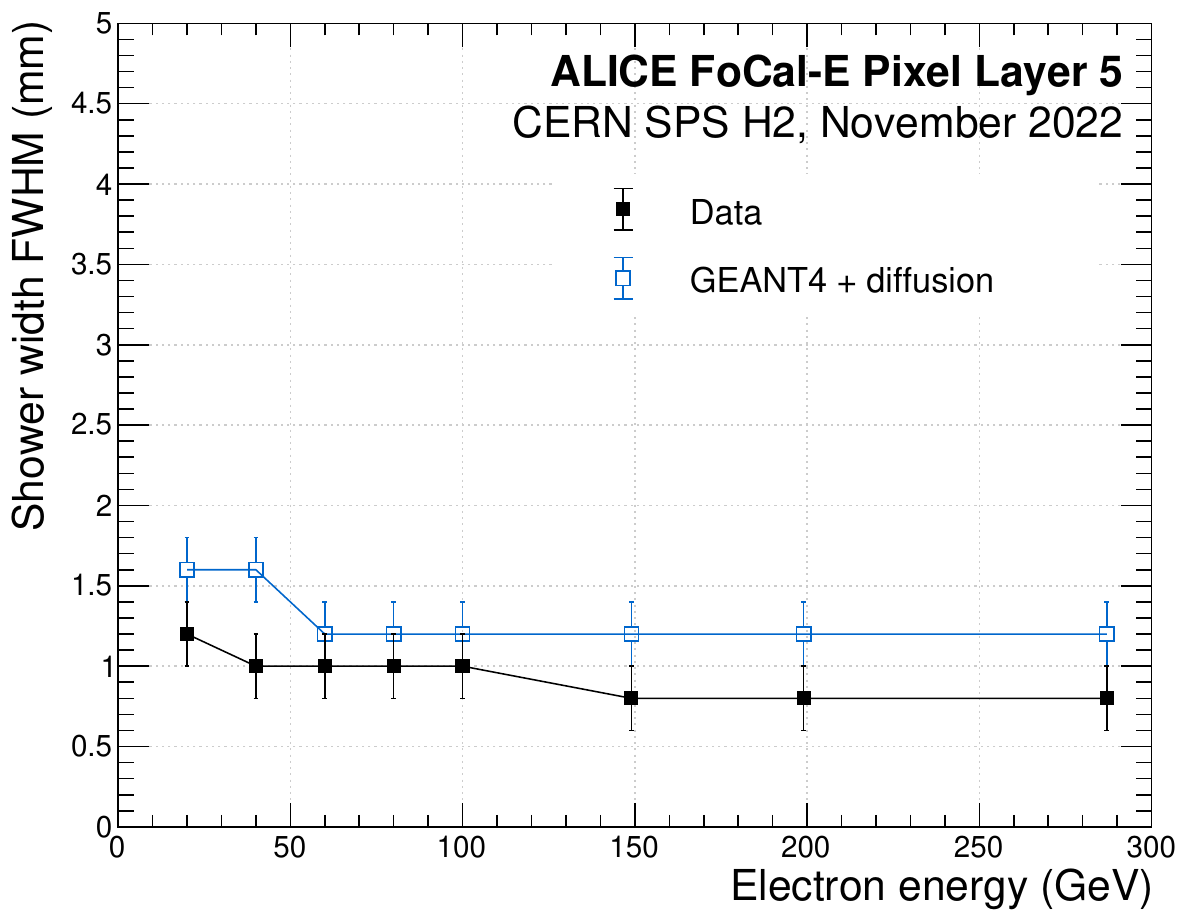}
\caption{\label{fig:pixel-swfwhm-L5-x}FoCal-E Pixel Layer 5: measured and simulated FWHM for the \textit{loose} (left) and \textit{stringent} (right) shower center method. The error bars indicate an uncertainty of 10\,\%.}
\end{figure}

\begin{figure}
\includegraphics[width=0.49\textwidth]{figures/pixel/showerwidth/pixel-swfwhm-L10-x-loose.pdf}
\includegraphics[width=0.49\textwidth]{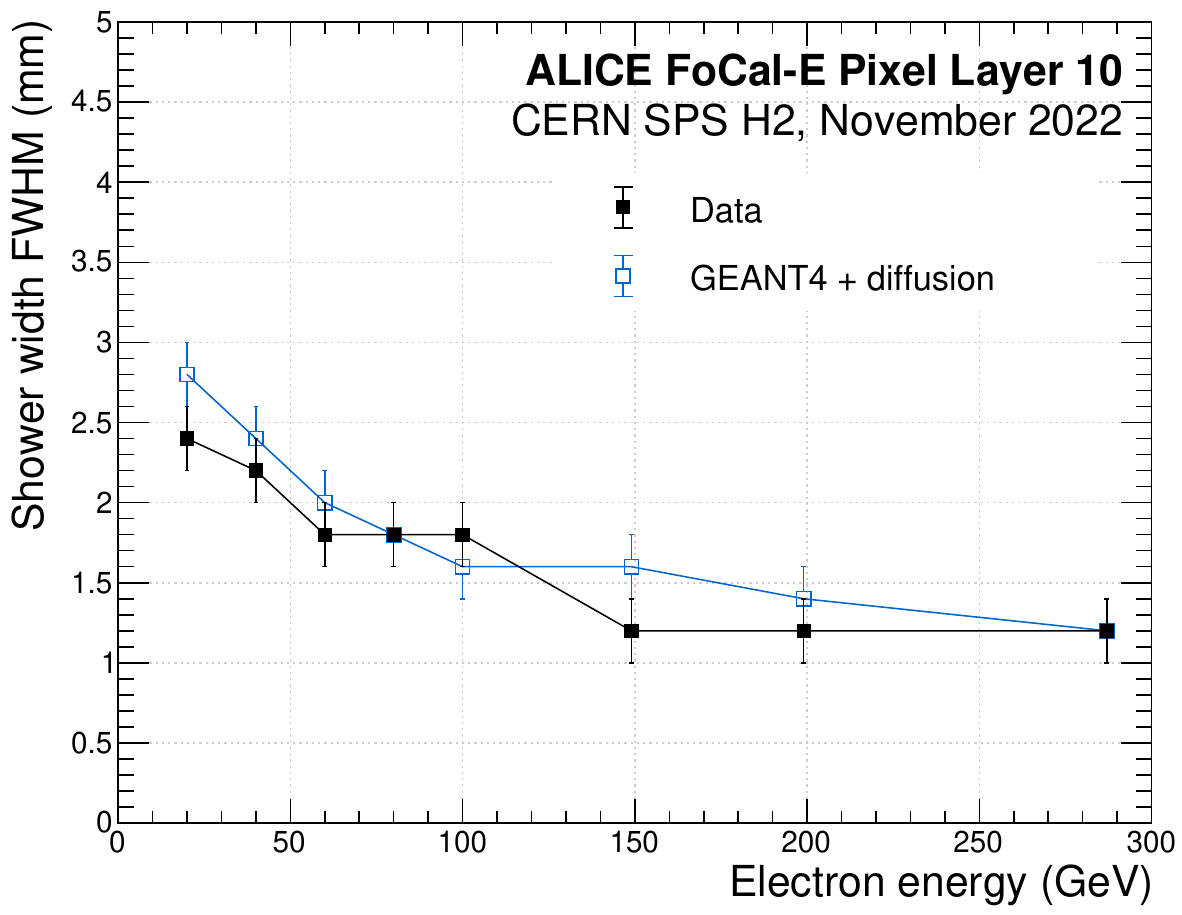}
\caption{\label{fig:pixel-swfwhm-L10-x}FoCal-E Pixel Layer 10: measured and simulated FWHM for the \textit{loose} (left) and \textit{stringent} (right) shower center method. The error bars indicate an uncertainty of 10\,\%.}
\end{figure}
The FWHM values are significantly smaller in layer 5 than in layer 10, which is expected because of the larger transverse spread at higher shower depths.
For the stringent shower center method in layer 5, we measure an FWHM of 2\mm{ }for electron energies of 20\GeV, and the FWHM drop with increasing energies down to 1.2\mm{ }for electron energies of 300\GeV.
For the loose shower center method, the FHWMs are significantly higher than for the stringent method, namely 3.2\mm{ }at 20\GeV{ }electron, and 1.6\mm{ }for 300\GeV{ }electrons.
Independently from the two methods, the measured FWHM values in layer 10 for higher energy electrons are measured to be in the order of $\approx 2.8$ to $3.6\mm$, which is about a factor of 2 higher than in layer 5.
For the stringent (loose) method the FWHM value increase towards lower electron energies, reach nearly 5\mm{ }(7\mm) at 20\GeV.\\
The simulation dataset - after having applied the cluster generation model - was analysed with the same criteria for the determination of the shower center like the data.
The simulated FWHM values are systematically higher than the data in layer 5, and lower in layer 10. However, the measured and simulated FWHM of the transverse shower widths are in general consistent within 20\,\%, or better.\\
Whether either the loose or the stringent method is applied for the determination of the shower center, has a direct effect on the prominence of the transverse shower peak.
While with the loose method outlier hits or potential noise hits might introduce fluctuations of the calculated shower center with respect to the real shower center, the stringent method rejects these contributions, emphasizing thus the high particle density core of the shower.
Other experiments (like e.g. \cite{Peitzmann:2022asy}) have circumvented the uncertainty, which originates from these methods, by making an independent position measurement of the incident electron before it enters the calorimeter.
\subsection{Characterization with an analytical model}
We attempt to describe the measured shower profiles with an analytic functional form with respect to the lateral distance $\Delta X = x - x_0$ from the shower center.
In general, this form depends on the calorimeter type and its detector geometry.
In particular, literature does not clearly indicate a generalized analytical form of transverse electromagnetic shower profiles
Hence, we attempt to describe the shower width with the ansatz presented in \cite{Grindhammer:1993kw}.
Here, the probability distribution function - with respect to radius $r$ from the shower center - is composed of two contributions, a core component, $f_C$ and a tail component, $f_T$, which are modulated with the distance constants $R_C$ and $R_T$, respectively:
\begin{align}
\label{eq:grindhammer_fr}
 \frac{d}{dr} \Nhits(r) = f(r) &= p f_C(r) + (1-p) f_T(r) \\
        &= p \frac{2r R_C^2}{(r^2 + R_C^2)^2} + (1-p) \frac{2r R_T^2}{(r^2 + R_T^2)^2} \, ,
\end{align}
where $p \in [0,1]$ denotes the fraction of particles which is carried by the core component of the shower.
In order to obtain lateral profiles, i.e. the probability density function along one axis in a cartesian coordinate system, we rewrite Eq.{ }\ref{eq:grindhammer_fr} with the surface element and assume azimuthal symmetry, thus eliminating the azimuthal angle $\phi$:
\begin{align}
\label{eq:grindhammer_fr}
 d \Nhits(r) = f(r) r dr  = s(x_{\parallel}) \; x_{\parallel} \; dx_{\parallel} \,
\end{align}
If we set $\Delta x = x-x_0 = x_{\parallel}$ and $x_0 = 0$, we can describe the measured shower shapes with the lateral profile
\begin{align}
 s( x) & = p \,x\,f_C( x)  + (1-p) \, x \, f_T( x)  \\
        &= p \, \frac{2 x^2 R_C^2}{( x^2 + R_C^2)^2} \; + \; (1-p) \, \frac{2  x^2 R_T^2}{( x^2 + R_T^2)^2} \, .
\end{align}
We describe the event-by-event uncertainty on $x_0$ by convoluting $s(x)$ with a simple Gaussian function
\begin{equation}
\tilde{g}(x) = N g(x) = N \exp{ \bigg( -  \frac{(x-x_{\text{sys}})^2}{2 \sigma_{x_0}^2}  \bigg) } \, ,
\end{equation}
where $N$ is the normalization constant, $x_{\text{sys}}$ describes potential systematic errors in the determination of the shower center, and $ \sigma_{x_0}$ accounts for event-by-event uncertainties on the shower center.
The fit function, which we implemented numerically, is given by
\begin{equation}
 t(x) = N g(x) * s(x)
\end{equation}
Figures \ref{fig:pixel-swidth-L5-x-stringent-log} to \ref{fig:pixel-swidth-L10-x-loose} show the function t(x) fitted to data and simulation.
The fit is stable and describes the data qualitatively very well.
However it does not provide a very accurate description since the $\chi^2 / \text{ndf}$ are high, and e.g.\; the peak in the shower center is not described well. \todo{ratio plots}
Figures \ref{fig:swparameters-L5-x-loose} to \ref{fig:swparameters-L10-x-stringent} show the fitted parameters, $R_C$, $R_T$, $\sigma_{x_{0}}$, and $p$, with respect to the electron energy.
\begin{figure}
\includegraphics[width=\textwidth]{figures/pixel/showerwidth/pixel-swparameters-L5-x-loose.pdf}
\caption{\label{fig:swparameters-L5-x-loose} FoCal-E pixel layer 5, \textit{loose} center method, fitted parameters $R_C$, $R_T$, $\sigma_{x_{0}}$, and $p$.}
\end{figure}
\begin{figure}
\includegraphics[width=\textwidth]{figures/pixel/showerwidth/pixel-swparameters-L5-x-stringent.pdf}
\caption{\label{fig:swparameters-L5-x-stringent} FoCal-E pixel layer 5, \textit{stringent} center method, fitted parameters $R_C$, $R_T$, $\sigma_{x_{0}}$, and $p$.}
\end{figure}
\begin{figure}
\includegraphics[width=\textwidth]{figures/pixel/showerwidth/pixel-swparameters-L10-x-loose.pdf}
\caption{\label{fig:swparameters-L10-x-loose} FoCal-E pixel layer 10, \textit{loose} center method, fitted parameters $R_C$, $R_T$, $\sigma_{x_{0}}$, and $p$.}
\end{figure}
\begin{figure} 
\includegraphics[width=\textwidth]{figures/pixel/showerwidth/pixel-swparameters-L10-x-stringent.pdf}
\caption{\label{fig:swparameters-L10-x-stringent} FoCal-E pixel layer 10, \textit{stringent} center method, fitted parameters $R_C$, $R_T$, $\sigma_{x_{0}}$, and $p$.}
\end{figure}

\fi
\ifgloss
\printglossary[type=\acronymtype]
\printglossary
\fi
\end{document}